\documentclass[acmsmall, nonacm]{acmart}

\renewcommand\footnotetextcopyrightpermission[1]{} 

\usepackage{graphicx}
\usepackage{wrapfig}
\usepackage{float}
\usepackage{subcaption}
\usepackage{multirow}
\usepackage[framemethod=tikz]{mdframed}
\usepackage{xcolor}

\definecolor{keybg}{RGB}{245,245,245}
\definecolor{keyborder}{RGB}{235,235,235}
\newmdenv[
    backgroundcolor=keybg,
    linecolor=keyborder,
    linewidth=0.6pt,
    roundcorner=2pt,
    innertopmargin=5pt,
    innerbottommargin=5pt,
    innerleftmargin=5pt,
    innerrightmargin=5pt,
    skipabove=6pt,
    skipbelow=6pt
]{keybox}

\newcounter{keynotectr}
\newcommand{\keynote}{\refstepcounter{keynotectr}\emph{Key takeaway (\thekeynotectr):- }
}

\newcommand{\parab}[1]{\vspace{0.03in}\noindent{\bf #1}}
\newcommand{\projectname}{\texttt{CosmicDancePro}}

\AtBeginDocument{%
  }

\begin{document}


\title{\projectname{} -- Measuring LEO satellite's orbital decay and network connectivity implications during solar storms}

\author{Suvam Basak}
\affiliation{
  \institution{Indian Institute of Technology Kanpur}
  \country{India}
}

\author{Amitangshu Pal}
\affiliation{
  \institution{Indian Institute of Technology Kanpur}
  \country{India}
}

\author{Debopam Bhattacherjee}
\affiliation{
  \institution{Microsoft Research India}
  \country{India}
}

\begin{abstract}

The May 2024 solar superstorm highlighted the vulnerability of rapidly expanding low Earth orbit (LEO) satellite networks to severe space weather events. 
To systematically evaluate LEO network resilience, we introduce an open-source tool, \projectname{}.
It enables a comprehensive analysis of the effects of solar storms in the LEO satellite network.
It integrates real-world multimodal datasets, including space weather measurements from several satellites, upper-atmospheric density conditions from data-driven and high-fidelity physics-based models, and LEO satellite trajectory and LEO network measurement traces to quantify orbital decay driven by enhanced atmospheric density and network connectivity degradation.
We utilize \projectname{} to analyze the Starlink constellation's behavior during two recent major solar storms. 
First, we identify the specific fleet management strategies Starlink adopts during the May 2024 solar superstorm and how they differ from its regular orbit-correction strategy. 
Second, we identify the mechanisms driving the previously unexplained `W'-shaped altitude variation pattern across orbital planes of LEO constellations. 
Finally, our network-layer analysis quantifies the connectivity degradation during these storms, revealing transient disruptions that include repetitive short-lived outages, reconfiguration latency spikes above 500 ms, up to 60\% increase in uplink loss, distorted diurnal latency patterns, and a 10+ Mbps drop in end-user data rates during storm peaks.

\end{abstract}



\maketitle


\section{Introduction}

The low Earth orbit (LEO) satellite networks are slowly emerging as an integral part of the global Internet infrastructure. 
More than one LEO satellite operator, specifically Starlink~\cite{starlinkWeb}, Eutelsat OneWeb~\cite{oneWebWeb}, and, more recently, Amazon Leo (formerly Kuiper)~\cite{amazonLeoWeb}, is competing to deploy hundreds of thousands of LEO satellites to serve the Internet from space, regardless of geographic challenges.
Starlink, as an early mover with SpaceX's competitive edge in reusable rockets, today has the largest fleet of 10K+ LEO satellites and plans to expand to up to 15K satellites~\cite{FCCStarlink15000}.
Not only that, as of today, Starlink is the only operator serving 10 million noncommercial customers across 150 countries outside the Airliner and maritime markets~\cite{StarlinkWikipedia}.
This development, on the one hand, provides relatively reliable Internet connectivity to regions of the Earth that are poorly connected~\cite{ma2023network}.
On the other hand, this development democratized access to satellite connectivity, allowing the networking research community to explore, analyze challenges, and propose solutions across several aspects to further improve such connectivity.
Therefore, over the last six years, we have observed increasing efforts to measure the performance of the Internet~\cite{michel2022first, kassem2022browser, ullah2025impact, ma2023network, bose2025investigating, StarlinkOneWayDelay, mohan2024multifaceted, jang2025geo} and the backbone operations~\cite{izhikevich2024democratizing, wang2024large, 10294034, 10623111, MakingSenseLEO} of the Starlink LEO satellite network.

In the last couple of years, particularly after the May 2024 solar superstorm, there has been growing interest~\cite{WeatheringSolarSuperstorm, DeepDiveSolarStorms, starlinkDirectCellService} in analyzing the implications of solar storms on the LEO satellite network.
Solar storms are space weather events. 
The space weather close to Earth is mainly driven by the Sun.
The Sun's active regions sometimes blast out a hot, dense cloud of charged particles into outer space at thousands of kilometers per second~\cite{chen2011coronal, webb2012coronal}.
When such a cloud of charged particles sweeps through the Earth, it creates severe disturbances in Earth's magnetic field, triggering a geomagnetic storm.
During such events, charged particles infiltrate through the weaker regions of the geomagnetic field, creating operational challenges for space infrastructure in Earth's orbit.
The implications of such a solar storm for satellites depend on their orbits~\cite{SatelliteAnomaliesBook}.
A satellite in geostationary (GEO) orbit, at 35,786 km above Earth, experiences extreme radiation exposure.
This could permanently damage the critical electronic component~\cite{schultz2012revealing}, leading to a complete mission failure.
A satellite in LEO, such as Starlink, experiences relatively limited radiation exposure compared to a GEO satellite because it is deployed below the Van Allen radiation belts~\cite{SatelliteAnomaliesBook}.
The primary challenge for the LEO satellite is the enhanced atmospheric density resulting from thermosphere expansion driven by the absorption of extreme ultraviolet and X-ray radiation and by charged-particle interactions~\cite{ashruf2024loss}.
This enhanced atmospheric density causes high atmospheric drag in LEO~\cite{berger2023thermosphere}.
Failing to counteract properly could cause a LEO satellite to re-enter Earth with unintended orbital decay.
In 2022, for similar reasons, Starlink lost 38 satellites due to a minor solar storm immediately after the launch of 49 satellites~\cite{berger2023thermosphere, guarnieri2023norad}.

Authors in~\cite{basak2024cosmicdance} built \emph{CosmicDance}, which illustrated the orbital decay of a Starlink satellite immediately after high-intensity solar events.
After the May 2024 solar superstorm, a few works~\cite{basak2025investigation, WeatheringSolarSuperstorm, DeepDiveSolarStorms}
using RIPE Atlas probes, investigated Starlink network connectivity during the solar storm.
They observed increased latency and packet loss during such events.
Further, authors in~\cite{DeepDiveSolarStorms} while investigating the spatial characteristics of LEO constellation altitude decay during the May 2024 solar superstorm, identify a `W' shaped altitude variation pattern across the consecutive orbits.
However, they do not completely explain the key driver of this phenomenon.
Altogether, none of the existing work provides a thorough end-to-end investigation of this aspect of LEO networking.

In this paper, we present \projectname{}, an advanced extension of \emph{CosmicDance}~\cite{basak2024cosmicdance} that enables a full-spectrum investigation of such events.
\projectname{} includes: 
(i) tracing solar storms using data sourced from multiple satellites, 
(ii) correlating LEO satellite trajectory change with atmospheric density enhancement using data-driven high-fidelity physics-based simulations, 
(iii) investigating LEO satellite fleet management and operational strategy, 
(iv) LEO network connectivity implications in the Starlink segment, as well as end-user perceived Internet experience using network measurement datasets. 
Using \projectname{}, we extensively analyze two recent major solar storms: (i) the May 2024 solar superstorm, and (ii) the October 2024 solar storm.

In our analysis of the May 2024 solar superstorm, we find the maximum upper-atmospheric density increased by up to 20 times above the long-term baseline.
Starlink responded almost in real time to counteract orbital decay.
Almost 30-40\% satellites in each shell are raised to 100s of meters above their actual altitude to compensate for the increased decay rate until atmospheric conditions normalize.
In contrast, in a regular scenario, Starlink satellites are periodically orbit-corrected every 2nd or 3rd day.
Additionally, by correlating the satellite trajectory with atmospheric conditions, we show that the main driver of that previously identified `W' pattern in altitude variation~\cite{DeepDiveSolarStorms} is the geospatial distribution of atmospheric density between the day and night sides.
This pattern is not a solar storm-specific artifact.
It persists throughout the year and shifts continuously in synchronization with Earth's position around the Sun. 
During an intense space weather event, enhanced atmospheric density significantly amplified this pattern.
We also examine the network connectivity implications during these physical maneuvers by Starlink satellites.
Our investigation of the network aspect not only shows latency and loss inflation, but also reveals distortion in the Starlink diurnal latency pattern during the solar superstorm.
The Starlink uplink is more sensitive to such events than the downlink, resulting in significantly higher packet loss of 10-60\%.
During the peak of the solar superstorm in May, we observed repetitive, short-lived outages of 10s of seconds and reconfiguration latency spikes up to 500+ ms, significantly higher than on regular days.
We also explored publicly available user-initiated speed test records and found a degradation in data rate of approximately 10s of Mbps across the majority of regions.
In these speed test records, we also observed increases in latency, jitter, and packet loss. 
However, we need more systematic measurements to quantify how the interactive application experience will degrade. 

In summary in this paper, our main contributions are as follows:
\begin{enumerate}
    \item We reveal Starlink fleet management strategies in both regular operation and during solar storms.
    \item We illustrate the driving mechanism of the previously identified `W' pattern altitude variation across the orbits of LEO constellations. 
    \item We reveal inflation and distortion in the diurnal latency pattern during the solar storm. Multiple transient outages and abrupt reconfiguration latency spikes during the peak of a solar storm.
    The Starlink uplink is more susceptible to space weather events than the downlink. 
    Degradation in end-user data rate during the solar storm.
    \item We open-sourced \projectname{}~\cite{CosmicDanceProSourceCode} to enable the community to further explore and investigate past and future solar events.
\end{enumerate}

The remaining part of the paper is organized as follows: 
In~\S{\ref{sec:background}} we give a brief background of space weather events, their implication on ground and space-based infrastructure, and the motivation behind this paper.
In~\S{\ref{sec:relatedWorks}} we give an overview of all related work in this area.
In~\S{\ref{sec:cosmicDancePro}}, we introduce \projectname{} and describe the datasets it uses and their sources.
Then, using \projectname{}, we dive deep into two major solar storms in~\S{\ref{sec:solarStorm}}.
Analyze the LEO satellite trajectory change in~\S{\ref{sec:decayMeasurements}}.
Their implication in connectivity is discussed in~\S{\ref{sec:networkMeasurement}}.
In~\S{\ref{sec:limitations}} we point out a few limitations.
Finally, we conclude the paper in~\S{\ref{sec:conclusion}}.

\section{Background}
\label{sec:background}

In this section, we provide a brief overview of space weather events and their probable implications for today's global connectivity infrastructure.

\subsection{Space weather}

The weather in space near Earth is mainly influenced by the Sun.
The Sun continuously emits charged particles (protons, electrons, etc.), or plasma, into outer space, known as the solar wind~\cite{nasaWhatSolar}. 
Earth's magnetosphere deflects the stream of these charged particles and funnels them through the "polar cusps". 
The collision of these charged particles with the upper atmosphere creates eye-pleasing aurora (northern and southern lights)~\cite{AuroraWikipedia} in the skies of higher latitudes, while extreme cases of solar events, such as Solar flares~\cite{dennis1988solar, sweet1969mechanisms} and Coronal Mass Ejections~\cite{chen2011coronal, webb2012coronal}, could pose a serious risk to space- and ground-based infrastructure~\cite{RoyalAcademyReport}.

\subsubsection{Extreme solar events:}

\emph{Solar flares}~\cite{dennis1988solar, sweet1969mechanisms}, intense bursts of Extreme UltraViolet (EUV)~\cite{ashruf2024loss,vourlidas2018euv} and X-ray~\cite{woods2012extreme} radiation from the Sun's active regions, can reach Earth in 8 minutes at the speed of light and cause temporary high frequency radio blackouts as well as navigation hazards for aircraft~\cite{airbusAirbusUpdate} and maritime vessels. 
A \emph{Coronal Mass Ejection (CME)}~\cite{chen2011coronal, webb2012coronal}, a massive explosion in the corona, the outermost layer of the Sun's atmosphere, blasts out a dense, hot cloud of plasma into outer space, with a strong magnetic field that travels at 200 to 3,000 km/s. 
This could smash into Earth within 1 to 4 days after the eruption, triggering a geomagnetic storm, which is a severe disturbance in Earth's magnetosphere. 
On one hand, such variation in Earth's magnetic field can produce Geomagnetically Induced Current (GIC)~\cite{GIC} as high as 100-130 Amps, capable of taking out ground-based strategic infrastructure like the national power grid~\cite{powergrid, RoyalAcademyReport}, submarine cables connecting continents~\cite{internetapocalypse}. 
On the other hand, charged particles create operational hazards to the space-borne infrastructure. 
Satellites at a higher altitude experience a radiation surge that might cause Single Event Effects (SEE), bit flip, surface charging, solar panel degradation, etc.~\cite {SatelliteAnomaliesBook}. 
All of which could lead to a temporary outage or permanent mission failure due to loss of a critical subsystem during such an event~\cite{schultz2012revealing}.
At lower altitudes, LEO satellites are somewhat shielded from direct radiation by Earth's Van Allen belts~\cite{SatelliteAnomaliesBook}, which results in different operational challenges. 
The collision of charged particles with gas molecules heats the Earth's upper atmosphere (thermosphere), causing it to expand up to 100s of km~\cite{berger2023thermosphere}. 
This sudden increase in atmospheric density causes high orbital drag (atmospheric friction) to LEO satellites, thus initiating unintended orbital decay~\cite{ashruf2024loss}. 
Therefore, unpreparedness for such a severe solar event could lead to the premature reentry of LEO satellites or, in a worst of the worst-case scenario, an in-orbit collision, which could initiate Kessler Syndrome~\cite{kesslerSyndrome} and block access to space for several years.

\subsubsection{Possibility of such extreme solar events in the near future:}

\begin{figure}
    \centering
    \begin{subfigure}[t]{0.50\columnwidth}
        \centering
        \includegraphics[width=\columnwidth, keepaspectratio]{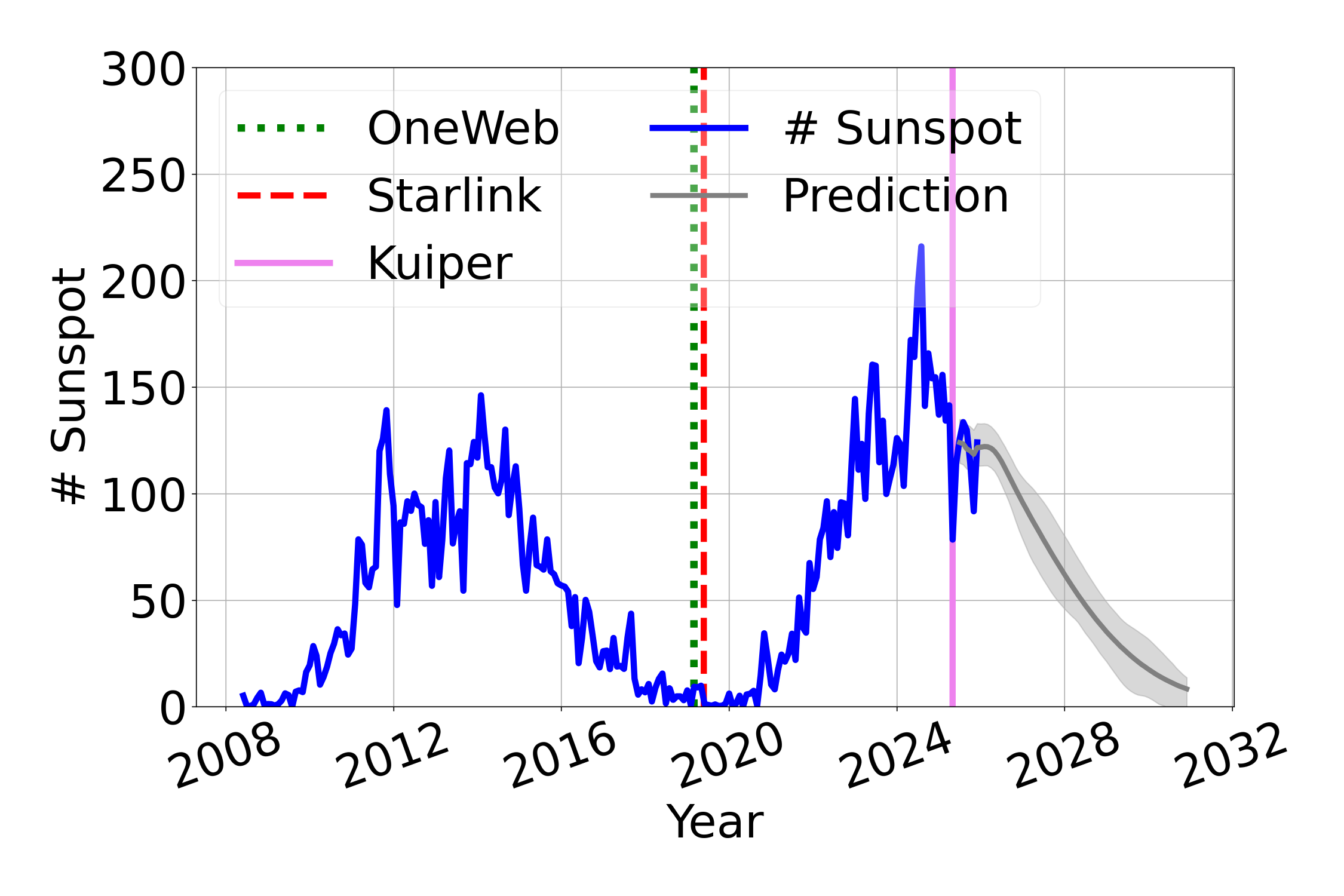}
        \caption{}
    \end{subfigure}%
    \hfill
    \begin{subfigure}[t]{0.50\columnwidth}
        \centering
        \includegraphics[width=\columnwidth, keepaspectratio]{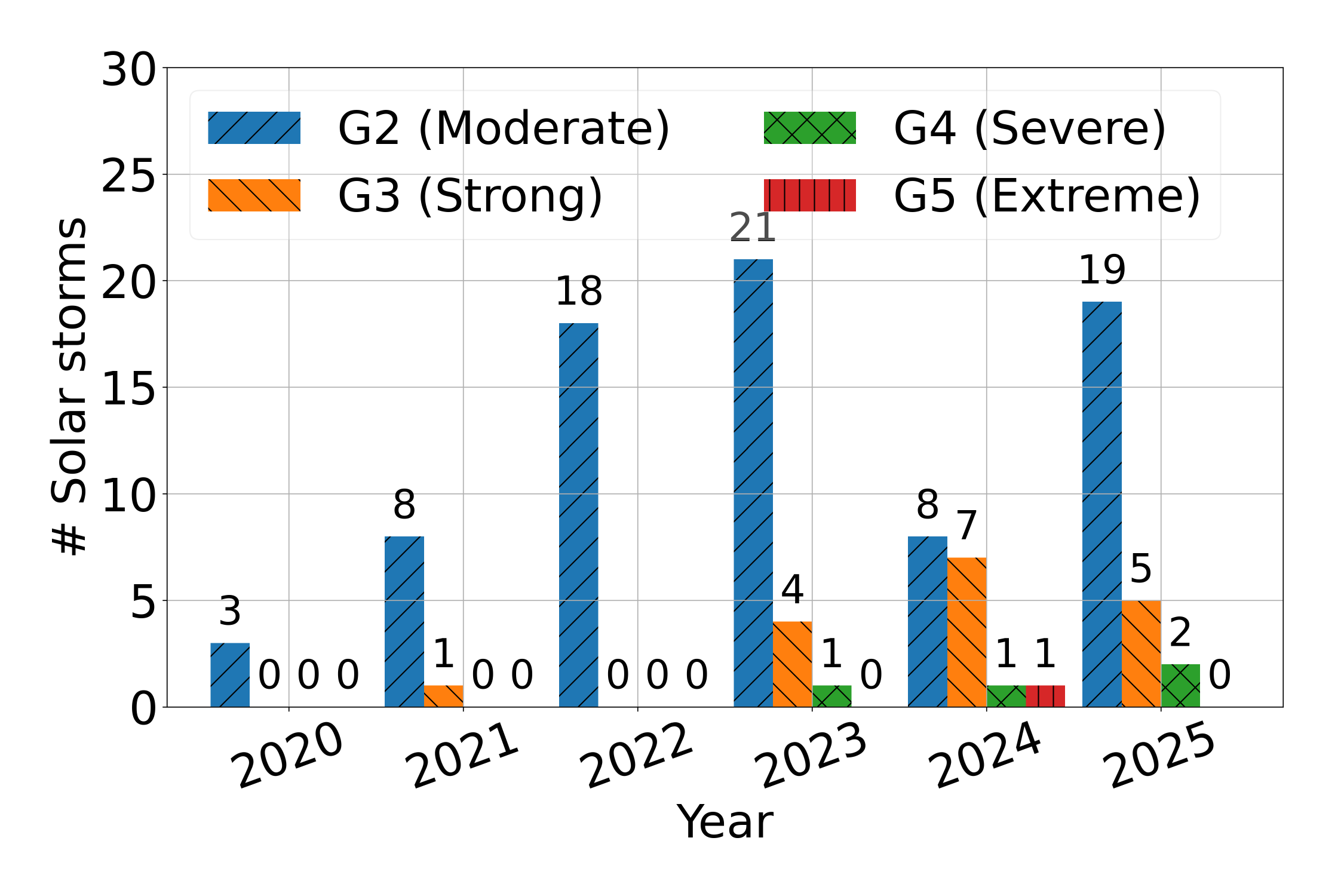}
        \caption{}
    \end{subfigure}%
    \caption{(a) Solar cycle progression and the launch date of the LEO satellite constellations (b) Solar storm statistics from January 2020 to December 2025.}
    \label{fig:solarCycleAndStormStats}
\end{figure}

The Sun advances through a maximum and minimum of its 11-year magnetic activity cycle, known as the solar cycle~\cite{SolarCycle}. 
The progression of the solar cycle is typically characterized by the number of sunspots.
Sunspots are the regions of intense magnetic field concentration, which are the primary indicators of Active Regions, the potential sources of solar flares and CMEs. 
As shown in Fig.~\ref{fig:solarCycleAndStormStats}(a), the Sun is currently at the declining phase after the maxima of solar cycle 25. 
Thus, in Fig.~\ref{fig:solarCycleAndStormStats}(b), there has been an increased number of severe (G4) and extreme (G5) categories of geomagnetic storms in the last couple of years compared to the previous years (2020 to 2022). 
These categories are the strongest level as per the G-scale classification system -- G1 (minor): below -30 nanoTesla (nT), G2 (moderate): below -50 nT, G3 (strong): below -100 nT, G4 (severe): below 200 nT, G5 (extreme): below -350 nT.) described by the National Oceanic and Atmospheric Administration (NOAA), based on intensity and potential impacts of such events~\cite{NOAAGScale}.

Moreover, in the long term, this 11-year solar cycle goes through a variation up to 4$\times$ called the Gleissberg cycle in the range of centuries (behaves like "amplitude modulation" over 80 to 100 years)~\cite{GleissbergCycle1, GleissbergCycle2}. 
Recent research~\cite{rodriguez2024we} suggests that the minima of the current Gleissberg cycle occurred around 2020-2025. 
So, Sun is now moving towards the Gleissberg maxima of the 21st century. 
Which is likely to appear in solar cycle 28 (around 2055)~\cite{adams2025turnover}. 
Hence, the probability of a solar superstorm in the near future is relatively high.
Although forecasting or predicting an exact date and time of the next solar superstorm is extremely difficult, with today's real-time monitoring of the Sun from Lagrange Point 1 (L1)~\cite{nasaSWFOL1NASA, isroADITYAL1}, the possible forecasting period of CME arrival is 1-4 days, depending on the velocity.

\subsection{Threat to the 21st century's connected world}

At first glance, this might seem like an extremely rare event that could be overlooked. 
Therefore, we examine past solar events and their impact on infrastructure, drawing on publicly available sources, to assess the severity of extreme solar events.

\subsubsection{Consequences on ground-based infrastructure:}

The strongest geomagnetic storm recorded in the past was on September 1859 (in solar cycle 10), known as the Carrington Event, with a reported intensity of -1,800 nT~\cite{carrington1, carrington2, wikipediaCarringtonEvent}. 
It produced a severe GIC that disrupted the telegraph network of the day in North America and Europe. 
The operator of the telegram station experienced an electric shock and was also reported to have a fire. 
Similarly, in May 1921 (in solar cycle 15), the New York Railroad storm of intensity -907 nT caused fire in the New York Central Railroad station's signaling system~\cite{Railroad, wikipedia1921Geomagnetic}.
In October 2003 (after the peak of solar cycle 23), the Halloween Storm of intensity -401 nT generated high GIC that took out Malmö Power Grid in Sweden, leaving thousands of people in blackout for an hour~\cite{wikipedia2003Halloween, pulkkinen2005geomagnetic}. 
In solar cycle 25, voltage irregularities have been reported by multiple power grids during the recent May 2024 solar superstorm (also known as the Gannon storm) with an intensity of -412 nT~\cite{wikipedia2024Solar, SpaceNew2025}.
In New Zealand, Transpower proactively took down some of the subsystems as a precautionary measure against the May 2024 solar superstorm~\cite{transpowerTranspowerRestores}. 

Here, we only highlight a few severe events. 
However, in Table~\ref{tbl:groundInfraImpactSolarEvent}, we have compiled a list of publicly reported anomalies after solar storms, showing far-reaching consequences in various ground infrastructures, including telecommunication, GPS receivers, and radar.

\begin{table}
\centering
\footnotesize
\caption{A list of publicly reported ground infrastructure anomalies and their impacts due to solar events recorded in the past.}
\begin{tabular}{c|c|c|c}

\hline
\bf{Event date} & \bf{Anomalies} & \bf{Anticipated reason} & \bf{Socioeconomic impact} \\ 
\hline\hline


    \begin{tabular}{@{}c@{}} Gannon Storm\\May 10, 2024\\\cite{wikipedia2024Solar,SpaceNew2025,transpowerTranspowerRestores} \end{tabular} & 
    \begin{tabular}{@{}c@{}} Electrical system and \\ telecommunication \end{tabular} &
    \begin{tabular}{@{}c@{}} Multiple CMEs \end{tabular} & 
    \begin{tabular}{@{}c@{}} Experienced Irregularities, \\ precautionary measures (switching \\ off subsystems) prevented blackout.  \end{tabular} 
    \\ 

\hline
    \begin{tabular}{@{}c@{}} Nov 4, 2015\\\cite{marque2018solar} \end{tabular} & 
    \begin{tabular}{@{}c@{}} Surveillance radar\\Sweden \end{tabular} &
    \begin{tabular}{@{}c@{}} M3.7 solar flare caused\\Solar Radio Burst \end{tabular} & 
    \begin{tabular}{@{}c@{}} Sweden airspace remain \\ closed for 90 minutes \end{tabular} 
    \\ 

\hline
    \begin{tabular}{@{}c@{}} Dec 6, 2006\\\cite{cerruti2008effect} \end{tabular} & 
    \begin{tabular}{@{}c@{}} GPS receivers \end{tabular} &
    \begin{tabular}{@{}c@{}} X6.5 solar flare caused \\ radio blackout \end{tabular} & 
    \begin{tabular}{@{}c@{}} GPS receivers on the day side \\ lost the lock on GNSS satellites \end{tabular} 
    \\

\hline
    \begin{tabular}{@{}c@{}} Halloween Storms\\Oct 30, 2003\\\cite{wikipedia2003Halloween,pulkkinen2005geomagnetic} \end{tabular} & 
    \begin{tabular}{@{}c@{}} Malmö power grid, \\ Sydkraft  \end{tabular} &
    \begin{tabular}{@{}c@{}} GICs tripped \\ a 130 kV line \end{tabular} & 
    \begin{tabular}{@{}c@{}} Blackout of 1 hour \\ for 50,000 customers \end{tabular} 
    \\ 

\hline
    \begin{tabular}{@{}c@{}} Mar 13, 1989\\\cite{boteler201921st, RoyalAcademyReport, wikipediaMarch1989} \end{tabular} & 
    \begin{tabular}{@{}c@{}} Hydro-Québec \\ power grid \end{tabular} & 
    \begin{tabular}{@{}c@{}}  GICs caused \\ transformer collapse  \end{tabular} & 
    \begin{tabular}{@{}c@{}} Blackout of 9 hours for 6 million \\ cost above \$13.2 million CAD \end{tabular} 
    \\ 

\hline
    \begin{tabular}{@{}c@{}} Jul 13, 1982\\\cite{SwedishRailwayNetwork}\end{tabular} & 
    \begin{tabular}{@{}c@{}} Swedish railway \\ network \end{tabular} &
    \begin{tabular}{@{}c@{}} GIC of 100s of amps \end{tabular} & 
    \begin{tabular}{@{}c@{}} Signaling Errors and flickering \\ between red and green  \end{tabular} 
    \\ 

\hline
    \begin{tabular}{@{}c@{}} New York Railroad \\ Storm, May 13, 1921\\\cite{Railroad,wikipedia1921Geomagnetic}  \end{tabular} & 
    \begin{tabular}{@{}c@{}} Telegraph networks \\ electrical services \end{tabular} &
    \begin{tabular}{@{}c@{}}Back to back CMEs  \end{tabular} & 
    \begin{tabular}{@{}c@{}} Killed switching system in \\ New York Central Railroad \end{tabular} 
    \\ 

\hline
    \begin{tabular}{@{}c@{}} Carrington Event\\Sept 1, 1859\\\cite{carrington1,carrington2,wikipediaCarringtonEvent} \end{tabular} & 
    \begin{tabular}{@{}c@{}} Telegraph networks  \end{tabular} &
    \begin{tabular}{@{}c@{}} GICs caused by \\ massive CME \end{tabular} & 
    \begin{tabular}{@{}c@{}} Collapse of telegraph system due to fire \\ and operator experienced electric shock \end{tabular} 
    \\ 

\hline \hline
\end{tabular}

\label{tbl:groundInfraImpactSolarEvent}
\end{table}

\subsubsection{Consequences on space infrastructure:}

Satellites in space experience a much harsher environment compared to the ground. 
In addition to that, a radiation surge for a geostationary (GEO) and inflation in orbital drag for LEO often cause satellite anomalies.
It is worth mentioning that the majority of such anomalies are not reported publicly, as these assets are mostly operated by private and military agencies~\cite{schultz2012revealing, SatelliteAnomaliesBook}.
Despite that, we found a fairly large list of satellite anomalies in Table~\ref{tbl:spaceInfraImpactSolarEvent} caused by different solar activities over the last few decades. 
Notice in Table~\ref{tbl:spaceInfraImpactSolarEvent}, almost all the GEO (and higher altitude) satellite anomalies are caused by exposure to intense radiation~\cite{RoyalAcademyReport, bodeau2007killer, Telstar401, lam2012anik, shen2021evaluation, milsatmagazineMilsatMagazine}, while the LEO satellites are more prone to orbital decay and uncontrolled reentry caused by an increase in atmospheric expansion or orbital drag~\cite{archiveJAXAOperational, RoyalAcademyReport}. For instance, in February 2022, 38 out of 49 SpaceX's Starlink satellites reentered the atmosphere closely after a launch because of a moderate geomagnetic storm~\cite{baruah2024loss,berger2023thermosphere,guarnieri2023norad}.

\begin{table}
\centering
\footnotesize
\caption{A list of publicly reported space infrastructure anomalies and their impacts due to solar events recorded in the past.}
\begin{tabular}{c|c|c|c}

\hline
\bf{Event date} & \bf{Space asset} & \bf{Anticipated reason} & \bf{Socioeconomic impact} \\ 
\hline\hline

    \begin{tabular}{@{}c@{}} Feb 4, 2022\\\cite{baruah2024loss,berger2023thermosphere,guarnieri2023norad} \end{tabular} & 
    \begin{tabular}{@{}c@{}} 38 Starlink \\ LEO satellite \end{tabular} &
    \begin{tabular}{@{}c@{}} Sudden atmospheric drag spike due \\ to G2 class geomagnetic storm \end{tabular} & 
    \begin{tabular}{@{}c@{}} Tens of millions of dollars \end{tabular} 
    \\

\hline
    \begin{tabular}{@{}c@{}} Apr 5, 2010\\\cite{milsatmagazineMilsatMagazine} \end{tabular} & 
    \begin{tabular}{@{}c@{}} Galaxy 15 \\ GEO satellite \end{tabular} &
    \begin{tabular}{@{}c@{}} Telemetry and command control \\lost after electrostatic discharge \end{tabular} & 
    \begin{tabular}{@{}c@{}} Failed after 8 months of \\ autonomous operational \end{tabular} 
    \\ 
    
\hline
    \begin{tabular}{@{}c@{}} Halloween Storms\\Oct 30, 2003\\\cite{archiveJAXAOperational,RoyalAcademyReport} \end{tabular} & 
    \begin{tabular}{@{}c@{}} ADEOS-II \\ (Midori II) \\ LEO satellite \end{tabular} &
    \begin{tabular}{@{}c@{}} Solar panel failed \\to severe radiation \end{tabular} & 
    \begin{tabular}{@{}c@{}} Scientific satellite \\ worth \$70 billion Yen \end{tabular} 
    \\ 

\hline
    \begin{tabular}{@{}c@{}} Halloween Storms\\Oct 30, 2003 \end{tabular} & 
    \begin{tabular}{@{}c@{}} SOHO \\ L1 satellite \end{tabular} &
    \begin{tabular}{@{}c@{}} Instrument anomalies \\due to severe radiation \end{tabular} & 
    \begin{tabular}{@{}c@{}} Temporary outage \end{tabular} 
    \\ 
    
\hline
    \begin{tabular}{@{}c@{}} Bastille Day\\Jul 15, 2000\\\cite{AstroDEvent} \end{tabular} & 
    \begin{tabular}{@{}c@{}} ASCA (Astro-D) \\ LEO satellite \end{tabular} &
    \begin{tabular}{@{}c@{}} Started uncontrolled spin \\ after failure of mission \\ critical instruments \end{tabular} & 
    \begin{tabular}{@{}c@{}} Total loss of a \\ astronomy satellite \end{tabular} 
    \\ 
    
\hline
    \begin{tabular}{@{}c@{}} May 19, 1998\\\cite{RoyalAcademyReport,bodeau2007killer} \end{tabular} & 
    \begin{tabular}{@{}c@{}} Galaxy 4 \\ GEO Satellite \end{tabular} &
    \begin{tabular}{@{}c@{}} Failure of primary and backup \\ processor to solar radiation \end{tabular} & 
    \begin{tabular}{@{}c@{}} Around 40-45 million \\ pager outage \end{tabular} 
    \\ 

\hline
    \begin{tabular}{@{}c@{}} Jan 11, 1997\\\cite{Telstar401} \end{tabular} & 
    \begin{tabular}{@{}c@{}} Telstar 401 \\ GEO Satellite\end{tabular} &
    \begin{tabular}{@{}c@{}} Total loss due to intense \\ electrostatic discharge \end{tabular} & 
    \begin{tabular}{@{}c@{}} TV, phone call, \\ data service \end{tabular} 
    \\

\hline
    \begin{tabular}{@{}c@{}} Jan 20, 1994\\\cite{lam2012anik} \end{tabular} & 
    \begin{tabular}{@{}c@{}} Intelsat K \\ GEO Satellite \end{tabular} &
    \begin{tabular}{@{}c@{}} Temporary loss of altitude \\ control to electrostatic discharge \end{tabular} & 
    \begin{tabular}{@{}c@{}} Temporary outage \end{tabular} 
    \\ 

\hline
    \begin{tabular}{@{}c@{}} Jan 20, 1994\\\cite{lam2012anik} \end{tabular} & 
    \begin{tabular}{@{}c@{}} Anik E-1 \\ GEO Satellite \end{tabular} &
    \begin{tabular}{@{}c@{}} Primary stabilizer failed to \\ electrostatic discharge \end{tabular} & 
    \begin{tabular}{@{}c@{}} Outage for approx 8 hours,\\ switched to secondary \end{tabular} 
    \\ 

\hline
    \begin{tabular}{@{}c@{}} Jan 20, 1994\\\cite{lam2012anik} \end{tabular} & 
    \begin{tabular}{@{}c@{}} Anik E-2 \\ GEO Satellite \end{tabular} &
    \begin{tabular}{@{}c@{}} Both stabilizer failed to \\ electrostatic discharge \end{tabular} & 
    \begin{tabular}{@{}c@{}} Brought back to service \\ by GLACS~\footnote{Ground Loop Attitude Control System} in Jul, 1994 \end{tabular} 
    \\ 

\hline
    \begin{tabular}{@{}c@{}} Feb 8, 1986\\\cite{shen2021evaluation} \end{tabular} & 
    \begin{tabular}{@{}c@{}} MARECS-A \\ GEO Satellite \end{tabular} &
    \begin{tabular}{@{}c@{}} Electrostatic discharge and surface \\ charging led to satellite Anomaly \end{tabular} & 
    \begin{tabular}{@{}c@{}} Maritime communication \\disruption \end{tabular} 
    \\ 

\hline
    \begin{tabular}{@{}c@{}} Jul 11, 1979\\\cite{wikipediaSkylab,americaspaceKingsHorsesskylab}\end{tabular} & 
    \begin{tabular}{@{}c@{}} Skylab \\ LEO Space Station \end{tabular} &
    \begin{tabular}{@{}c@{}} Uncontrolled reentry caused \\ by rise of orbital drag \end{tabular} & 
    \begin{tabular}{@{}c@{}} -- \end{tabular} 
    \\ 


\hline \hline
\end{tabular}
\label{tbl:spaceInfraImpactSolarEvent}
\end{table}

\subsubsection{Why should the networking community bother?:}

The Internet is designed to be a decentralized entity. 
However, the underlying hardware infrastructure, i.e., routers, Internet Exchange Points (IXPs), and long-haul cables, is completely dependent on regional power grids. 
Further, regional power grids spanning large geographic areas are tightly coupled with Wide Area Synchronous Grids (WASGs)~\cite{epaUSGrid, wikipediaContinentalEurope, OneNationOneGrid, sappSouthernAfrican}. 
This means that 65\% of the Internet infrastructure is dependent on only 10 WASGs~\cite{jyothi2023characterizing}. 
Any failure of this centralized power source to a solar superstorm could lead to an Internet blackout over a large geographic area.
In the 21st century, given the role of Internet connectivity across various sectors of industry, a Carrington-scale event could result in economic losses in the order of trillions. A recent investigation of the terrestrial segment of the Internet in this context has revealed that 43\% of submarine cables, 43\% of IXPs, 38\% of routers, and 57\% of Autonomous Systems in higher latitudes are vulnerable to such a solar superstorm~\cite{internetapocalypse}. 

Today, Starlink and OneWeb, with 10,000+ and 600+ LEO satellites~\cite{spaceStarlinkSatellites, eutelsatHighspeedLowlatency}, respectively, are not only serving more than 9 million active users across 150+ countries~\cite{StarlinkWikipedia}.
Rather, after airlines, Starlink began offering direct-to-cell service~\cite{starlinkDirectCellService} too.
Amazon Leo (formerly Project Kuiper), currently with over 200+ LEO satellites~\cite{aboutamazonAmazonMission}, is likely to launch its service soon.
These networks in space are emerging as an integral part of the global Internet and future ubiquitous connectivity. 
As a result, LEO satellites are a recent consideration for future failover connectivity in the event of international undersea cable failures~\cite{LEONationalFailover}.
This makes it crucial to comprehensively study the resilience of these networks against solar events.
Therefore, in this work, we assess network connectivity and the characteristics of LEO satellites' orbital trajectories during recent solar storms using publicly available real-world datasets.

\section{Related works}
\label{sec:relatedWorks}

All previous efforts in the area of space weather, orbital decay of LEO satellites, and LEO network connectivity fall into the following four categories, which we summarize below.

\parab{Investigation of solar events} - 
The impact of solar events on the Earth has always been an active area of research~\cite{boteler201921st, carrington1, carrington2, Railroad, pulkkinen2005geomagnetic}. After the recent May 2024 solar superstorm, authors in~\cite{tulasi2024super, zakharenkova2025unveiling} have analyzed the spatial impact of the storm on Earth's upper atmosphere (thermosphere and ionosphere) and magnetosphere.

\parab{LEO satellites' orbital decay} - 
In~\cite{oliveira2019satellite}, the author used atmospheric density data from two LEO satellites -- CHAMP and GRACE to investigate how solar activity inflicts orbital decay, and reveals inaccuracies in the prominent thermosphere empirical density model JB2008~\cite{nasaJB20082008}.
In February 2022, after the reentry of 38 Starlink satellites from their staging orbit, authors in~\cite{miteva2023space} conducted a comparative analysis of Starlink satellites' reentry and historical failures, highlighting how extreme solar events impact the spacecraft. 
Further, in~\cite{berger2023thermosphere, baruah2024loss}, authors investigated the same incident using empirical models (NRLMSIS, JB08, and HASDM), showing atmospheric density increased 20\% to 30\% at their staging altitude, which is catastrophic for satellite operation~\cite{berger2023thermosphere}. 
While authors in~\cite{baruah2024loss} claimed that Starlink's satellite design choices, i.e., a low mass-to-area ratio, are also a crucial factor in this incident. 
In contrast, a recent work~\cite{basak2024cosmicdance} developed a data-driven tool to investigate the trajectory changes of LEO satellites after solar events and illustrated the orbital decay of many Starlink satellites following high-intensity solar activity. 

After the May 2024 solar superstorm, an investigation~\cite{doi:10.2514/1.A36164} revealed that the thermosphere density increased up to six times the baseline, and forecast intensity had been underestimated. This led to unplanned station-keeping maneuvers across many satellites.
Another work~\cite{ashruf2024loss} illustrated the May 2024 solar superstorm, and the precondition of solar superstorm increased the decay rate of the Starlink decommissioned satellite, which resulted in the early reentry of the 12 satellite.
A similar observation is also reported in~\cite{oliveira202510} during the October 2024 solar storm.

\parab{LEO network measurements} - 
Starlink, being the first consumer-facing LEO satellite constellation network with near worldwide availability, has been an active area of network measurements for the last five years. 
After the launch, authors in~\cite{michel2022first} measured Starlink's latency, throughput, and packet loss for months. 
Their measurements revealed that the web-browsing experience over Starlink is nearly as good as that of wired connections. 
However, under heavy load, performance can fluctuate due to packet loss and queuing delays.
While authors in~\cite{kassem2022browser} developed a browser extension and deployed Raspberry Pis in multiple countries to capture Starlink's performance. 
Their study revealed that network performance also depends on geographical location and weather conditions. Modern congestion control protocols, such as BBR, are effective at managing frequent packet losses of up to 50\%~\cite {kassem2022browser}.
Later, a dedicated study~\cite{ullah2025impact} of the weather impact on Starlink performance showed rain drops both up and down link throughput roughly by 52\% and 38\%, respectively. 
Cloud cover also degrades throughput, while RTT remains unchanged~\cite{ullah2025impact}.
Measuring Starlink at a global scale is challenging. 
In~\cite {izhikevich2024democratizing}, the author proposed and evaluated a novel methodology that exploits Internet-exposed services to democratize Starlink network measurement at a global scale. 
Furthermore, analysis of Starlink performance across 34 countries~\cite{mohan2024multifaceted} has shown that Starlink’s service is competitive with terrestrial cellular networks. 
Their experiment also revealed periodic variations in Starlink's throughput and latency, which impact real-time applications such as gaming and video conferencing.
A series of systematic measurements in~\cite{ma2023network} evaluated the reliability of Starlink service across diverse geographic locations, including remote areas. 
Another line of effort~\cite{10.1145/3696348.3696879} evaluated the Content Delivery Networks (CDNs) over Starlink, and results~\cite{bose2025investigating} showed Starlink’s network operation often leads to DNS mislocalization, thus forcing traffic to distant servers that inflates latency beyond 200 ms.
Authors in~\cite{StarlinkOneWayDelay, wang2024large, 10294034, 10623111, mohan2024multifaceted, izhikevich2024democratizing} dives deeper into Starlink operation. 
Their work uncovers Starlink's diurnal latency pattern~\cite{StarlinkOneWayDelay}, scheduling or one way delay characteristics~\cite{StarlinkOneWayDelay}, globally synchronized periodic 15-second reconfiguration~\cite{StarlinkOneWayDelay, 10623111, mohan2024multifaceted, izhikevich2024democratizing}, backbone network topology (i.e., Points of Presence (PoPs) and their internal connectivity)~\cite{wang2024large}, latency difference between residential, business services, and inter-satellite link usage~\cite{10294034, 10623111}. 
A recent work~\cite{jang2025geo} also evaluated the Starlink in-flight connectivity against traditional GEO satellites' connectivity.

\parab{Network measurement against solar storms} - 
Being a nontraditional area, only one work~\cite{internetapocalypse} has explored terrestrial networks against solar superstorms, and found that 43\% of submarine cables, 43\% of IXPs, and 38\% of routers at northern latitudes are vulnerable to such events. 
Only a few studies~\cite{WeatheringSolarSuperstorm, basak2025investigation, DeepDiveSolarStorms, fadaei2026comprehensive} explored Starlink's performance during solar storms.
Their analysis using RIPE Atlas~\cite{WeatheringSolarSuperstorm, basak2025investigation} and other network measurements~\cite{fadaei2026comprehensive} has shown connectivity degradation at higher latitudes and sun-facing orbits experienced more impact~\cite{DeepDiveSolarStorms}.
As opposed to these existing contributions, our paper proposes an integrated tool set that allows end-to-end full-spectrum analysis, which includes:
(i) dissecting a solar storm from solar eruption to thermosphere expansion, 
(ii) measuring and quantifying the shell-wise orbital decay of LEO satellites and their geospatial nature,
(iii) fine-grain LEO network connectivity implication, and 
(iv) end-user Internet degradation experience during such events.

\section{Building \projectname{}}
\label{sec:cosmicDancePro}

To study the impact of solar storms on both LEO satellite trajectories and end-user Internet connectivity, a comprehensive system is required that can integrate and process heterogeneous datasets, including space weather data, satellite tracking information, and network measurement data across multiple formats.
Prior to the May 2024 solar superstorm, \emph{CosmicDance}~\cite{basak2024cosmicdance} introduced a data-driven framework focused exclusively on satellite orbital decay immediately after solar activity. However, it does not allow exploration of solar storms, the geospatial pattern of orbital decay, and their effects on Internet performance.
Following the May 2024 event, subsequent studies~\cite{WeatheringSolarSuperstorm, basak2025investigation, DeepDiveSolarStorms} used built-in RIPE Atlas measurements with \emph{CosmicDance}’s orbital analysis to derive preliminary insights. 
These efforts remain limited in scope, as they rely on loosely coupled analyses and lack an integrated framework for joint reasoning across domains.

In this work, we present \projectname{}~\cite{CosmicDanceProSourceCode}, a substantial extension of \emph{CosmicDance}, designed to enable deeper, end-to-end investigation of solar storm impacts. \projectname{} addresses key limitations of prior work, including:
(i) eliminating the need for manual filtering of orbit-raising maneuvers, and
(ii) incorporating datasets for tracking decommissioned satellites.
(ii) mapping satellite decay with upper atmospheric density conditions and more.
More importantly, \projectname{} enables systematic exploration of potential cause–and–effect relationships across multiple layers, including solar eruptions, magnetospheric disturbances, upper atmospheric density inflation, LEO orbital decay, and resulting network performance degradation.
The system is built on Polars, a high-performance data processing library implemented in Rust, allowing efficient handling of datasets on the order of hundreds of gigabytes on a single machine. 
In addition, \projectname{} provides high-level APIs for querying atmospheric density at arbitrary latitude, longitude, altitude, and timestamp, without requiring full in-memory loading of large-scale geospatial time-series datasets.

\subsection{Datasets}

Here, we provide an overview of the data sources used in \projectname{} to trace solar storms, atmospheric density inflation, and their implications on LEO satellites' trajectory and network connectivity.

\subsubsection{Solar storms:}

We use the following datasets to analyze solar storms and their impact on the upper atmosphere.

\parab{Solar eruptions} - 
\projectname{} leverages SunPy~\cite{sunpy_community2020} to collect one-second cadence time series of X-ray emissions readings from Extreme Ultraviolet and X-ray Irradiance Sensors (EXIS) on GOES-18~\cite{noaaGOES18Launch}.
GOES-18 is a space and Earth weather monitoring satellite operated by NOAA in geostationary orbit at 137.2$^{\circ}$ west above the Pacific Ocean. 
A sudden spike in X-ray irradiance readings on GOES-18 indicates a solar flare from the Sun.

\parab{CME arrival} - 
CMEs are mostly preceded by the strongest solar flares; however, this does not always occur at the same time.
CMEs are highly directional. 
Hence, they move more often towards outer space in a completely different direction (not towards the Earth). 
So, in \projectname{} we leverage Deep Space Climate Observatory (DSCOVR)~\cite{noaaDSCOVRDeep} at L1 to collect solar wind velocity, density, temperature, and direction of the Interplanetary Magnetic Field (IMF) time series in one-minute cadence. 
Additionally, \projectname{} collects measurements from GOES-18's~\cite{noaaGOES18Launch} magnetometer too.

\parab{Geomagnetic activity} - 
To acquire the intensity of a geomagnetic storm, \projectname{} relies on the Disturbance Storm Time (Dst) index. 
Dst index is a measurement of Earth's geomagnetic disturbance from four geomagnetic observatories located at Kakioka, Honolulu,  San Juan, and Hermanus~\cite{WorldDataCenterGeomagnetism}. 
The World Data Center for Geomagnetism compiles geomagnetic field data from the aforementioned observatory network and distributes it hourly. 
\projectname{} collects Dst index in raw format, then parses and exports a timeseries of geomagnetic disturbance into a \texttt{CSV} file.

\parab{Atmospheric conditions} -
For upper atmospheric conditions, in \projectname{}, we utilize physics-based high-fidelity simulations using Thermosphere-Ionosphere-Electrodynamics General Circulation Model (TIE-GCM 2.0)~\cite{TIEGCM}.
It takes real-world solar radiation measurements as input and produces a high-dimensional timeseries of upper-atmospheric conditions.
Specifically, in this context, it provides atmospheric density observations at different altitudes across a 5°$\times$5° latitude-longitude grid at a 20-minute cadence.
\projectname{} provides a high-level API for looking up the density value at any given timestamp, latitude, longitude, and elevation without loading large datasets in memory.

\subsubsection{Satellite tracking:}
 
Private corporations like LeoLabs~\cite{LEOLabsRadars} and the US Space Force’s Combined Space Operations Center (CSpOC)~\cite{CSPOC} use their Space Surveillance Network (i.e., ground-based radar systems~\cite{LEOLabsRadars} and optical sensors) to track all Earth orbiting objects with a unique identification number (NORAD ID). 
After orbit determination of such objects, they release a new TLE~\cite{TLEs}, typically every 8-12 hours, which is distributed via Space-Track~\cite{spacetrack}. 

\parab{Two-Line Element (TLE)} - 
A TLE is a standard data format that encodes orbital elements -- inclination, Right Ascension of the Ascending Node (RAAN), eccentricity, argument of perigee, mean anomaly, and mean motion, together uniquely describe an object's orbit and the position of the object in the orbit. 
Additionally, it includes the satellite catalog number (NORAD ID), epoch, orbital drag coefficient, and higher-order derivatives of mean motion, which allows predicting the object's position in orbit at any arbitrary time using perturbation models (such as SGP4~\cite{vallado2006revisiting, SPACETRACKREPORTNO3}).

In \projectname{}, we inherit the script from prior work~\cite{basak2024cosmicdance} to download all satellite TLEs for a given NORAD ID and time window using Space-Track's APIs~\cite{spacetrack} while adapting to the request rate limit. 
In addition, we address two drawbacks of the prior work~\cite{basak2024cosmicdance}.

First, prior work relies solely on CelesTrak~\cite{celestrakCelesTrak} to acquire the satellite's NORAD IDs of different LEO constellations. 
Hence, it misses the TLEs of 100s of decommissioned satellites~\cite{starlinkDeorbitingSats} when investigating past solar events.
Therefore, in \projectname{}, we source the NORAD IDs from a comprehensive satellite catalog database maintained in Jonathan's Space Pages~\cite{planet4589JonathanapossSpace}.

\begin{figure}
    \centering
    \begin{subfigure}[t]{0.50\columnwidth}
        \centering
        \includegraphics[width=\columnwidth, keepaspectratio]{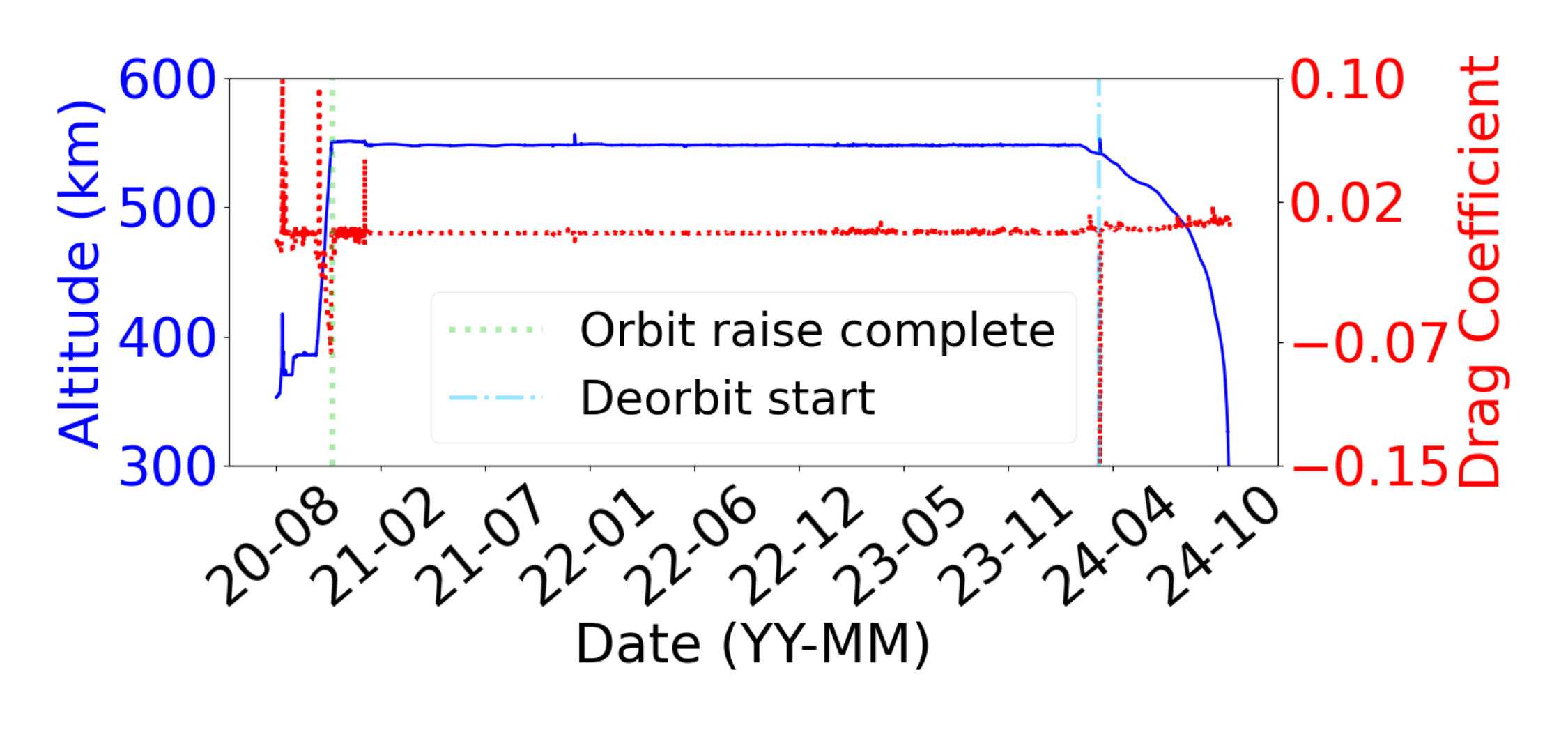}
        \caption{Starlink NORAD ID: 46149}
    \end{subfigure}%
    \hfill
    \begin{subfigure}[t]{0.50\columnwidth}
        \centering
        \includegraphics[width=\columnwidth, keepaspectratio]{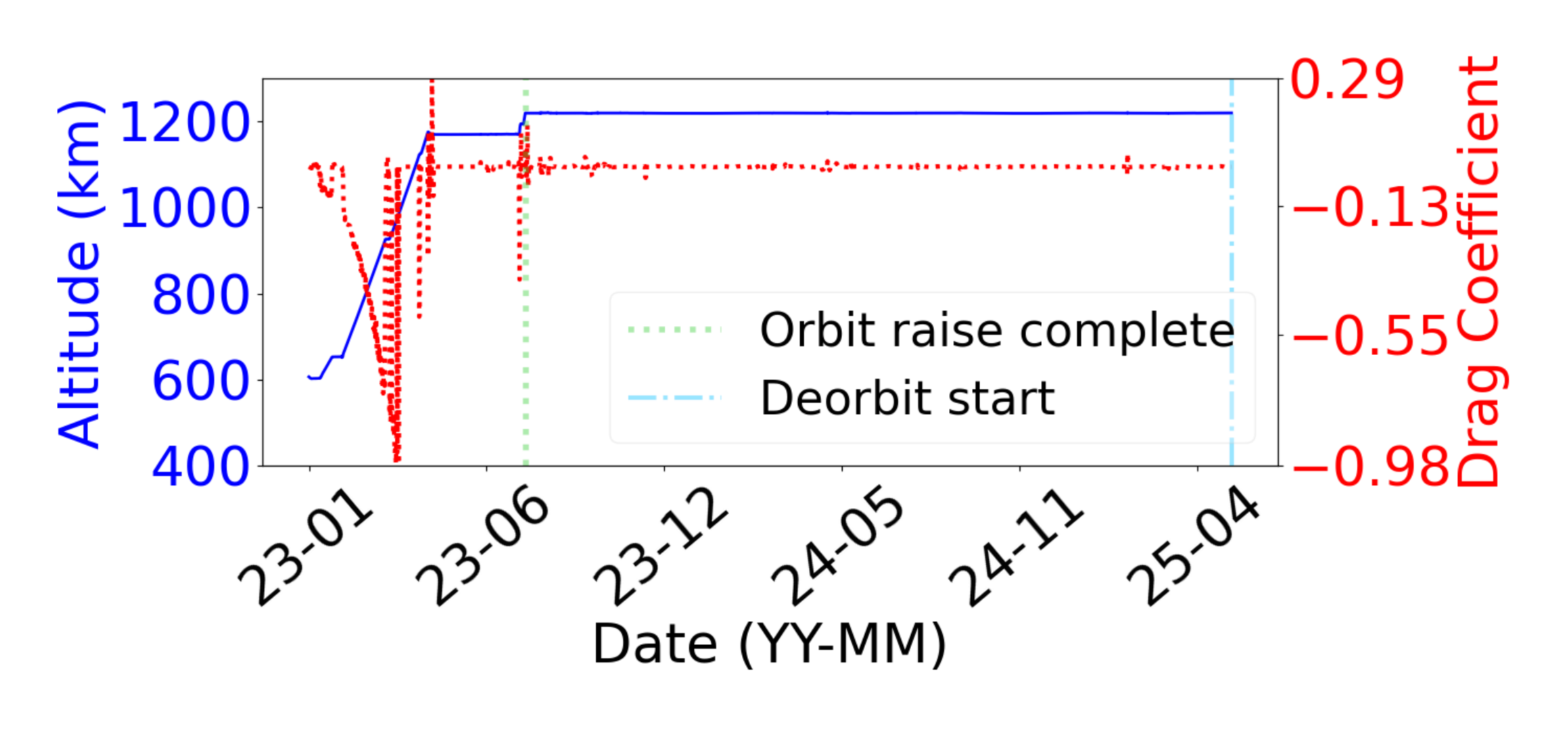}
        \caption{OneWeb NORAD ID: 55176}
    \end{subfigure}%
    \caption{Detecting the orbit raise complete and deorbit start with (a) Starlink and (b) OneWeb's TLE timeseries.}
    \label{fig:TLEclean}
\end{figure}

Second, the satellite launch vehicle injects the LEO satellites into a low-altitude staging orbit, typically at an altitude of around 350 km. 
After initial testing and screening, the LEO satellite uses its thrusters to ascend to an operational orbit of 500 km or more. 
These orbit-raising maneuvers are performed in multiple phases and take a variable period of time (a few weeks to a few months), as shown with a Starlink and OneWeb satellites in Fig.~\ref{fig:TLEclean}.
Prior work requires manual intervention to trim TLEs during orbit-raising phases, which is not scalable, given that Starlink is now operating over 10,000+ satellites and plans to have over 15,000 in total~\cite{FCCStarlink15000}. 
In \projectname{}, we automatically trim out TLEs of orbit-raising and de-orbiting periods to extract LEO satellites' operational period only. 
We achieve this with a four-step procedure:

\begin{enumerate}
    \item Compute the median altitude from all the  TLEs.
    Any satellite with a sufficiently long operational period will have operational altitude as the median altitude.
    Then, from the beginning as well as from the end of the timeseries, find the first TLE that encounters the median altitude.
    \item Consider the TLEs between these two to compute the altitude threshold: $\mu_h-\sigma$, where $\mu_h$ is the mean altitude and $\sigma$ the first standard deviation of altitude.
    \item Now consider the entire timeseries to find the TLE that encounters the altitude threshold at both the beginning and the end. 
    The epochs of these two TLE provide approximate timestamps for the completion of the orbit raise and the start of de-orbiting.
    \item Set the orbit raise complete and the de-orbiting start timestamp to the epoch of TLEs having maximum drag coefficient within a day window of approximate timestamp.
\end{enumerate}

This approach accurately detects the operational period of any LEO satellite that has spent a sufficiently long time in orbit.
The detected orbit raise is complete, and the deorbit start with this approach is shown using two vertical dotted lines with Starlink in Fig.~\ref{fig:TLEclean}(a) and OneWeb in Fig.~\ref{fig:TLEclean}(b) satellite. 
In our analysis, we consider only satellites with an operational period of more than one year.
Note that this is an essential step because, for a given investigating time window, a newly launched satellite could perform an orbit raise, or an older satellite might start de-orbiting, leading to an incorrect interpretation.

\subsubsection{LEO Internet connectivity:}

For connectivity \projectname{} uses four publicly available LEO network measurement datasets.

\parab{LENS} - High-cadence (up to 10 ms resolution) network latency traces over Starlink were collected in collaboration with a handful of Starlink users~\cite{LENSdatasets}. 
The measurements capture latency between the user terminal (dish) and the Starlink Point of Presence (PoP) at \texttt{100.64.0.1}, which is a single IP hop encompassing the Starlink satellite(s) segment and the associated ground-station-to-PoP terrestrial segment.
Latency traces are obtained using two methods. 
(i) ICMP-based \texttt{ping} for round-trip time (RTT), and (ii) UDP-based \texttt{IRTT} for end-to-end one-way delay. 
For this study, we collect both \texttt{ping} and \texttt{IRTT} traces for May and October 2024. 
This includes measurements from 6 and 10 user terminals in May and October, distributed across the US, Canada, and Europe.

\parab{M-LAB (NDT)} - The Measurement Lab (M-Lab)~\cite{MLAB_SIGCOMM_COMM_REVIEW} host network performance records of the end user initiated speed test~\footnote{https://speed.measurementlab.net/} in publicly accessible Google BigQuery.
This leverages the Network Diagnostic Tool (NDT)~\cite{ndt} to measure end-to-end network connectivity (transport capacity and latency) via a single TCP connection to the nearest available measurement server, selected from a globally distributed infrastructure of physical and virtual nodes~\cite{ndtSelectsServers}.
For this study, we acquire the speed test records from two BigQuery tables \texttt{ndt5} (uses CUBIC) and \texttt{ndt7} (uses BBR if available).
Specifically, we extract mean throughput, minimum latency, packet loss, and client and server geolocation metadata filtered by Starlink Autonomous System Number (AS14593)~\cite{StarlinkASN}.
In total, we use 600K+ (36K+) and 1M+ (22K+) \texttt{ndt7} (\texttt{ndt5}) speed test results across 80+ (50+) countries over the Starlink network between 3rd to 13th May and 1st to 13th October in 2024.

\parab{Cloudflare AIM} - Cloudflare Aggregated Internet Measurement (AIM) is also an end-user–initiated speed test~\footnote{https://speed.cloudflare.com/}. 
When resolving the hostname, Cloudflare leverages anycast routing to direct requests to the nearest Cloudflare edge CDN.
Unlike traditional speed tests, Cloudflare AIM does not attempt to saturate the link capacity~\cite{cloudflareHowTestWorks}. 
Instead, it transmits a sequence of small data blocks over HTTP/3 QUIC (when available), thereby better mimicking real-world network usage conditions and capturing user-perceived application-level performance.
These measurements are publicly accessible via M-Lab Google BigQuery too~\cite{cloudflareMeasuringNetwork,measurementlabMLabProviding}. 
Similar to M-Lab (NDT), we extract records, including uplink and downlink throughput, latency, packet loss, jitter, and client and server geolocation metadata, filtered by the Starlink ASN~\cite{StarlinkASN}.
In total, we analyze 2K+ and 4K+ AIM speed test results collected over the Starlink network between 3rd to 13th May and 1st to 13th October 2024 across 60+ countries.

\parab{RIPE Atlas} - RIPE Atlas probes~\cite{ripeAtlasProbes} are small hardware devices or software agents deployed worldwide. 
This distributed network of probes functions as a swarm of sensors, continuously measuring Internet connectivity worldwide.
In this study, we leverage the RIPE Atlas REST API~\cite{ripeAPIreference} in \projectname{} and use the Starlink ASN~\cite{StarlinkASN} to identify 91 probe IDs hosted behind Starlink user terminals across 23 countries. 
We then collect measurements from the built-in periodic \texttt{ping} tests~\cite{ripeBuiltinMeasurements} to all DNS root servers over both IPv4 and IPv6 from, during two time periods in May and October 2024.

In the following sections, we use these datasets with \projectname{} to dive deep into solar storms and their implications for LEO satellites' orbital trajectories and network connectivity.

\section{Dissecting solar storms}
\label{sec:solarStorm}

\begin{figure}
    \centering
    \begin{subfigure}[t]{0.60\columnwidth}
        \centering
        \includegraphics[height=11cm, keepaspectratio]{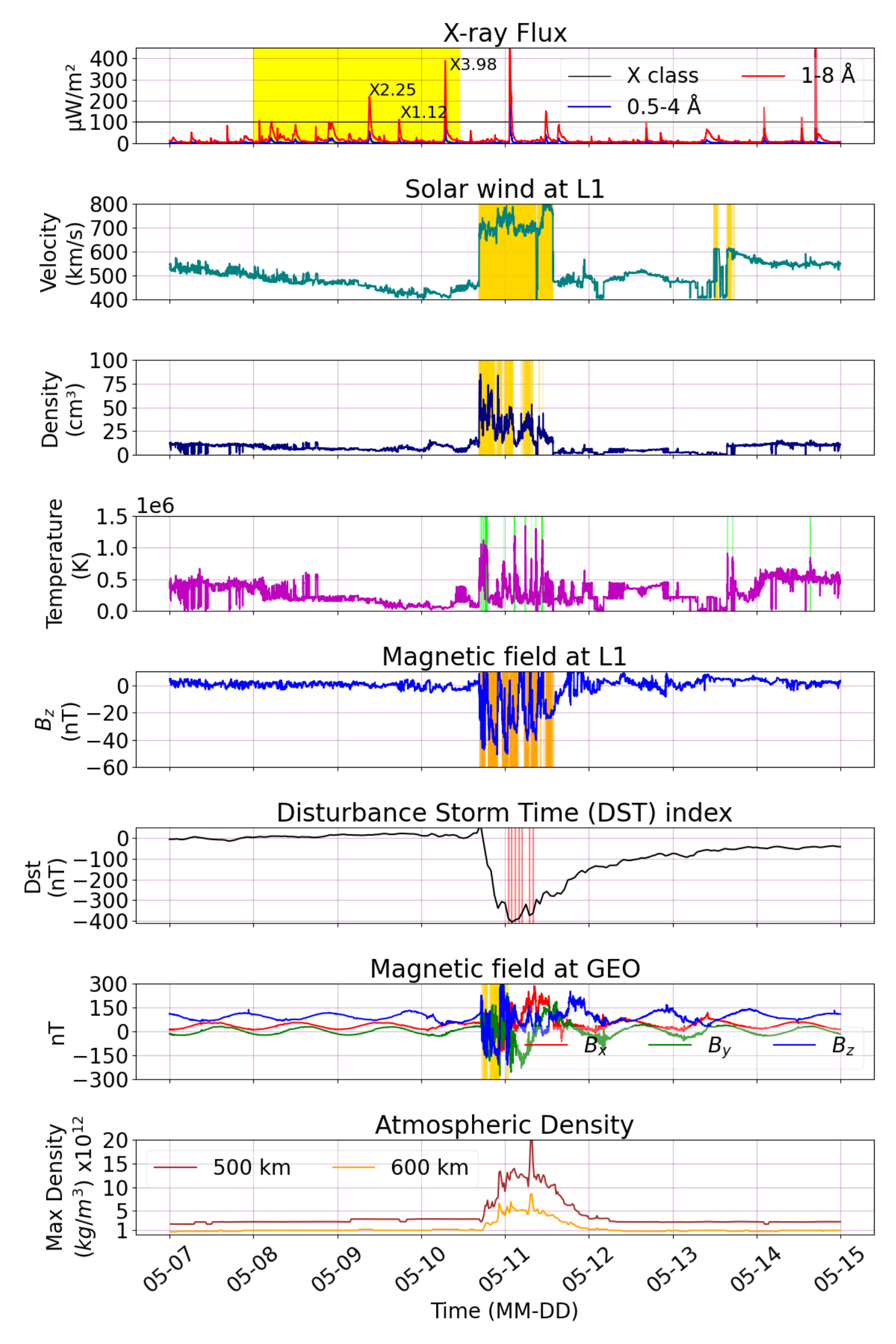}
        \caption{May 2024 solar superstorm}
    \end{subfigure}%
    \hfill
    \begin{subfigure}[t]{0.40\columnwidth}
        \centering
        \includegraphics[height=11cm, keepaspectratio]{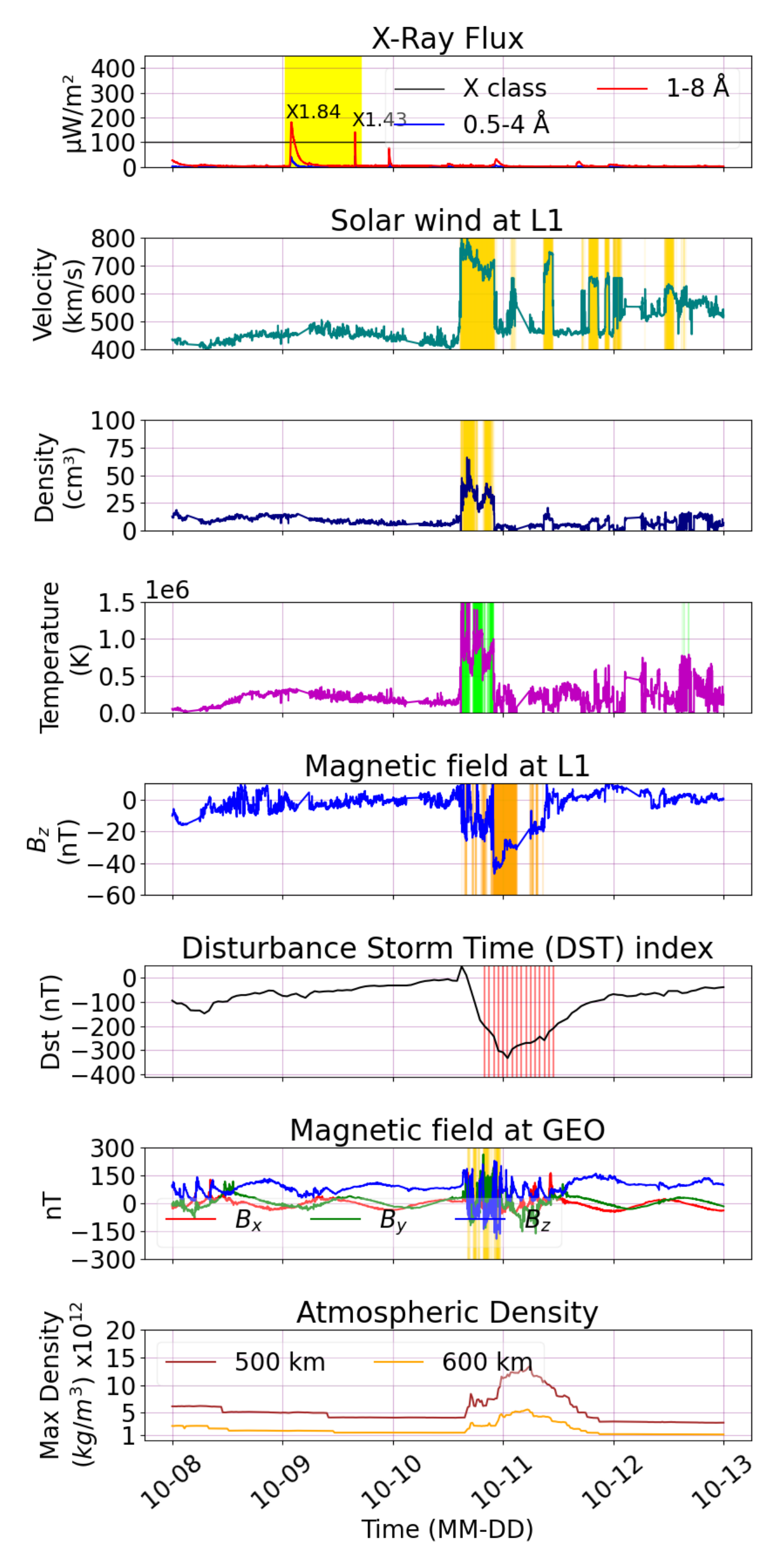}
        \caption{October 2024 solar severe storm}
    \end{subfigure}%
    \caption{Dissecting two major solar events (a) May 2024 solar superstorm (NOAA G5) and (b) October 2024 solar severe storm (NOAA G4), illustrating phase-by-phase events from solar eruption to LEO operational hazards.}
    \label{fig:deepDriveSolarStorm}
\end{figure}

We start by tracing solar storms with \projectname{} to provide an end-to-end understanding of how eruptions reach Earth and pose operational hazards to LEO satellites.
We use the Dst indices of the last five years, from January 2020 (following the launch of Starlink operational satellites on November 11, 2019) to December 2025, to list all geomagnetic storms in Table~\ref{tbl:listSolarStorm} having intensity of at least NOAA G3 ($\leq$-100 nT).
From this, we pick the two strongest geomagnetic storms for our analysis. 
First is the May 2024 solar superstorm, also known as the Gannon storm (named after physicist Jennifer Gannon) or 2024 Mother's Day solar storm~\cite{wikipedia2024Solar}, with an intensity of -406 nT. 
This is considered to be the strongest geomagnetic storm of the 21st century as of now. 
The second is the October 2024 NOAA G4-class geomagnetic storm, with an intensity of -333 nT, the second strongest since the launch of Starlink.
In Fig.~\ref{fig:deepDriveSolarStorm}, we merge all space weather datasets acquired from the aforementioned satellites and observatories to deep dive into these two events.

\begin{table}
\centering
\footnotesize
\caption{All the G3 to G5 scale geomagnetic storms from January 2020 to December 2025.}
\label{tbl:listSolarStorm}
\begin{tabular}{c|c|c||c|c|c}
\hline
NOAA Scale                                                              & Event date & Peak intensity & NOAA Scale                                                             & Event date & Peak intensity \\ \hline\hline
\multirow{11}{*}{\begin{tabular}[c]{@{}c@{}}G3\\ (Strong)\end{tabular}} & 2021-11-04 & -105 nT        & \multirow{6}{*}{\begin{tabular}[c]{@{}c@{}}G3\\ (Strong)\end{tabular}} & 2024-11-09 & -101 nT        \\ \cline{2-3} \cline{5-6} 
                                                                        & 2023-02-27 & -132 nT        &                                                                        & 2025-04-16 & -138 nT        \\ \cline{2-3} \cline{5-6} 
                                                                        & 2023-03-24 & -163 nT        &                                                                        & 2025-06-01 & -119 nT        \\ \cline{2-3} \cline{5-6} 
                                                                        & 2023-11-05 & -163 nT        &                                                                        & 2025-06-13 & -101 nT        \\ \cline{2-3} \cline{5-6} 
                                                                        & 2023-12-01 & -108 nT        &                                                                        & 2025-09-30 & -106 nT        \\ \cline{2-3} \cline{5-6} 
                                                                        & 2024-03-03 & -112 nT        &                                                                        & 2025-11-06 & -116 nT        \\ \cline{2-6} 
                                                                        & 2024-03-24 & -128 nT        & \multirow{4}{*}{\begin{tabular}[c]{@{}c@{}}G4\\ (Severe)\end{tabular}} & 2023-04-24 & -213 nT        \\ \cline{2-3} \cline{5-6} 
                                                                        & 2024-04-19 & -117 nT        &                                                                        & 2024-10-11 & -333 nT        \\ \cline{2-3} \cline{5-6} 
                                                                        & 2024-06-28 & -107 nT        &                                                                        & 2025-01-01 & -210 nT        \\ \cline{2-3} \cline{5-6} 
                                                                        & 2024-08-12 & -188 nT        &                                                                        & 2025-11-12 & -217 nT        \\ \cline{2-6} 
                                                                        & 2024-09-12 & -188 nT        & \begin{tabular}[c]{@{}c@{}}G5\\ (Extreme)\end{tabular}                 & 2024-05-11 & -406 nT        \\ \hline\hline
\end{tabular}
\end{table}

\parab{Gannon storm} - 
The origin of this solar superstorm is multiple back-to-back eruptions from the Sun's active region AR13664 between 8th and 10th May, 2024.
These eruptions are captured in X-ray flux on the top panel of Fig.~\ref{fig:deepDriveSolarStorm}(a). 
Notice six-seven X-class flares (irradiance $\geq$10$^{-4}$W/m$^2$) in the highlighted time span in yellow, where the last three are the strongest: X2.25, X1.12, and X3.98 class flares.
These slow CMEs, followed by the fast CMEs, collided in space and merged to form the strongest composite magnetic field as they moved towards the Earth.
Notice the next four panels. 
The first radiation shock wave arrived at L1 around 16:30 UTC on 10th May, 2024, resulting in a significant increase in solar wind velocity (up to 800 km/s), plasma density, and temperature, accompanied by strong southward ($B_z$) interplanetary magnetic field disturbances reaching up to -50 nT. 
Around 17:15 UTC on 10th May, 2024, this shock wave of solar wind and interplanetary magnetic field struck the Earth's magnetosphere, triggering a push inward, causing a decline in the Dst index. 
By 2:30 UTC on 11th May, 2024, the Dst index fell below -400 nT, and the storm was classified as a NOAA G5 geomagnetic storm. 
The intensity remained below -100 nT till the next morning, 9:00 UTC on 12th May, 2024.
Strong disturbances in the magmatic field after 17:30 UTC on 10th May, 2024, have also been recorded in geostationary orbit, as shown in the next panel.
As charged particles began hitting the upper atmosphere, it began to heat up and expand. 
As a result, notice that the bottom panel shows the maximum atmospheric density at altitudes of 500 and 600 km (the envelope where Starlink mostly operates) began increasing around 18:30 UTC on 10th May, 2024. 
This density inflation peaked at 20 times the baseline around 7:00 UTC on 11th May, 2024, 4 hours after the recording peak Dst index of -406 nT. 
Which was recovered after 12th May, 2024.

\parab{October 2024 storm} - 
Similarly, the origin of this solar storm is a solar eruption from AR3848, approximately at 9:00 UTC on 9th October, 2024. 
Again, this is recorded as a spike in X-ray radiation of an X1.84 class flare in the top panel in Fig.~\ref{fig:deepDriveSolarStorm}(b). 
The impact of this eruption arrived at L1 around 15:00 UTC on 10th October, 2024. 
Hence, a sudden spike in solar wind velocity, plasma density, and temperature, as well as in southward interplanetary magnetic field disturbances in the next four panels.
Notice, as compared to the Gannon storm in Fig.~\ref{fig:deepDriveSolarStorm}(a), the time span of the surge is shorter, and the magnitude of the southward interplanetary magnetic field disturbance and plasma density is also relatively smaller.
Therefore, the pressure on Earth's magnetosphere during this eruption did not exceed -350 nT, and it is classified as a NOAA G4 geomagnetic storm.
At 16:00 UTC on 10th October, 2024, a magnetic field disturbance at geostationary altitude has also been observed. 
Then, the atmospheric density increased up to 13 times above the baseline at around 5:00 UTC on 11th October, 2024, four hours after the peak of the Dst index of -333 nT.
Which returned to normal by 12th October, 2024.

\begin{keybox}
\keynote 
A longer duration of high-speed solar wind with a strong southward pointing magnetic field induces stronger solar storms, leading to higher upper atmospheric density inflation, recorded up to 20 and 13 times above the baseline during the Gannon and October 2024 solar storms, respectively. 
\end{keybox}

\section{Measuring the implications on LEO satellite's trajectory}
\label{sec:decayMeasurements}

In this section, we investigate operational hazards posed by these two solar storms for LEO satellites at different altitudes and inclinations. 
For this, we use the satellites' TLE from the Starlink and OneWeb constellations. 
In the prior discussion of the solar storm analysis, we saw that atmospheric density inflation is much higher at lower altitudes than at higher ones. 
That means satellites orbiting at different altitudes are expected to experience different intensities of orbital drag and cannot be compared with one another.
A large constellation like Starlink deploys its satellites in multiple shells.
Where each shell is a batch of satellites orbiting at similar altitudes and inclinations. 
Hence, to accurately measure the impact, first, we need to separate the satellites into their respective shells.

\subsection{Segregating shells of LEO constellations}

We use the Starlink and OneWeb shell configurations (altitude and inclination) from Table 1 in~\cite{basak2025leocraft}, which are compiled from FCC filings. 
To segregate the shells, we first compute the mean altitude and inclination of each satellite using TLEs with epochs within the solar storm window. 
Then, we compare the mean altitude and inclination with officially reported shell altitudes and inclination. 
We consider a given satellite to belong to that shell if the altitude and inclination differences fall below a threshold. 
Also, note that there are some caveats. 
For instance, in Starlink's TLEs, we found that shells' altitudes in real deployment might differ up to a few 10s of km.
We have made these adjustments in our implementation after exploring the datasets.
For our measurements, we use the solar storm windows from 10th May to 15th May 2024 and from 9th October to 14th October 2024. And altitude thresholds are 7 km, and inclination thresholds are kept between 0.15$^{\circ}$ to 0.5$^{\circ}$.
In Table~\ref{tbl:listShellSegregation}, we have tabulated the shells configuration and satellites per shell during the May 2024 solar superstorm and October 2024 solar storm.

\begin{table}
\centering
\footnotesize
\caption{The number of satellites is segregated in each shell of Starlink and OneWeb for both solar storms. (+x) indicates the difference observed in the TLEs relative to the mentioned values in the FCC filling.}
\label{tbl:listShellSegregation}
\begin{tabular}{c|c|c|c|c}
\hline
Constellation & Altitude (km) & Inclination ($^{\circ}$) & \# Satellites (May'24) & \# Satellites (Oct'24) \\ 
\hline\hline

OneWeb & 1,200 & 87.9 & 372 & 372 \\
\hline
\multirow{5}{*}{Starlink} & 530 (+36) & 43 & 1,335 & 1,332 \\
                        \cline{2-5} 
                          & 540 (+6) & 53.2 & 1,528 & 1,499 \\
                        \cline{2-5} 
                          & 550 (+4) & 53 & 1,210 & 1,296 \\
                        \cline{2-5} 
                          & 570 (+9) & 70 & 390 & 392 \\
                        \cline{2-5} 
                          & 560 (+10) & 97.6 (+0.1) & 232 & 233 \\
\hline\hline
\end{tabular}
\end{table}

\subsection{Exploring satellite decay characteristics}

We select four Starlink satellites from four shells to examine the nature of orbital decay at different altitudes and inclinations. 
These selected satellites are,

\begin{enumerate}
    \item NORAD 55434 at 579.22 km altitude and 70$^{\circ}$ inclination.
    
    \item NORAD 53482 at 570.12 km altitude and 97.7$^{\circ}$ inclination.
    
    \item NORAD 57508 at 566.26 km altitude and 43$^{\circ}$ inclination.
    
    \item NORAD 44973 at 551.47 km altitude and 53$^{\circ}$ inclination.
\end{enumerate}

\begin{figure}
    \centering
    \begin{subfigure}[t]{0.5\columnwidth}
        \centering
        \includegraphics[width=\columnwidth, keepaspectratio]{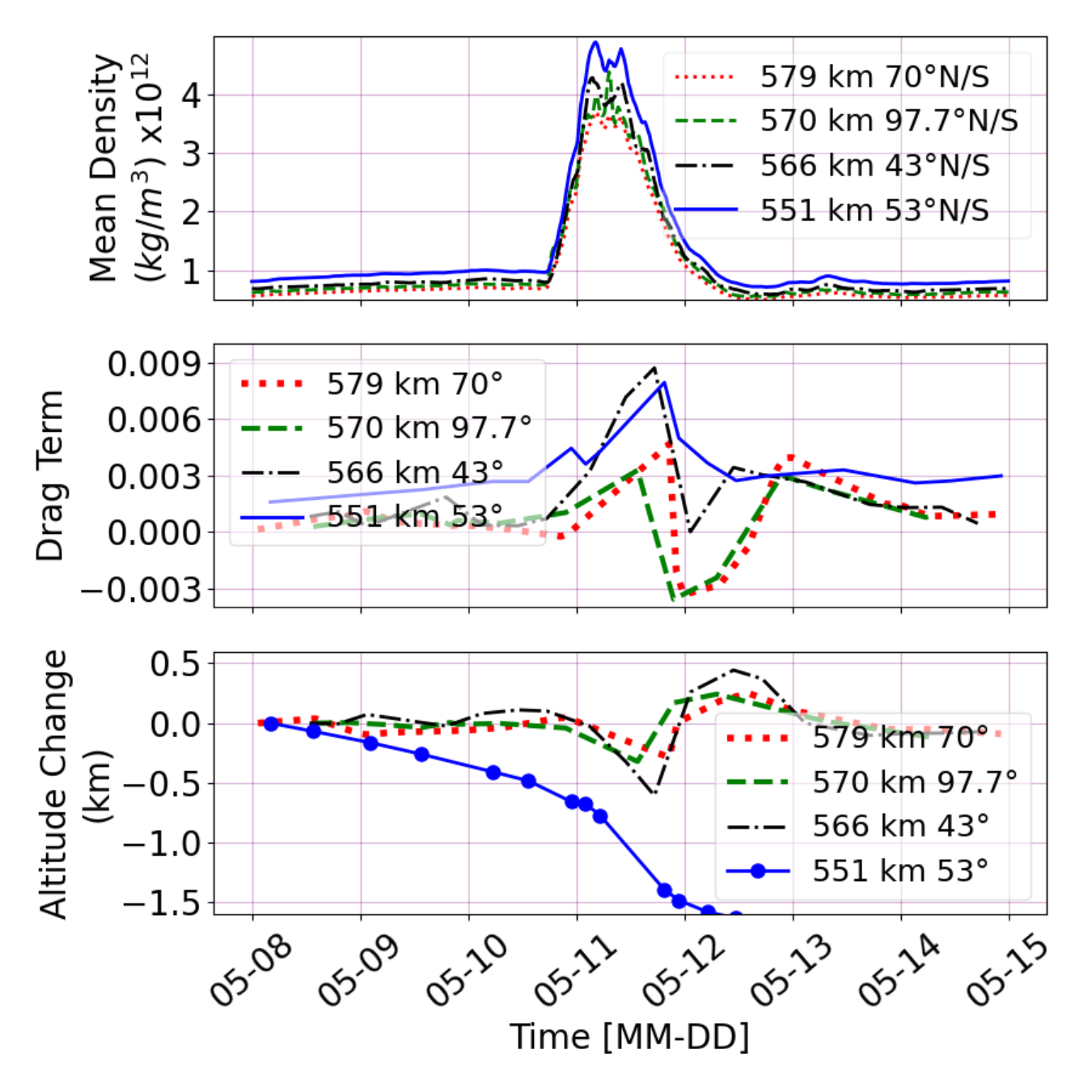}
        \caption{May 2024 solar superstorm}
    \end{subfigure}%
    \hfill
    \begin{subfigure}[t]{0.5\columnwidth}
        \centering
        \includegraphics[width=\columnwidth, keepaspectratio]{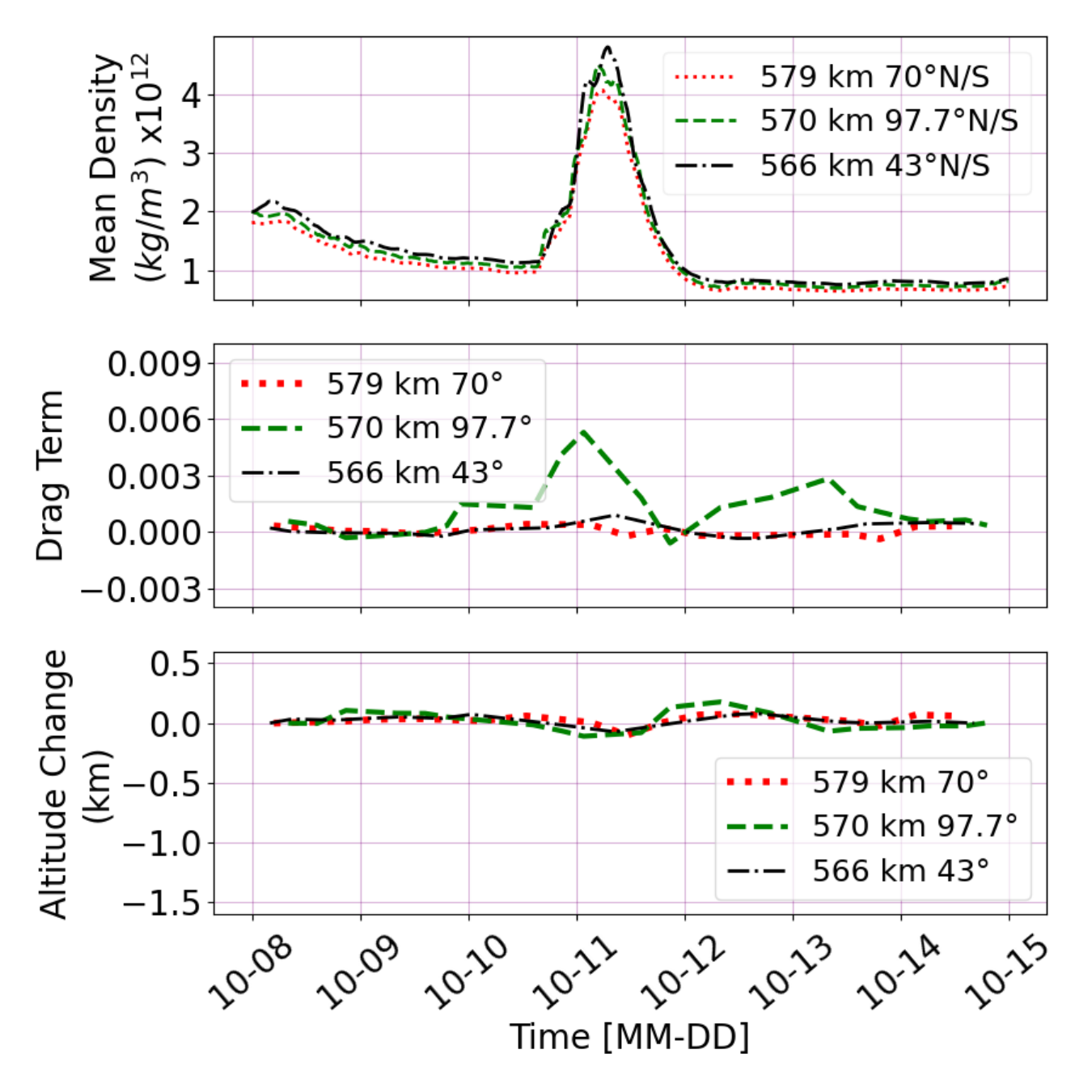}
        \caption{October 2024 solar severe storm}
    \end{subfigure}%
    \caption{Cherry-picked one satellite from each shell of Starlink to demonstrate the nature of orbital decay at different altitudes and inclinations during the solar storm. One satellite is decommissioned, so its orbit is never corrected.}
    \label{fig:decayNatureTimeseries}
\end{figure}

We extract atmospheric density from the TIE-GCM simulation for each satellite’s operational region. 
At a given timestamp, the mean atmospheric density is calculated by averaging density values at the satellite’s operational altitude, over all longitudes, and within the latitude band bounded by $\pm i^{\circ}$, where i represents the satellite’s inclination.
In Fig.~\ref{fig:decayNatureTimeseries}, we plot a stacked timeseries.
Where, at the top, we have the mean atmospheric densities.
In the middle and at the bottom, we show the drag coefficient and the observed altitude decay relative to the first day of this observation window, calculated from the TLEs of these satellites.

\parab{Gannon storm} - 
Notice in Fig.~\ref{fig:decayNatureTimeseries}(a), on 11th May 2024, during the peak of the solar superstorm, as the atmospheric density increases, the TLEs of NOARD 55434 (579 km), 53482 (570 km), and 57508 (566 km) report 5, 3, and 8 times the orbital drag over the baseline.
Consequently, loses the altitude up to 332, 318, and 716 meters within a day.
After a few hours, in the timeseries, notice the immediate steep drop in drag, followed by an increase in the satellite's altitude.
This is an orbit-correction maneuver to counter unexpected orbital decay during the solar storm.
Starlink in the post-storm report~\cite{StarlinkResponse} to FCC public notice~\cite{FCCPublicNotice} also mentions that the satellite's onboard thruster responded in real time when the satellite experienced an orbital drag increase up to 5 times.
Also, observe that this orbit correction raises the satellites to a slightly higher altitude of up to 500 meters than their long-term operational altitude to countermeasure the decay rate till the atmospheric conditions normalize.
Satellites at lower altitudes experience higher peak atmospheric density. 
However, given irregularities between consecutive TLE epochs and other unknowns, such as when the thrusters were activated, we cannot systematically compare the three factors: satellite altitude, atmospheric density inflation at that altitude, and the corresponding decay rate.

Another interesting observation in this figure is that the satellite with NORAD 44973 (551 km) began decaying well before the start of the May 2024 solar storm. 
This has never been orbit-corrected; consequently, it reentered the atmosphere on 11th September 2025 (not shown in this figure).
This satellite was launched on 7th January 2020.
Therefore, likely to be decommissioned around the first week of May 2024 after approximately 5 years of operational life.
Assuming this satellite has not yet lost its maneuvering capabilities, Starlink is using atmospheric conditions to expedite the de-orbiting of older satellites. 
The key observation here is the markers on this satellite's altitude change that indicate the TLE epochs. 
When the satellite starts decaying faster, almost 1 km per day, we see a longer delay in TLE release. 
If this delay is due to tracking challenges caused by rapid trajectory changes, it is affecting the conjunction assessment pipeline's ability to correctly predict the future conjunction event~\cite{doi:10.2514/1.A36164}.
This increases the risk of unforeseen close encounters between LEO satellites and in-orbit collisions in today's crowded LEO space.
To address this risk, Starlink is using an onboard autonomous collision-avoidance system~\cite{autonomousCollisionAvoidanceSystem} and developing a novel space situational awareness system, \emph{Stargaze}~\cite{stargaze}, for the LEO satellite operator community to minimize ground-based active operations.

\parab{October 2024 storm} - 
In Fig.~\ref{fig:decayNatureTimeseries}(b), during the October 2024 solar storm, we see that these operational satellites NORAD 55434 (579 km), NORAD 53482 (570 km), and NORAD 57508 (566 km) experience a drop in altitude around 161, 95, and 146 meters, respectively.
This is close to the regular orbital decay rate in LEO, thus we do not see any large orbit correction or drastic change in Fig.~\ref{fig:decayNatureTimeseries}(b) compared to Fig.~\ref{fig:decayNatureTimeseries}(a).
This indicates that extreme-scale solar events, such as solar superstorms, can pose serious operational hazards in LEO, whereas lower-intensity solar storms have minor implications.

\begin{keybox}
\keynote 
Starlink does not appear to perform real-time orbit corrections for decommissioned satellites (assuming they have not run out of fuel), instead leveraging the solar storm to accelerate their de-orbiting.
\end{keybox}


\subsection{Decoding the operation of LEO constellations}

These constellations are built from hundreds of thousands of LEO satellites orbiting the Earth, maintaining a specific formation to provide uniform Internet connectivity across the globe.
So, here we look at the entire LEO constellation, try to decode their operational strategies, and assess their likely implications for their services.

\subsubsection{Quantifying the orbital decay:}
\begin{figure}
    \centering
    \begin{subfigure}[t]{0.30\columnwidth}
        \centering
        \includegraphics[width=\columnwidth, keepaspectratio]{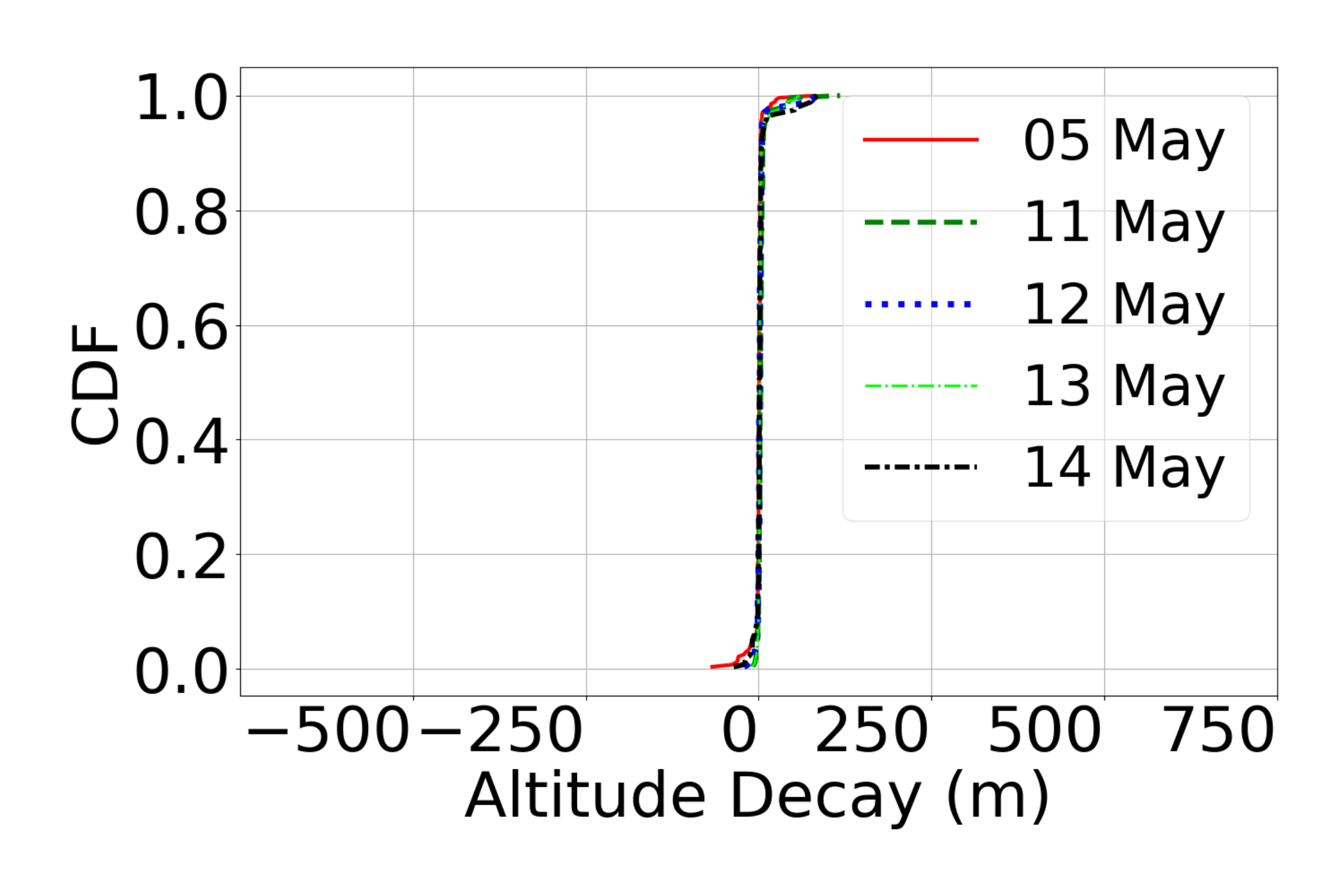}
        \caption{87.9$^\circ$, 1,200 km}
    \end{subfigure}%
    \hfill
    \begin{subfigure}[t]{0.30\columnwidth}
        \centering
        \includegraphics[width=\columnwidth, keepaspectratio]{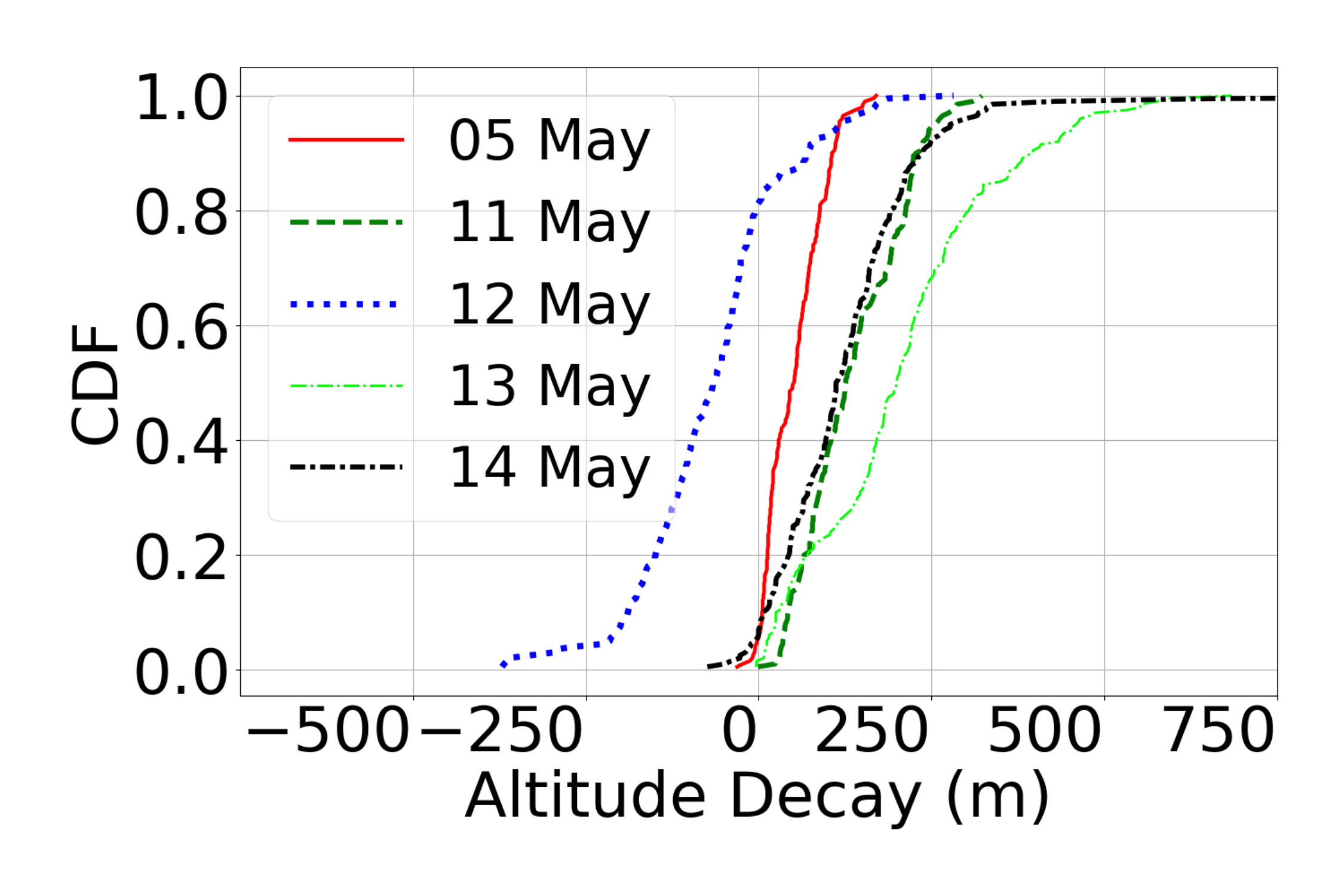}
        \caption{97.6$^\circ$, 560 km}
    \end{subfigure}%
    \hfill
    \begin{subfigure}[t]{0.30\columnwidth}
        \centering
        \includegraphics[width=\columnwidth, keepaspectratio]{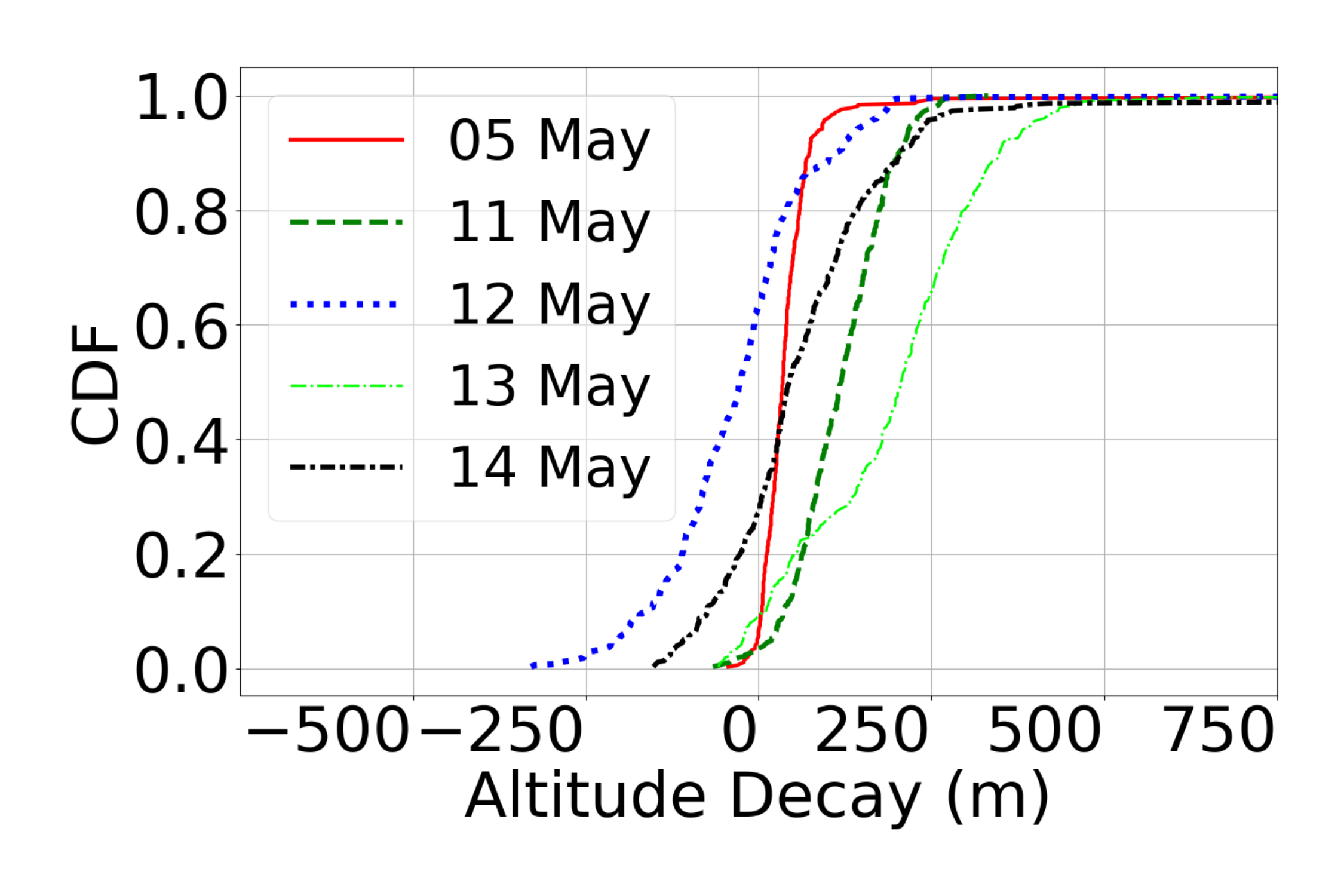}
        \caption{70$^\circ$, 570 km}
    \end{subfigure}%
    \hfill
    \begin{subfigure}[t]{0.30\columnwidth}
        \centering
        \includegraphics[width=\columnwidth, keepaspectratio]{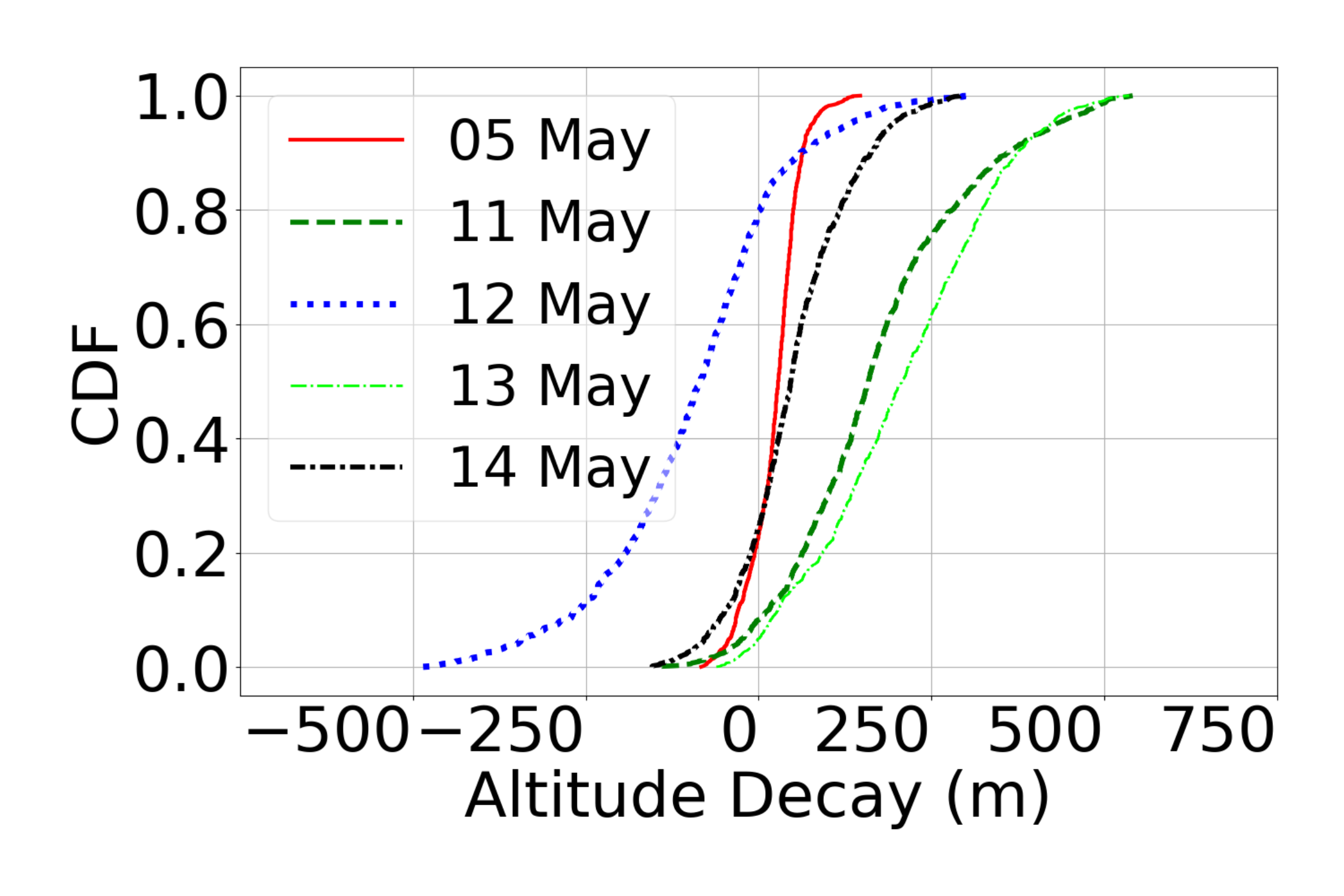}
        \caption{53$^\circ$, 550 km}
    \end{subfigure}%
    \hfill
    \begin{subfigure}[t]{0.30\columnwidth}
        \centering
        \includegraphics[width=\columnwidth, keepaspectratio]{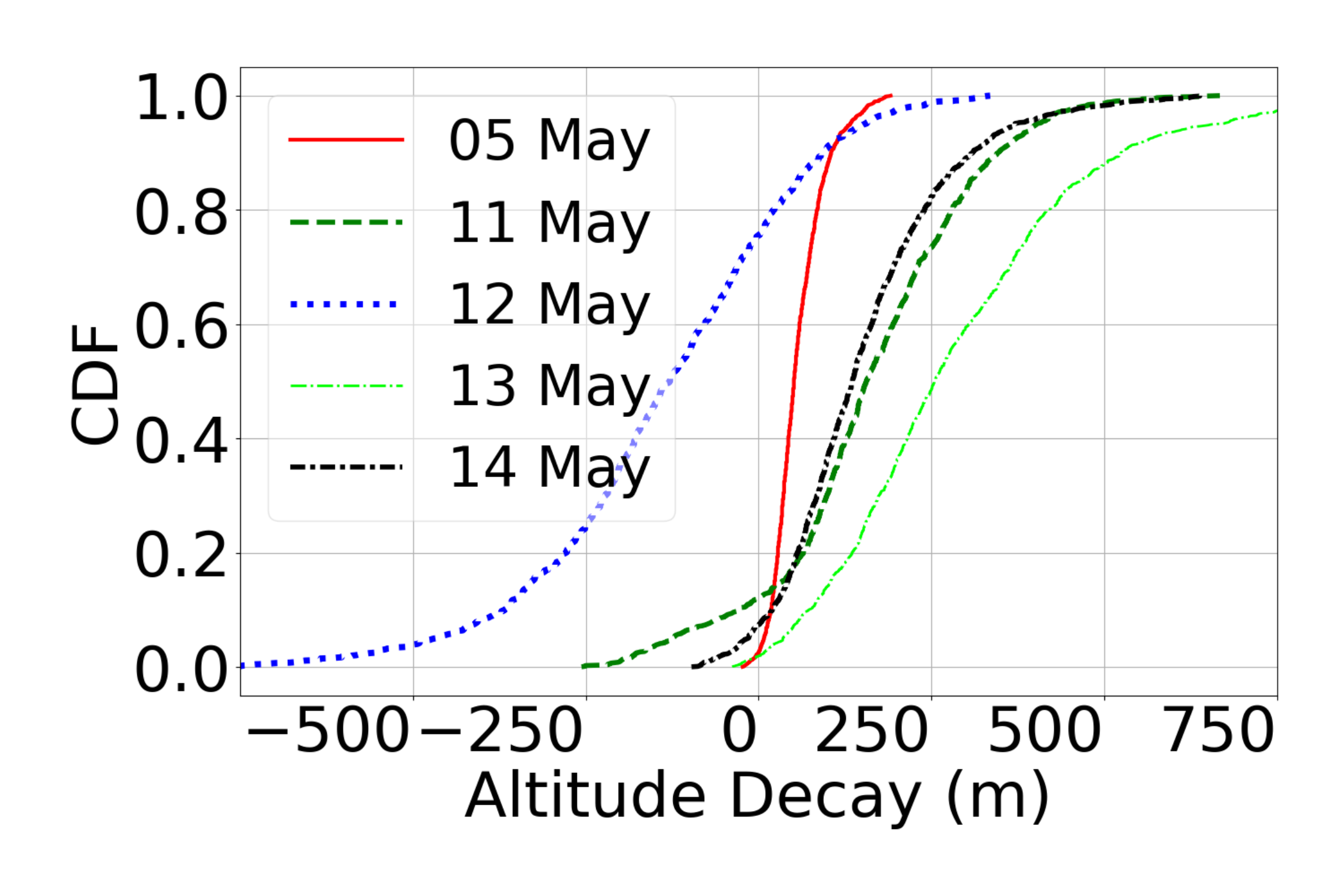}
        \caption{53.2$^\circ$, 540 km}
    \end{subfigure}%
    \hfill
    \begin{subfigure}[t]{0.30\columnwidth}
        \centering
        \includegraphics[width=\columnwidth, keepaspectratio]{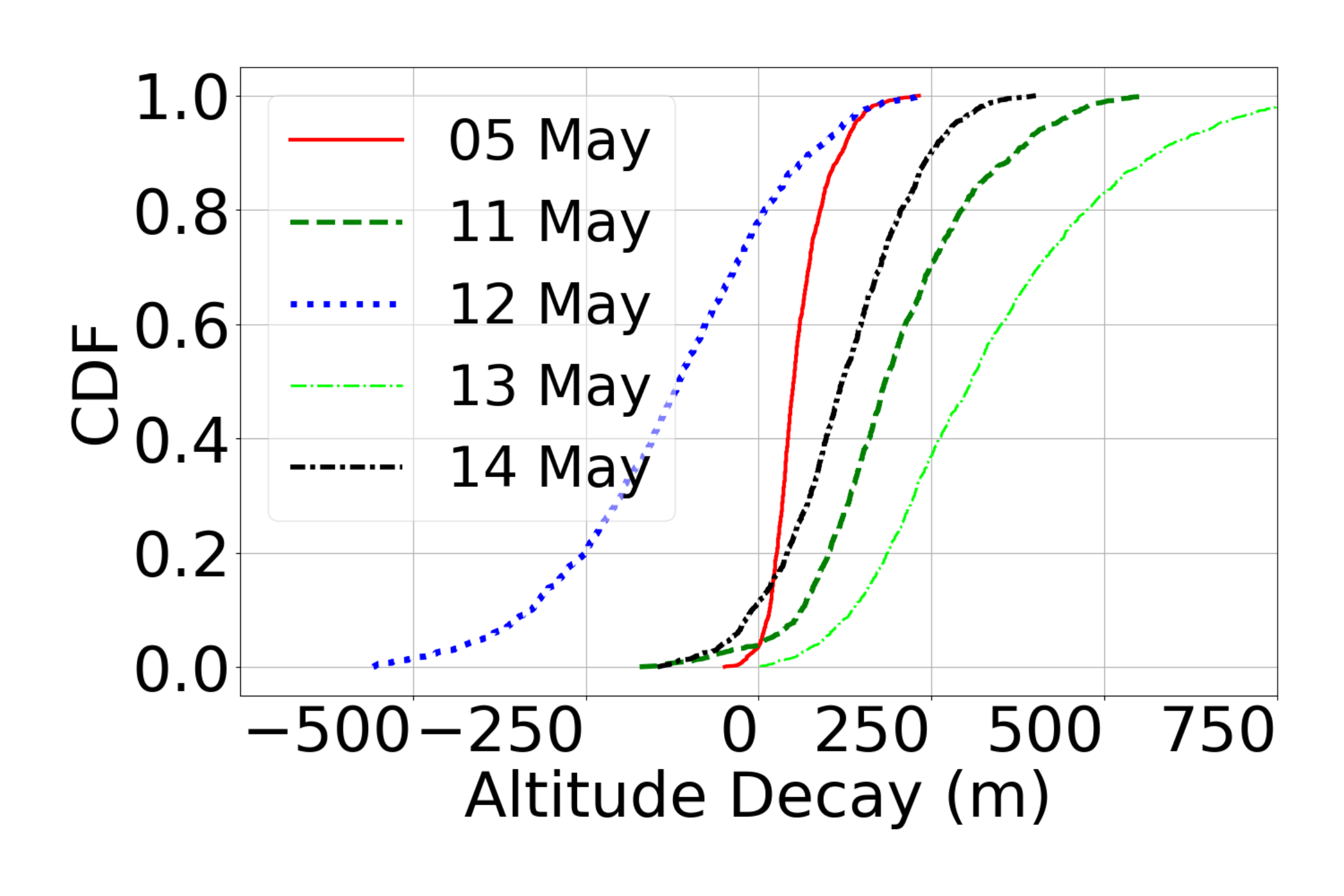}
        \caption{43$^\circ$, 530 km}
    \end{subfigure}%
    
    \caption{Shell-wise distribution of altitude decay compared to the last day in (a) OneWeb, and (b)-(f) Starlink constellation during May 2024 solar superstorm. OneWeb satellite at a higher altitude remains stable, Starlink satellite experiences significant altitude decay on 11th May, and then orbits are raised on 12th May 2024.}
    \label{fig:decayCDFMay24}
\end{figure}

\begin{figure}
    \centering
    \begin{subfigure}[t]{0.3\columnwidth}
        \centering
        \includegraphics[width=\columnwidth, keepaspectratio]{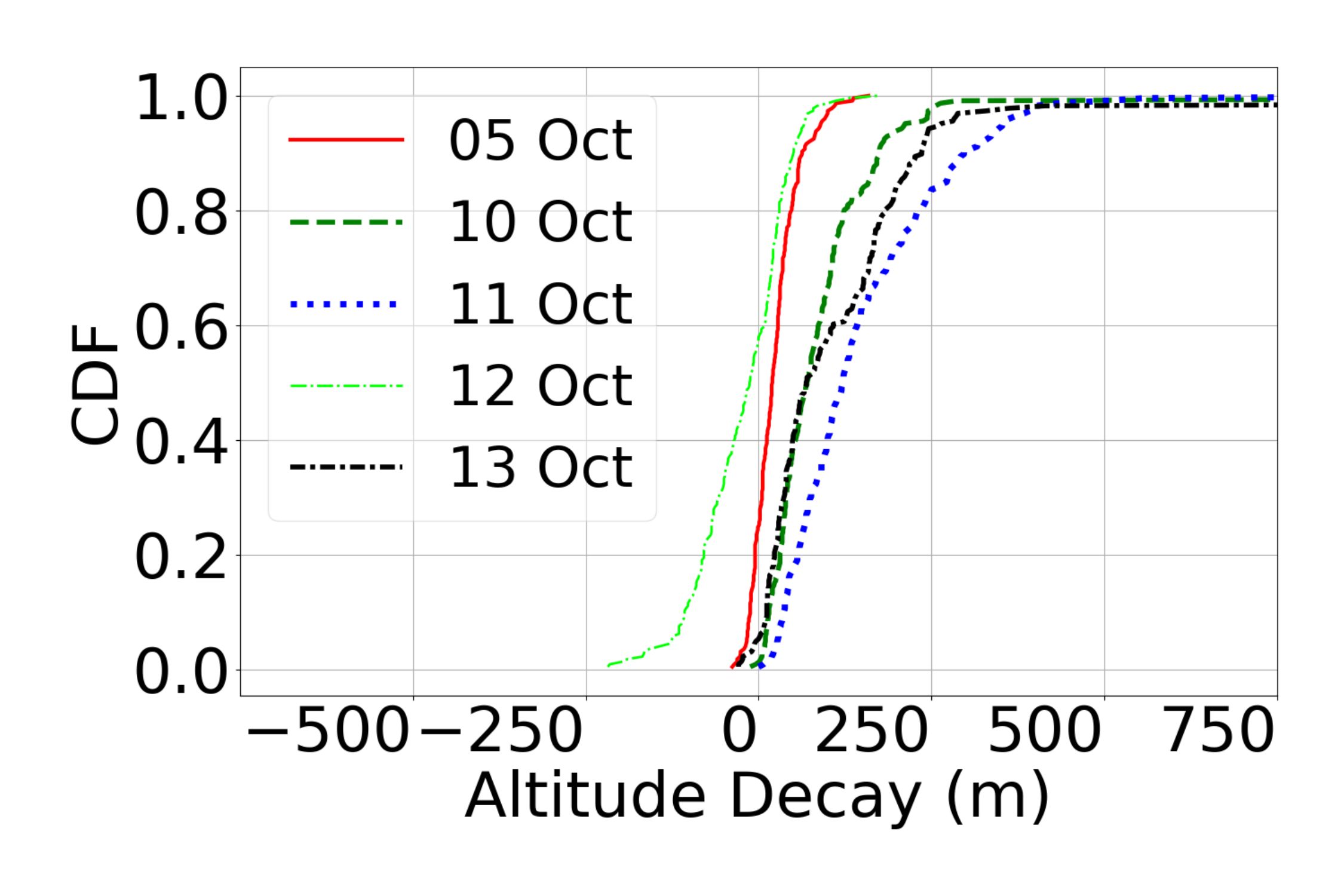}
        \caption{97.6$^\circ$, 560 km}
    \end{subfigure}%
    \hfill
    \begin{subfigure}[t]{0.3\columnwidth}
        \centering
        \includegraphics[width=\columnwidth, keepaspectratio]{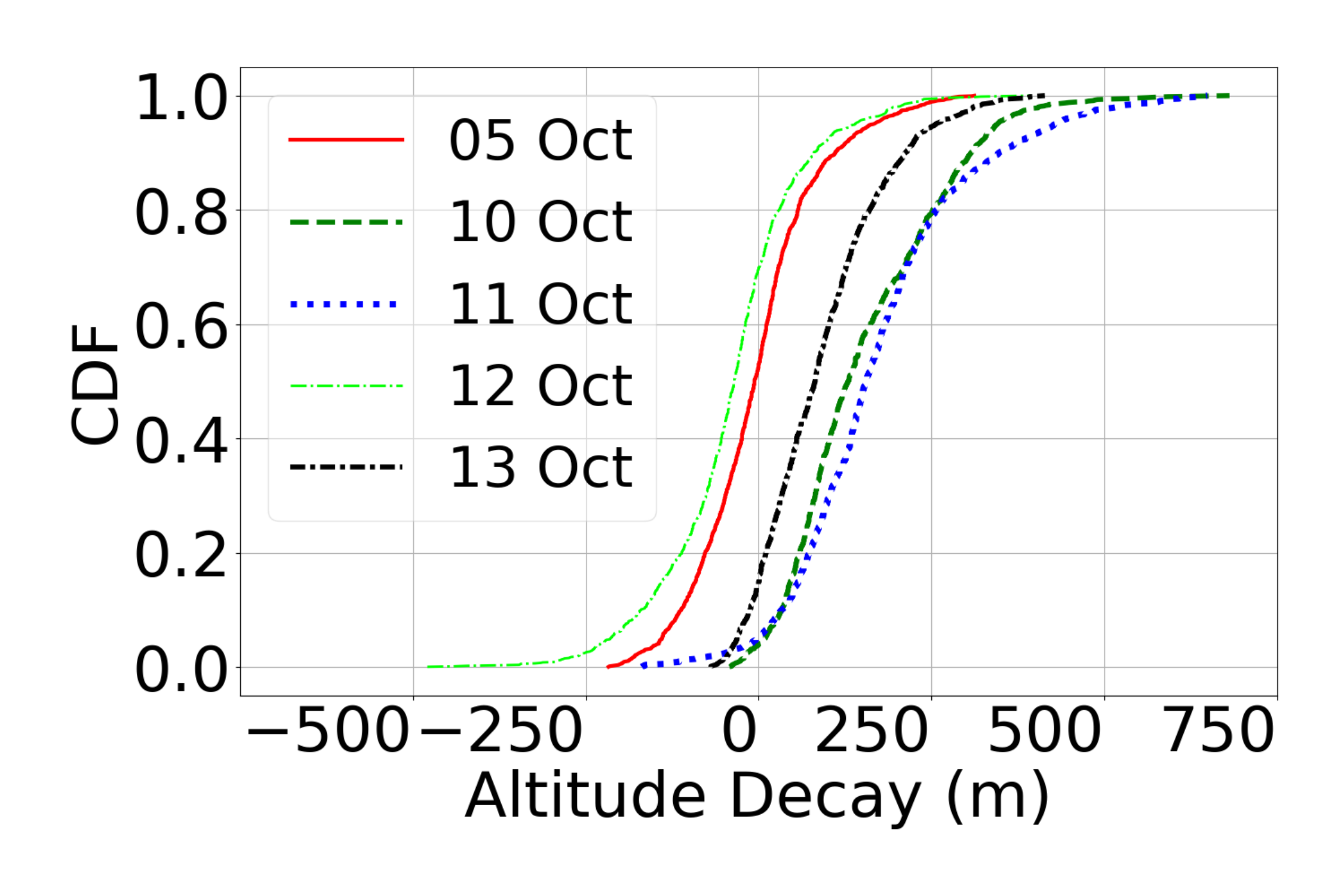}
        \caption{53$^\circ$, 550 km}
    \end{subfigure}%
    \hfill
    \begin{subfigure}[t]{0.3\columnwidth}
        \centering
        \includegraphics[width=\columnwidth, keepaspectratio]{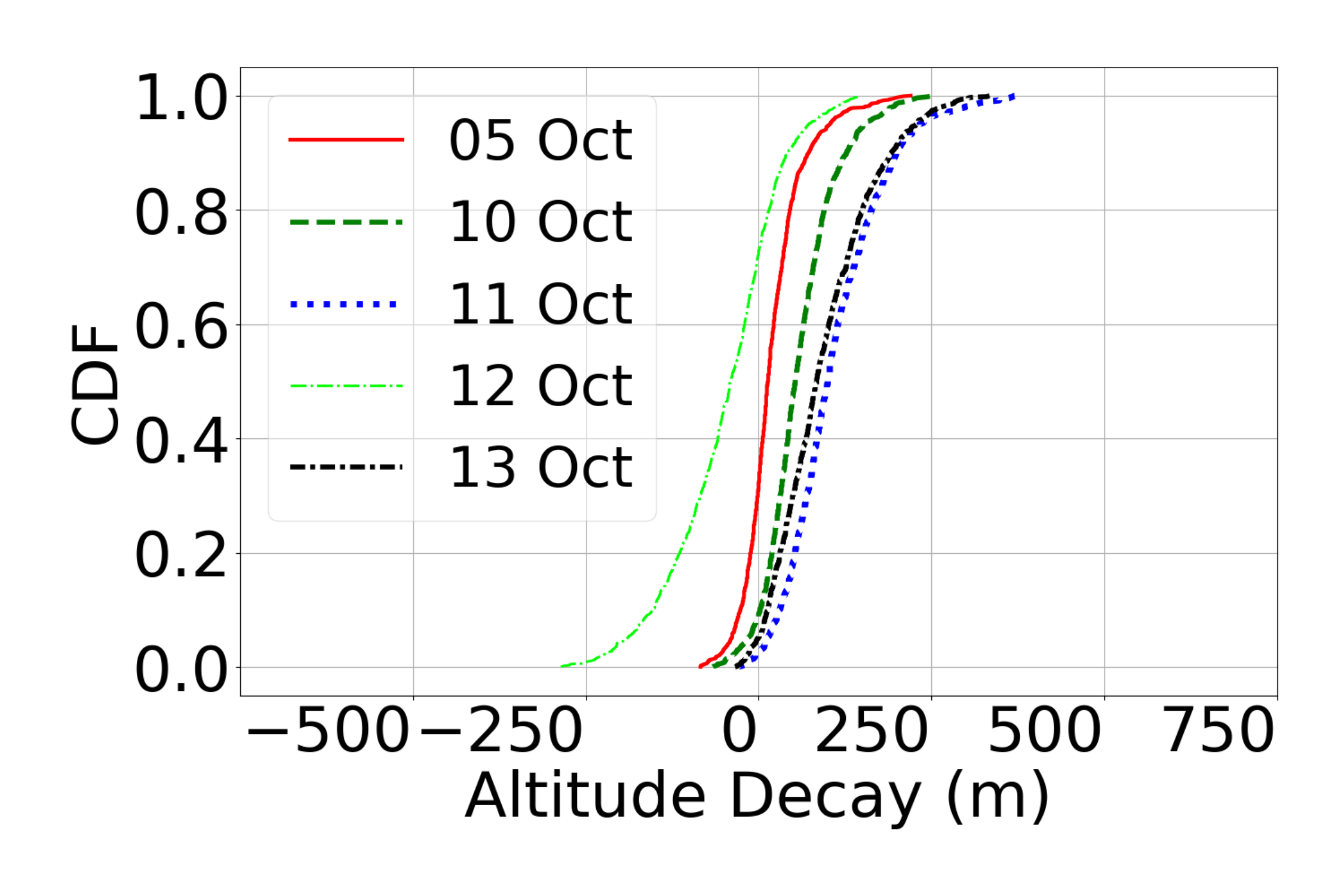}
        \caption{43$^\circ$, 530 km}
    \end{subfigure}%
    
    \caption{Starlink satellite altitude decayed during 10-11th October 2024 solar storm too. 
    However, Starlink did not respond in real time like the May 2024 solar superstorm.}
    \label{fig:decayCDFOct24}
\end{figure}

We quantify the overall altitude decay observed across the shells of OneWeb and Starlink constellations.
For this, we use TLEs to calculate decay as the difference in altitude over the previous day.
When more than one TLE is available, we take the maximum reported altitude from the previous day and the minimum from the current day to quantify the total change in altitude over the previous day.
We remove observations beyond the 99th and 0.01th percentiles to eliminate extreme outliers of a few kilometers, which are likely due to tracking errors rather than solar storm effects.
We plot the CDF of day-wise altitude decay during the May 2024 solar superstorm in Fig.~\ref{fig:decayCDFMay24} and the October 2024 solar storm in Fig.~\ref{fig:decayCDFOct24} for each shell tabulated in Table~\ref{tbl:listShellSegregation}.
Additionally, we plot the altitude decay for 5th May and 5th October as a baseline, since these two days have minimal solar activity intensity, with a Dst index between -39 and 21 nT. 
Note that, here, a positive value means the decay, i.e., altitude fall compared to the previous day, and a negative value means the opposite, i.e., orbit rise.

\parab{Gannon storm} - 
Notice the OneWeb's altitude decay CDFs in Fig.~\ref{fig:decayCDFMay24}(a).
The distribution of altitude decay on 5th May and from 11th to 14th May 2024 closely overlaps, with no noticeable change. 
This is because OneWeb satellites are deployed at 1,200 km, at much higher altitudes than those of other systems.
Hence, they encounter almost negligible differences in the atmospheric drag even during a solar superstorm.
In contrast, Starlink satellites are deployed at altitudes of 550-600 km. 
Which is more than twice as close to the Earth as OneWeb satellites.
As a result, in Fig.~\ref{fig:decayCDFMay24}(b)-(f), we see a large variation in satellite altitude due to much higher atmospheric resistance even in the quiet days.
Notice the distribution of 5th May. 
On a quiet day, we see a median decay up to 60 meters, and 95\% of satellites have not decayed beyond 150 meters across these five shells in Fig.~\ref{fig:decayCDFMay24}(b)-(f).
On 11th May 2024, during the solar superstorm, the decay distribution shifts significantly to the right.
We see median (95\%) decay increases to 126 (255), 120 (227), 160 (430), 156 (402), and 180 (423), respectively, which is up to 3 (2.8) times higher than on 5th May.
The next day, 12th May, we see a negative shift. 
The altitudes of approximately 80\% of the satellites increase up to 250 meters in higher altitude and higher inclination shells in Fig.~\ref{fig:decayCDFMay24}(b)-(c) and beyond 500 meters in lower altitude and lower inclination shells in Fig.~\ref{fig:decayCDFMay24}(d)-(f).
Further observe in all the shells in Fig.~\ref{fig:decayCDFMay24}(d)-(f), we see even higher altitude decay on 13th May compared to the 11th May. 
Then, from 14th May onward, atmospheric conditions start to normalize, so the distribution shifts towards the characteristics of 5th May. 

Altogether, this provides insight into Starlink's operations.
On 11th May, during the solar superstorm, when satellites started decaying rapidly, the onboard thruster kicked in at some point, so the distribution of altitude decay on 11th May remains relatively lower than on 13th May. 
Orbit raising continues until 12th May to proactively raise the orbit to a slightly higher altitude than its long-term operational altitude, compensating for the expected enhanced atmospheric density over the next few days. 
Therefore, we see shells with lower altitudes in Fig.~\ref{fig:decayCDFMay24}(d)-(f) have a large negative shift compared to shells with higher altitudes in Fig.~\ref{fig:decayCDFMay24}(b)-(c).
This can be noticed in Fig.~\ref{fig:decayNatureTimeseries}(a) too, where the satellite at 566 km altitude decays more, and then consecutively rises to a higher altitude than the other two satellites.

\parab{October 2024 storm} - 
After the October 2024 solar storm on 10th October, we observe a similar type of inflation at a much lower magnitude than the May 2024 solar superstorm, shown in three shells in Fig.~\ref{fig:decayCDFOct24} for brevity. 
Consequently, we do not observe a real-time orbit correction, but rather a gradual increase in altitude decay up to 11th October. 
On 12th October, we see an orbit rise in 60-65\% of satellites in each shell; however, not beyond 250 meters (excluding a few satellites in Fig.~\ref{fig:decayCDFOct24}(b)).
This seems to be scheduled orbit maintenance rather than a response to this solar storm, as we discuss in the following subsections.

\begin{keybox}
\keynote 
During solar superstorm, Starlink satellites counteract orbital decay in real time and raise the satellite altitude 100s of meters above the long-term operational altitude to account for the enhanced decay rate until the upper atmosphere normalizes back to regular conditions.
\end{keybox}

\subsubsection{Starlink fleet management:}
Starlink aims to be a global Internet service provider, currently serving 150+ countries~\cite{starlinkWeb} and maritime customers~\cite{starlinkMaritime} too.
A Starlink user terminal connects to a LEO satellite via a line-of-sight radio link, and the satellite relays the user traffic to a gateway ground station within its coverage. 
When user terminals and a gateway ground station are not available within the footprint of the same satellite, traffic is routed via inter-satellite laser links (ISLs) to the nearest or assigned gateway ground station~\cite{izhikevich2024democratizing, mohan2024multifaceted, bose2025investigating}. 
Therefore, the Starlink satellite constellation must maintain a specific ISL topology.
Any network-agnostic maneuvers could cause a temporary ISLs drop due to misalignment in extreme narrow laser beam divergence~\cite{StarlinkSelfDriving}.
Then how does Starlink manage the largest constellation of a few thousand satellites with a network disruption of less than 1 minute during a solar superstorm~\cite{StarlinkResponse}? 

\parab{Maneuver detection using TLEs is challenging} - 
There exists a long line of work that focuses on detecting satellite maneuvers using the TLEs~\cite{kelecy2007satellite, mukundan2021simplified, lemmens2014two}. 
Given that the TLEs are (i) often mixed with significant noise and (ii) the epochs between two consecutive TLEs are highly inconsistent (vary from 90 minutes to 33 hours in our Starlink's TLE datasets). 
Detecting a satellite's maneuvers with 100\% accuracy is extremely challenging~\cite{kelecy2007satellite}.
What makes this even worse is Starlink's electric propulsion system~\cite{starlinkTechnology}. 
Unlike conventional chemical propulsion, Starlink's electric propulsion system generates low thrust, i.e., a gentle push that slowly accelerates the satellites~\cite{de2025control}. 
As a result, we do not observe any clearly distinguishable values from a regular variation in the TLE time series. 
An alternative is to leverage the ephemeris datasets released by Starlink~\cite{starlinkOperators}, which forecast the satellite's position for the next 3 days at 1-minute intervals; however, unlike TLE, older ephemeris is not available.

\begin{figure}
    \centering
    \begin{subfigure}[t]{0.50\columnwidth}
        \centering
        \includegraphics[width=\columnwidth, keepaspectratio]{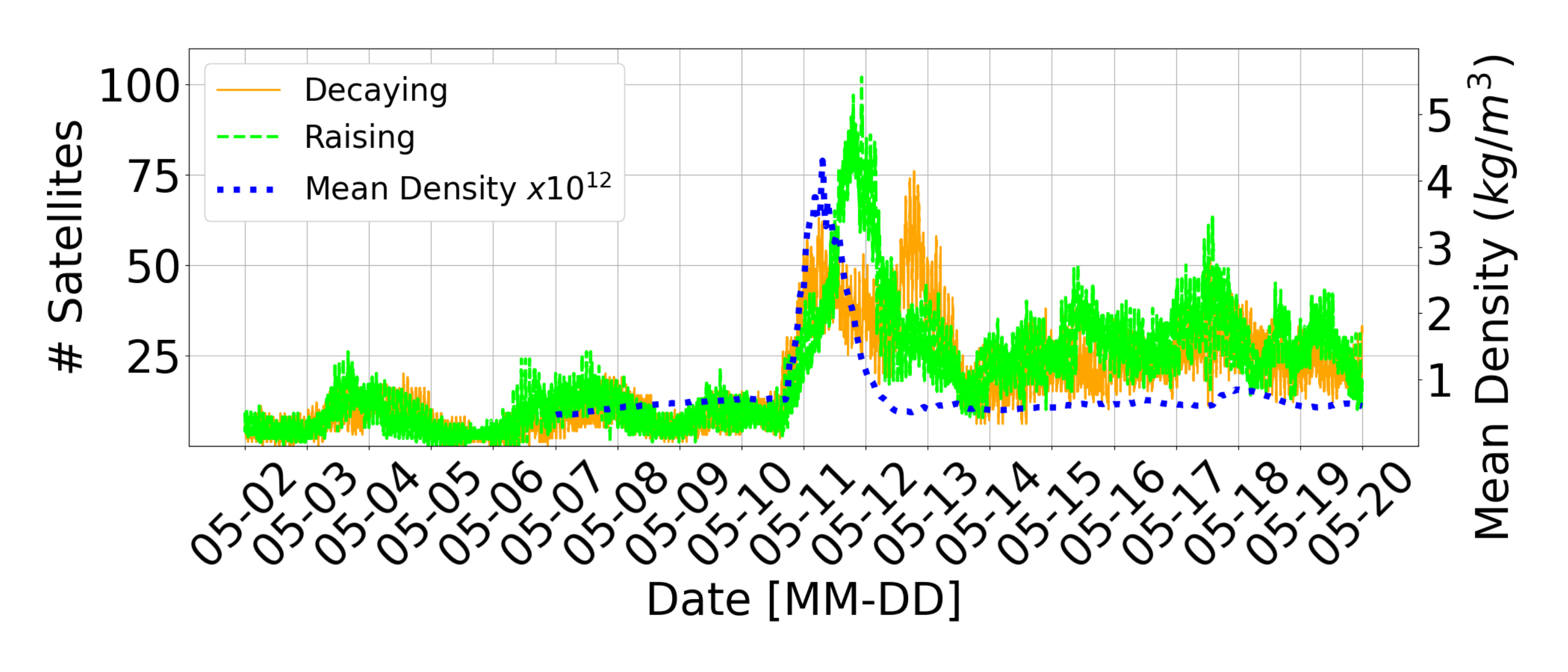}
        \caption{97.6$^{\circ}$, 560 km, during May'24}
    \end{subfigure}%
    \hfill
    \begin{subfigure}[t]{0.50\columnwidth}
        \centering
        \includegraphics[width=\columnwidth, keepaspectratio]{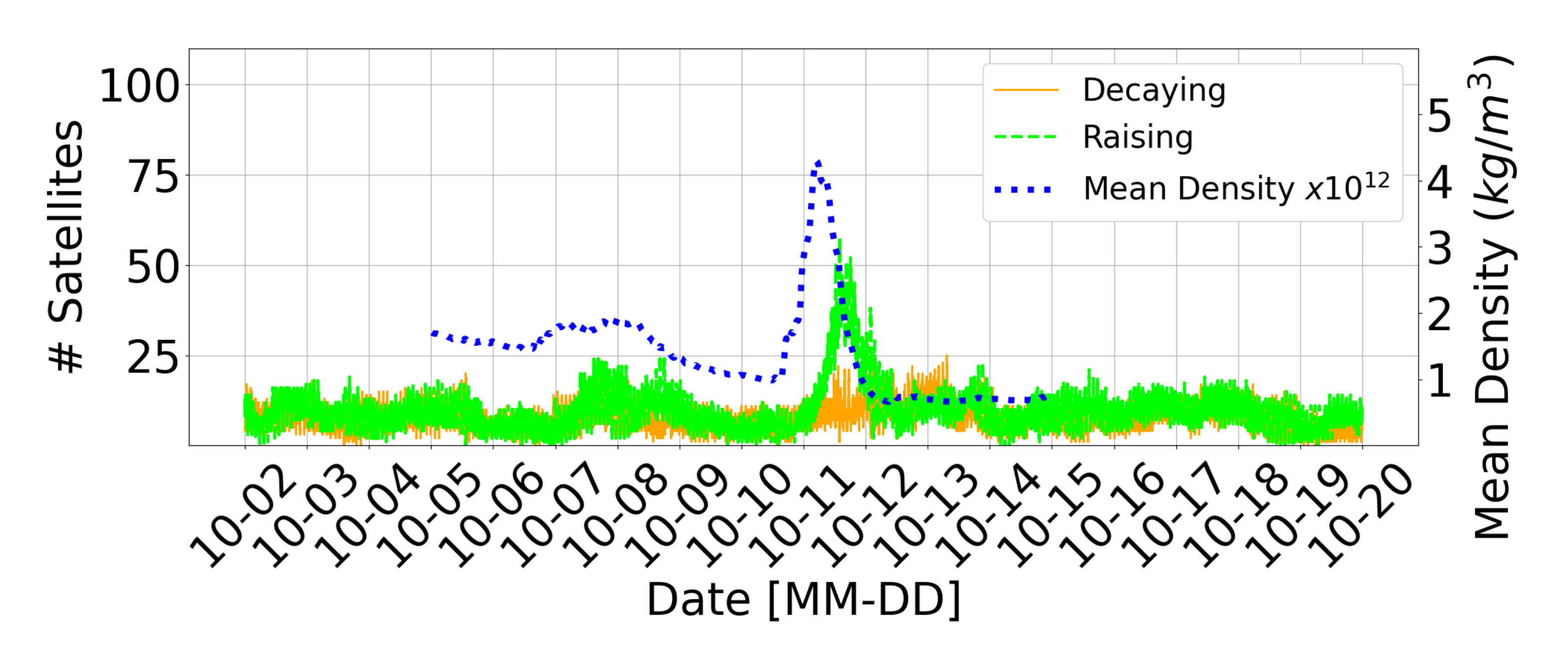}
        \caption{97.6$^{\circ}$, 560 km, during Oct'24}
    \end{subfigure}%
    \hfill
    \begin{subfigure}[t]{0.50\columnwidth}
        \centering
        \includegraphics[width=\columnwidth, keepaspectratio]{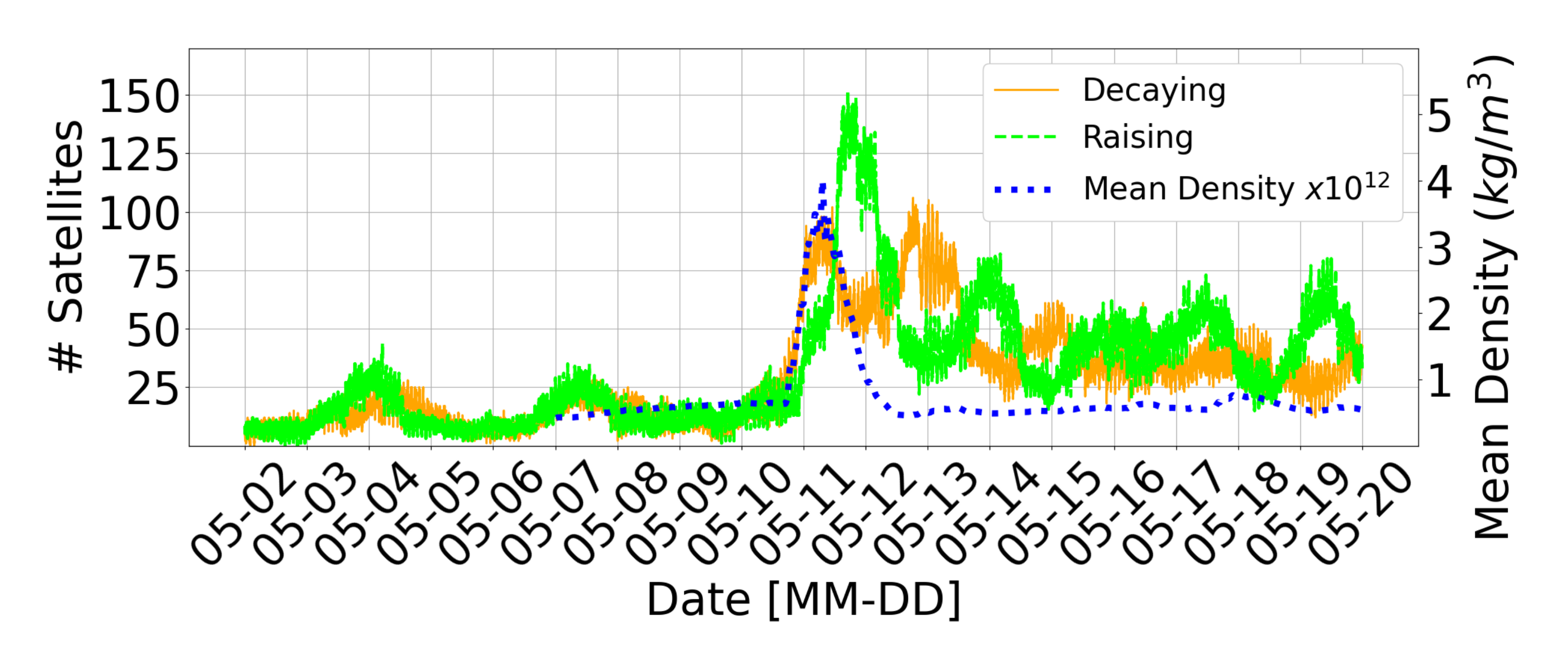}
        \caption{70$^{\circ}$, 570 km, during May'24}
    \end{subfigure}%
    \hfill
    \begin{subfigure}[t]{0.50\columnwidth}
        \centering
        \includegraphics[width=\columnwidth, keepaspectratio]{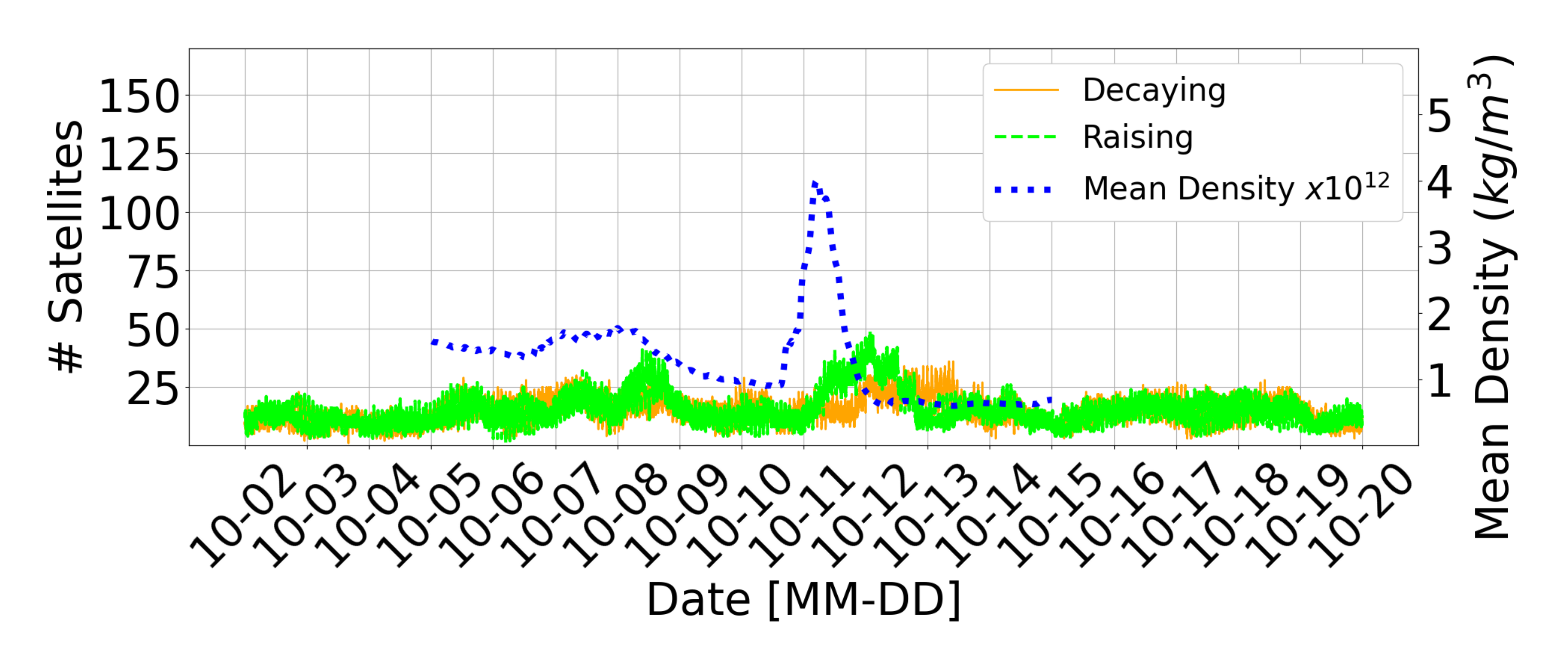}
        \caption{70$^{\circ}$, 570 km, during Oct'24}
    \end{subfigure}%
    \hfill
    \begin{subfigure}[t]{0.50\columnwidth}
        \centering
        \includegraphics[width=\columnwidth, keepaspectratio]{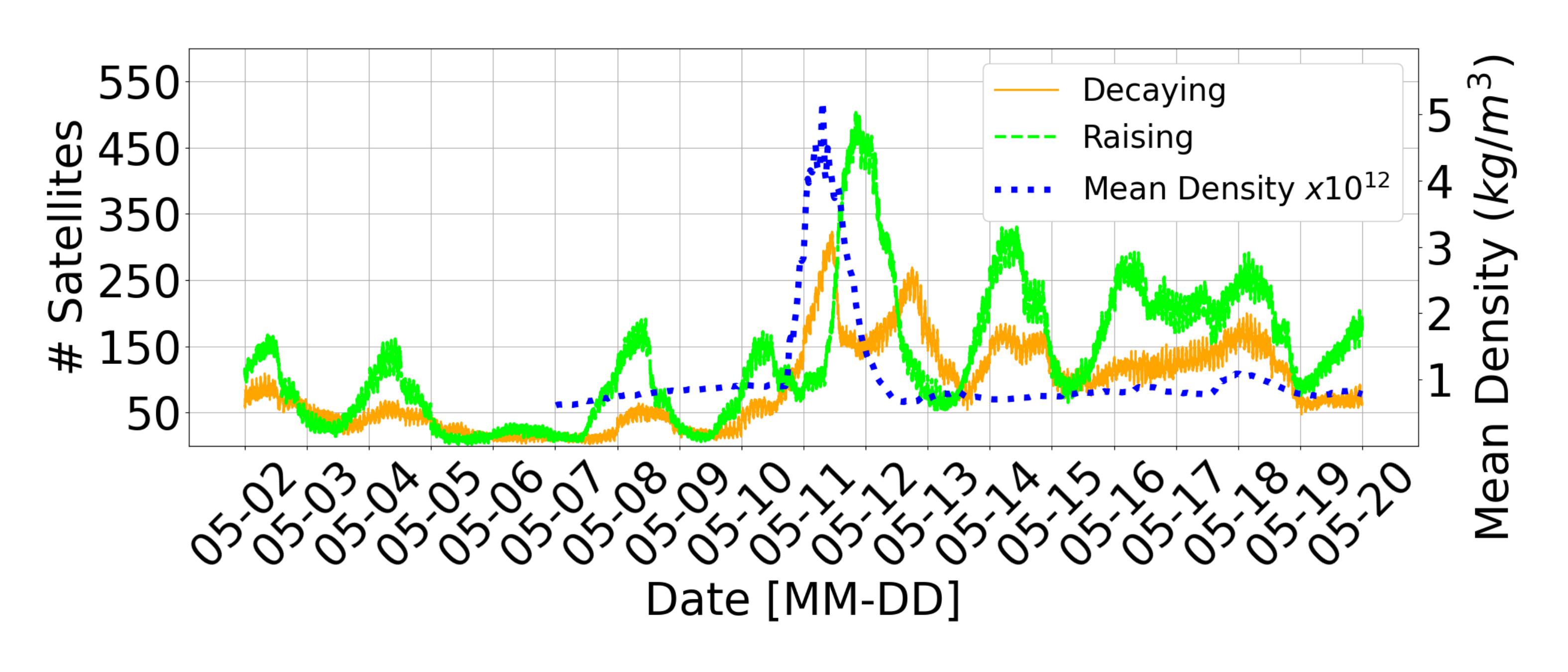}
        \caption{53$^{\circ}$, 550 km, during May'24}
    \end{subfigure}%
    \hfill
    \begin{subfigure}[t]{0.50\columnwidth}
        \centering
        \includegraphics[width=\columnwidth, keepaspectratio]{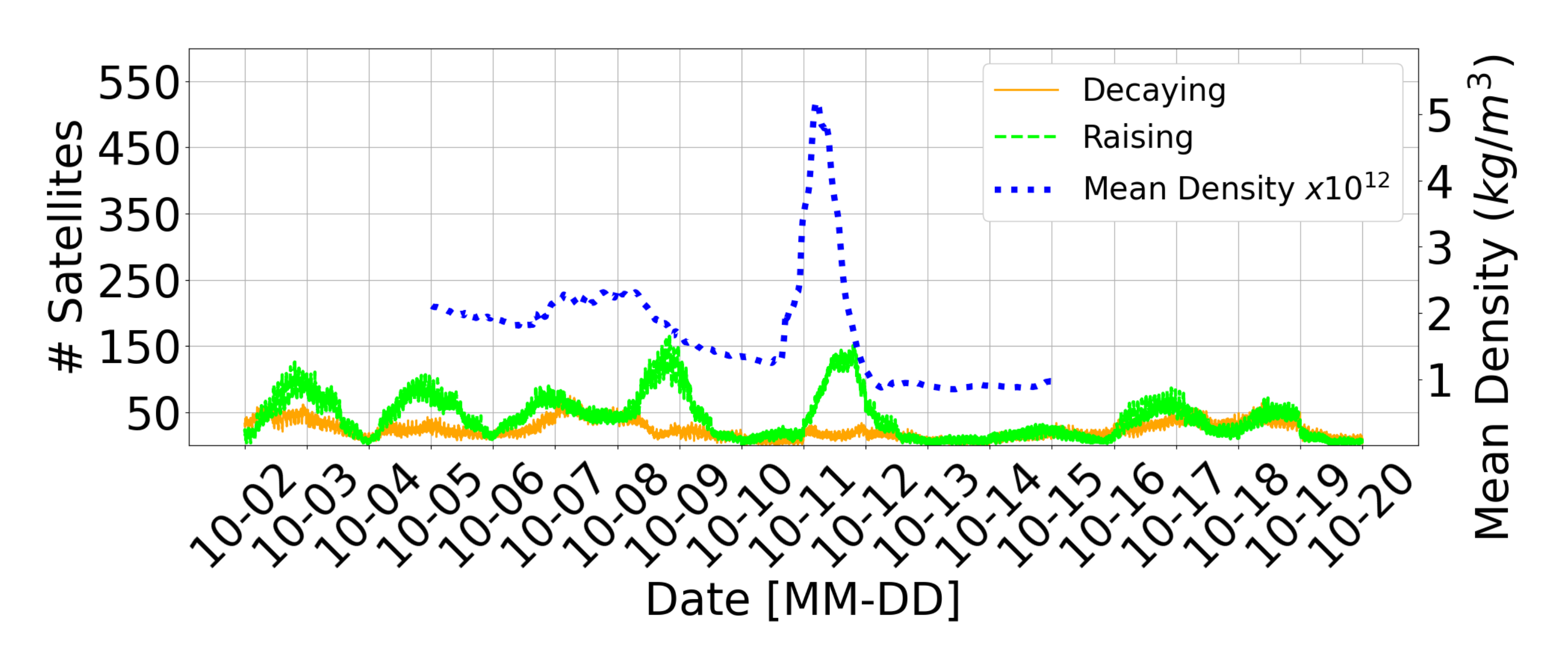}
        \caption{53$^{\circ}$, 550 km, during Oct'24}
    \end{subfigure}%
    \hfill
    \begin{subfigure}[t]{0.50\columnwidth}
        \centering
        \includegraphics[width=\columnwidth, keepaspectratio]{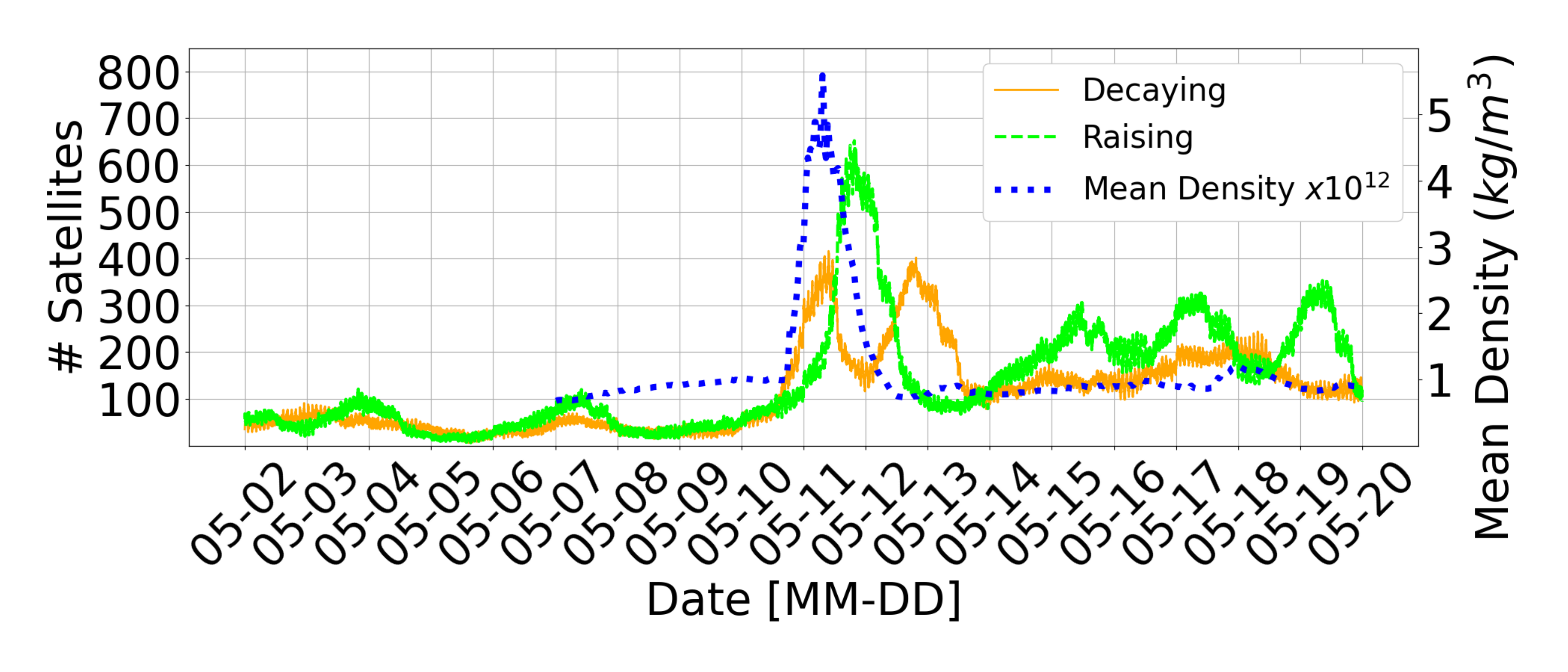}
        \caption{53.2$^{\circ}$, 540 km, during May'24}
    \end{subfigure}%
    \hfill
    \begin{subfigure}[t]{0.50\columnwidth}
        \centering
        \includegraphics[width=\columnwidth, keepaspectratio]{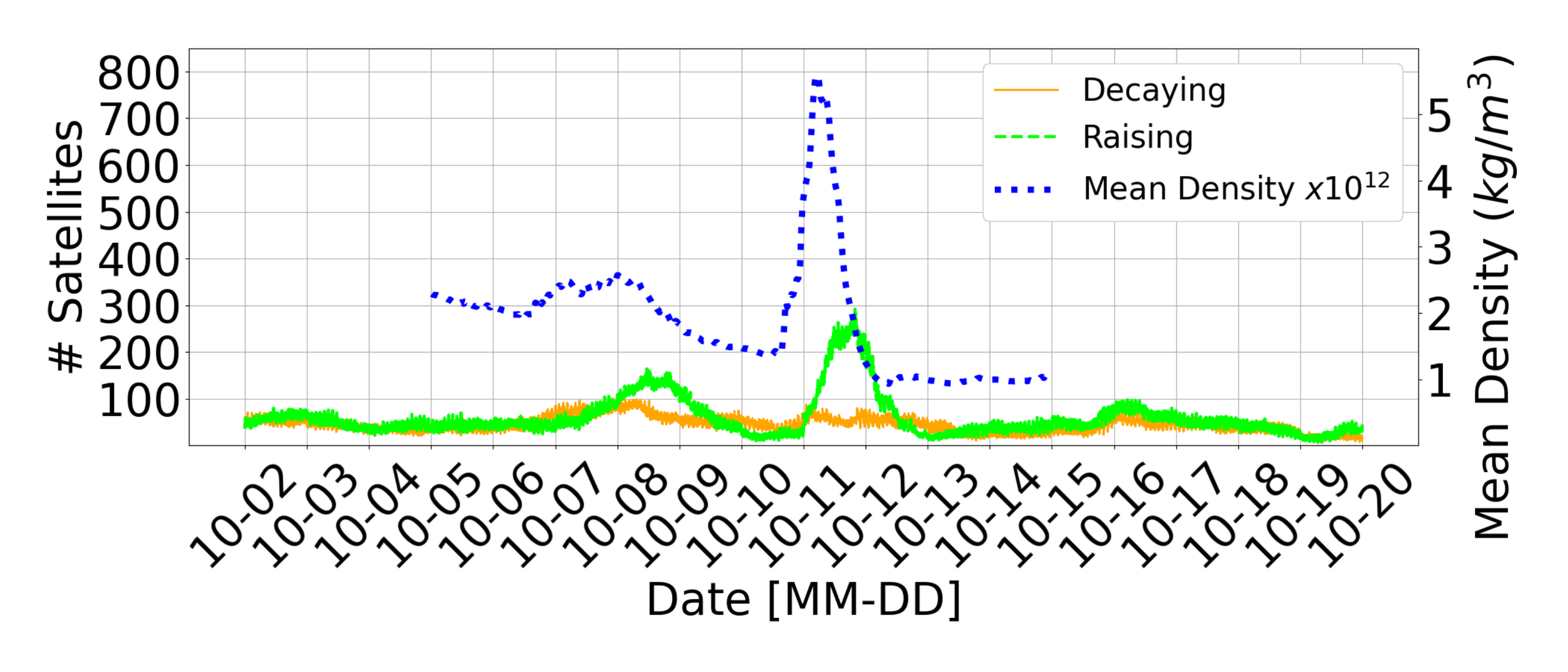}
        \caption{53.2$^{\circ}$, 540 km, during Oct'24}
    \end{subfigure}%
    \hfill
    \begin{subfigure}[t]{0.50\columnwidth}
        \centering
        \includegraphics[width=\columnwidth, keepaspectratio]{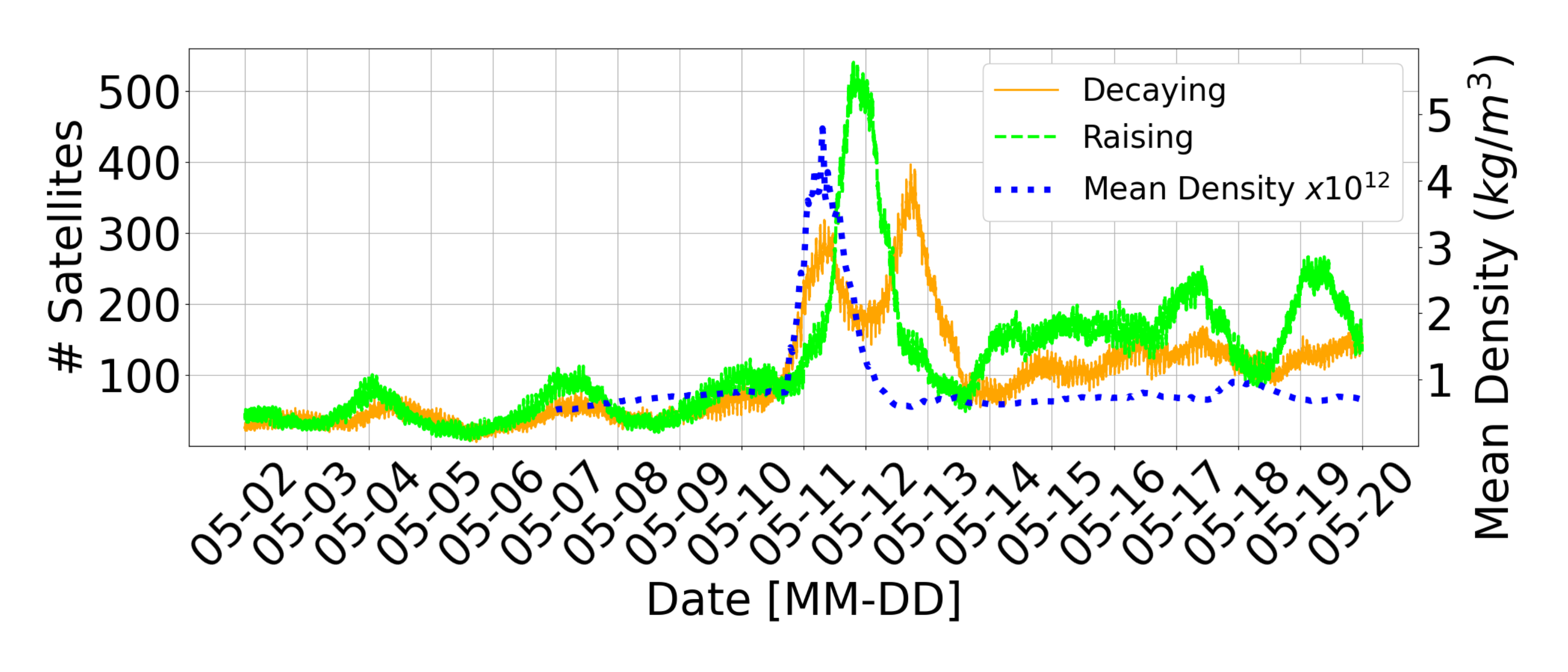}
        \caption{43$^{\circ}$, 530 km, during May'24}
    \end{subfigure}%
    \hfill
    \begin{subfigure}[t]{0.50\columnwidth}
        \centering
        \includegraphics[width=\columnwidth, keepaspectratio]{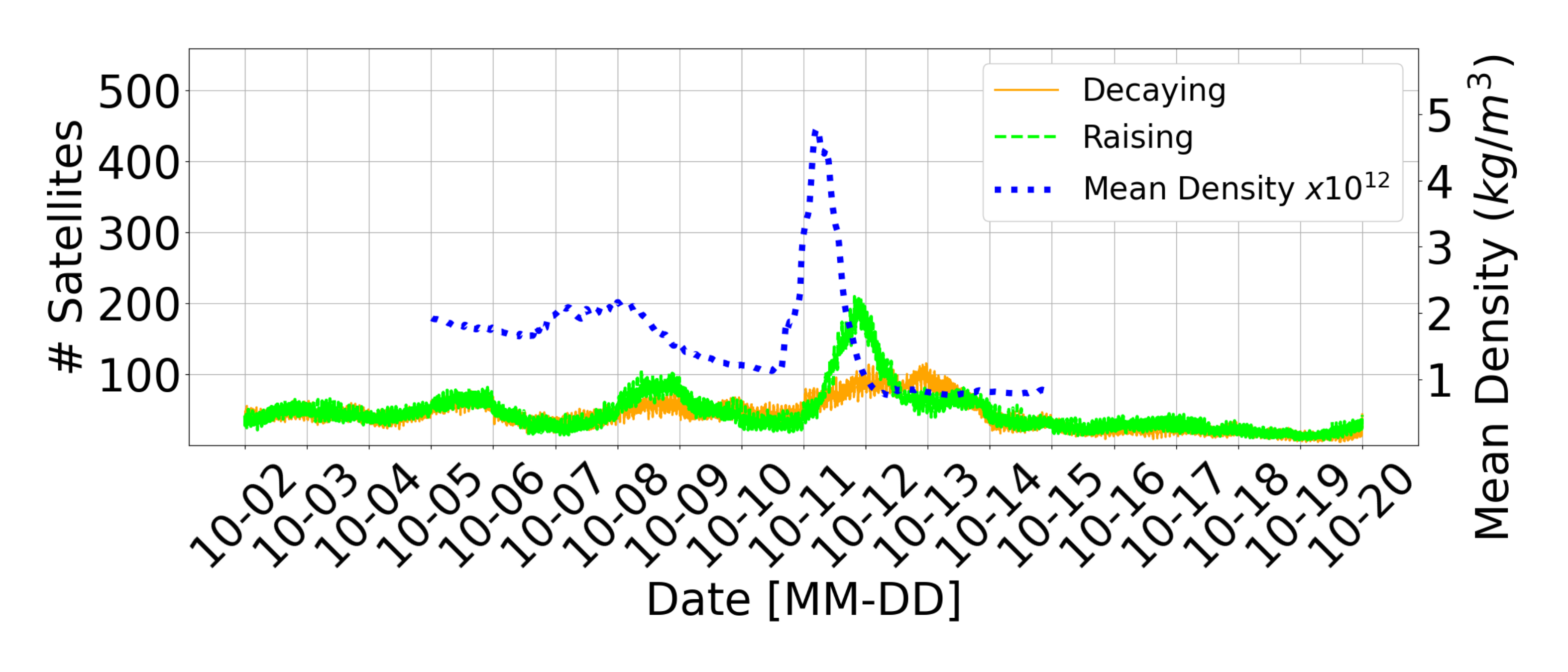}
        \caption{43$^{\circ}$, 530, during km Oct'24}
    \end{subfigure}%
    \hfill
    
    \caption{Counting the number of Starlink satellites per shell altitude decaying (solid orange line) and raising (dashed green line) and corresponding mean atmospheric density in that shell altitude (dotted blue line) over time. 
    In May 2024 (left side), the solar superstorm increased atmospheric density, and Starlink satellites began decaying immediately. 
    Starlink responded with a countermeasure within hours and started rising in altitude. 
    In October 2024 (right side), the response is not in real time.}
    \label{fig:StarlinkFleetManagement}
\end{figure}

\parab{Leveraging orbit propagator} - 
We overcome this roadblock using \texttt{skyfield}~\cite{rhodesmillSkyfieldx2014}. 
It predicts the satellite position at any arbitrary time using the SGP4~\cite{vallado2006revisiting, SPACETRACKREPORTNO3} orbit propagator with a given TLE. 
In our measurement, we use two consecutive TLEs of each Starlink satellite, let's say $TLE$ and $TLE-1$.
Then predict the satellite's altitude at any given time $t$ as $h_{TLE}^{t}$ and $h_{TLE-1}^{t}$ respectively.
In an ideal LEO operational scenario, if no maneuver is performed in between these two $TLE$ and $TLE-1$, then $h_{TLE}^{t} < h_{TLE-1}^{t}$ due to increase in decay rate after a solar storm; otherwise, $h_{TLE}^{t} \geq h_{TLE-1}^{t}$ possible only if orbit is raised. 
However, in the real world, Earth's oblate shape, atmospheric conditions, and other interplanetary factors contribute to satellite trajectory deviations, and TLE also contains noise. 
For that, similar to prior work~\cite{kelecy2007satellite}, we use statistics of regular altitude variation as the trajectory deviation margin and flag a satellite as decaying or rising only if the altitude difference exceeds this margin. 
Therefore, in each Starlink shell, we first calculate the mean ($\mu$) and standard deviation ($\sigma$) of the altitude differences ($\Delta h^t = h_{TLE}^{t} - h_{TLE-1}^{t}$) from a time window of a few quiet days. 
Then we flag a satellite that is decaying or rising at time $t$, if the altitude difference ($\Delta h^t$) is beyond $\mu\pm2\sigma$.
Note that here our intention is not to detect all satellite maneuvers with 100\% accuracy. 
Rather, approximately quantify irregular altitude falls and rises to decode any existing trends, which reveal a large satellite fleet management strategy in quiet and storm time.
Using this approach, we count the number of satellite altitude rises and falls over time for each Starlink shell and plot in Fig.~\ref{fig:StarlinkFleetManagement} (left X-axis). 
We also plot the mean atmospheric density over time at the respective shells' operational altitudes (right X-axis).
All the figures on the left are measurements during the Gannon storm, while the figures on the right are measurements during the October 2024 storm for each Starlink shell. 

\parab{Gannon storm} - 
A week ahead of the solar superstorm, notice a periodic increase in the number of satellites rising after 3rd May, 6th May, and 8th May across almost all the shells. 
This clearly indicates that Starlink employs a batch-wise orbit maintenance strategy in regular operation, i.e., a batch of Starlink satellites is simultaneously raised to maintain the operational orbit.
Another prior work~\cite{liu2024maneuver} that analyzed the ephemeris of a few Starlink satellites also reported a similar pattern of orbit correction.
This means this observation with \projectname{} is accurate.
More importantly, achieved using easily accessible course TLEs rather than high-cadence ephemeris.
Late on 10th May, as the solar storm commences, the mean atmospheric density increases exponentially to 4-6 times over the baseline within a few hours.
Consequently, we see a large spike in the number of satellites that started a rapid decay. 
Then, within an hour, we see a huge spike in the number of satellites rising in all the shells.
Quantitatively, in Fig.~\ref{fig:StarlinkFleetManagement}(a) up to 100 (43\%), Fig.~\ref{fig:StarlinkFleetManagement}(c) up to 150 (38\%), Fig.~\ref{fig:StarlinkFleetManagement}(e) up to 450 (37\%), Fig.~\ref{fig:StarlinkFleetManagement}(g) up to 650 (42\%), and Fig.~\ref{fig:StarlinkFleetManagement}(i) up to 510 (38\%) satellites are rising simultaneously during the solar superstorm.
After the solar superstorm, from 13th May, periodic batch-wise orbit maintenance is not visible, and it looks like a post-storm orbit correction over the next few weeks.
This analysis does not make it clear whether this post-event correction is autonomous or requires operator intervention.

\parab{October 2024 storm} - 
In October 2024, we see two solar storms, one on 11th October and, before that, on 7th October, with relatively lower intensity, so the atmospheric density inflation is much lower.
Unlike a solar superstorm, we do not notice any immediate change in the number of satellites decaying or rising. 
However, notice an increase in the number of satellites undergoing orbit correction after 8th and 11th October. 
This indicates that a relatively lower-intensity solar storm (i) does not increase the satellite decay rate to the point where the satellite requires an instantaneous countermeasure, (ii) in our measurement, the decay rate increase of such magnitude is buried under regular variation, thus not clearly visible.

Altogether, it gives insight into the Starlink constellation's autonomous operation.
After a solar storm, post-storm orbit correction is sufficient to maintain a topological formation.
In exceptional cases of intense solar storms, the onboard thrusters kick in to counteract the rapid decay. 
Once the atmospheric conditions normalize, the constellation goes through a post-event topological formation correction over a period of time. 
From a networking perspective, simultaneous orbit-raising of a large batch of satellites will maintain relative positioning, enabling satellites to sustain ISL connectivity during the operation.
Activating the onboard thrusters will introduce vibration, which is likely to degrade connectivity due to jitter in the laser beam pointing.

\begin{keybox}
\keynote 
Starlink satellites undergo batch-wise orbit corrections once every 2-3 days.
During a solar superstorm, when satellites start decaying rapidly, typically 35-45\% satellites per shell perform simultaneous orbit-raising followed by a post-storm course correction.
\end{keybox}

\subsection{Geospatial characteristics of LEO satellite decay}

So far, we have discussed that the Starlink satellite trajectory during a solar storm gives insight into the Starlink fleet operations.
This does not shed light on the geospatial aspect of satellite trajectory change.
This aspect is explored in a recent work~\cite{DeepDiveSolarStorms} investigating the LEO satellite network during solar storms, which identified a `W' shaped altitude change pattern across the orbits of the Starlink constellation.
Their work does not explain the reason behind this pattern.
In this work, using \projectname{}, we dive deeper into this mysterious geospatial pattern to find the root cause of this.

\subsubsection{`W' pattern in altitude change:}

Authors in~\cite{DeepDiveSolarStorms} have demonstrated systematic differences in altitude decay responses in satellites belonging to different orbits during the May 2024 superstorm. 
Their analysis shows that satellites in an orbit position inline between the Sun and Earth experience significantly larger absolute altitude changes over one day.
Whereas satellites positioned approximately orthogonal to this direction (around $\pm90^{\circ}$ phase offset) exhibit comparatively stable behavior. 
When all the orbital planes are aggregated in order, these variations form a `W'-shaped profile in absolute altitude change.
They attribute this pattern primarily to differences in solar radiation exposure across the orbital planes, suggesting that satellites closer to the Sun experience stronger perturbations, while those farther away are less affected~\cite{DeepDiveSolarStorms}. 
Note that, unlike high altitude GEO satellites, LEO satellites operate below the inner Van Allen belt and are largely shielded from direct extreme radiation exposure except for two anomalous regions: the South Atlantic Anomaly and the polar region.
As we have already demonstrated in our prior discussion, the dominant driver of orbital decay in LEO is atmospheric drag, which is modulated by thermospheric density variations rather than direct solar proximity.
The average distance between the Sun and Earth is approximately 150 million km.
A LEO satellite at 550 km altitude while orbiting Earth will have distance variation with the Sun up to 0.009\% of 150 million km.
Which we believe is insufficient to produce meaningful differences in radiation exposure. 
This highlights a gap in understanding the key underlying mechanism behind the observed pattern. 
Specifically, we find that the following four question is still remain to be answered:

\begin{enumerate}
    \item Is this `W' pattern an extreme solar storm specific phenomenon?
    \item What is the key driver of systematic variation in altitude changes across orbital planes?
    \item How does orbital orientation relative to the Sun influence this orbital dynamics?
    \item Is it a universal LEO Trait?
\end{enumerate}

\subsubsection{Decoding the root cause behind `W' pattern:}

To investigate the root cause of the observed `W'-shaped pattern, we perform a year-long analysis across orbital planes.
For each satellite within a shell, we collect all available TLEs for a given day and compute the daily altitude change as the difference between the maximum and minimum semi-major axes reported in those TLEs. 
This captures a satellite's net altitude change within a day.
We then propagate satellite positions (latitude, longitude, elevation) using TLEs at 1-second granularity to obtain fine-grained orbital trajectories. 
Using these positions, we compute (i) the duration of solar exposure and (ii) the atmospheric density encountered along the orbit around the Earth. 
We use the atmospheric condition datasets to obtain density by mapping satellite positions to the nearest grid point.
Since density measurements are available at a 20-minute cadence, satellite positions are temporally aligned to the closest available snapshot.

\parab{The pattern exists throughout the year} - 
In Fig.~\ref{fig:WpatternShiftOverYear}, we summarize these measurements in one day per month for a one-year period. 
For May and October, we show the peak solar storm day (i.e., the 11th day of the month), while for all other months we show the first day.
Notice that the x-axis in these figures represents the RAAN, ordering orbital planes. 
RAAN is the angular term that denotes where a satellite intersects the Earth's equator while moving towards the geographic north.
Therefore, all satellites in the same orbit will have the same or very similar RAAN values.
Using this, we plot the left y-axis, which shows the distribution of daily altitude change experienced by the satellite across 72 orbital planes. 
On the right side, using a blue y-axis, we show the maximum difference in atmospheric density encountered by the satellites while orbiting the Earth in that orbital plane.
Using another red y-axis beside that, the median of sunlight exposure duration on a day in that orbital plane.
From this summary of one-year measurements, we make the following observations:

\begin{figure}
    \centering
    \begin{subfigure}[t]{0.5\columnwidth}
        \centering
        \includegraphics[width=\columnwidth, keepaspectratio]{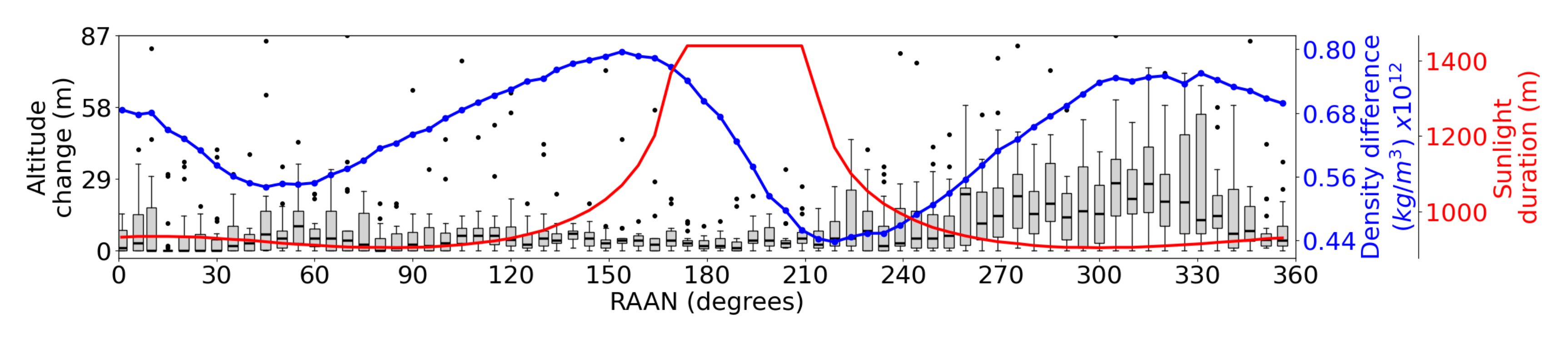}
        \caption{1st January, 2024}
    \end{subfigure}%
    \begin{subfigure}[t]{0.5\columnwidth}
        \centering
        \includegraphics[width=\columnwidth, keepaspectratio]{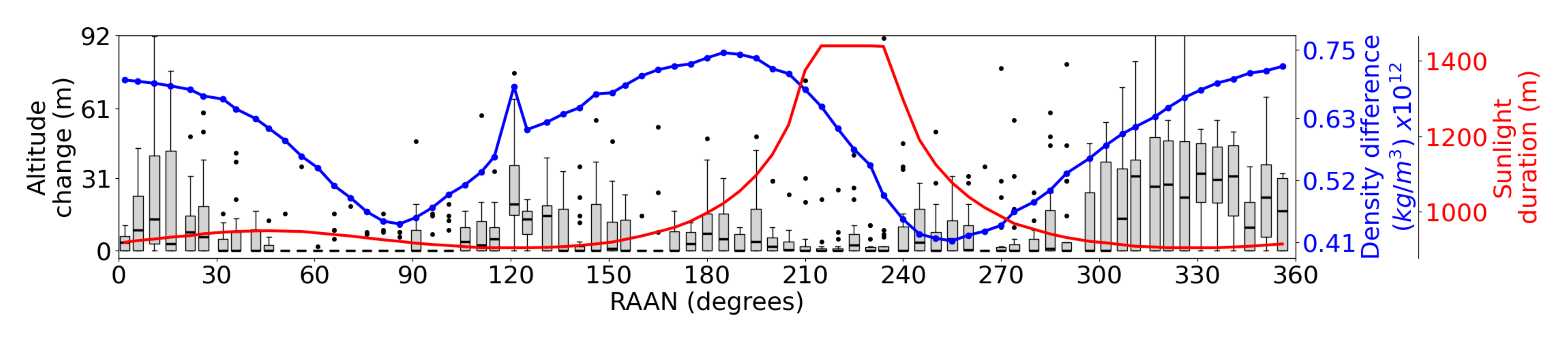}
        \caption{1st February, 2024}
    \end{subfigure}%
    \hfill
    \begin{subfigure}[t]{0.5\columnwidth}
        \centering
        \includegraphics[width=\columnwidth, keepaspectratio]{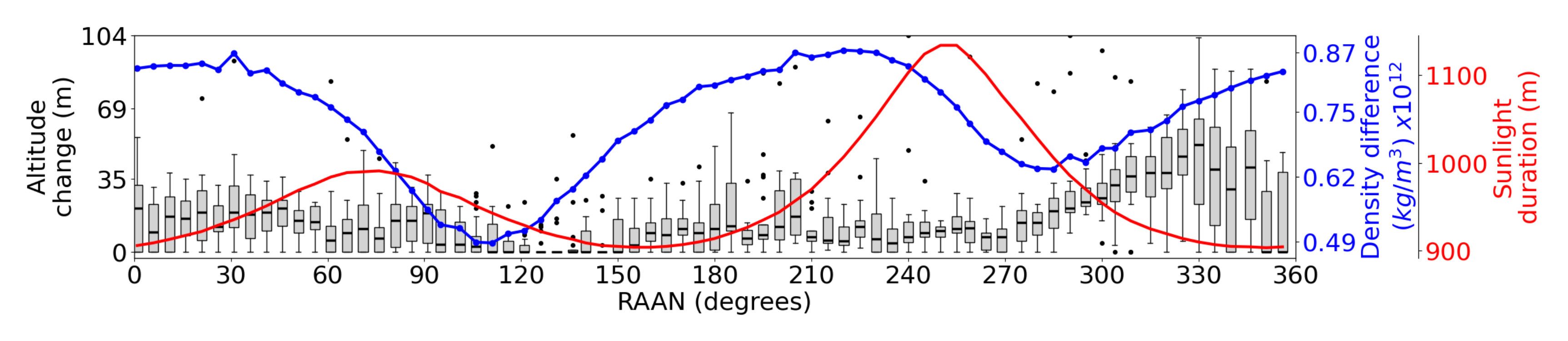}
        \caption{1st March, 2024}
    \end{subfigure}%
    \begin{subfigure}[t]{0.5\columnwidth}
        \centering
        \includegraphics[width=\columnwidth, keepaspectratio]{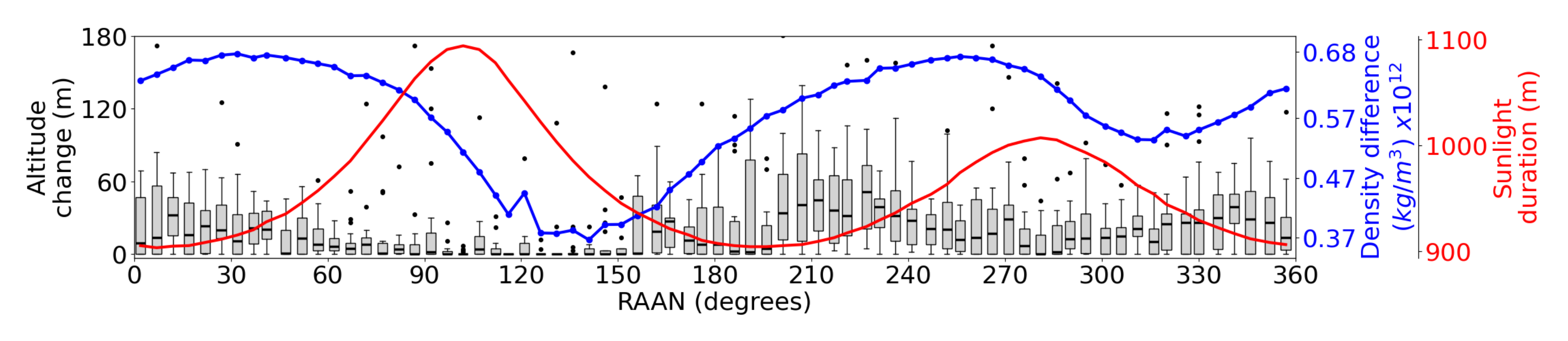}
        \caption{1st April, 2024}
    \end{subfigure}%
    \hfill
    \begin{subfigure}[t]{0.5\columnwidth}
        \centering
        \includegraphics[width=\columnwidth, keepaspectratio]{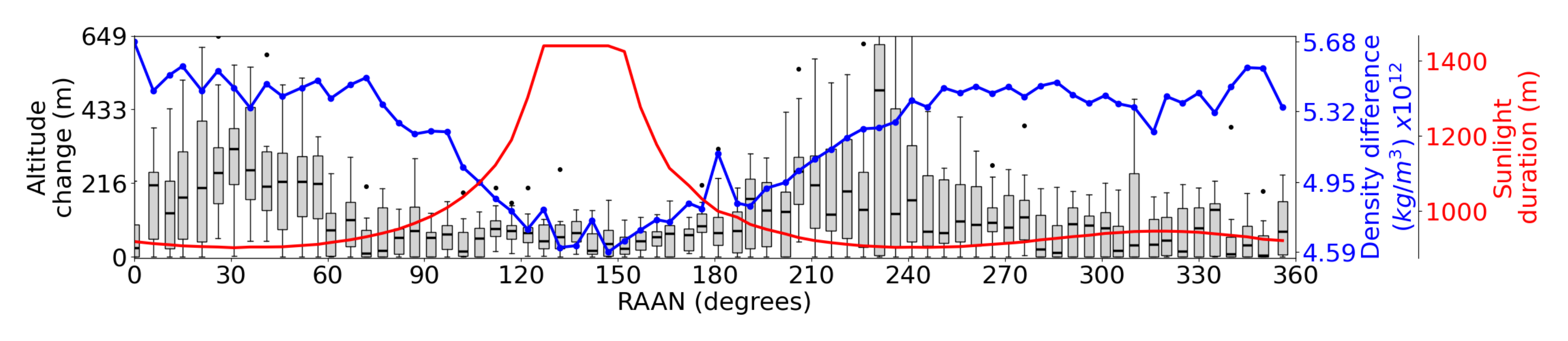}
        \caption{11th May, 2024 (Solar superstorm)}
    \end{subfigure}%
    \begin{subfigure}[t]{0.5\columnwidth}
        \centering
        \includegraphics[width=\columnwidth, keepaspectratio]{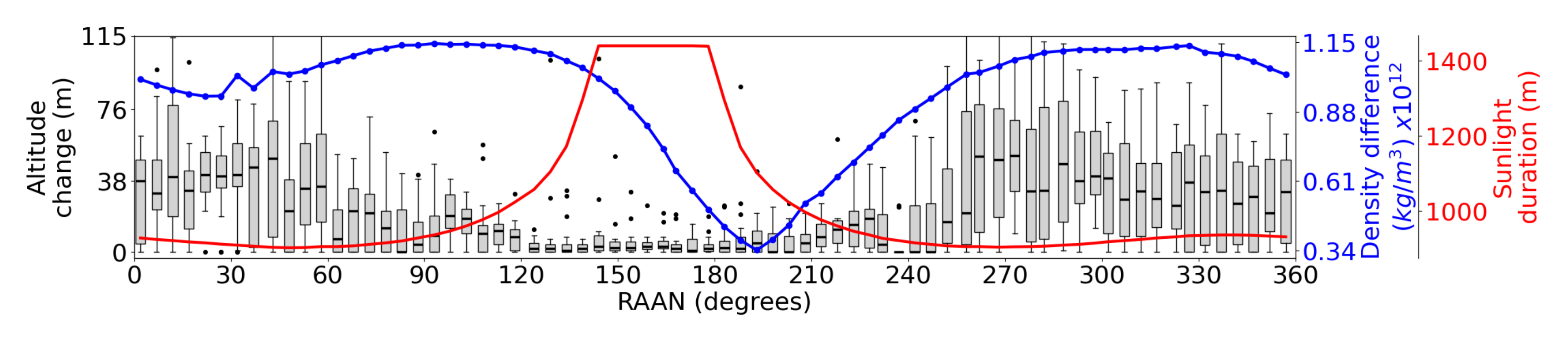}
        \caption{1st June, 2024}
    \end{subfigure}%
    \hfill
    \begin{subfigure}[t]{0.5\columnwidth}
        \centering
        \includegraphics[width=\columnwidth, keepaspectratio]{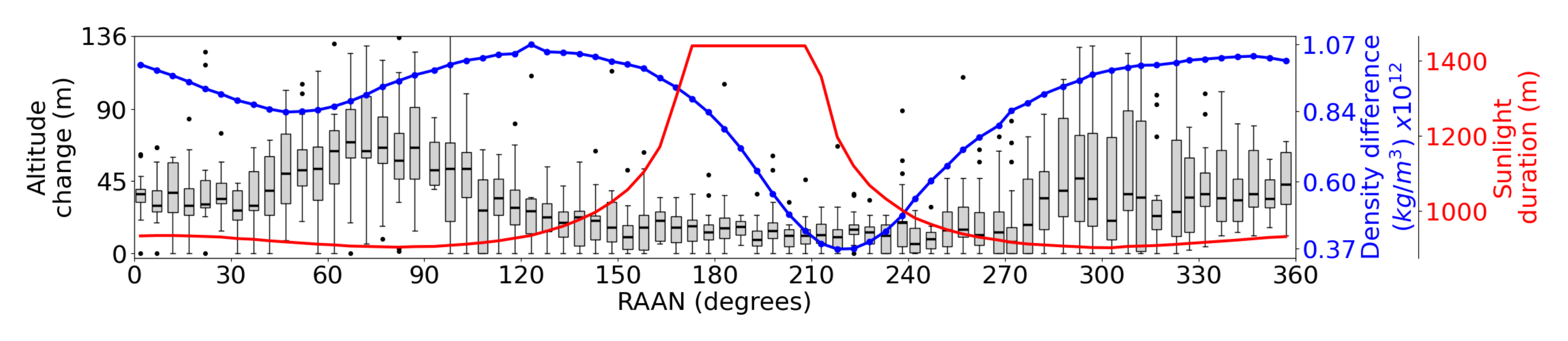}
        \caption{1st July, 2024}
    \end{subfigure}%
    \begin{subfigure}[t]{0.5\columnwidth}
        \centering
        \includegraphics[width=\columnwidth, keepaspectratio]{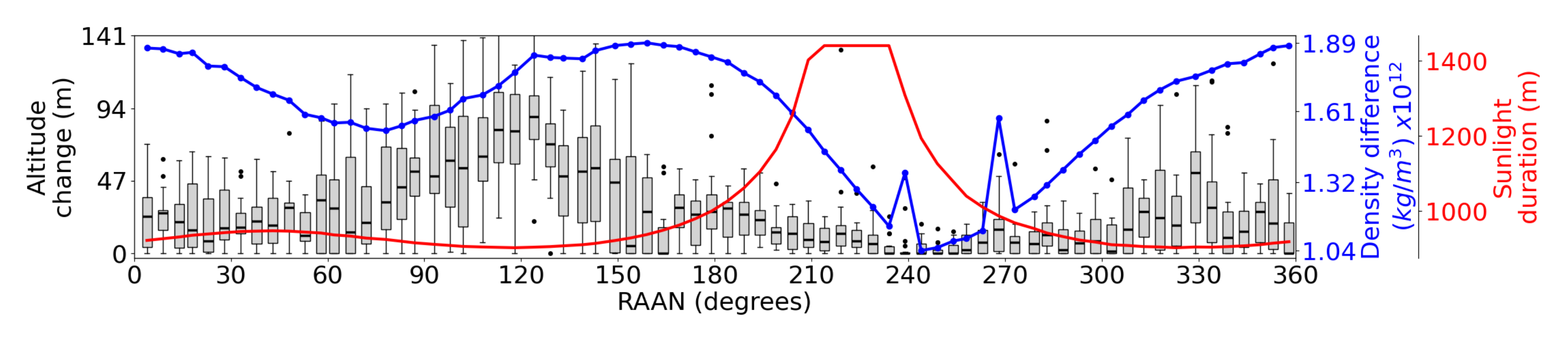}
        \caption{1st August, 2024}
    \end{subfigure}%
    \hfill
    \begin{subfigure}[t]{0.5\columnwidth}
        \centering
        \includegraphics[width=\columnwidth, keepaspectratio]{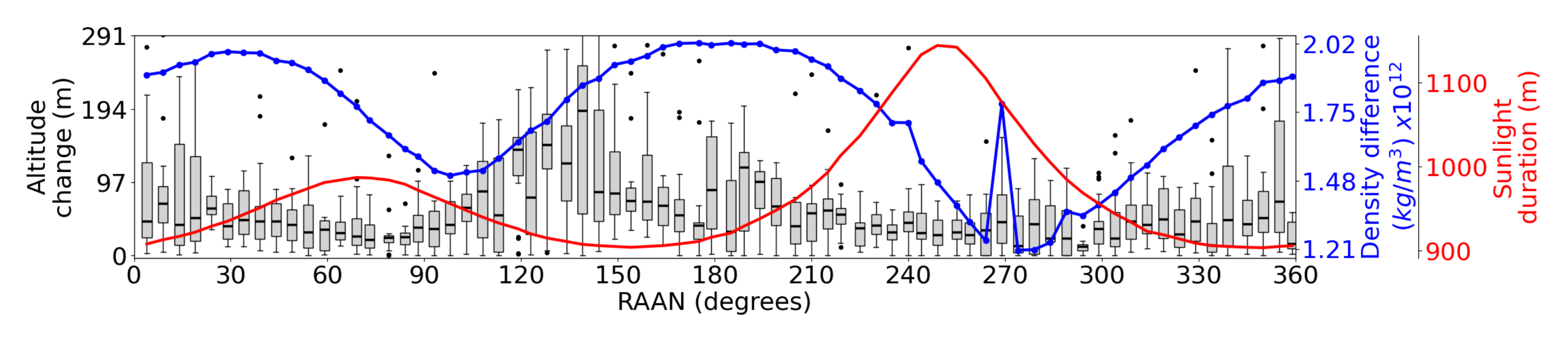}
        \caption{1st September, 2024}
    \end{subfigure}%
    \begin{subfigure}[t]{0.5\columnwidth}
        \centering
        \includegraphics[width=\columnwidth, keepaspectratio]{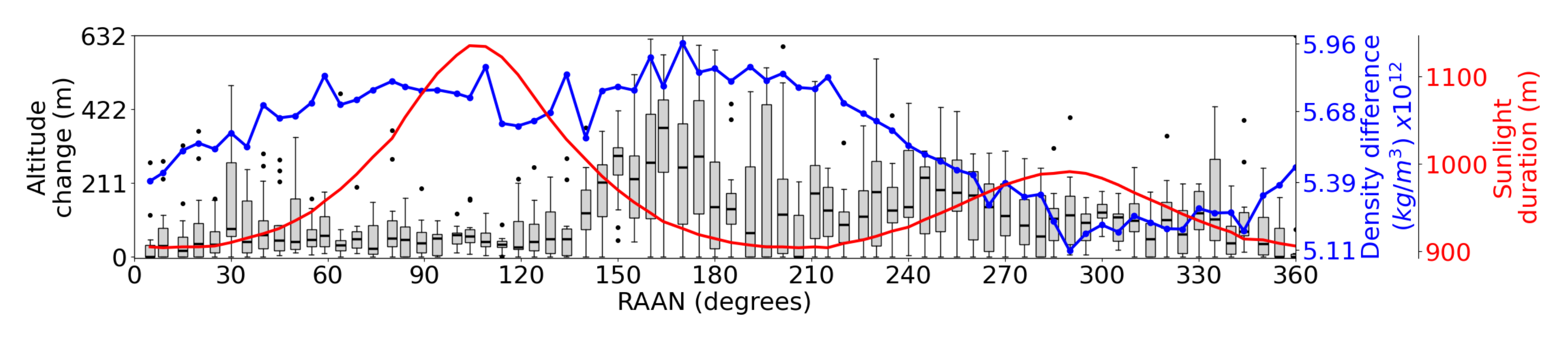}
        \caption{11th October, 2024 (Solarstorm)}
    \end{subfigure}%
    \hfill
    \begin{subfigure}[t]{0.5\columnwidth}
        \centering
        \includegraphics[width=\columnwidth, keepaspectratio]{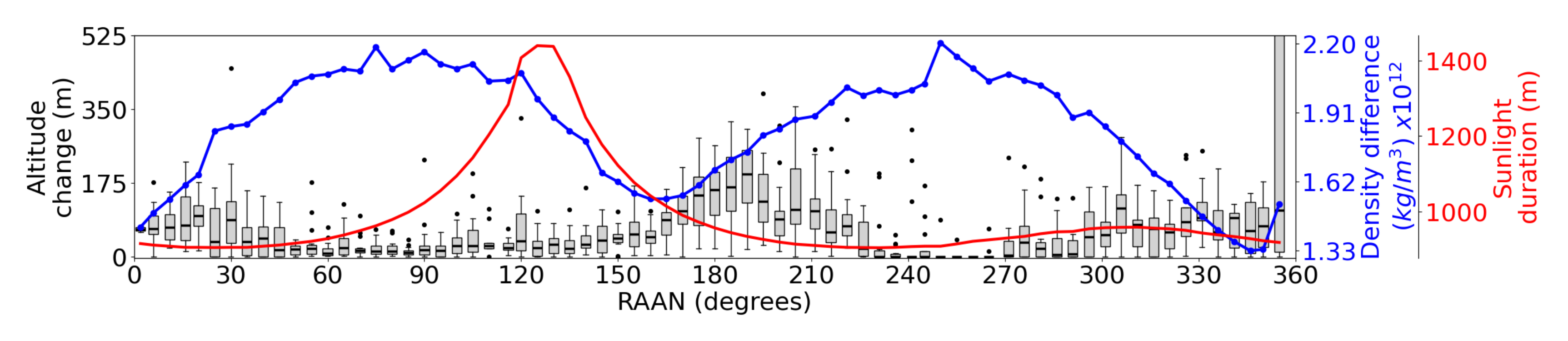}
        \caption{1st November, 2024}
    \end{subfigure}%
    \begin{subfigure}[t]{0.5\columnwidth}
        \centering
        \includegraphics[width=\columnwidth, keepaspectratio]{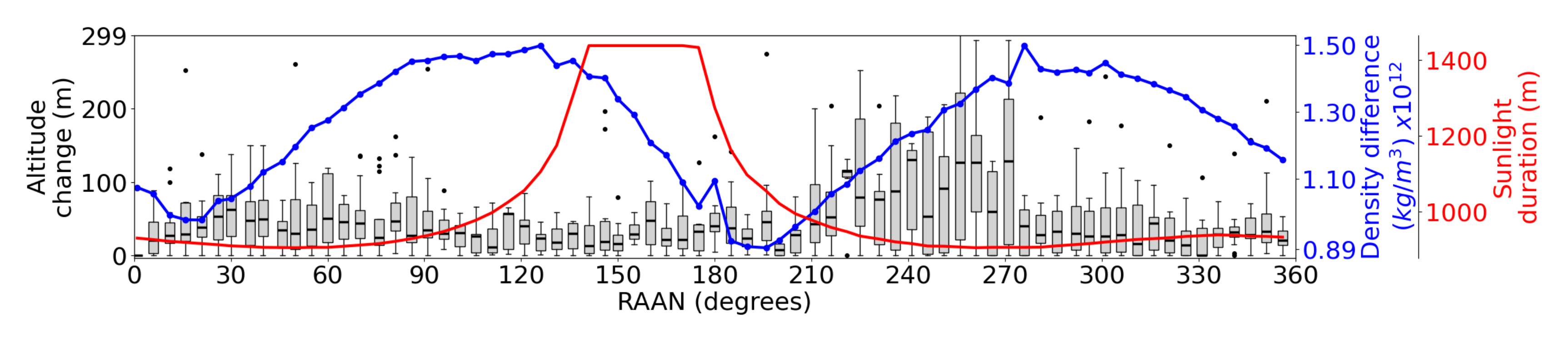}
        \caption{1st December, 2024}
    \end{subfigure}%

    \caption{The distribution of Starlink satellite orbital altitude changes (left y-axis), along with the maximum atmospheric density change encountered (right y-axis, blue) and median sunlight exposure (right y-axis, red), follows a similar pattern throughout the year. 
    This pattern gradually shifts to the right, with transitions in March–April and September–October.
    During periods of solar storms in May and October, the magnitude of altitude and atmospheric density variation is significantly higher than under baseline conditions.}
    \label{fig:WpatternShiftOverYear}
\end{figure}

\begin{table}
\centering
\caption{Range of RAAN degrees where significantly high and low altitude variations are observed per day over the span of one year.}
\footnotesize
\begin{tabular}{c|cc|cc}
\hline
\textbf{Month} 
& \multicolumn{2}{c|}{\textbf{High variation RAANs}} 
& \multicolumn{2}{c}{\textbf{Low variation RAANs}} \\
\cline{2-5}
& \textbf{Range 1} & \textbf{Range 2} 
& \textbf{Range 1} & \textbf{Range 2} \\
\hline\hline
January   & 240--330 & 30--90  & 150--200 & 350--0   \\
February  & 300--360 & 120--150 & 60--90   & 210--240 \\
March     & 330--360 & 180--210 & 120--150 & 250--270 \\
April     & 210--240 & 0--30    & 90--150  & 290--310 \\
May       & 180--270 & 0--60    & 120--150 & 330--360 \\
June      & 270--300 & 30--60   & 120--180 & 330--0   \\
July      & 290--330 & 60--90   & 180--240 & 0--30    \\
August    & 90--150  & 310--340 & 210--260 & 0--60    \\
September & 120--150 & 350--30  & 60--90   & 240--270 \\
October   & 150--210 & 30--60   & 90--120  & 360--30  \\
November  & 170--240 & 0--30    & 60--90   & 240--270 \\
December  & 240--270 & 30--90   & 180--210 & 360--20  \\
\hline\hline
\end{tabular}
\label{tbl:variationRanges}

\end{table}

\begin{enumerate}
    \item This `W'-shaped pattern in altitude variation exists throughout the year, including days without any significant solar activity, observes the other months in Fig.~\ref{fig:WpatternShiftOverYear} except May and October.
    We summarized RAAN degrees of high and low variation orbital planes in Table~\ref{tbl:variationRanges}.
    Also, notice that the magnitude of change increases significantly during solar storms, reaching up to 700 meters per day in May and October, while in non-storm periods, it is typically within 150 meters per day, and still exhibits a `W' structure. 
    This indicates that the pattern is not solely induced by extreme solar events, but reflects an underlying systematic phenomenon. 

    \item Altitude variation across orbital planes strongly correlates with the atmospheric density difference encountered along the orbit around the Earth. 
    Orbital planes with high altitude variance show a larger difference in atmospheric density. 
    In contrast, sunlight exposure duration shows an inverse relationship with altitude variation.
    Orbits with longer cumulative sunlight exposure tend to exhibit lower altitude changes.

    \item The spatial structure of the pattern evolves continuously over the year, with two prominent transition periods, one in March–April and another in September–October. 
    Between these transitions, the `W'-pattern shifts rightward progressively over the RAAN degree. 
    This indicates a tight correlation among the orientations of the Sun, Earth, and the orbital planes.
\end{enumerate}

\begin{figure}
    \centering
    \begin{subfigure}[t]{0.5\columnwidth}
        \centering
        \includegraphics[width=\columnwidth, keepaspectratio]{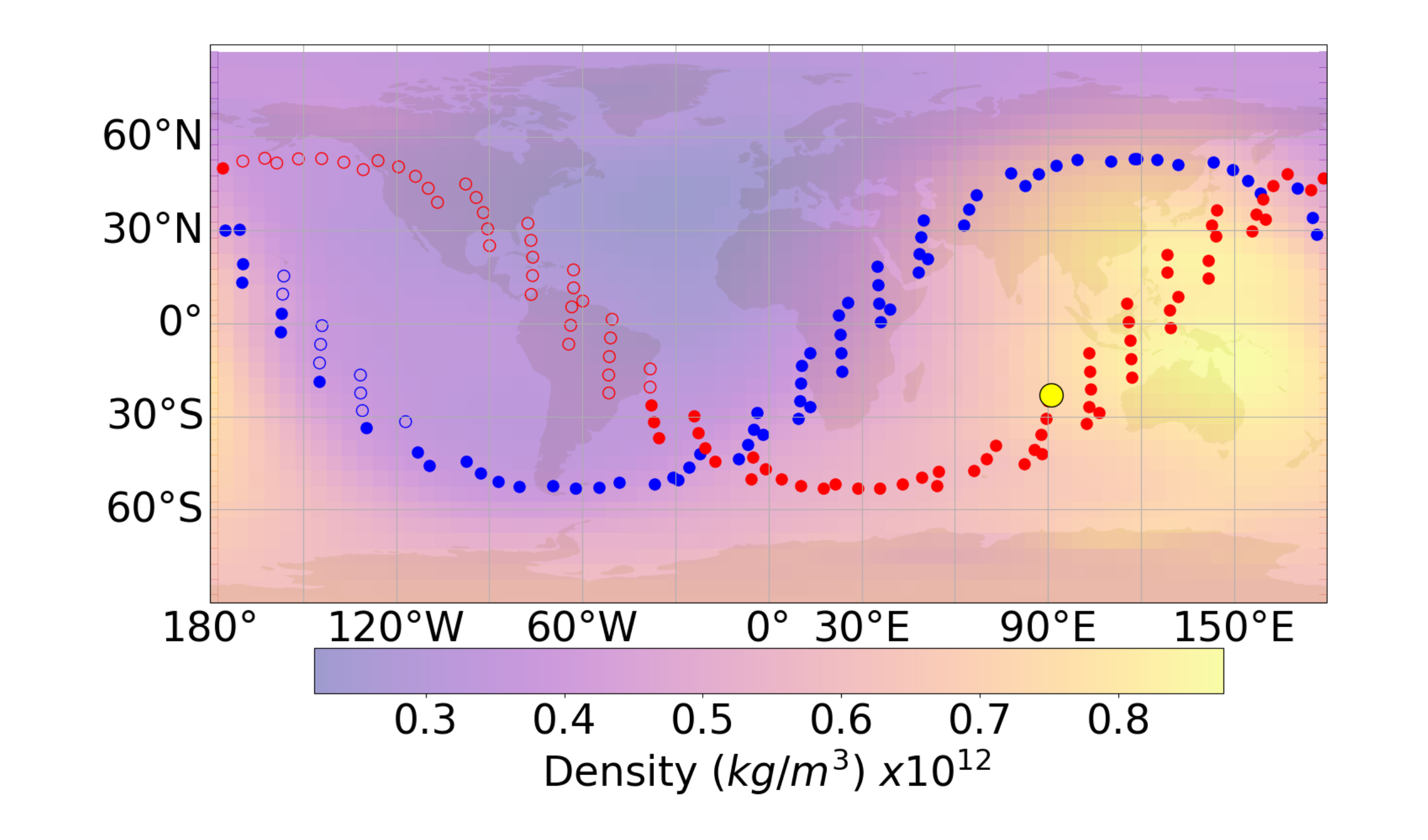}
        \caption{At 06:00 UTC, 1st January, 2024}
    \end{subfigure}%
    \hfill
    \begin{subfigure}[t]{0.5\columnwidth}
        \centering
        \includegraphics[width=\columnwidth, keepaspectratio]{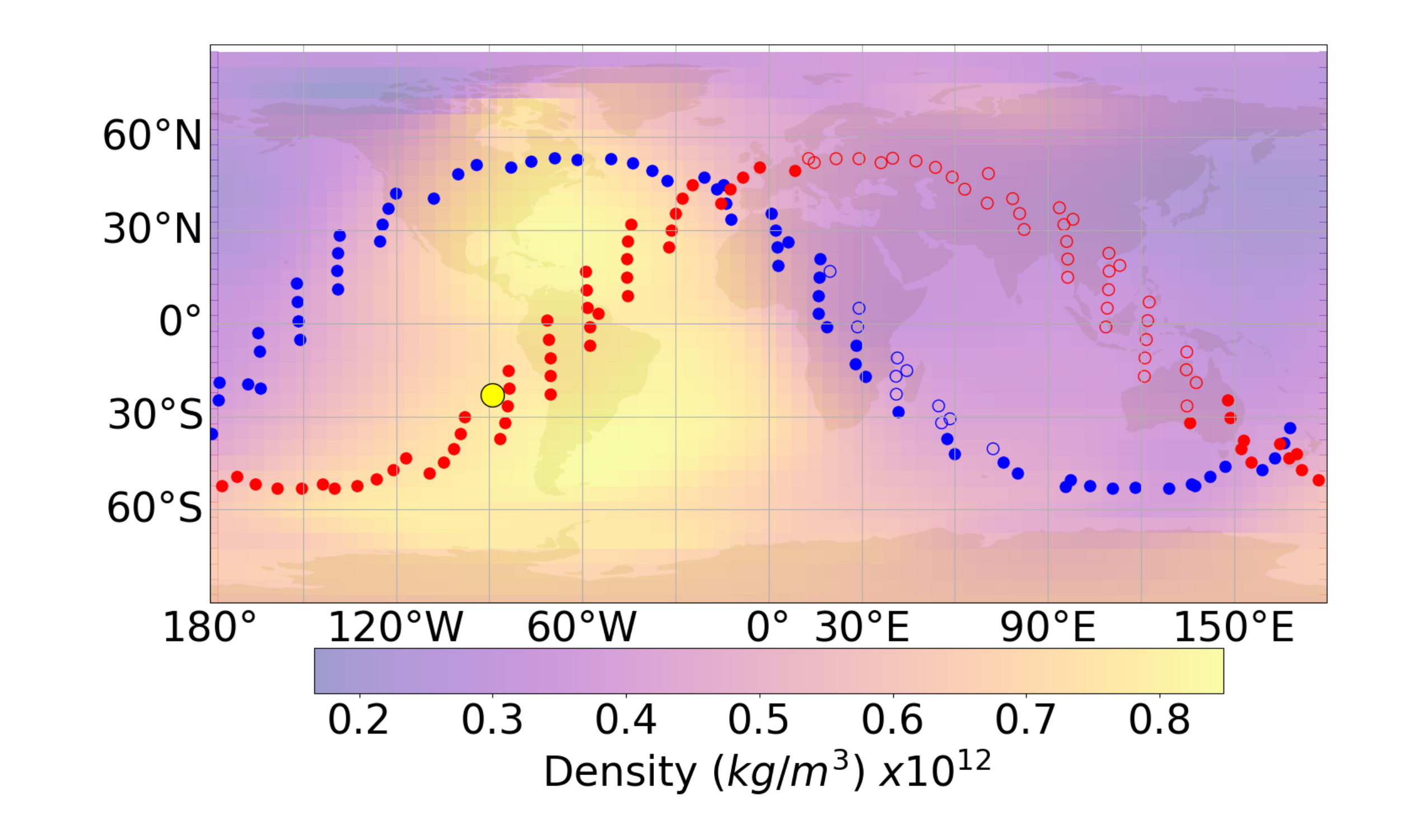}
        \caption{At 18:00 UTC, 1st January, 2024}
    \end{subfigure}%

    \caption{Color mesh in background represents the global atmospheric density at 550 km altitude, showing that daytime density is about four times higher than nighttime levels. 
    Red markers indicate satellites passing through regions of maximum and minimum density, leading to significant altitude variations over the course of a day. 
    Blue markers represent satellites traveling along dawn–dusk paths, where atmospheric density remains nearly constant, resulting in minimal altitude changes.}
    \label{fig:geospatialViewJan}
\end{figure}

\parab{Decoding the `W' pattern} - 
To further understand the origin of the observed pattern, we analyze the geospatial states of satellite and atmospheric density conditions at their altitude. 
In Fig.~\ref{fig:geospatialViewJan}, we visualize this state on 1st January at two timestamps, 06:00 UTC and 18:00 UTC, respectively. 
The background color in these figures represents atmospheric density at the Starlink shell’s operational altitude of 550 km, while the yellow marker denotes the subsolar point. 
Satellite positions at that particular timestamp from a few orbital planes are marked using the red and blue circles.
Where red circles correspond to satellites in five consecutive orbital planes closest to RAAN $300^\circ$ that experience the maximum altitude changes, notice in Fig.~\ref{fig:geospatialViewJan}(a).
Similarly, the blue circles correspond to satellites in five consecutive orbital planes, with the RAAN at $180^\circ$ experiencing the minimum altitude change. 
Further, the solid and hollow red/blue circles indicate that satellites are in sunlight and Earth’s shadow, respectively.

At 06:00 UTC in Fig.~\ref{fig:geospatialViewJan}(a), notice a strong day–night asymmetry in atmospheric density.
The dayside shows significantly higher density than the nightside, up to a factor of four at the same altitude. 
This gradient directly impacts the drag experienced by satellites along their orbit.
Notice the satellites in the red orbital group traverse through regions spanning the subsolar point (maximum density) and the midnight sector (minimum density) within a single orbit. 
As a result, these satellites encounter large density variations along their orbits around Earth.
This leads to varying drag forces.
Consequently, larger cumulative altitude changes over a day are reflected in TLE-derived measurements.
In contrast, satellites in the blue orbital group remain close to the terminator (dawn–dusk line), where atmospheric density remains relatively stable due to the transition between day and night conditions. This results in near-constant drag along the orbit around Earth, producing minimal altitude variation across consecutive TLEs.
This behavior is not transient. 
The same spatial situation persists 12 hours later, notice the Fig.~\ref{fig:geospatialViewJan}(b). 
As shown earlier in Fig.~\ref{fig:WpatternShiftOverYear}, it remains consistent throughout the year.

These figures also explain the inverse relationship between sunlight exposure and altitude variation. 
Satellites in dawn–dusk aligned orbital planes remain illuminated for most of their orbital period, whereas satellites in subsolar aligned orbital planes undergo extended eclipse phases, typically 30–40\% of their orbital period. 
Observe this illustrated in Fig.~\ref{fig:geospatialViewJan}(a), satellites in blue orbits are predominantly sunlit except a few between 120$^{\circ}$-170$^{\circ}$W, while a large fraction of satellites in red orbits are in Earth’s shadow, notice between 30$^{\circ}$-180$^{\circ}$W.

\begin{figure}
    \centering
    \begin{subfigure}[t]{0.5\columnwidth}
        \centering
        \includegraphics[width=\columnwidth, keepaspectratio]{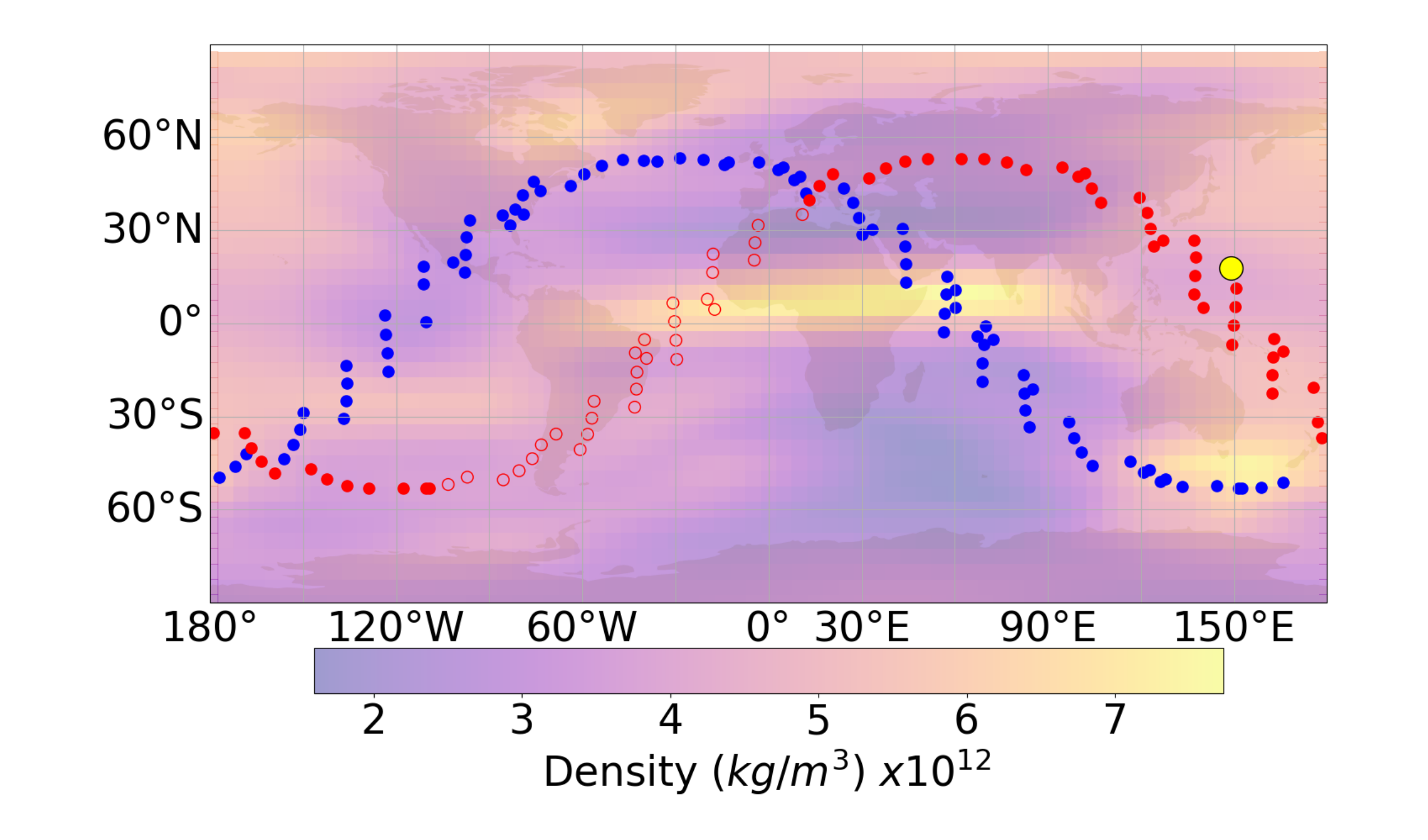}
        \caption{At 02:00 UTC, 11th May, 2024}
    \end{subfigure}%
    \hfill
    \begin{subfigure}[t]{0.5\columnwidth}
        \centering
        \includegraphics[width=\columnwidth, keepaspectratio]{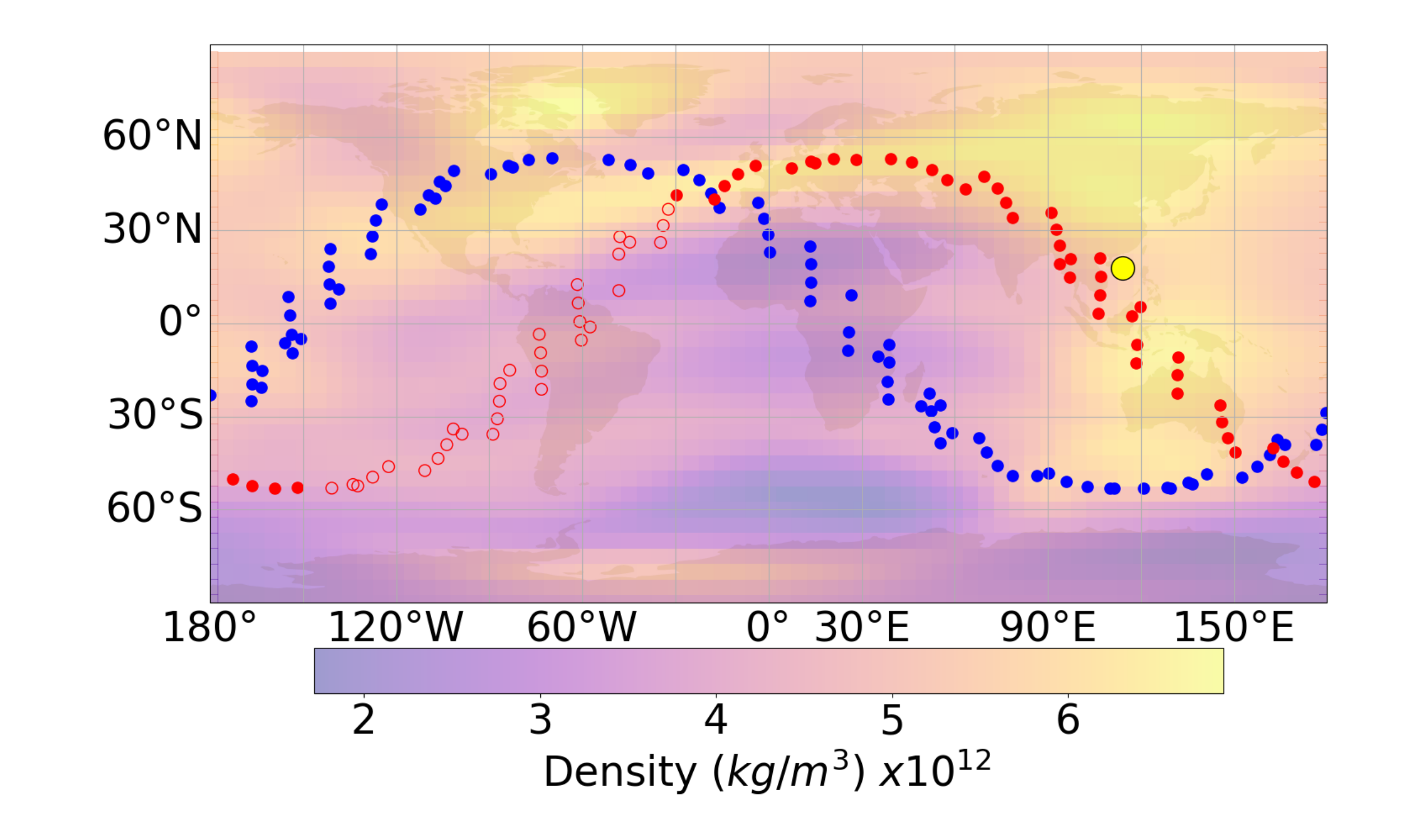}
        \caption{At 04:20 UTC, 11th May, 2024}
    \end{subfigure}%
    \hfill
    \begin{subfigure}[t]{0.5\columnwidth}
        \centering
        \includegraphics[width=\columnwidth, keepaspectratio]{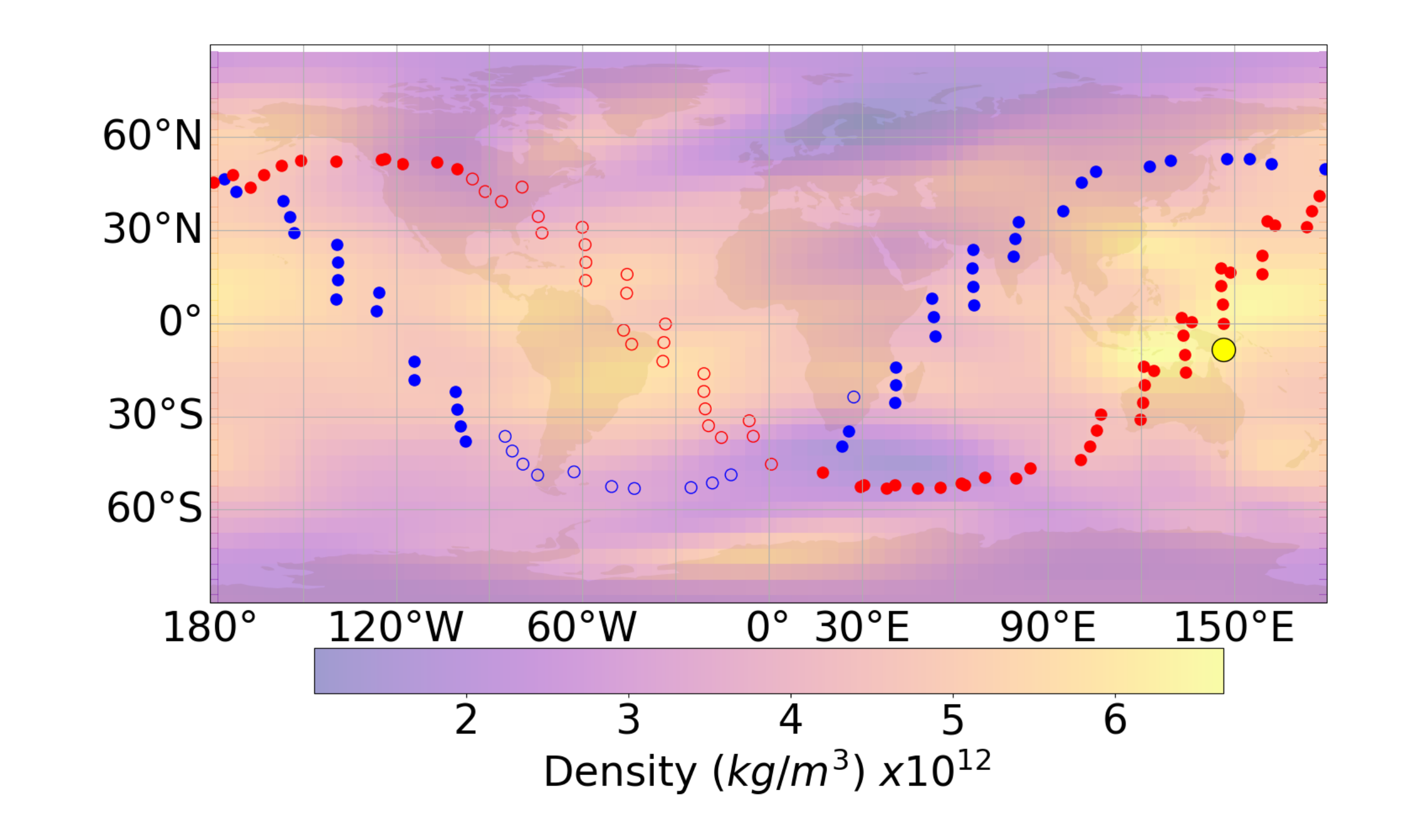}
        \caption{At 02:00 UTC, 11th Oct, 2024}
    \end{subfigure}%
    \hfill
    \begin{subfigure}[t]{0.5\columnwidth}
        \centering
        \includegraphics[width=\columnwidth, keepaspectratio]{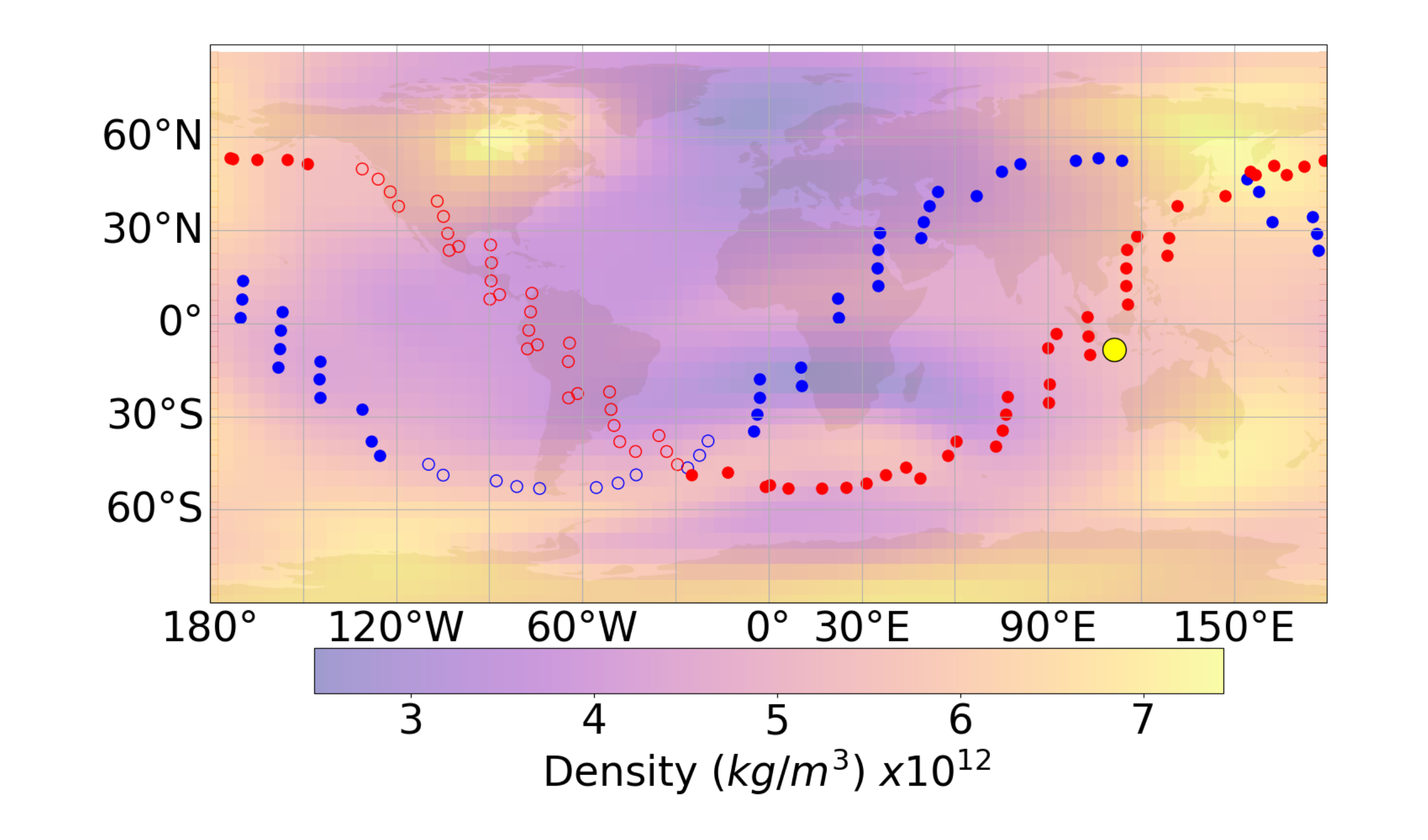}
        \caption{At 04:20 UTC, 11th Oct, 2024}
    \end{subfigure}%

    \caption{Color mesh in background represents dynamics of spatiotemporal global atmospheric density distribution at 550 km altitude during (a)-(b) May 2024 solar superstorm and (c)-(d) October 2024 solar storm. 
    The overall atmospheric density increased by up to 8 times compared to typical conditions. 
    Despite that, overall dayside density remains about 4 times higher than nightside density, so satellite altitude variations become larger in magnitude, while the underlying pattern of behavior remains unchanged.}
    \label{fig:geospatialViewStorm}
\end{figure}

\parab{What happens during a solar storm?} - 
During solar storms, the spatial distribution of atmospheric density exhibits a strong temporal and geographic heterogeneity, rather than a uniform global increase. 
To illustrate this behavior, in Fig.~\ref{fig:geospatialViewStorm}, we present two snapshots from the May 2024 solar superstorm and the October 2024 solar storm.
At the onset of the solar storm, the atmospheric density inflation is concentrated in the polar regions.
As the storm evolves, we observe high-density patches propagate towards the equator like waves, forming a non-systematic dynamic temporal density distribution. 
In Fig.~\ref{fig:geospatialViewStorm}(a) at 02:00 UTC on 11th May, we can observe a high-density patch originating in the polar region moving towards the equator in the nightside hemisphere.
All the states in Fig.~\ref{fig:geospatialViewStorm} illustrate this complex scenario of non-systematic dynamic temporal density distribution in atmospheric conditions during the solar storm.
Despite these complex dynamics of localized high-density patches, the overall day–night asymmetry in atmospheric density remains preserved. 
As shown in the snapshot at 04:20 UTC for both events, the dayside continues to exhibit overall higher density than the nightside. 
Consequently, the key driver of `W' pattern formation in altitude changes remains intact even during storm conditions, as reflected in Fig.~\ref{fig:WpatternShiftOverYear}(e) and (j).
However, the absolute magnitude of atmospheric density increases substantially during the storm, with typical values rising by a factor of 6–8 compared to quiet periods. 
This amplification directly translates into stronger drag forces, resulting in a corresponding increase in altitude decay from approximately 150 meters under nominal conditions to up to 700 meters during the solar storm.

\parab{Decoding pattern transition} - The Earth revolves around the Sun at an angular rate of $0.986^\circ$ per day, which can be approximated as $1^\circ$ per day for simplicity. 
Therefore, the relative angular rotation of Earth around the Sun is roughly $30^\circ$ per month. 
Notice this value is consistent with Fig.~\ref{fig:WpatternShiftOverYear}, where the `W' pattern shows a systematic rightward shift of $30^\circ$ in RAAN across successive months.
In addition to this longitudinal shift, the latitudinal position of the subsolar point varies over the year due to the Earth's axial tilt of approximately $23.5^\circ$. 
As a result, the subsolar point migrates between $23.5^\circ$N in summer and $23.5^\circ$S in winter over an annual cycle. 
This seasonal movement is reflected in the geospatial visualizations, where the subsolar point appears in the southern hemisphere in Fig.~\ref{fig:geospatialViewJan}(a)-(b), while it shifts to the northern hemisphere in Fig.~\ref{fig:geospatialViewStorm}(a)-(b).
The transition of the subsolar point across the equator occurs around the equinox periods (March–April and September–October). 
These transitions correspond directly to the shifts observed in the structure of the `W'-pattern in Fig.~\ref{fig:WpatternShiftOverYear}(c)–(d) and (i)–(j), indicating that the evolution of the pattern is tightly coupled to seasonal changes in Sun–Earth geometry.

\parab{Universal LEO Trait?} - The LEO altitude range is generally considered to be 160-2,000 km. 
All the Starlink operational shells are within a range of 540–570 km.
This `W' pattern altitude variation is visible in all the Starlink shells. 
The pattern becomes more pronounced in shells with a larger number of orbital planes, whereas higher-inclination shells (e.g., $97.6^\circ$), which contain fewer orbits, exhibit a weaker manifestation of the pattern.
In the prior work~\cite{DeepDiveSolarStorms}, this was already shown. 

\begin{figure}
    \centering
    \begin{subfigure}[t]{0.33\columnwidth}
        \centering
        \includegraphics[width=\columnwidth, keepaspectratio]{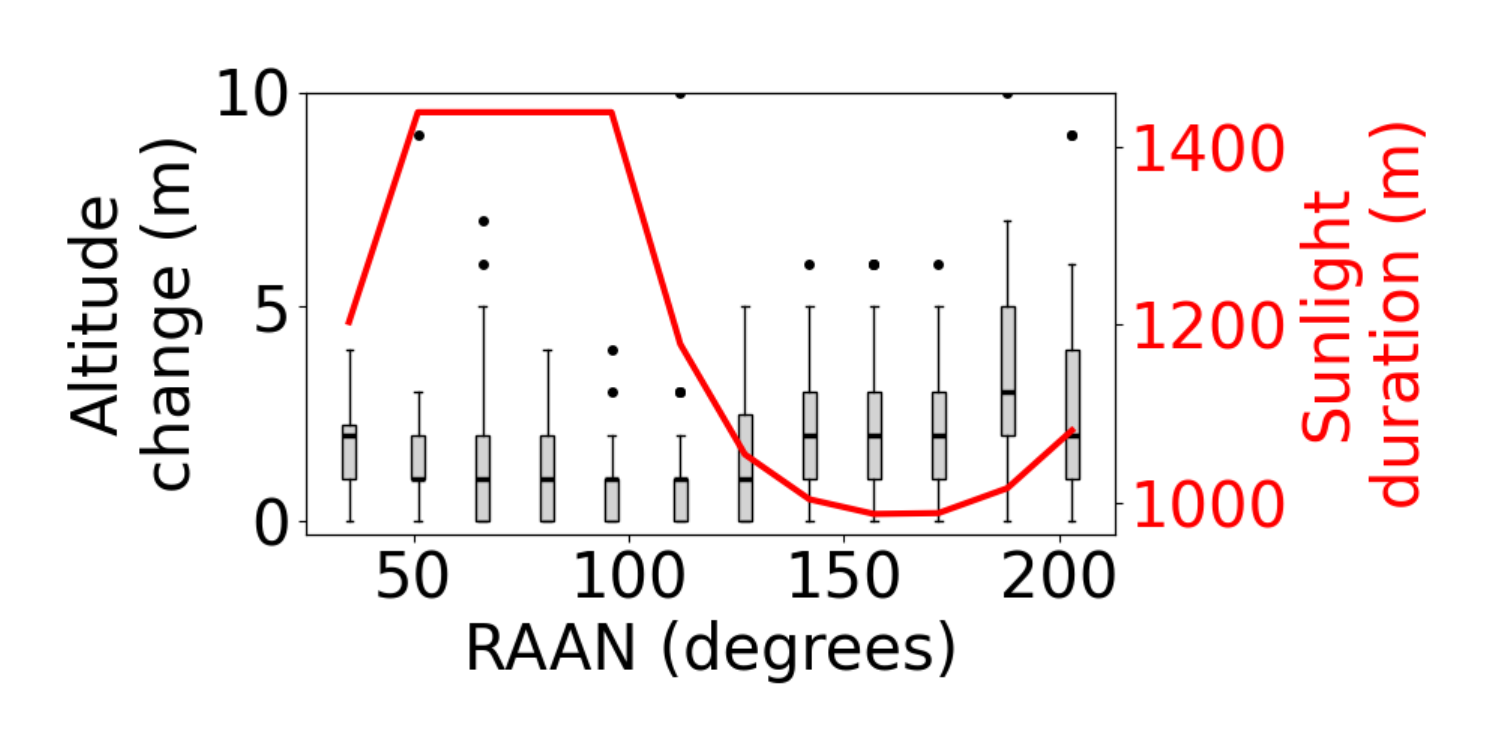}
        \caption{1st March, 2024}
    \end{subfigure}%
    \begin{subfigure}[t]{0.33\columnwidth}
        \centering
        \includegraphics[width=\columnwidth, keepaspectratio]{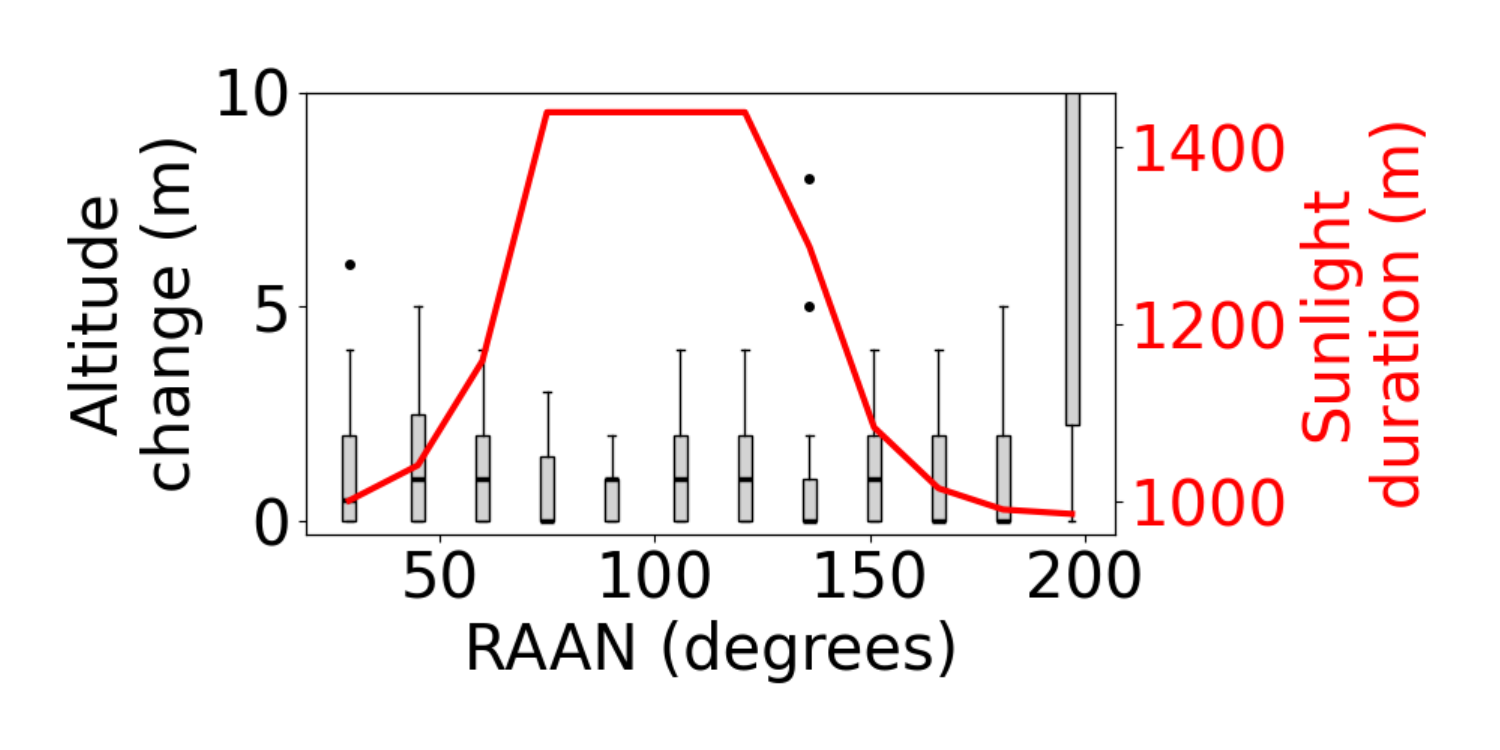}
        \caption{1st April, 2024}
    \end{subfigure}%
    \begin{subfigure}[t]{0.33\columnwidth}
        \centering
        \includegraphics[width=\columnwidth, keepaspectratio]{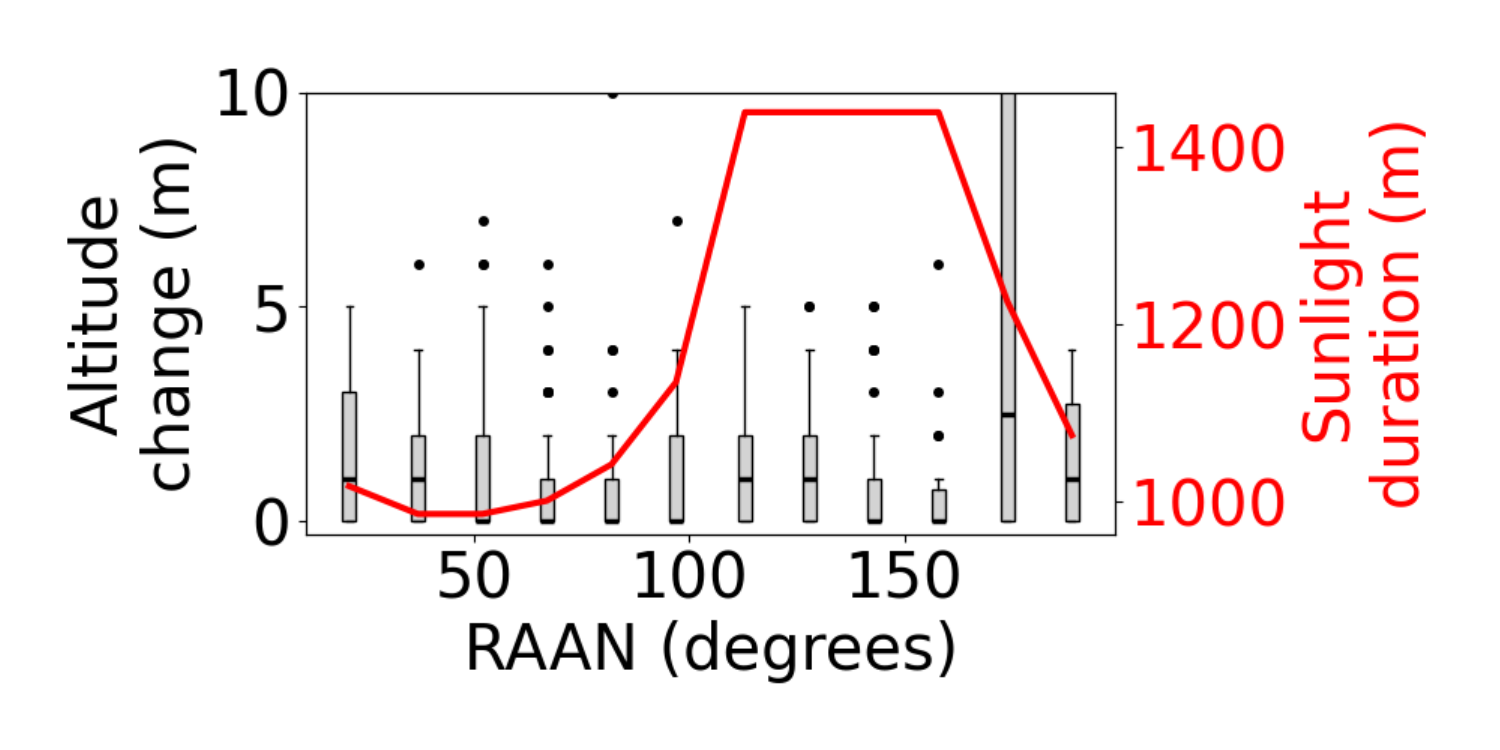}
        \caption{11th May, 2024}
    \end{subfigure}%
    
    \caption{Altitude variation of OneWeb satellites at 1,200 km shows that the `W' pattern nearly disappears. 
    The difference in altitude change is only about 2–3 meters in (a) March and (b) April. 
    Even during the May 2024 solar superstorm (c), there is no noticeable difference.}
    \label{fig:ValidateWpatternOneWeb}
\end{figure}

To examine whether this behavior generalizes across LEO constellations at different altitudes, we analyze the OneWeb constellation, which operates at 1,200 km, roughly twice the altitude of Starlink.
In Fig.~\ref{fig:ValidateWpatternOneWeb}, we present the corresponding analysis. 
Note that OneWeb satellites are deployed in Walker-star configuration~\cite{SurreyPhDThesis, basak2025leocraft} with 12 orbital planes uniformly across $180^\circ$ of RAAN degree.
Also, we do not account for atmospheric density data in this analysis, as the TIE-GCM model~\cite{TIEGCM} is limited to altitudes below 1,000 km. 
At higher altitudes, the atmosphere becomes too thin to manifest general fluid-dynamic properties.
Consequently, Fig.~\ref{fig:ValidateWpatternOneWeb} shows marginal altitude differences on the order of 2–3 meters across orbital planes. 
Furthermore, even during the May 2024 solar superstorm in Fig.~\ref{fig:ValidateWpatternOneWeb}(c), no significant amplification is observed. 
This indicates that, at higher LEO altitudes, atmospheric drag is too weak to induce the structured variations seen in lower-altitude constellations, resulting in relatively stable orbital behavior.

\begin{figure}
    \centering
    \begin{subfigure}[t]{0.6\columnwidth}
        \centering
        \includegraphics[height=4cm, keepaspectratio]{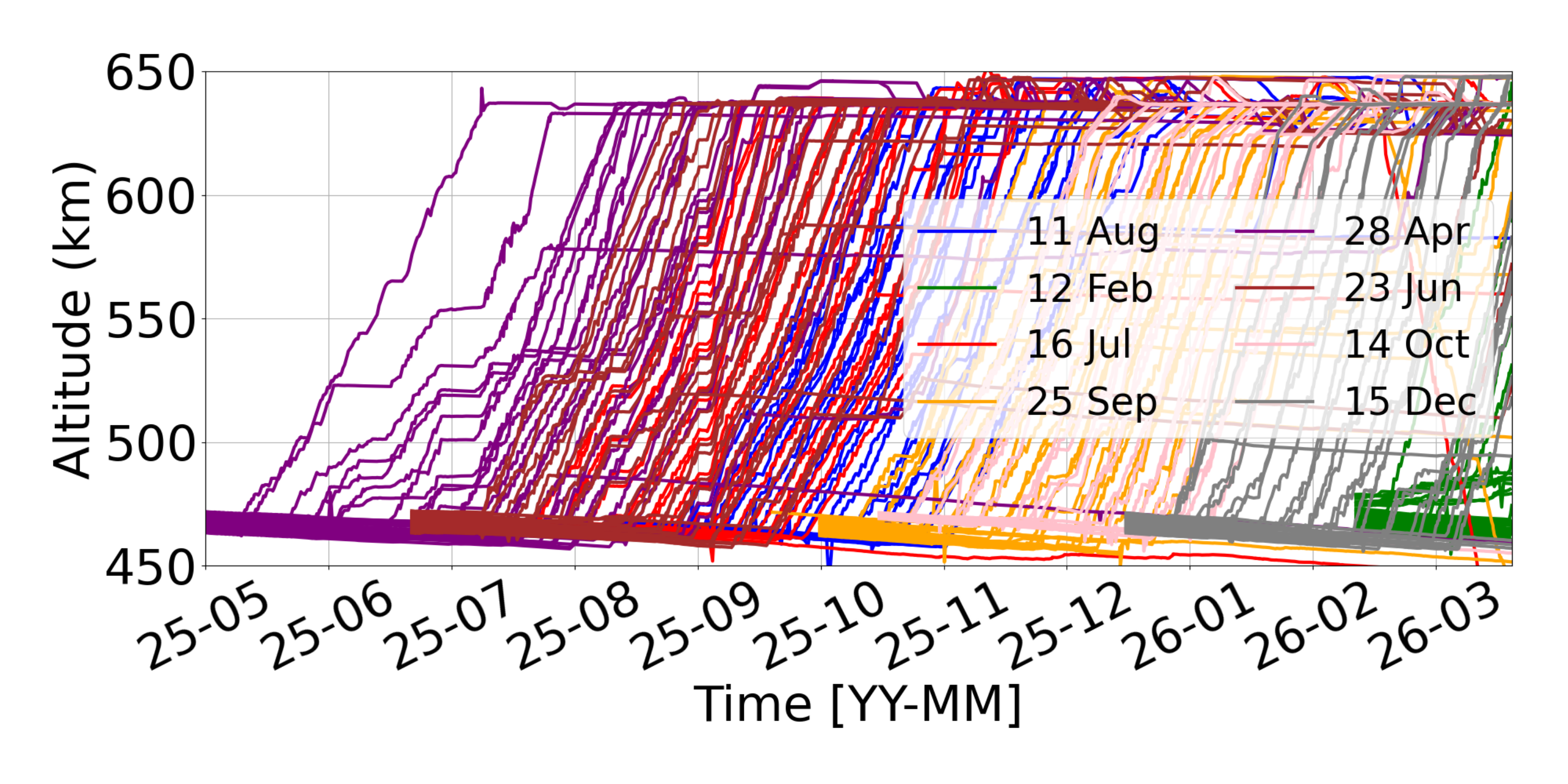}
        \caption{Launch batch-wise orbit raise since the first launch.}
    \end{subfigure}%
    \hfill
    \begin{subfigure}[t]{0.4\columnwidth}
        \centering
        \includegraphics[height=3.9cm, keepaspectratio]{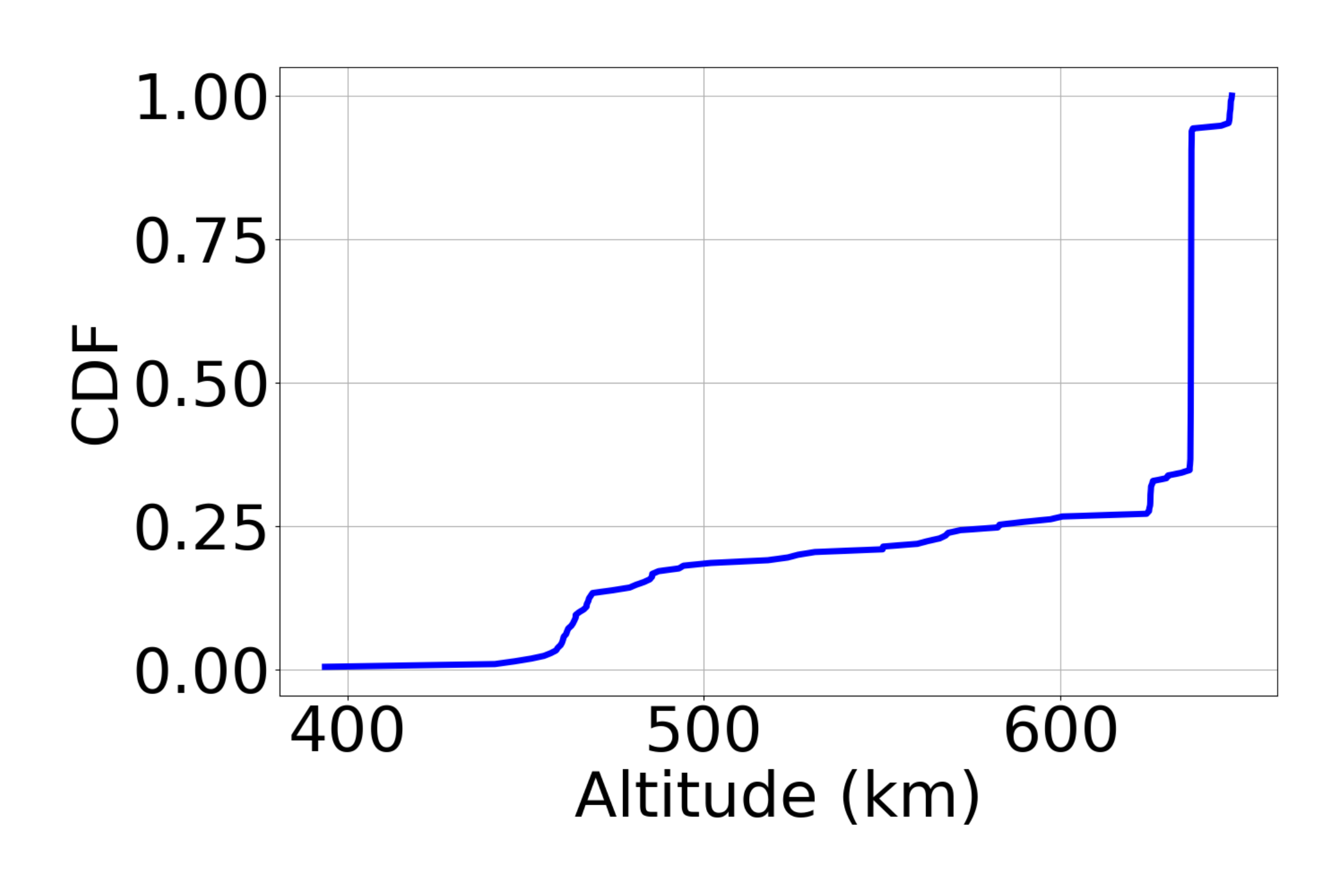}
        \caption{Satellite's altitude at this current state.}
    \end{subfigure}%
    
    \caption{Amazon Leo currently has 212 satellites in orbit (a) going through the orbit raise maneuver, (b) at this current phase, around 40\% of satellites are yet to reach the designated orbit. }
    \label{fig:kuiperCurrentStatus}
\end{figure}

Between these regimes, the Amazon Leo constellation is being deployed at altitudes ranging from 590 to 650 km~\cite{basak2025leocraft}, directly above the Starlink shells. 
However, the current deployment is of 212 satellites, with most still in the orbit-raising phase. 
In Fig.~\ref{fig:kuiperCurrentStatus}, we show the altitude change since the first launch and the distribution of all the satellites at the current state, indicating that only a fraction have reached operational altitude. 
As a result, the current state is insufficient to evaluate the presence or absence of the `W'-pattern in this constellation.
There are Earth observation satellites operating at altitudes around 800 km. 
Again, these satellites are typically placed in sun-synchronous orbits with a narrow range of RAAN. 
This limited angular coverage prevents meaningful analysis of cross-orbital variation, making them unsuitable for validating the observed pattern.

\begin{keybox}
\keynote 
The root cause of the `W' pattern in altitude variation (up to 150 meters) across the orbits is the geospatiality of atmospheric density distribution between day and night. 
This pattern exists throughout the year and continuously shifts in synchronization (1$^{\circ}$/day) with Earth's rotation around the Sun. 
During the solar storm, this pattern amplifies significantly, reaching up to 700 meters.
\end{keybox}

\section{Measuring the implications on network connectivity}
\label{sec:networkMeasurement}

In this section, we examine connectivity during the solar storm.
We use the real-world network measurement datasets discussed above to compare how Starlink connectivity characteristics change from the pre-storm period and how these implications are perceived by end users.
We start with a short overview of Starlink's LEO network operation.

\subsection{Starlink network operation}

All Starlink users access the Starlink LEO satellite network through a user terminal (dish). 
The terminal determines its location using GPS and waits for a beacon signal from nearby Starlink satellites. 
After authentication, it automatically orients itself in the optimal direction~\cite{10623111}. 
It then receives satellite scheduling information.
A centralized controller assigns specific satellites to each user terminal~\cite{MakingSenseLEO}.
The user terminal uses a phased-array antenna to electronically steer transmissions toward fast-moving satellites~\cite{10623111}.  
Each satellite locally manages medium access.
Allocates channel time to user terminals within its coverage area in a round-robin manner~\cite{MakingSenseLEO}.
Traffic from end devices is transmitted via the user terminal to the satellite using a Ku-band radio link. 
The satellite routes this traffic through space (if no gateway available nearby) and downlinks it to a nearby or assigned gateway ground station using a Ka-band radio link~\cite{mohan2024multifaceted, izhikevich2024democratizing}.
Every 15 seconds, a global scheduler reevaluates network conditions and reallocates resources, optimizing the mapping between user terminals, satellites, and gateway ground stations~\cite{mohan2024multifaceted, izhikevich2024democratizing, MakingSenseLEO, 10623111}.
From the gateway ground station, user traffic is routed to the user’s home PoP using Multi-Protocol Label Switching (MPLS), regardless of the location of the gateway ground station, which handles the downlink~\cite{10623111,izhikevich2024democratizing}.
At the PoP, a Carrier-Grade Network Address Translation system assigns a public IP address for egress traffic to the public Internet~\cite{izhikevich2024democratizing}.

\parab{What does it mean for our analysis of network measurement datasets?} - 
Starlink PoPs are mostly co-located with major Internet Exchange Points and CDN edge infrastructure~\cite{bose2025investigating}.
This enables direct peering with large ISPs and content providers.
As a result, the measurement endpoints of M-Lab, Cloudflare, and the DNS servers are highly likely to be geographically close to Starlink PoPs (assuming no geolocation errors~\cite{bose2025investigating}).
This proximity effectively captures the end-user experience of Starlink connectivity, with minimal artifacts introduced by the long terrestrial segment to the measurement endpoints. 
Complementing this, LENS offers high-cadence latency measurements between a device behind the user terminal and the home PoP, enabling fine-grained, sub-second analysis. 
Therefore, in the following, we investigate transient disruptions and short-lived latency variations using LENS datasets, and then examine the end-user Internet experience using M-LAB, Cloudflare speed tests, and RIPE Atlas probes.

\subsection{Latency characteristics}

Using LENS's ICMP-based \texttt{ping} measurements, we analyze how latency behavior changes during a solar storm compared to the pre-storm days.

\subsubsection{Exploring the latency change during solar storms:}

\begin{figure}
    \centering
    \begin{subfigure}[t]{0.50\columnwidth}
        \centering
        \includegraphics[width=\columnwidth, keepaspectratio]{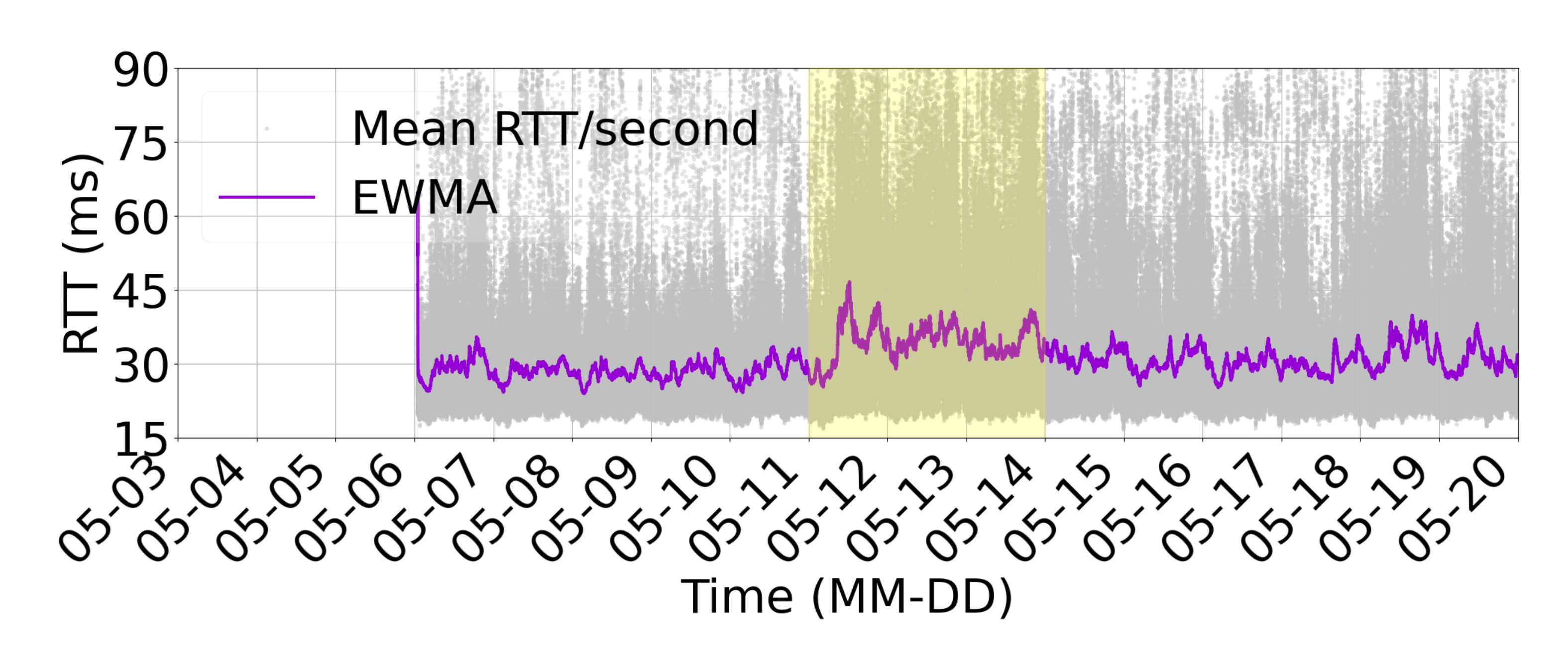}
        \caption{Frankfurt, Germany}
    \end{subfigure}%
    \hfill
    \begin{subfigure}[t]{0.50\columnwidth}
        \centering
        \includegraphics[width=\columnwidth, keepaspectratio]{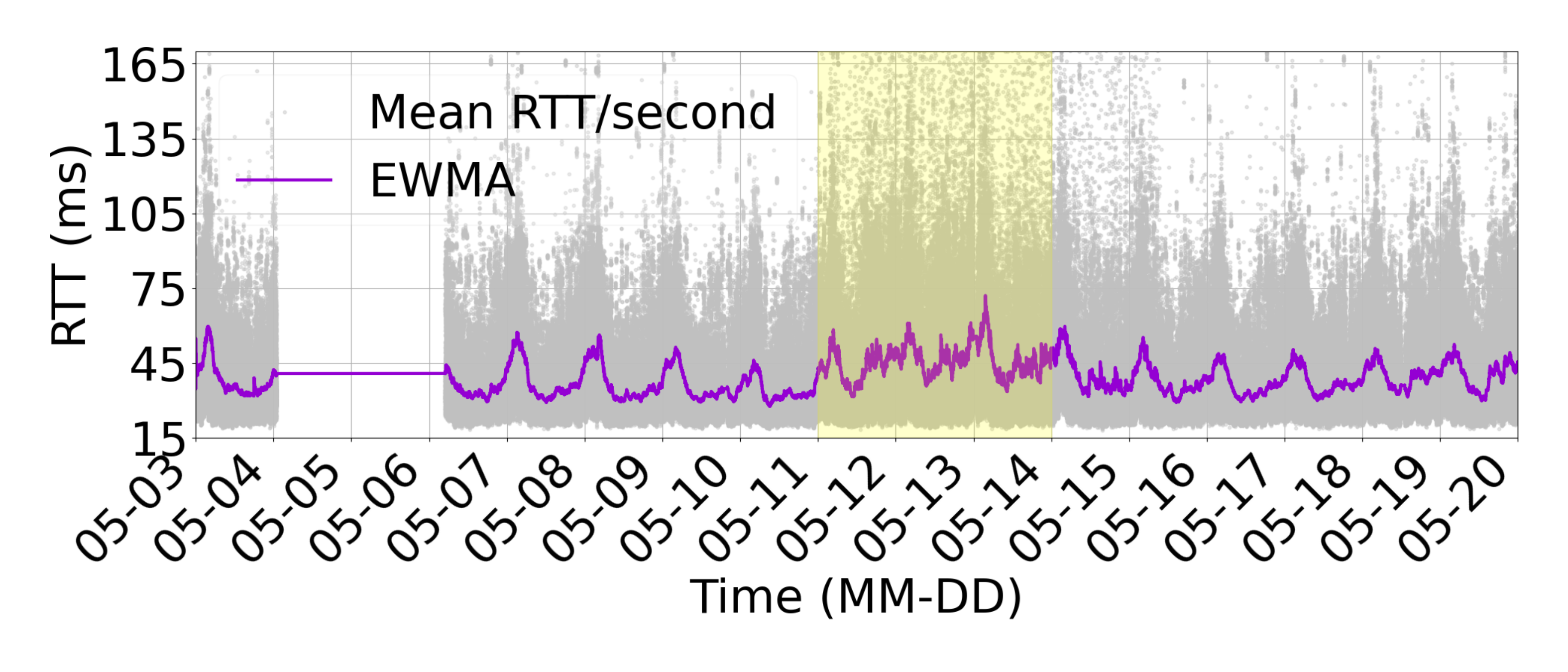}
        \caption{Vancouver, Canada}
    \end{subfigure}%
    \hfill
    \begin{subfigure}[t]{0.50\columnwidth}
        \centering
        \includegraphics[width=\columnwidth, keepaspectratio]{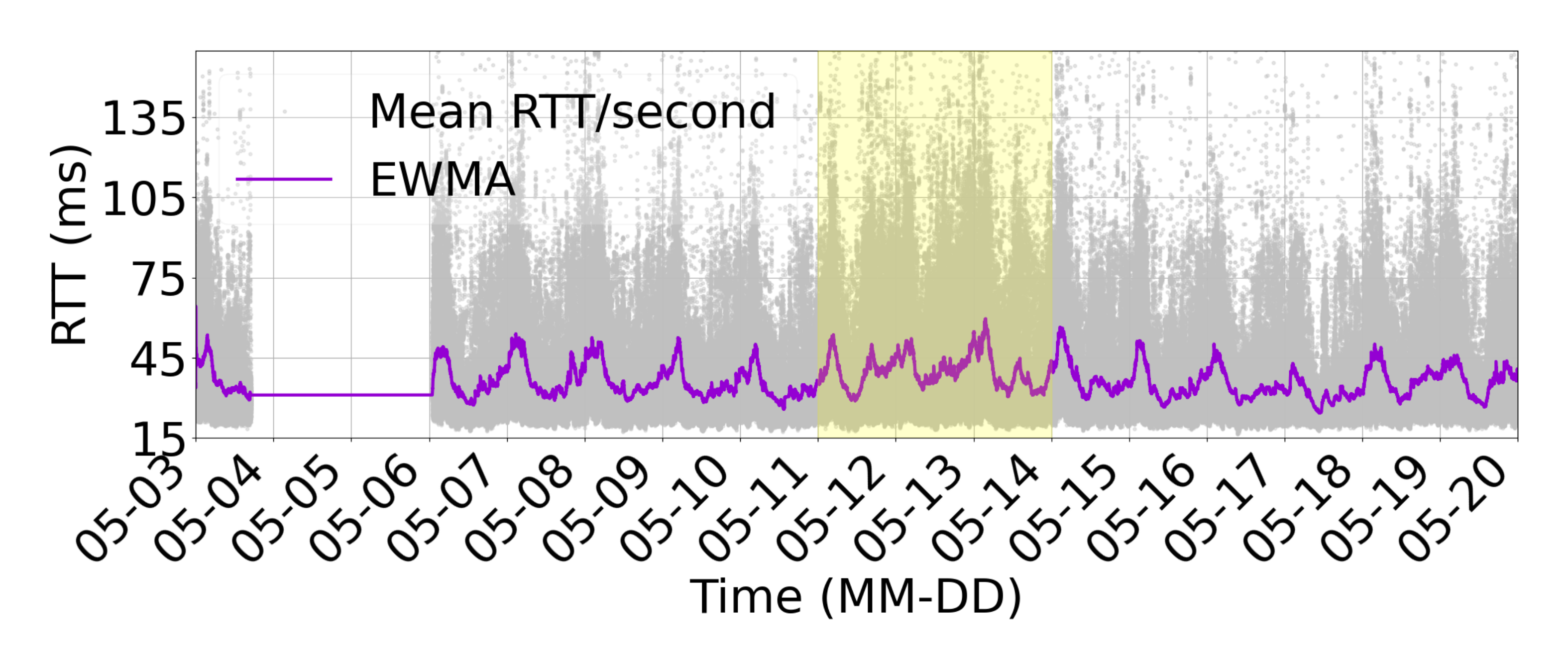}
        \caption{Victoria, Canada}
    \end{subfigure}%
    \hfill
    \begin{subfigure}[t]{0.50\columnwidth}
        \centering
        \includegraphics[width=\columnwidth, keepaspectratio]{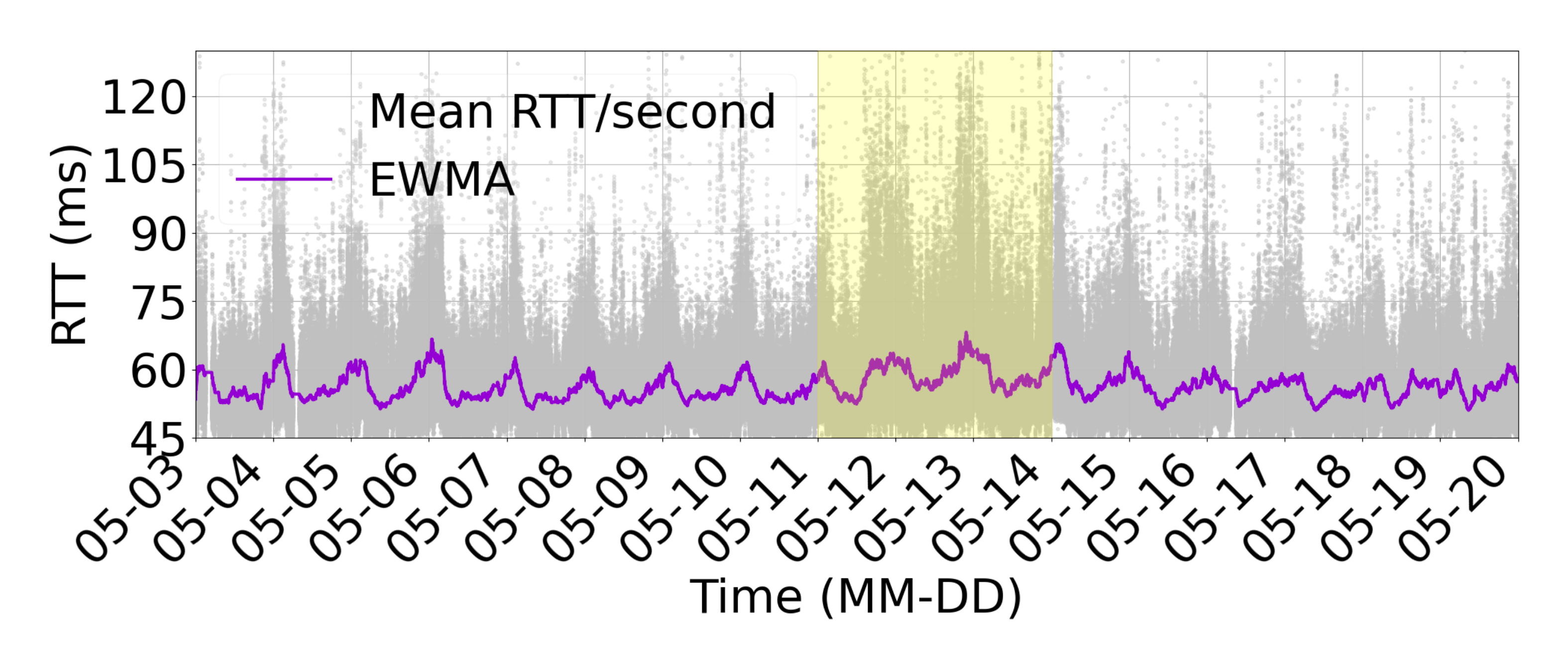}
        \caption{Denver, US}
    \end{subfigure}%
    \hfill
    \begin{subfigure}[t]{0.50\columnwidth}
        \centering
        \includegraphics[width=\columnwidth, keepaspectratio]{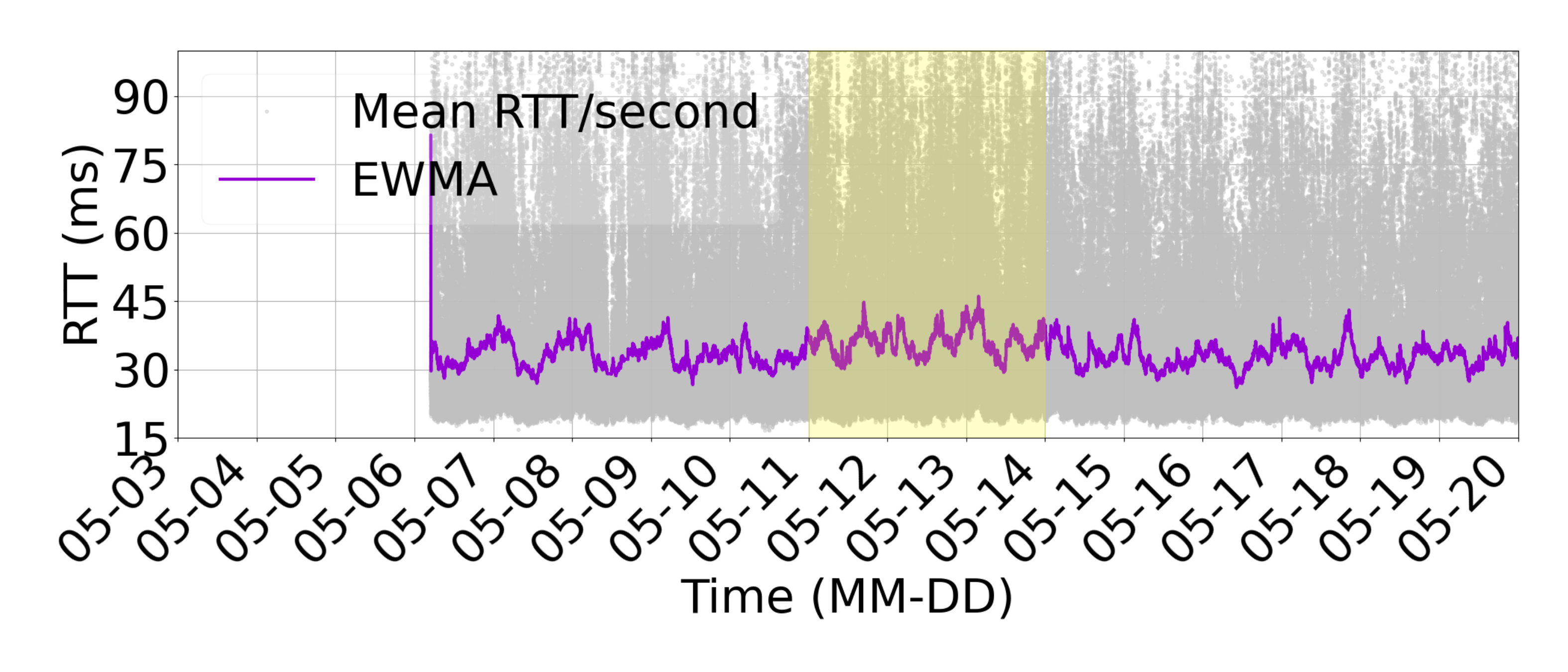}
        \caption{Seattle UT-1, US}
    \end{subfigure}%
    \hfill
    \begin{subfigure}[t]{0.50\columnwidth}
        \centering
        \includegraphics[width=\columnwidth, keepaspectratio]{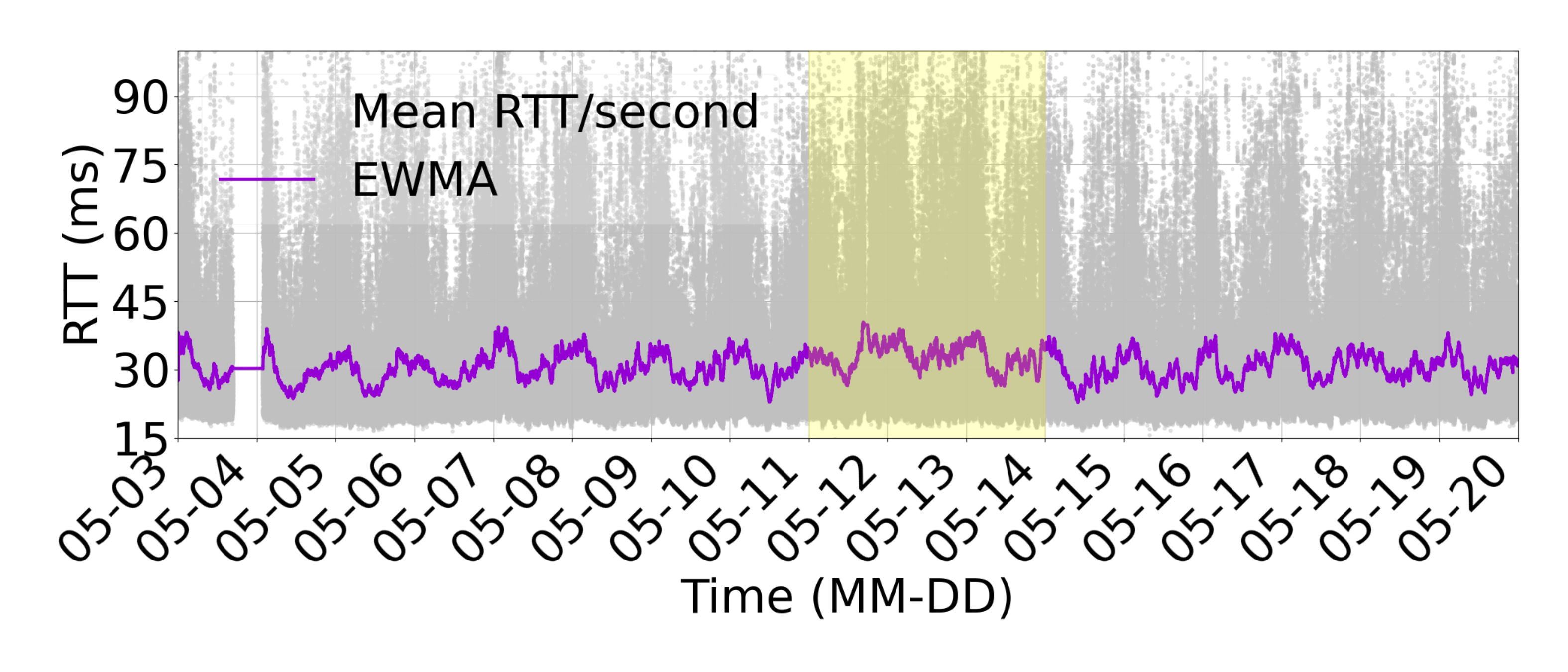}
        \caption{Seattle UT-2, US}
    \end{subfigure}%

    \caption{Time series of mean RTT per second (gray dots), and Exponentially Weighted Moving Average (EWMA) (purple line) from six vantage points from (a) Germany, (b)-(c) Canada, and (d)-(f) the US during the May 2024 solar superstorm, showing the inflation and distortion in diurnal latency patterns during the event.}
    \label{fig:TSpingRTTMay24}
\end{figure}

\begin{figure}
    \centering
    \begin{subfigure}[t]{0.50\columnwidth}
        \centering
        \includegraphics[width=\columnwidth, keepaspectratio]{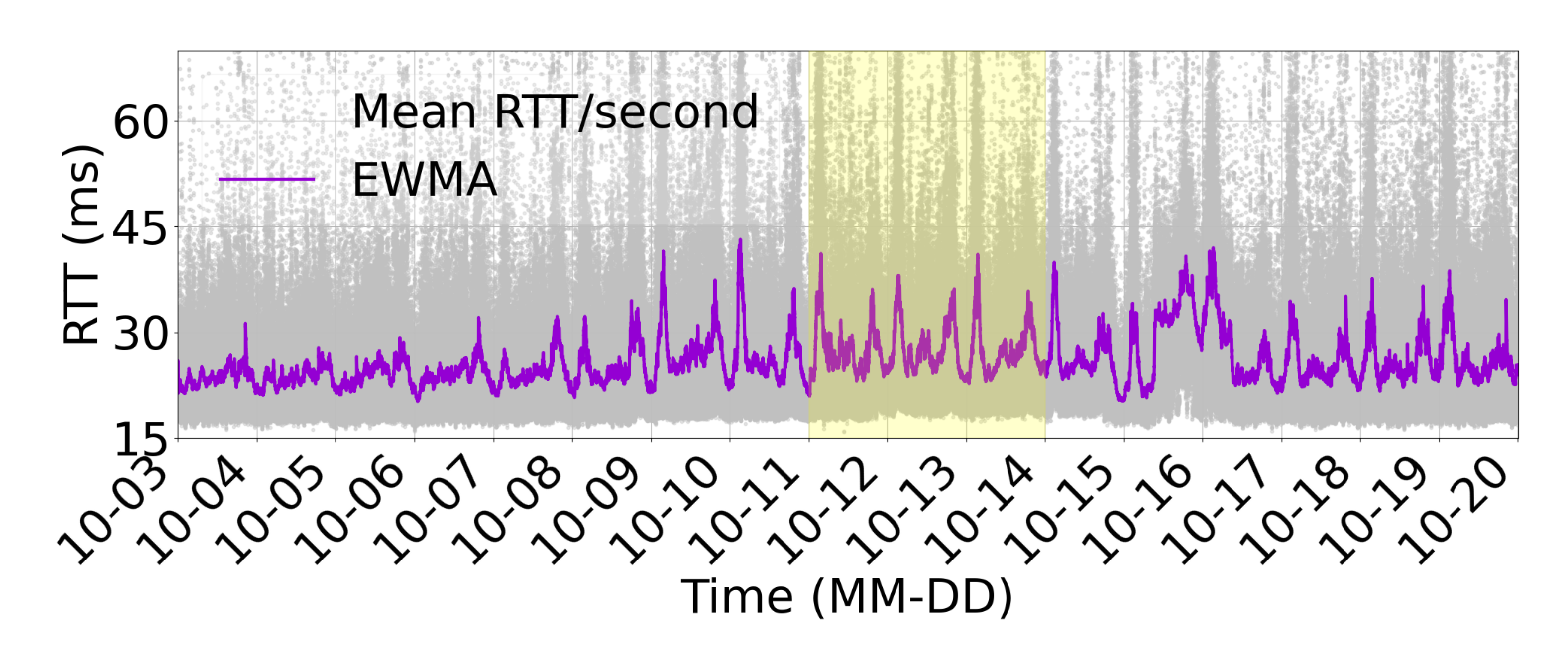}
        \caption{Bruhl, Germany}
    \end{subfigure}%
    \hfill
     \begin{subfigure}[t]{0.50\columnwidth}
        \centering
        \includegraphics[width=\columnwidth, keepaspectratio]{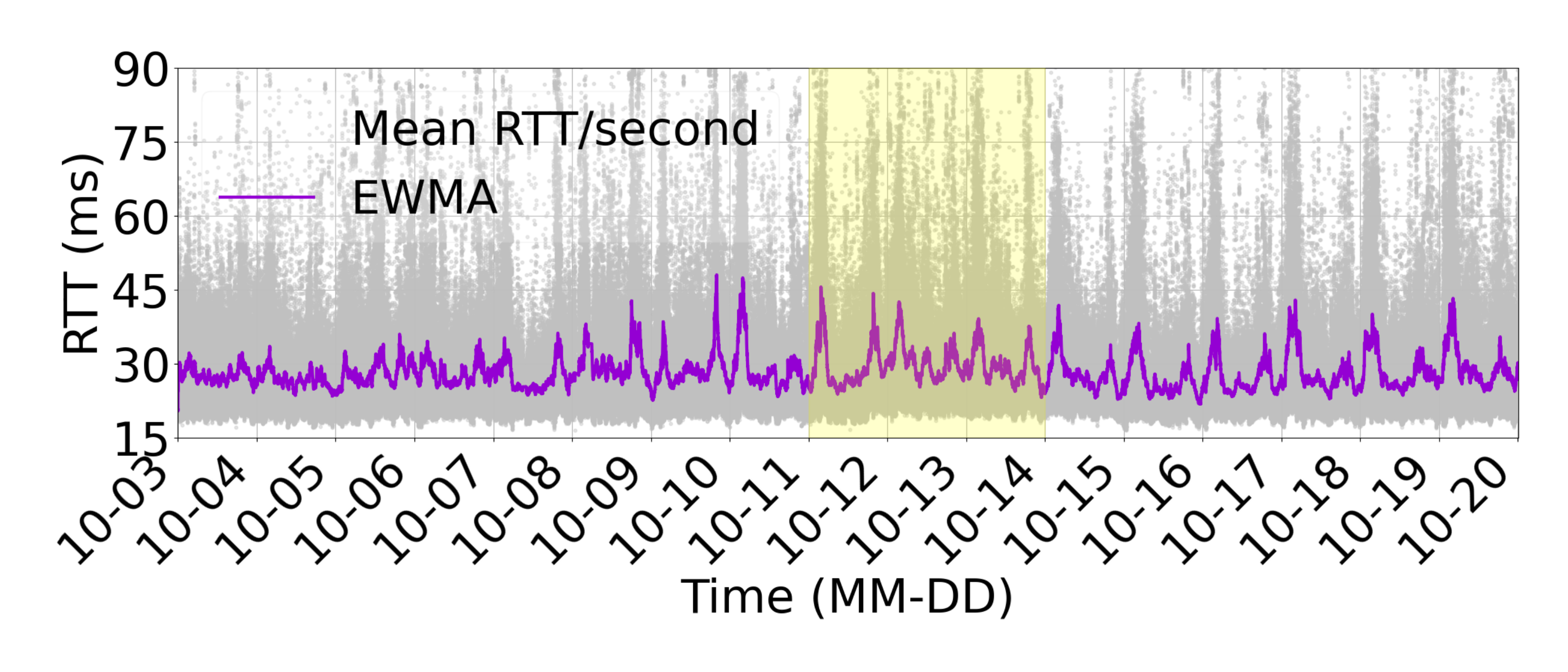}
        \caption{Frankfurt, Germany}
    \end{subfigure}%
    \hfill
    \begin{subfigure}[t]{0.50\columnwidth}
        \centering
        \includegraphics[width=\columnwidth, keepaspectratio]{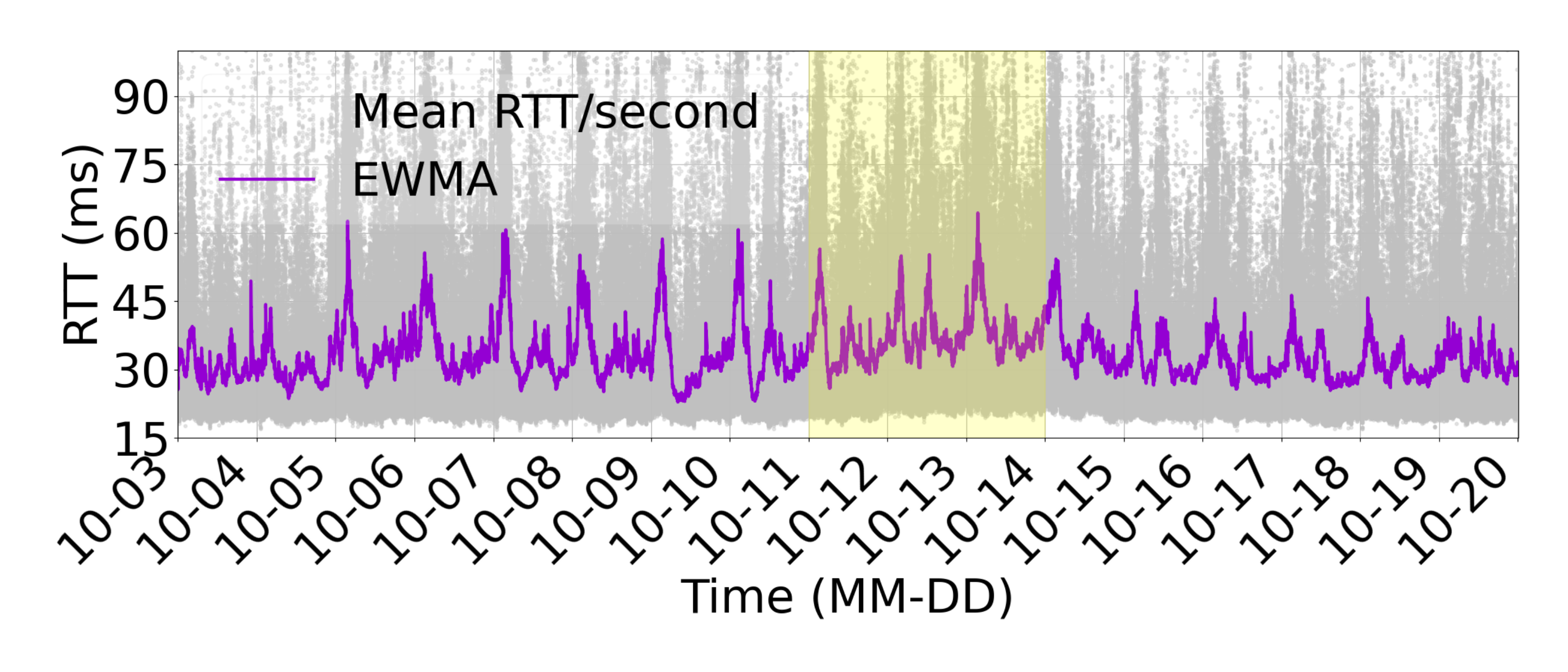}
        \caption{Victoria, Canada}
    \end{subfigure}%
    \hfill
    \begin{subfigure}[t]{0.50\columnwidth}
        \centering
        \includegraphics[width=\columnwidth, keepaspectratio]{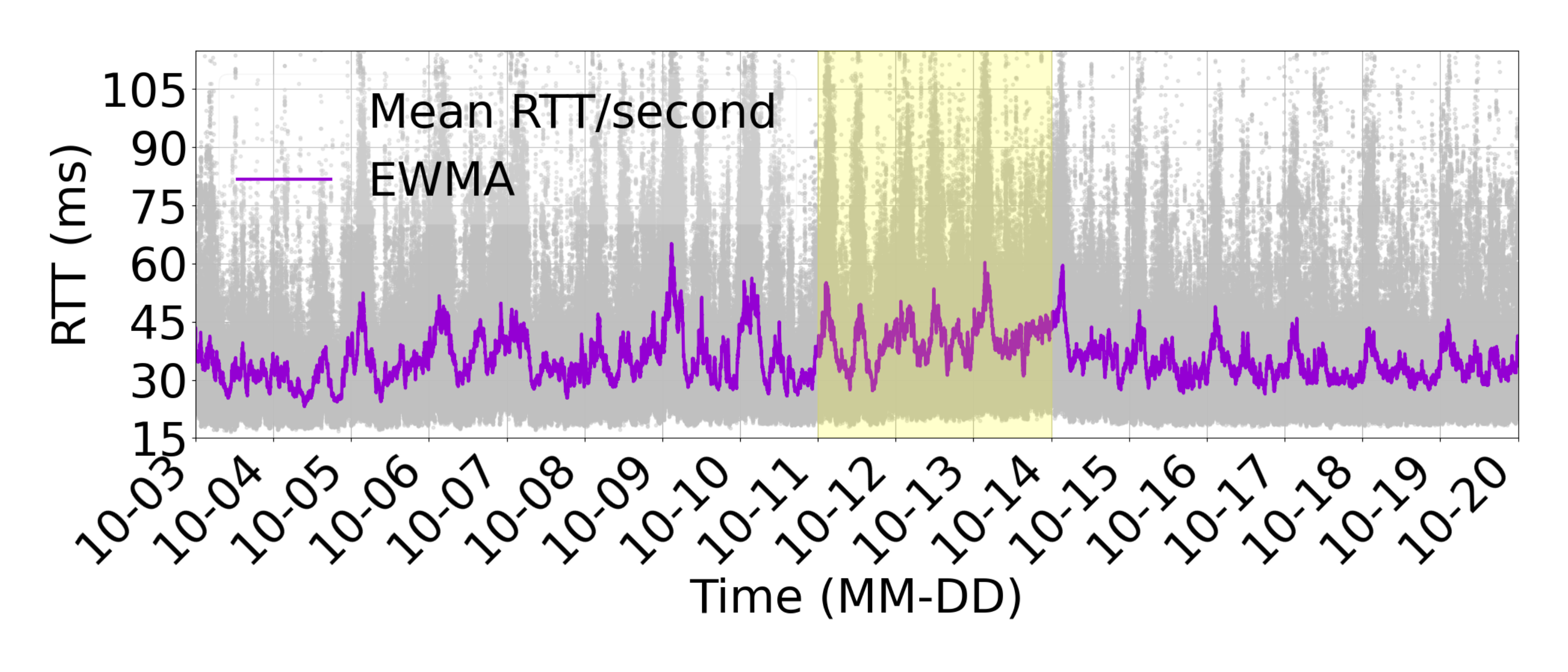}
        \caption{Calgary, Canada}
    \end{subfigure}%
    \hfill
    \begin{subfigure}[t]{0.50\columnwidth}
        \centering
        \includegraphics[width=\columnwidth, keepaspectratio]{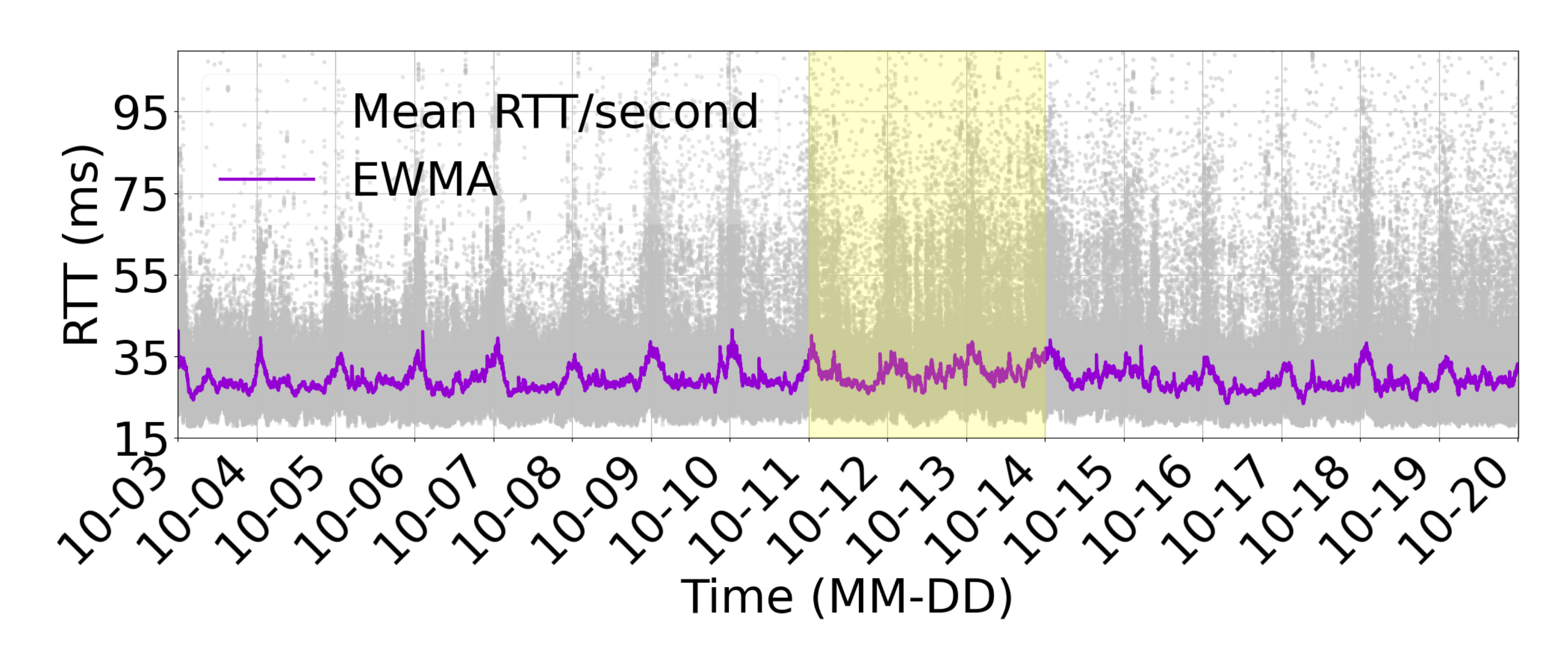}
        \caption{Ottawa IPv4, Canada}
    \end{subfigure}%
    \hfill
    \begin{subfigure}[t]{0.50\columnwidth}
        \centering
        \includegraphics[width=\columnwidth, keepaspectratio]{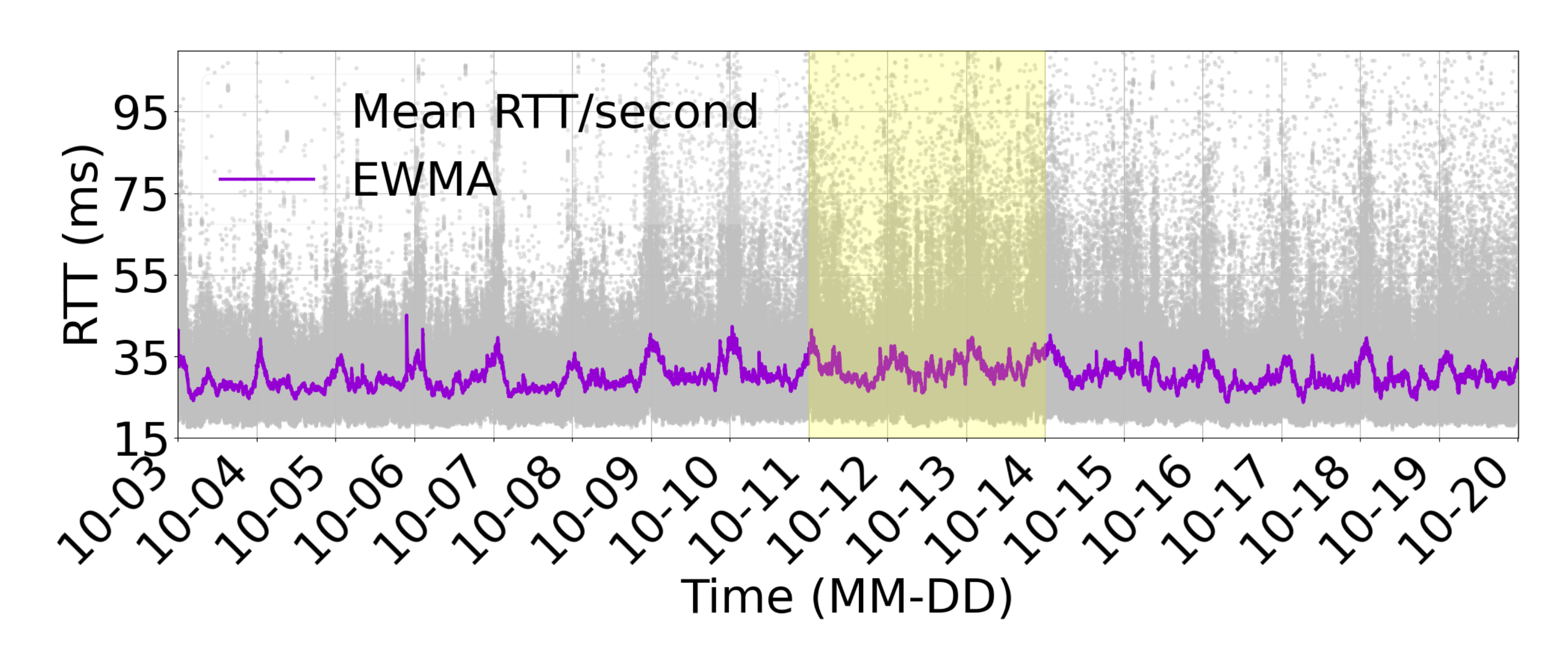}
        \caption{Ottawa IPv6, Canada}
    \end{subfigure}%
    \hfill
    \begin{subfigure}[t]{0.50\columnwidth}
        \centering
        \includegraphics[width=\columnwidth, keepaspectratio]{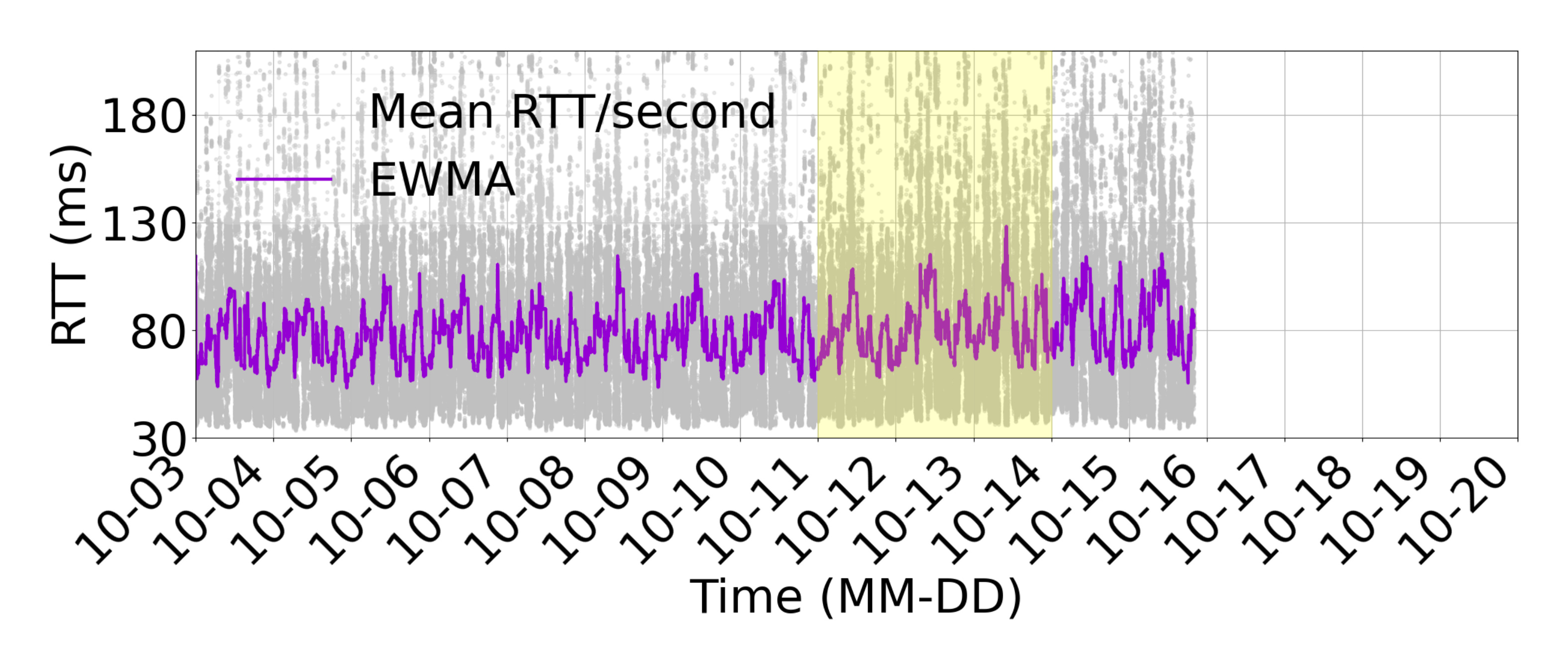}
        \caption{Ulukhaktok, Canada}
    \end{subfigure}%
    \hfill
    \begin{subfigure}[t]{0.50\columnwidth}
        \centering
        \includegraphics[width=\columnwidth, keepaspectratio]{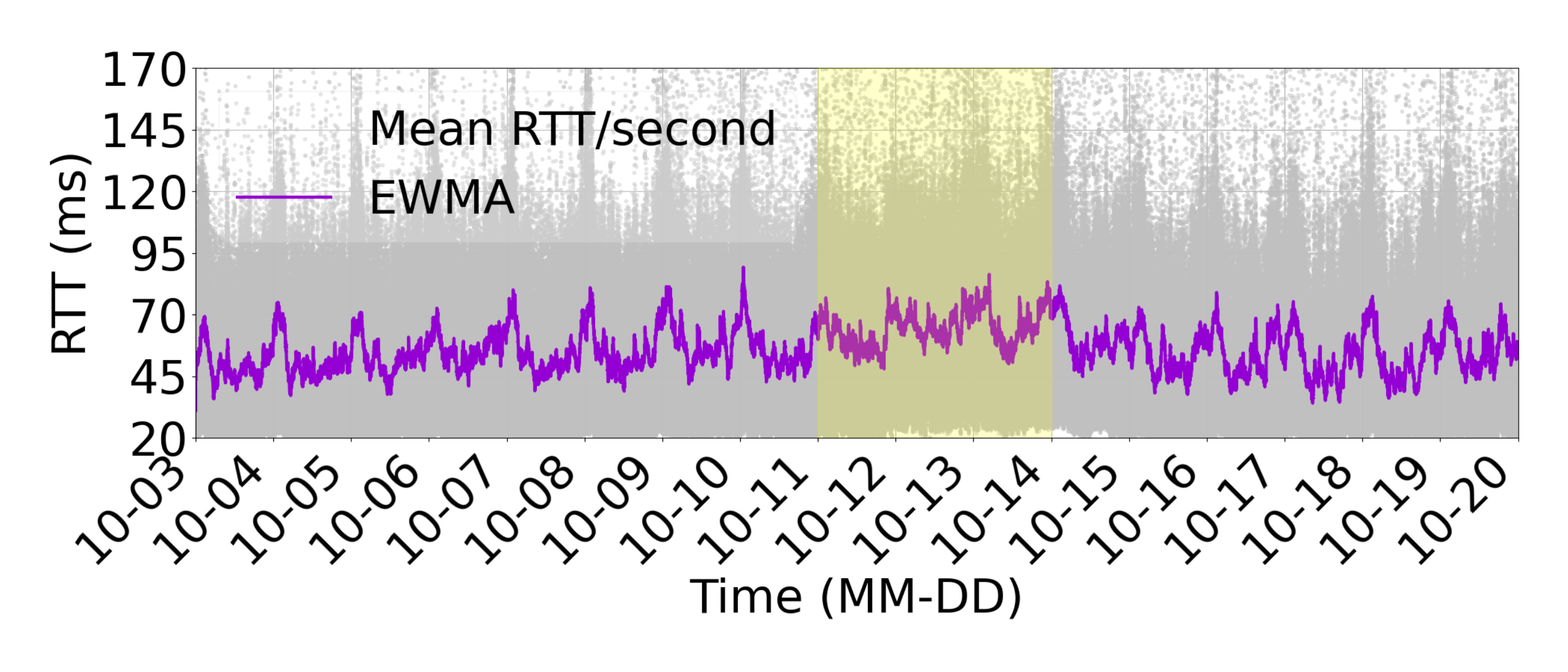}
        \caption{Dallas, US}
    \end{subfigure}%
    \hfill
    \begin{subfigure}[t]{0.50\columnwidth}
        \centering
        \includegraphics[width=\columnwidth, keepaspectratio]{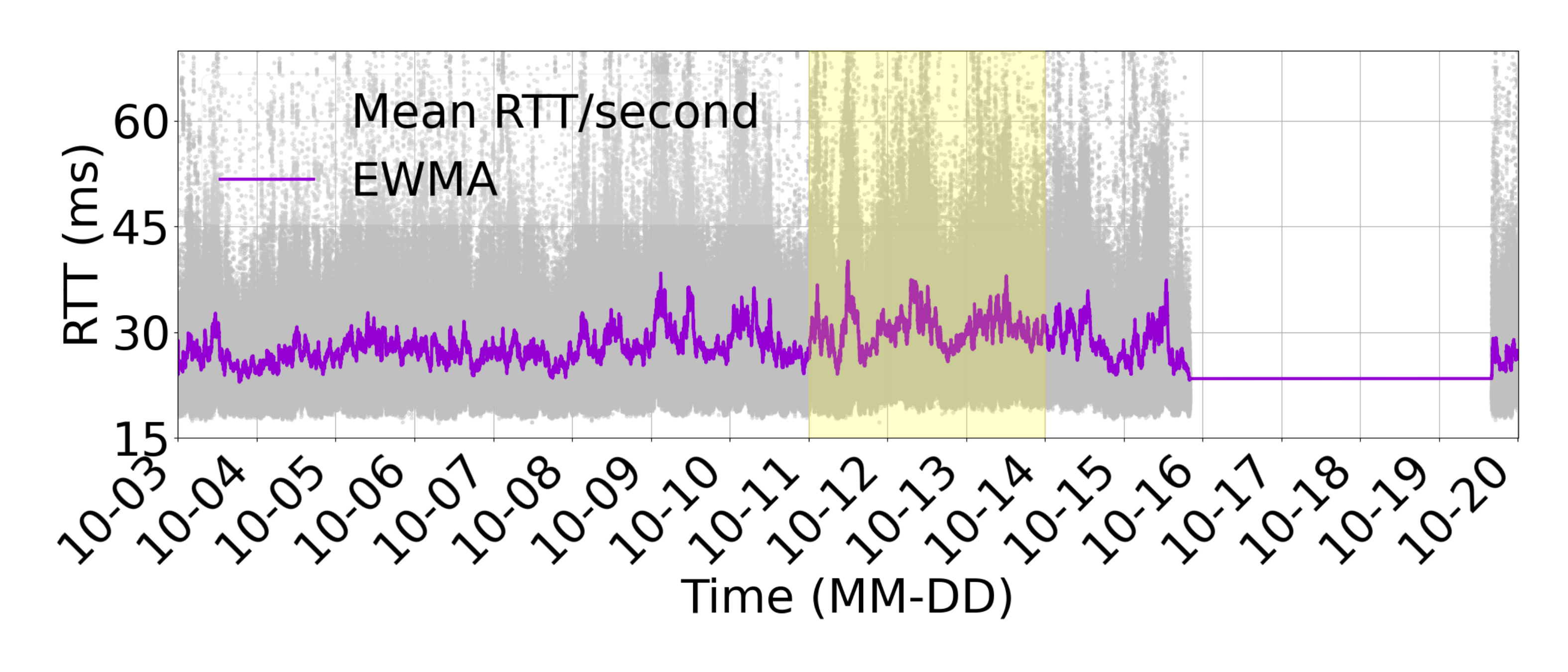}
        \caption{Salt Lake City, US}
    \end{subfigure}%
    \hfill
    \begin{subfigure}[t]{0.50\columnwidth}
        \centering
        \includegraphics[width=\columnwidth, keepaspectratio]{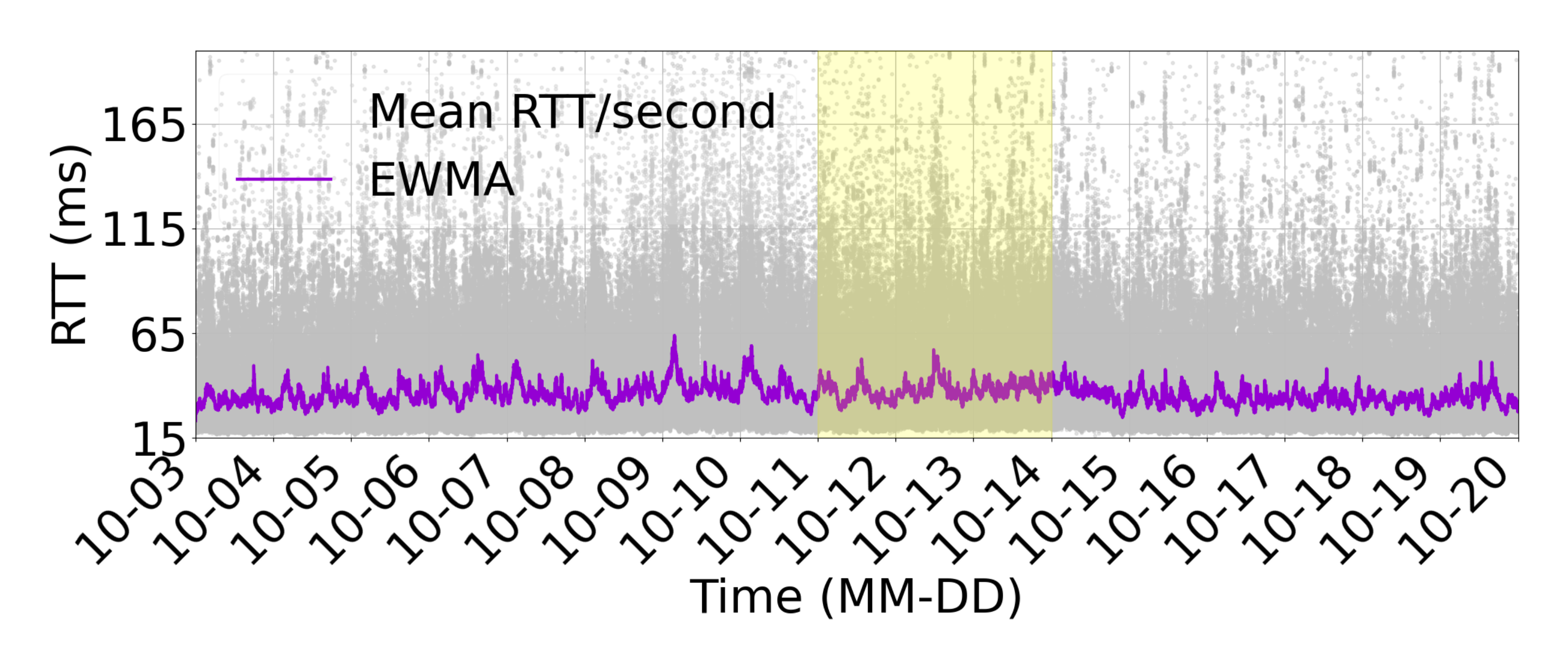}
        \caption{seattle, US}
    \end{subfigure}%

    \caption{Time series of mean RTT per second (gray dots), and Exponentially Weighted Moving Average (EWMA) (purple line) from ten vantage points from (a)-(b) Germany, (c)-(g) Canada, and (h)-(j) the US during the October 2024 solar storm, showing the minor inflation and repetitive spike in diurnal latency patterns during the event.}
    
    \label{fig:TSpingRTTOct24}
\end{figure}

We take the 10 ms cadence ping measurements and aggregate them to the mean RTT per second. 
Then we calculate the Exponentially Weighted Moving Average (EWMA) of the mean RTT per second to observe any meaningful shift in the latency pattern while suppressing the baseline noise.
We use an empirically decided span of 30 minutes (half-life of approximately 623 seconds) for EWMA calculation.
In Fig.~\ref{fig:TSpingRTTMay24}, we plot a timeseries of RTT from 3rd May (7 days before) to 20th May (7 days after) for the May 2024 solar superstorm.
The gray points in these figures are the mean RTT observations, and the violet line shows the EWMA of the mean RTTs.
We also highlight the three-day window in yellow when a large number of orbital decays and rises are observed.
This broader view of 17 days in Fig.~\ref{fig:TSpingRTTMay24} shows a typical diurnal latency pattern in Starlink connectivity, and how this pattern changes during the solar storm.
After 11th May, during the solar superstorm, we can observe a subtle upward shift in the latency pattern in Frankfurt and Vancouver, as noticed in the highlighted region in Fig.~\ref{fig:TSpingRTTMay24}(a)-(b).
In Victoria, Denver, and Seattle, the upward shift is not strong enough.
However, we observe a distortion in the RTT pattern compared to the pre-storm window.

In Fig.\ref{fig:TSpingRTTOct24} we plot the same from 3rd October to 20th October to look at the October 2024 solar storm.
Here, we observe a series of large consecutive spikes in the EWMA line reaching up to 2$\times$ over the baseline RTT measurements from Bruhl, Frankfurt, Victoria, and Calgary. 
These spikes started after 5th October, another G2-class (Moderate) solar storm, before the 11th October G4-class solar storm.
In Fig.~\ref{fig:StarlinkFleetManagement}, we can observe mean atmospheric density inflation after 6th October, consequently an increased number of satellite orbits rising after 7th October.
We do not observe any change at a similar magnitude in the latency pattern from any vantage point in the US. 
Interestingly, Ottawa and Ulukhaktok seem to be unaffected, too, despite being in Canada.
The distances between Frankfurt and Bruhl are 196 km, and Victoria and Calgary are 1,000 km, respectively.
The distance from Victoria to Ottawa and Ulukhaktok is more than 4,000 km. 
This seems to be a localized anomaly. 
However, we do not see any noticeable change in RTT from Seattle despite being close to Victoria and sharing the same PoP~\cite{10623111}. 
This indicates a localized network implication originating somewhere between the user terminal and the gateway ground station or satellite link segment.

\subsubsection{Zooming into diurnal latency pattern:}

\begin{figure}
    \centering
    \begin{subfigure}[t]{0.50\columnwidth}
        \centering
        \includegraphics[height=3cm, keepaspectratio]{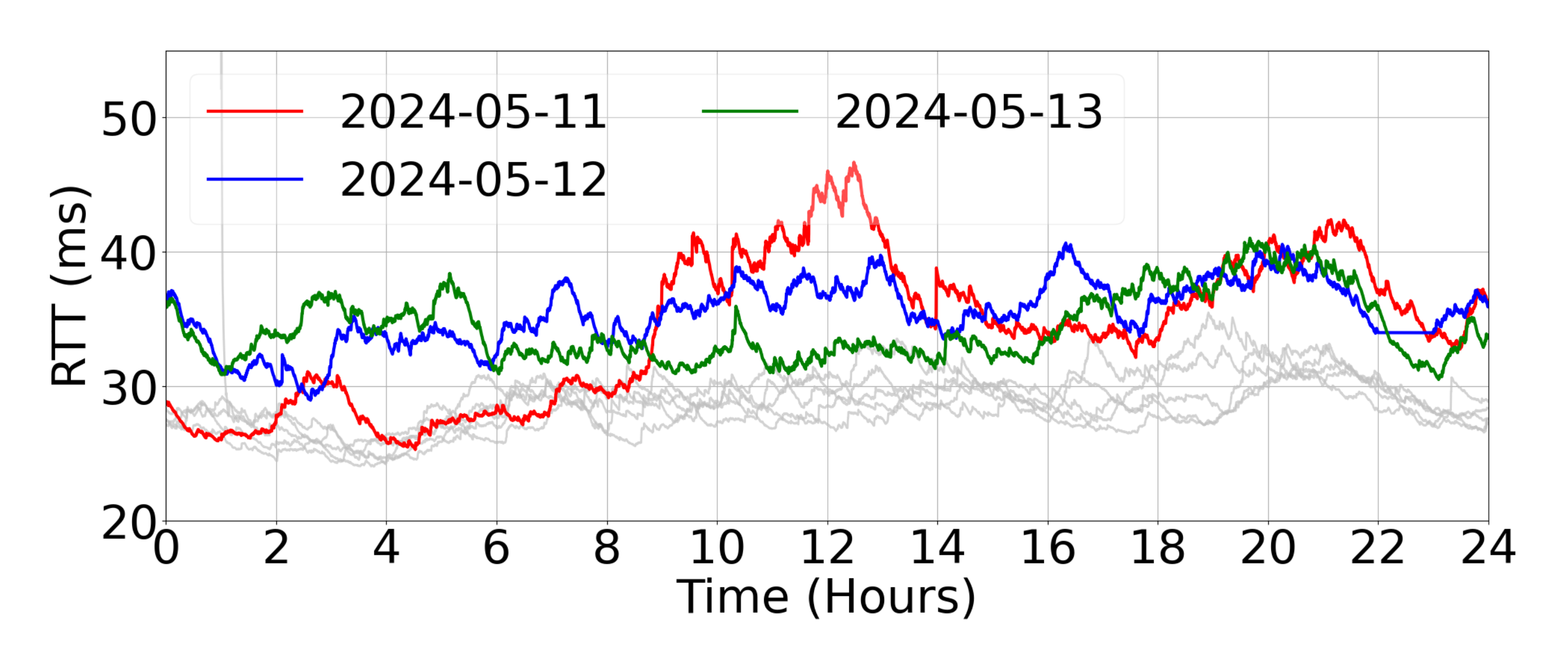}
        \caption{Frankfurt, Germany}
    \end{subfigure}%
    \hfill
    \begin{subfigure}[t]{0.50\columnwidth}
        \centering
        \includegraphics[height=3cm, keepaspectratio]{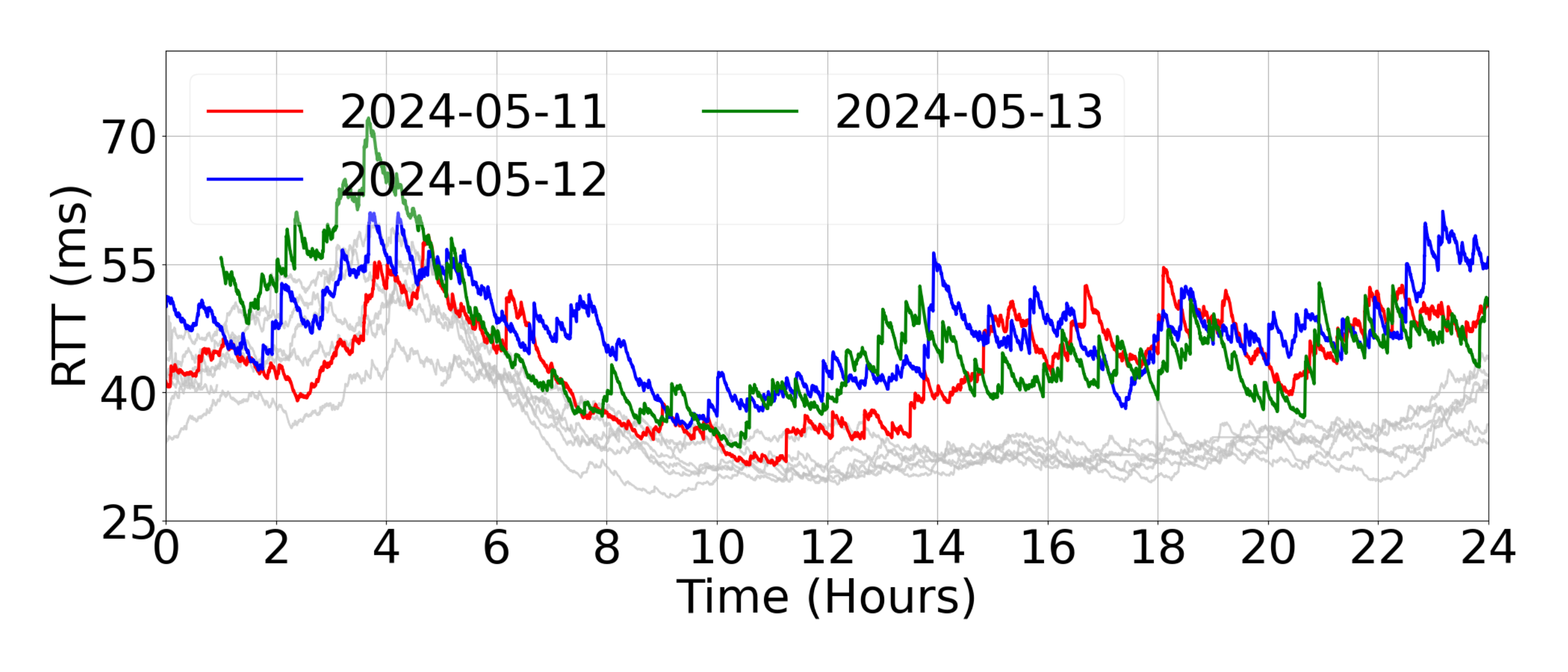}
        \caption{Vancouver, Canada}
    \end{subfigure}%
    \hfill
    \begin{subfigure}[t]{0.50\columnwidth}
        \centering
        \includegraphics[height=3cm, keepaspectratio]{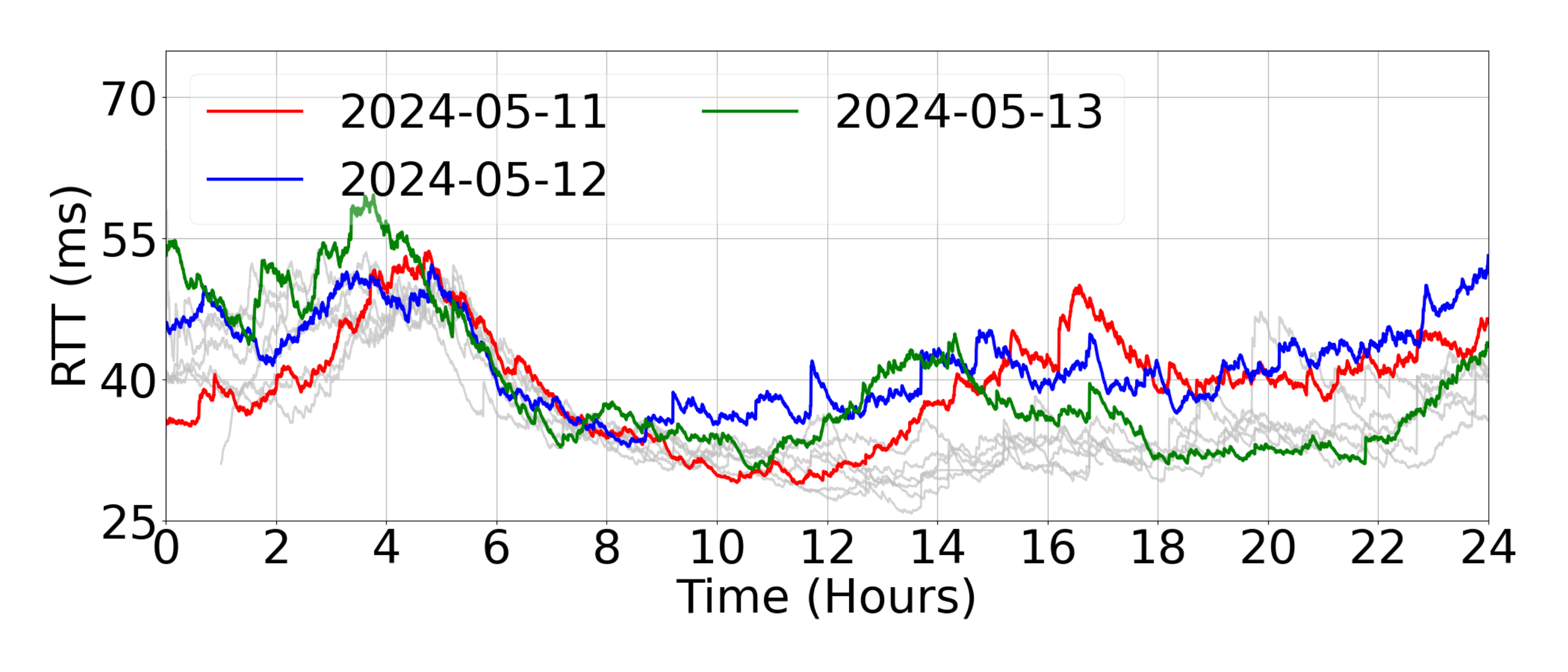}
        \caption{Victoria, Canada}
    \end{subfigure}%
    \hfill
    \begin{subfigure}[t]{0.50\columnwidth}
        \centering
        \includegraphics[height=3cm, keepaspectratio]{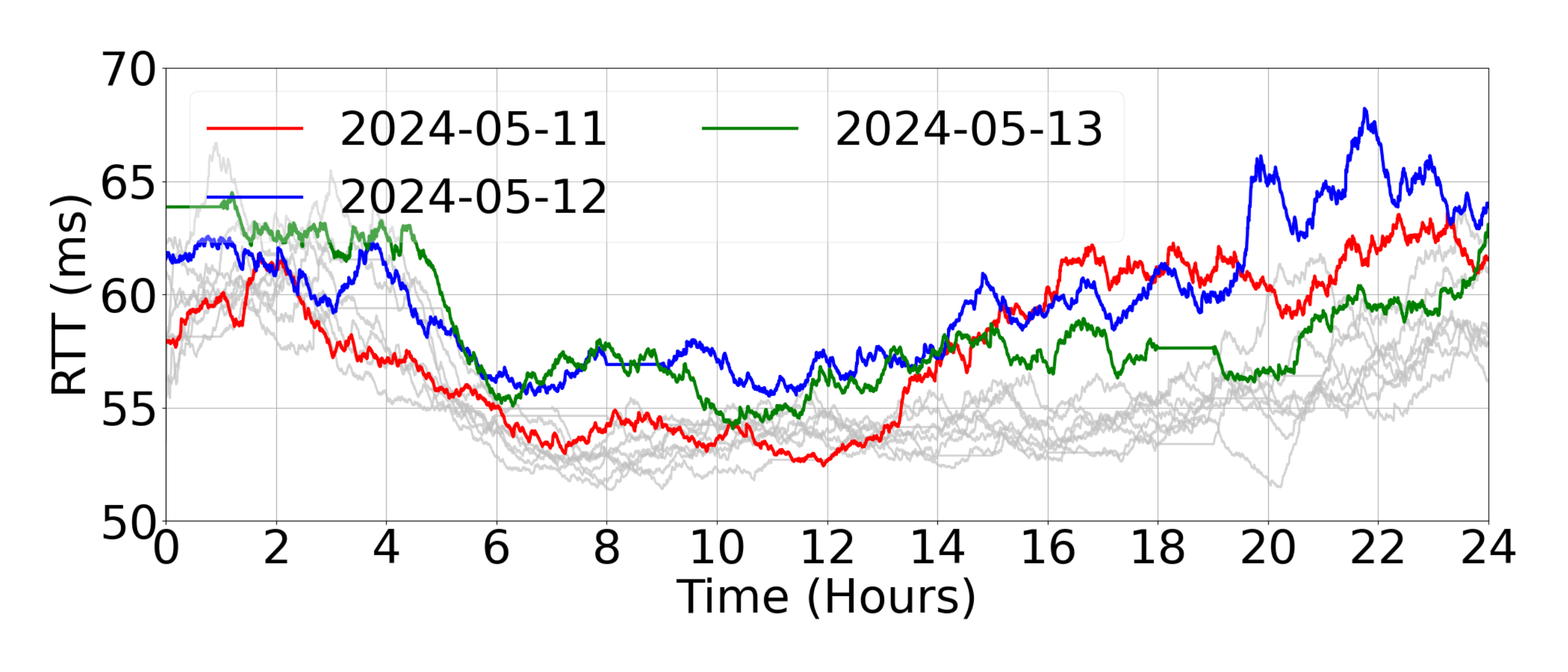}
        \caption{Denver, US}
    \end{subfigure}%
    \hfill
    \begin{subfigure}[t]{0.50\columnwidth}
        \centering
        \includegraphics[height=3cm, keepaspectratio]{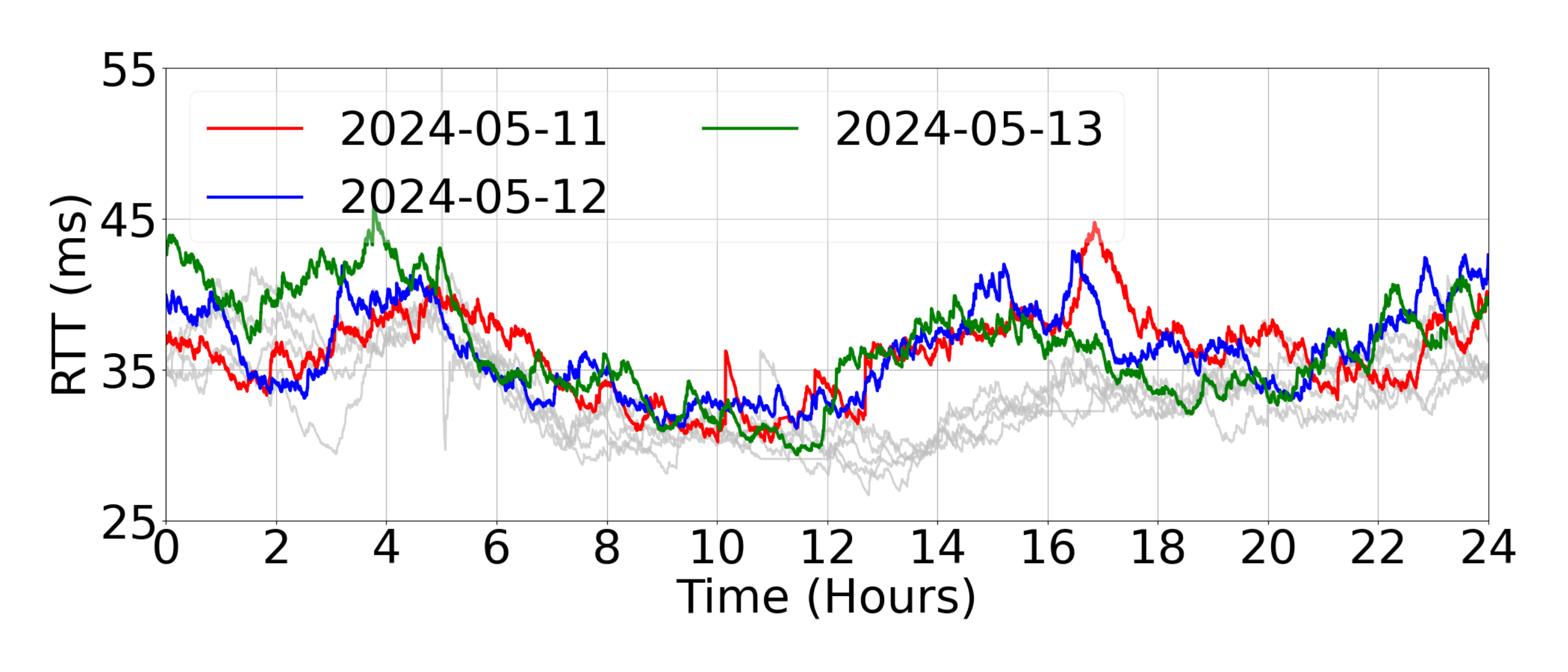}
        \caption{Seattle UT-1, US}
    \end{subfigure}%
    \hfill
    \begin{subfigure}[t]{0.50\columnwidth}
        \centering
        \includegraphics[height=3cm, keepaspectratio]{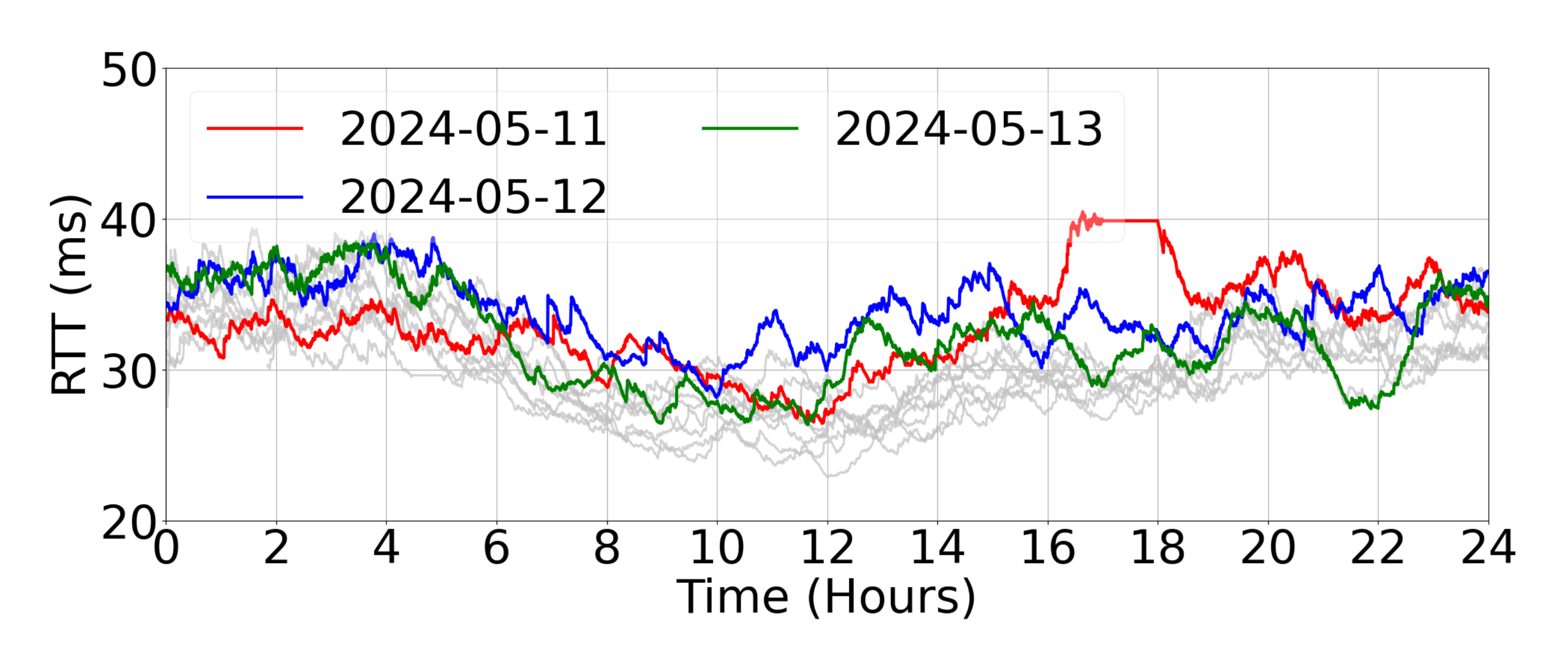}
        \caption{Seattle UT-2, US}
    \end{subfigure}%

    \caption{Zooming into day-wise latency pattern before the solar storm (gray lines) and during the solar superstorm (colored line) reveals that latency started increasing at (a) 8th hour, (b)-(c) 12th hour, (d) 14th hour, (e) 12th hour, and (f) 12th hour on 11th May. In a few places, latency remains high for the next few days.}
    \label{fig:TSpingRTTdiurnalMay24}
\end{figure}

\begin{figure}
    \centering
    \begin{subfigure}[t]{0.50\columnwidth}
        \centering
        \includegraphics[height=3cm, keepaspectratio]{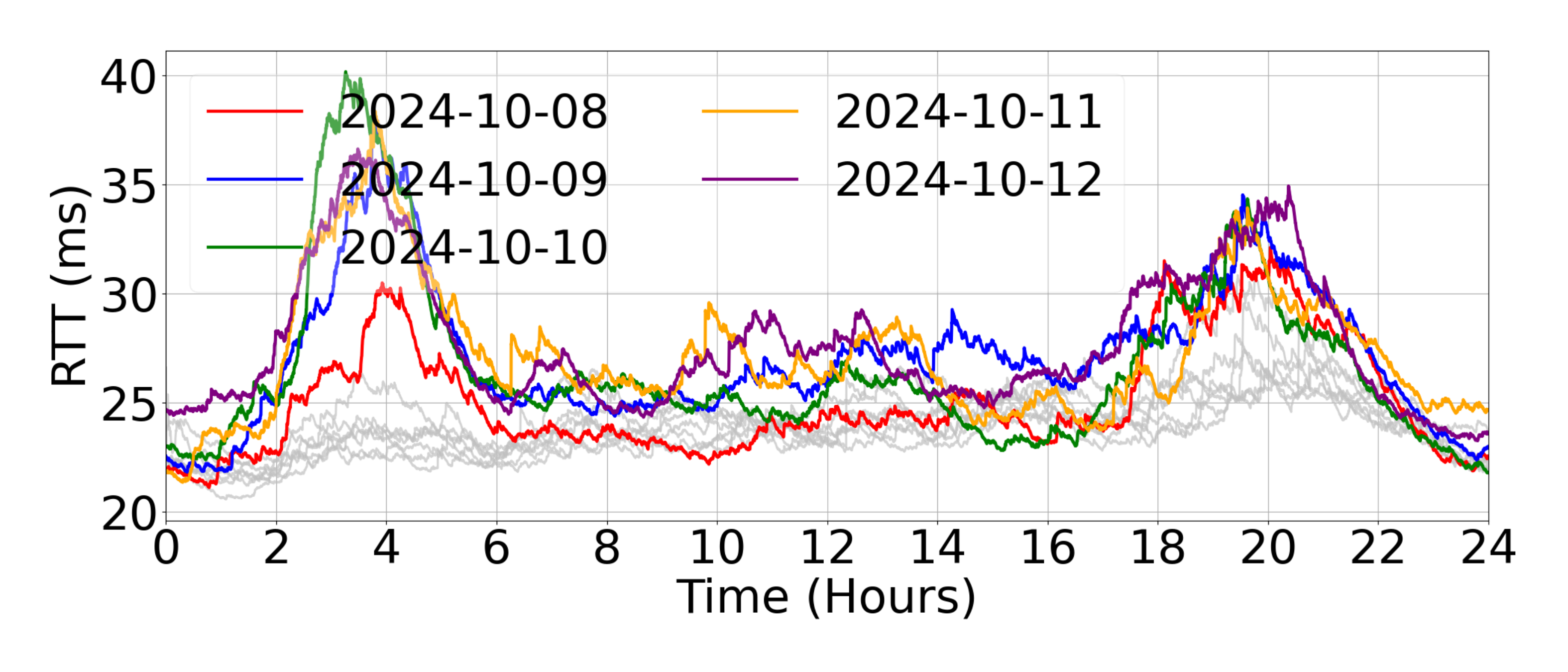}
        \caption{Bruhl, Germany}
    \end{subfigure}%
    \hfill
     \begin{subfigure}[t]{0.50\columnwidth}
        \centering
        \includegraphics[height=3cm, keepaspectratio]{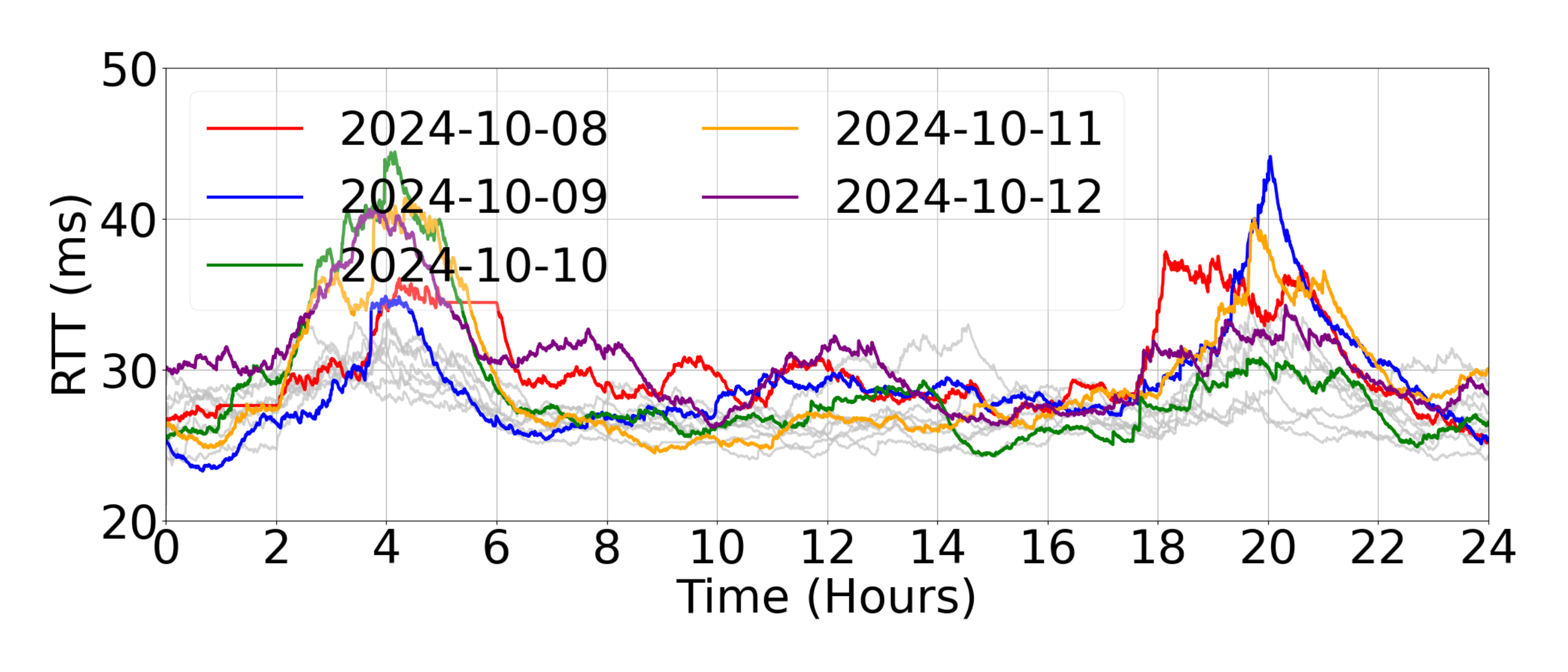}
        \caption{Frankfurt, Germany}
    \end{subfigure}%
    \hfill
    \begin{subfigure}[t]{0.50\columnwidth}
        \centering
        \includegraphics[height=3cm, keepaspectratio]{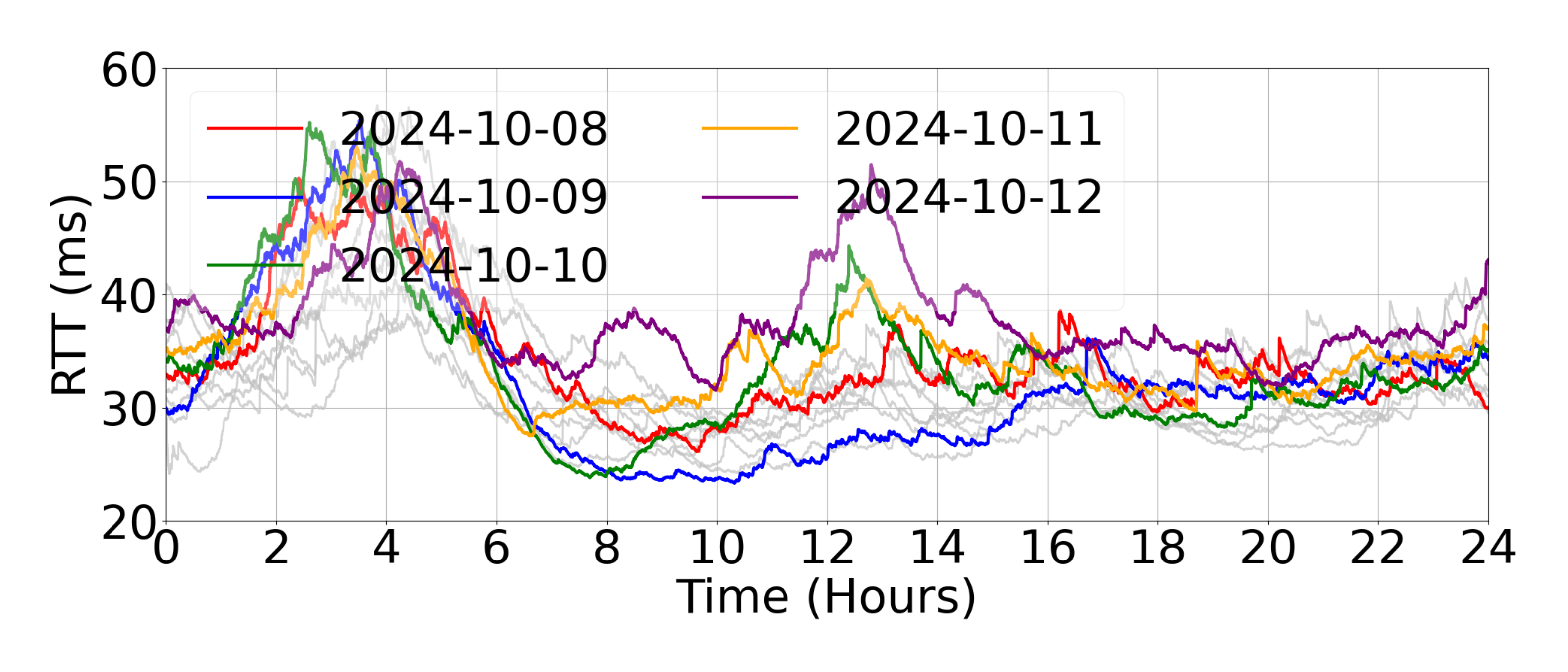}
        \caption{Victoria, Canada}
    \end{subfigure}%
    \hfill
    \begin{subfigure}[t]{0.50\columnwidth}
        \centering
        \includegraphics[height=3cm, keepaspectratio]{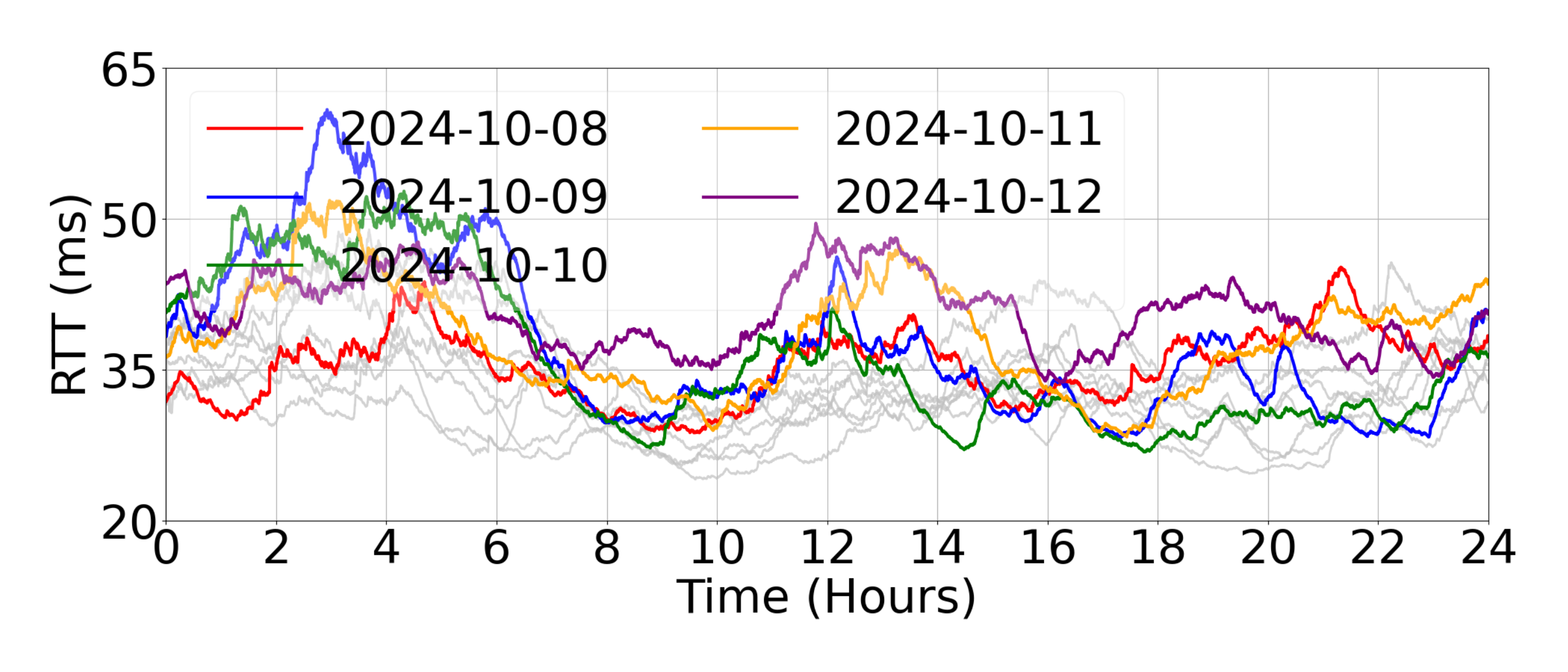}
        \caption{Calgary, Canada}
    \end{subfigure}%
    \hfill
    \begin{subfigure}[t]{0.50\columnwidth}
        \centering
        \includegraphics[height=3cm, keepaspectratio]{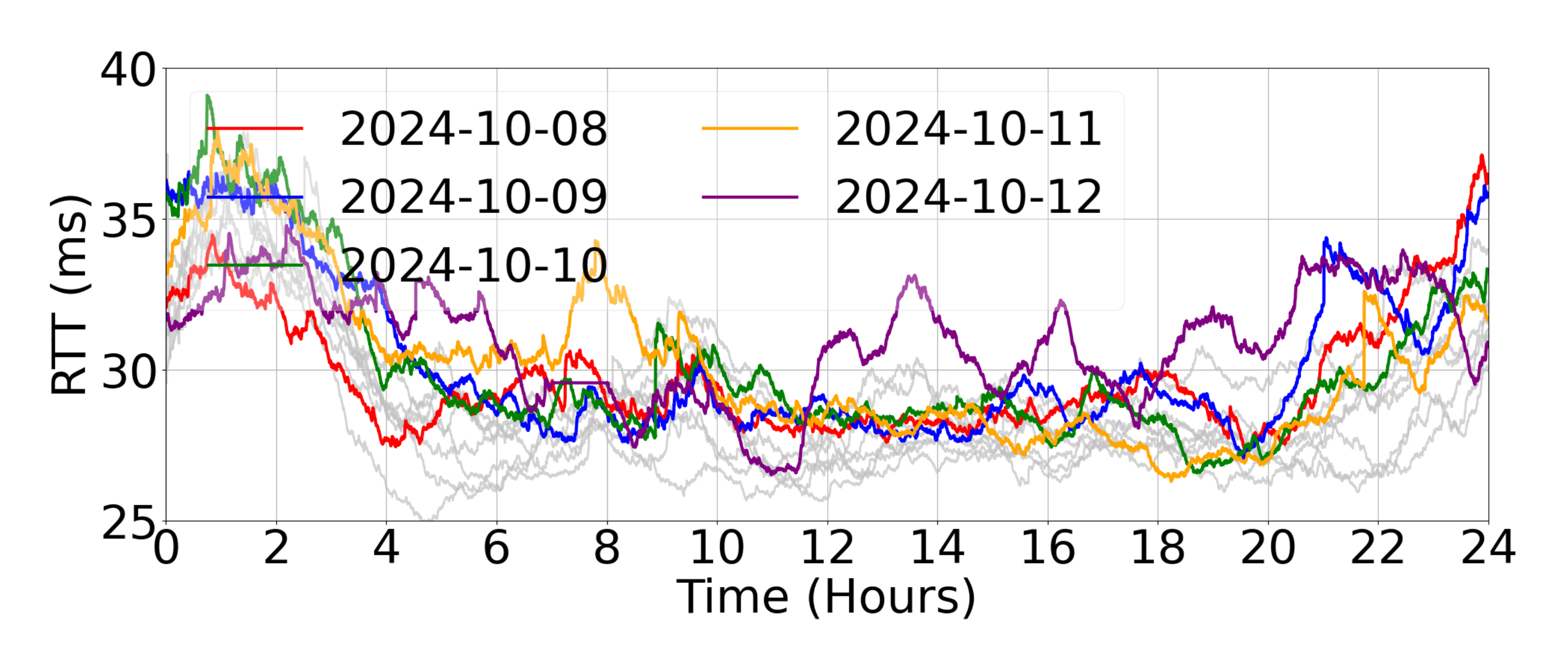}
        \caption{Ottawa IPv4, Canada}
    \end{subfigure}%
    \hfill
    \begin{subfigure}[t]{0.50\columnwidth}
        \centering
        \includegraphics[height=3cm, keepaspectratio]{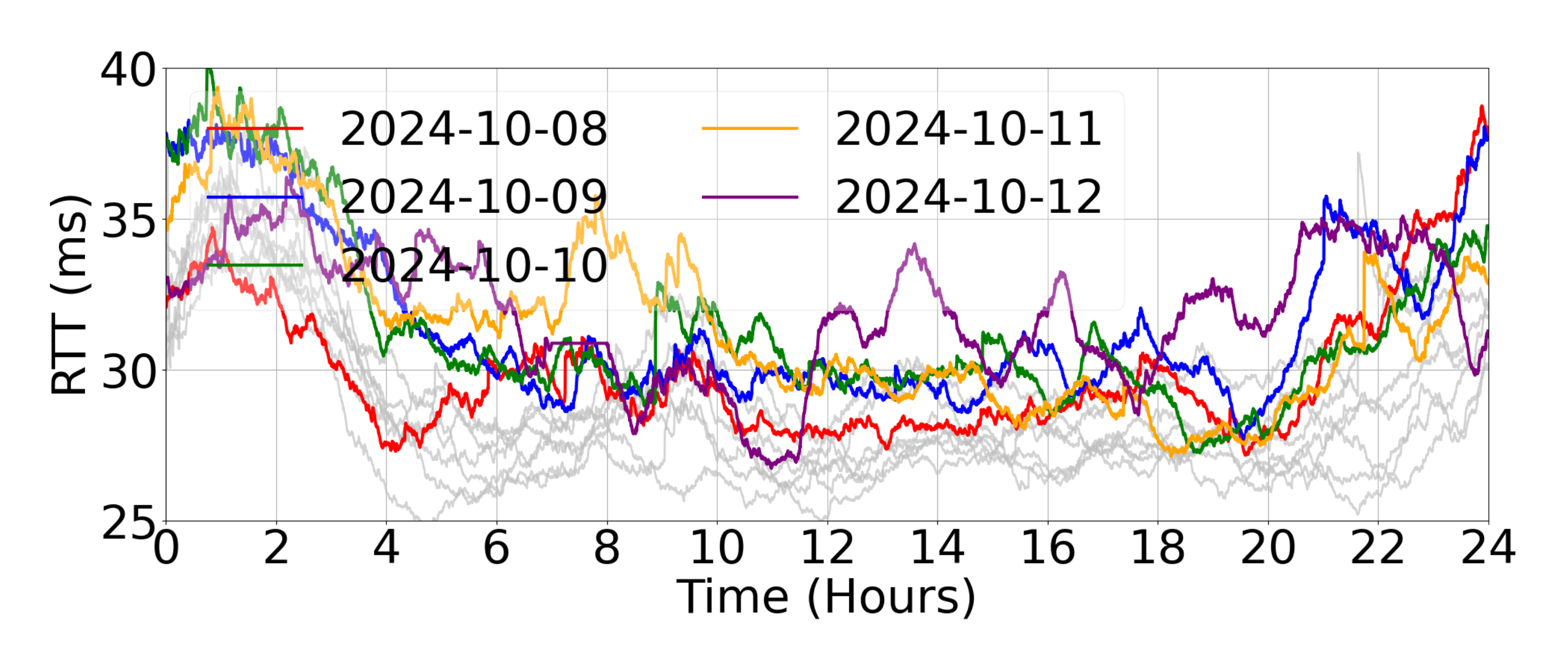}
        \caption{Ottawa IPv6, Canada}
    \end{subfigure}%
    \hfill
    \begin{subfigure}[t]{0.50\columnwidth}
        \centering
        \includegraphics[height=3cm, keepaspectratio]{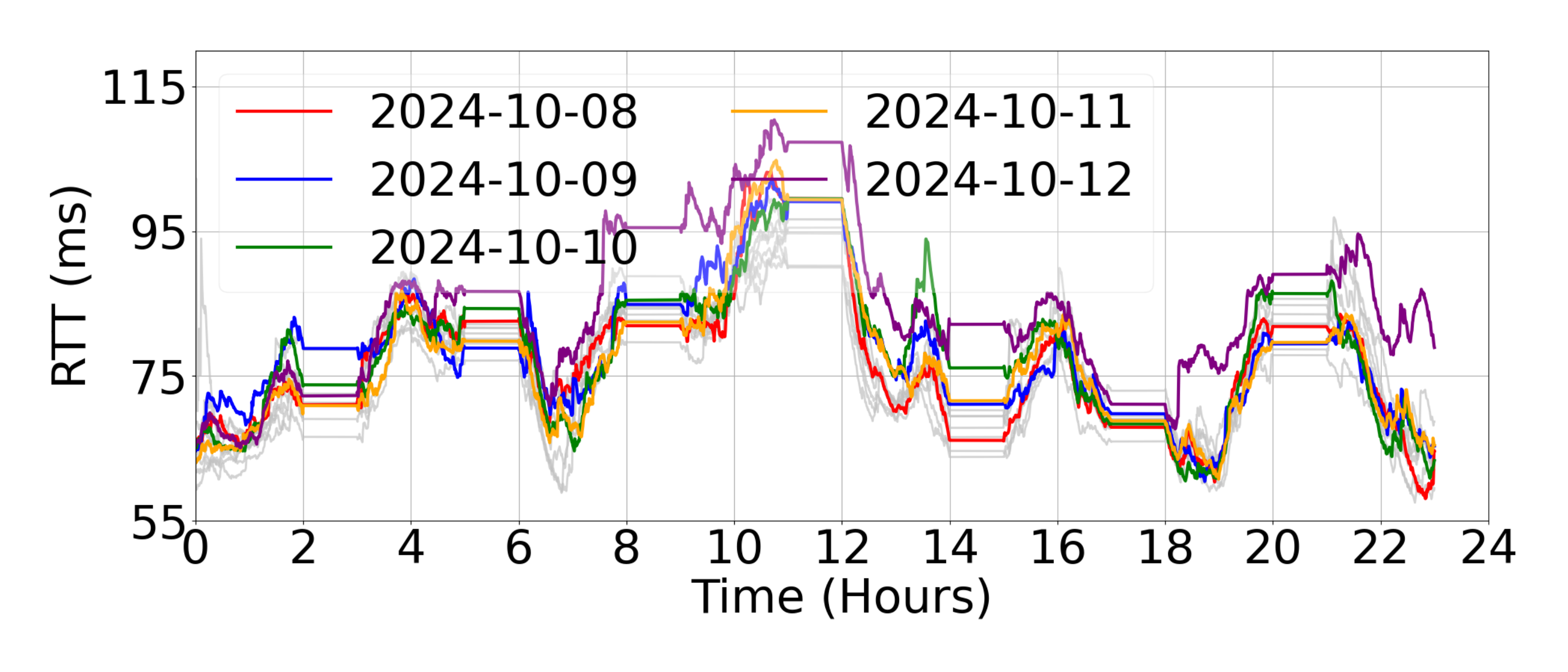}
        \caption{Ulukhaktok, Canada}
    \end{subfigure}%
    \hfill
    \begin{subfigure}[t]{0.50\columnwidth}
        \centering
        \includegraphics[height=3cm, keepaspectratio]{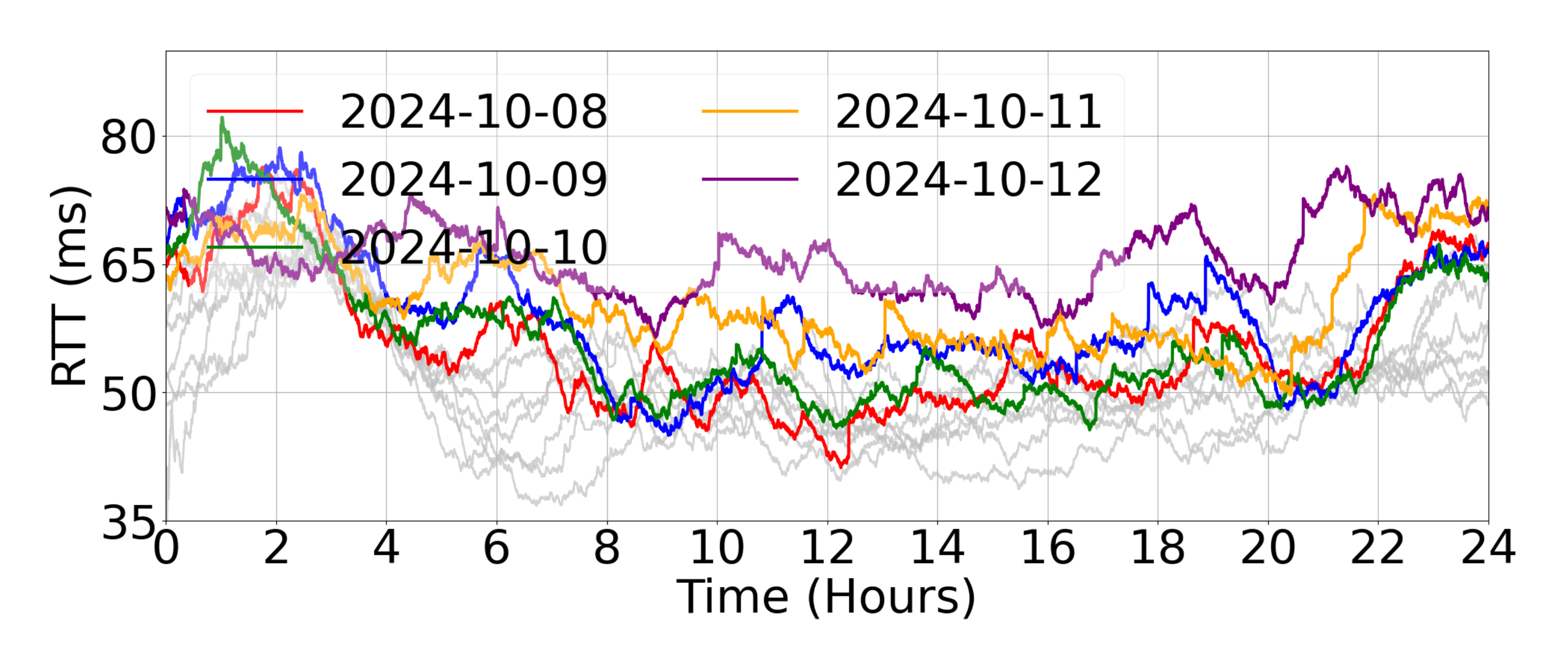}
        \caption{Dallas, US}
    \end{subfigure}%
    \hfill
    \begin{subfigure}[t]{0.50\columnwidth}
        \centering
        \includegraphics[height=3cm, keepaspectratio]{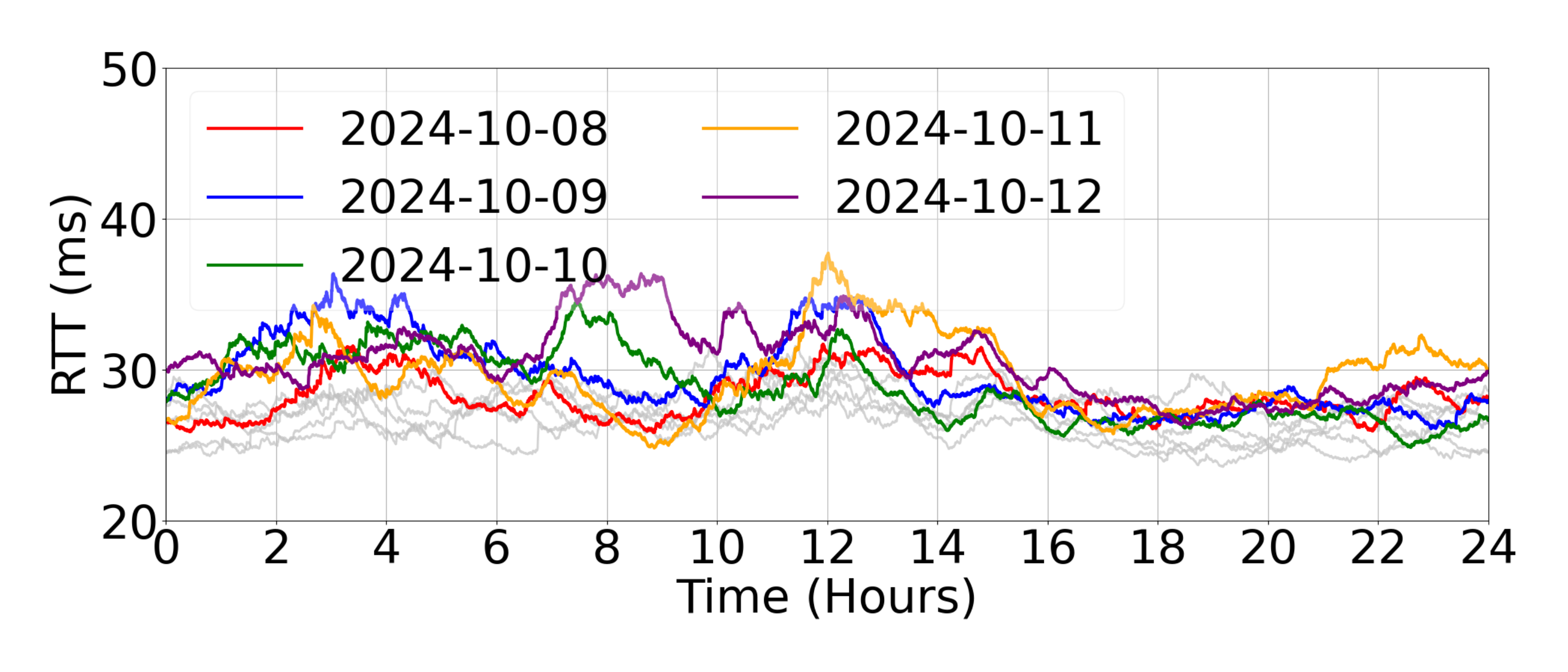}
        \caption{Salt Lake City, US}
    \end{subfigure}%
    \hfill
    \begin{subfigure}[t]{0.50\columnwidth}
        \centering
        \includegraphics[height=3cm, keepaspectratio]{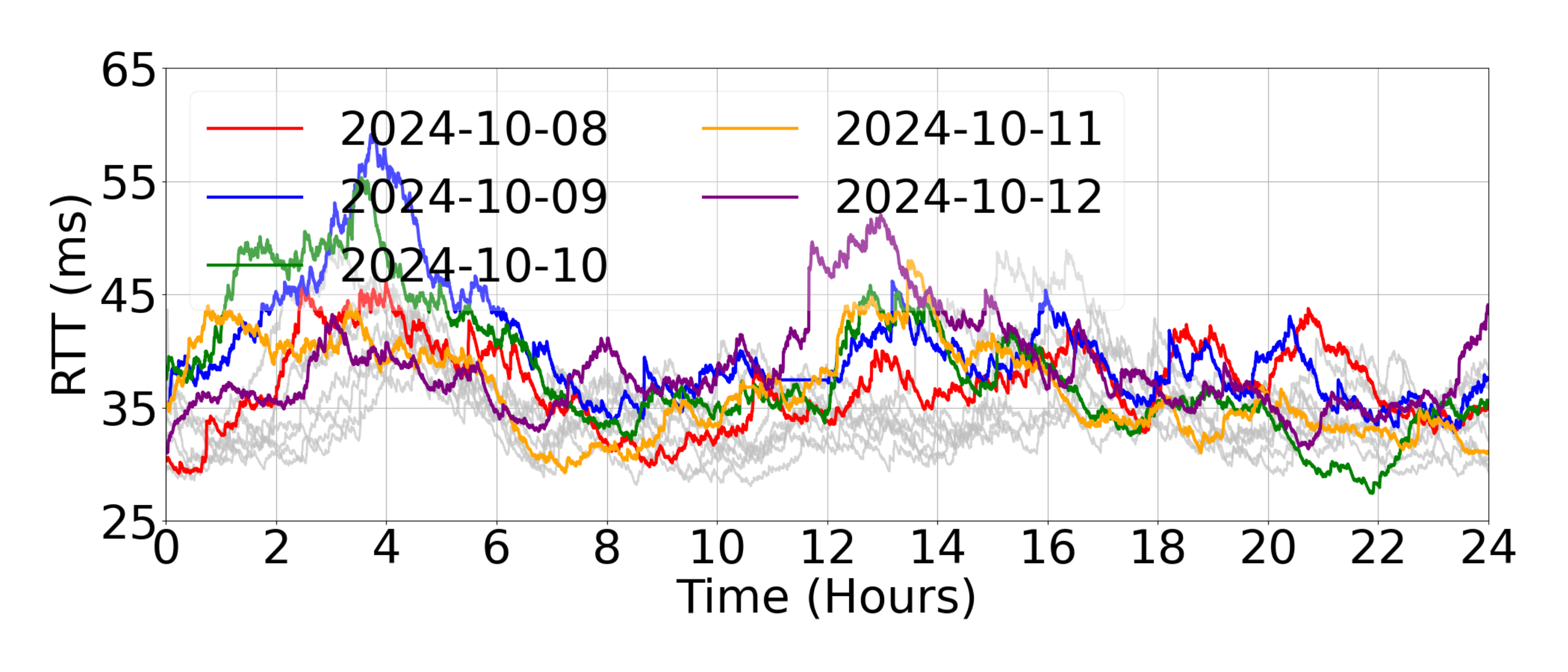}
        \caption{Seattle, US}
    \end{subfigure}%

    \caption{Zooming into day-wise latency pattern before the solar storm (gray lines) and during the solar October storm (colored line) reveals that latency spikes in broader window view are hours-long inflated latency in diurnal trends.}
    \label{fig:TSpingRTTdiurnalOct24}
\end{figure}

A common observation during the May 2024 solar storms is a subtle upshift and distortion of the diurnal cycle.
During the October 2024 solar storms, a short-lived latency spike occurred in certain areas of Germany and Canada. 
So, here we zoom in on that diurnal latency pattern from all these vantage points to observe the nature of the anomaly compared to pre-storm times.
For that, we plot the day-wise EWMA of the mean RTT per second in Fig.~\ref{fig:TSpingRTTdiurnalMay24} from 3rd to 13th May 2024.
In these figures, each line represents one day, with the X-axis showing time (24 hours) and the Y-axis showing RTT.
All the gray lines are the baseline from 3rd to 10th May, while the colored lines highlight the solar superstorm days.
Notice in Frankfurt, the latency on 11th May starts increasing at the 8th hour, reaching 50\% above the baseline over the next 4 hours.
The latency remains consistently high over the next few days.
Whereas latency from two vantage points in Canada, Vancouver and Victoria, shows latency inflation starting almost simultaneously, at 12th hour on 11th May.
However, unlike Frankfurt, we see latency inflation persist only during the hours when RTT was minimal on pre-storm days.
Notice 12th to 24th hours in Vancouver and 12th to 18th in Victoria.
Also, the scale of latency inflation in Vancouver is larger, up to 70\% over the baseline, while in Victoria it is up to 40\% over the baseline.
In Seattle, US, we see similar characteristics.
Inflation in latency after the 12th hour, when latency was minimal on pre-storm days.
The window of inflation is much shorter in Seattle, UT-1, from the 12th to 18th hours on 11th-12th May, from the 12th to 16th hours on 13th May.
In Seattle, UT-2, inflation is not as strong, except for a 2-hour span from 16th to 18th on 11th May.

In Fig.~\ref{fig:TSpingRTTdiurnalOct24} we show the diurnal latency pattern during the October 2024 storms.
Note that here we use the pre-storm window from 1st to 7th October as the baseline, represented by the gray lines.
The window from 8th to 12th October of storm time in the color line, accounting for increased satellite decay and rise (Fig.~\ref{fig:StarlinkFleetManagement}) after both G2- and G4-class solar storms.
Notice the similar diurnal latency pattern from two vantage points in Germany.
These two latency spikes per day we saw previously in Fig.~\ref{fig:TSpingRTTOct24}(a)-(b) are 4-hour-long (2nd to 6th and 18th to 22nd hours) inflation in latency on each day.
This starts on 8th October, after the G2-class solar storm, and on 10th October reaches a maximum of 60\% in Bruhl and 50\% in Frankfurt, above the pre-storm baseline during a G4-class storm. 
In contrast, the diurnal latency pattern from five vantage points in Canada shows no significant anomalies during the G2-class solar storm.
Because the latency spikes at Victoria and Calgary started on 5th May, the pre-storm and post-storm latency from 2nd to 6th hours overlap in these figures. 
However, we observe another latency inflation window between the 12th and 14th hours on 12th October, following the G4-class solar storm across all locations in Canada.
In Ulukhaktok, notice multiple 1-hour+ long straight lines at the same time every day. 
We speculate this is due to missing observations.
A reasonable explanation seems to be an outage caused by Starlink's limited coverage in the Arctic region.
In the US, we do not see a strong impact from the G2-class solar storm, except in Seattle, where it is strongest from the 2nd hour to the 5th hour on the 9th and 10th of May. 
On 12th October, we observe short-lived latency inflation between the 6th and 10th hours in Salt Lake City and between the 12th and 14th hours in Seattle, similar to what was observed in Canada.
Dallas is an exception. 
On 12th October, after 4th hours, latency increased and remained high for the next few days.

\subsubsection{Quantifying latency inflation:}

\begin{figure}
    \centering
    \begin{subfigure}[t]{0.30\columnwidth}
        \centering
        \includegraphics[height=2.9cm, keepaspectratio]{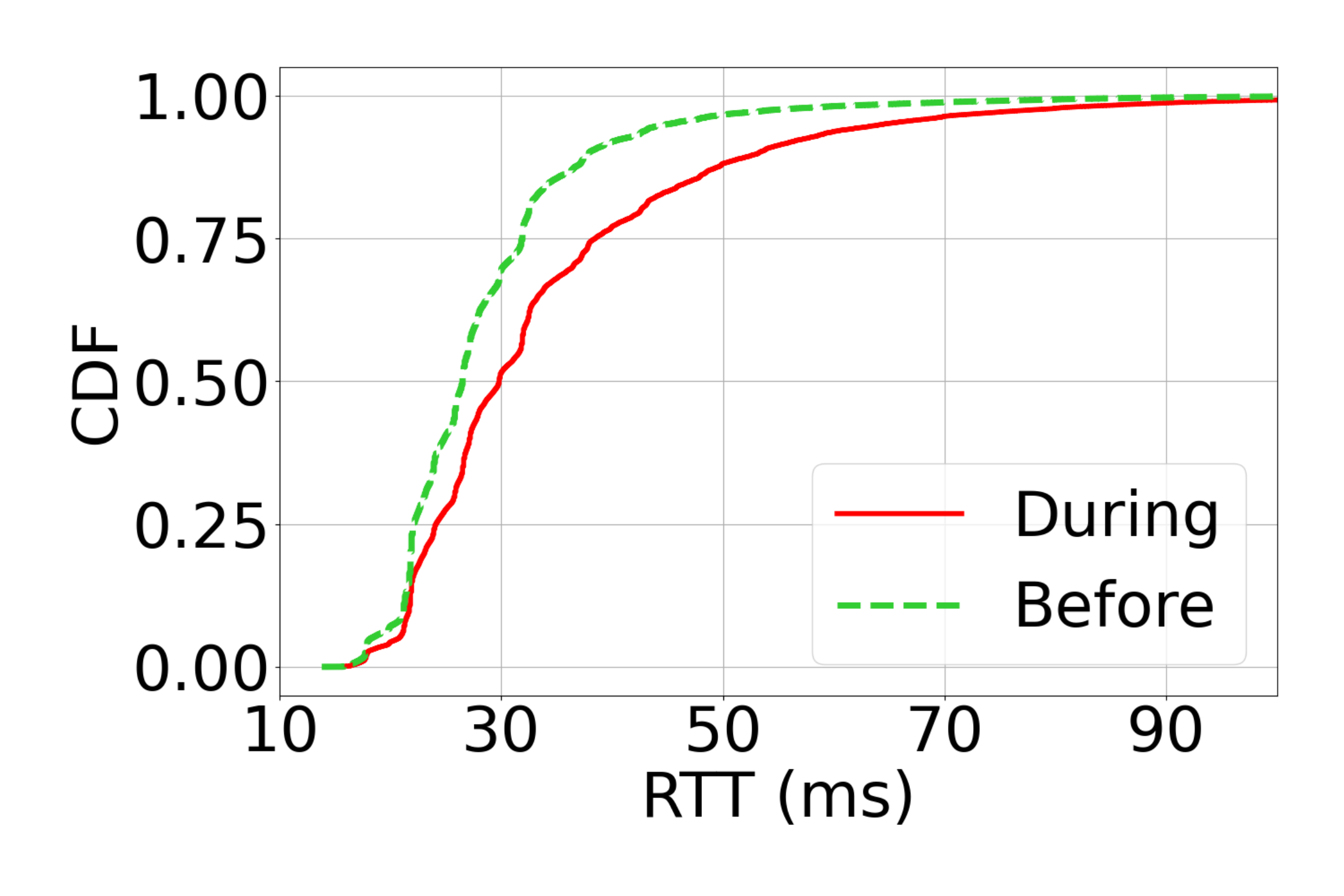}
        \caption{Frankfurt, Germany}
    \end{subfigure}%
    \hfill
    \begin{subfigure}[t]{0.30\columnwidth}
        \centering
        \includegraphics[height=2.9cm, keepaspectratio]{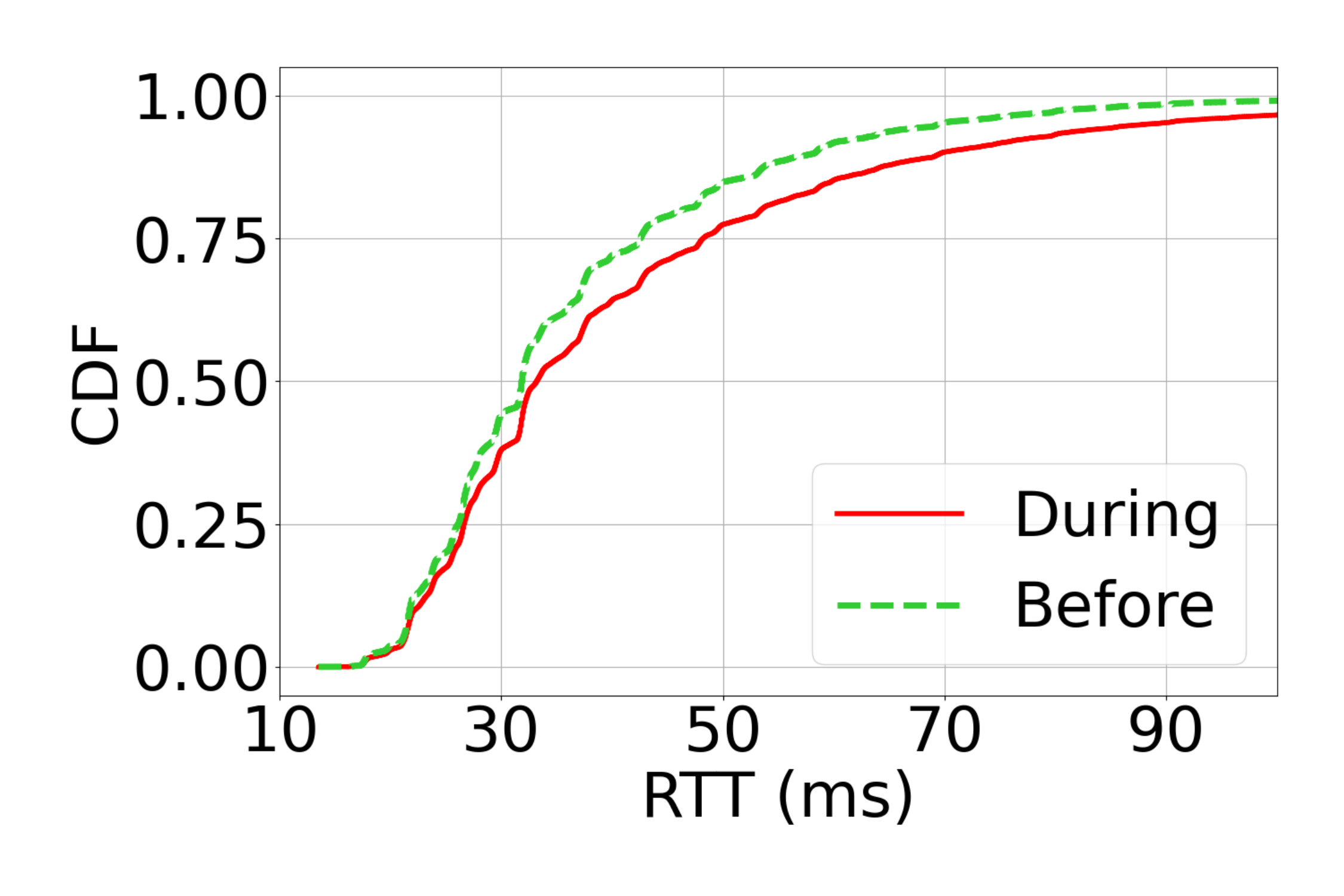}
        \caption{Vancouver, Canada}
    \end{subfigure}%
    \hfill
    \begin{subfigure}[t]{0.30\columnwidth}
        \centering
        \includegraphics[height=2.9cm, keepaspectratio]{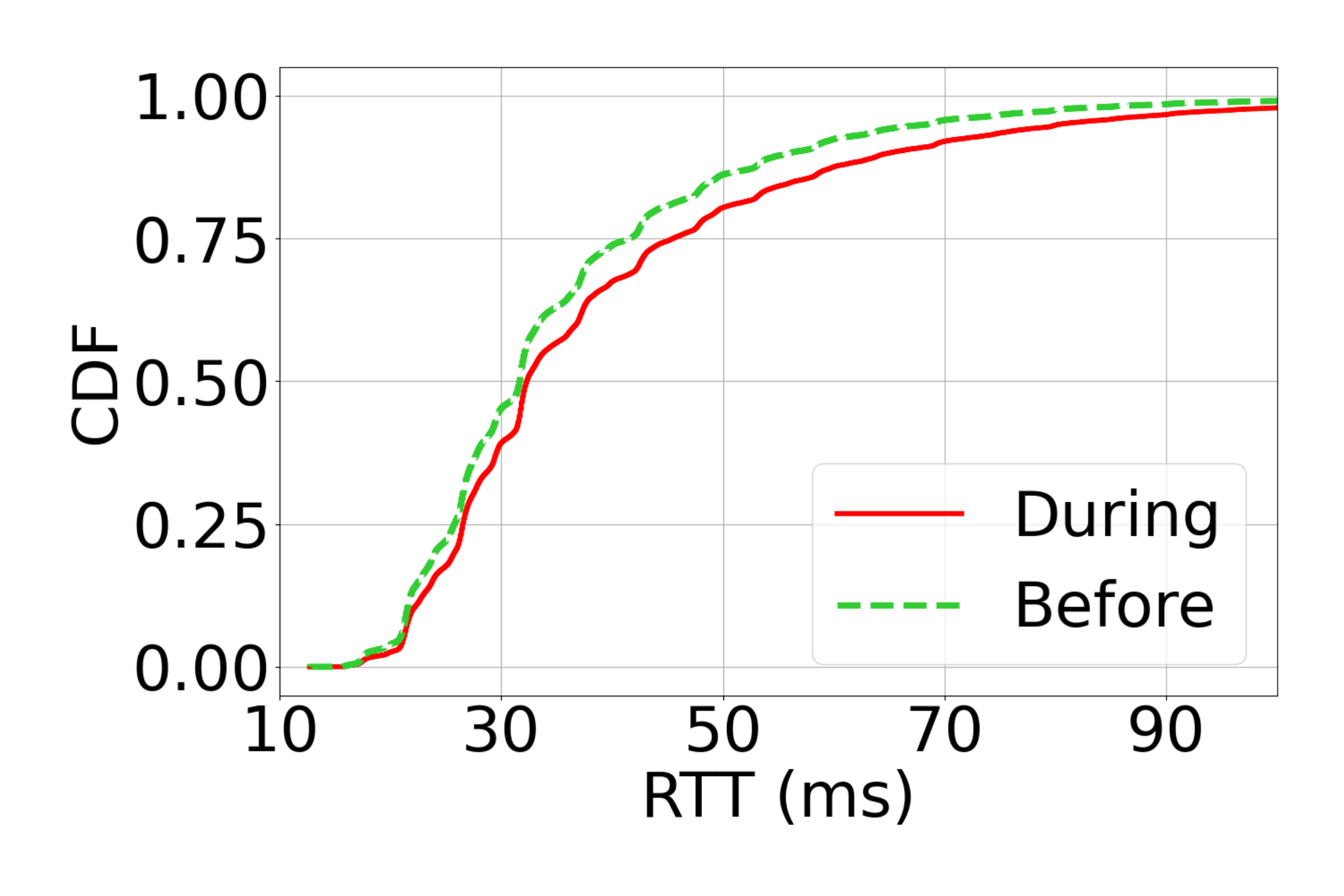}
        \caption{Victoria, Canada}
    \end{subfigure}%
    \hfill
    \begin{subfigure}[t]{0.30\columnwidth}
        \centering
        \includegraphics[height=2.9cm, keepaspectratio]{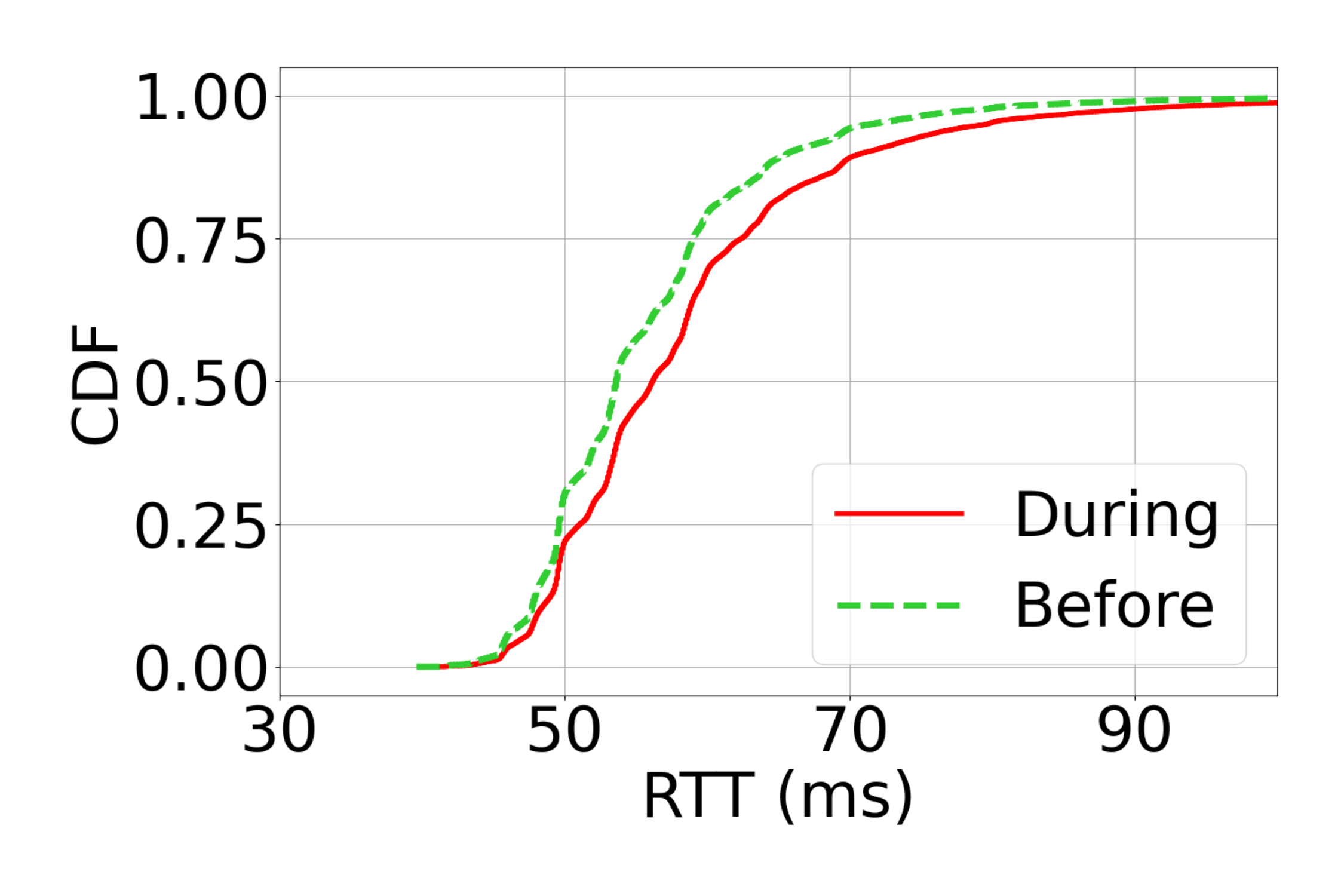}
        \caption{Denver, US}
    \end{subfigure}%
    \hfill
    \begin{subfigure}[t]{0.30\columnwidth}
        \centering
        \includegraphics[height=2.9cm, keepaspectratio]{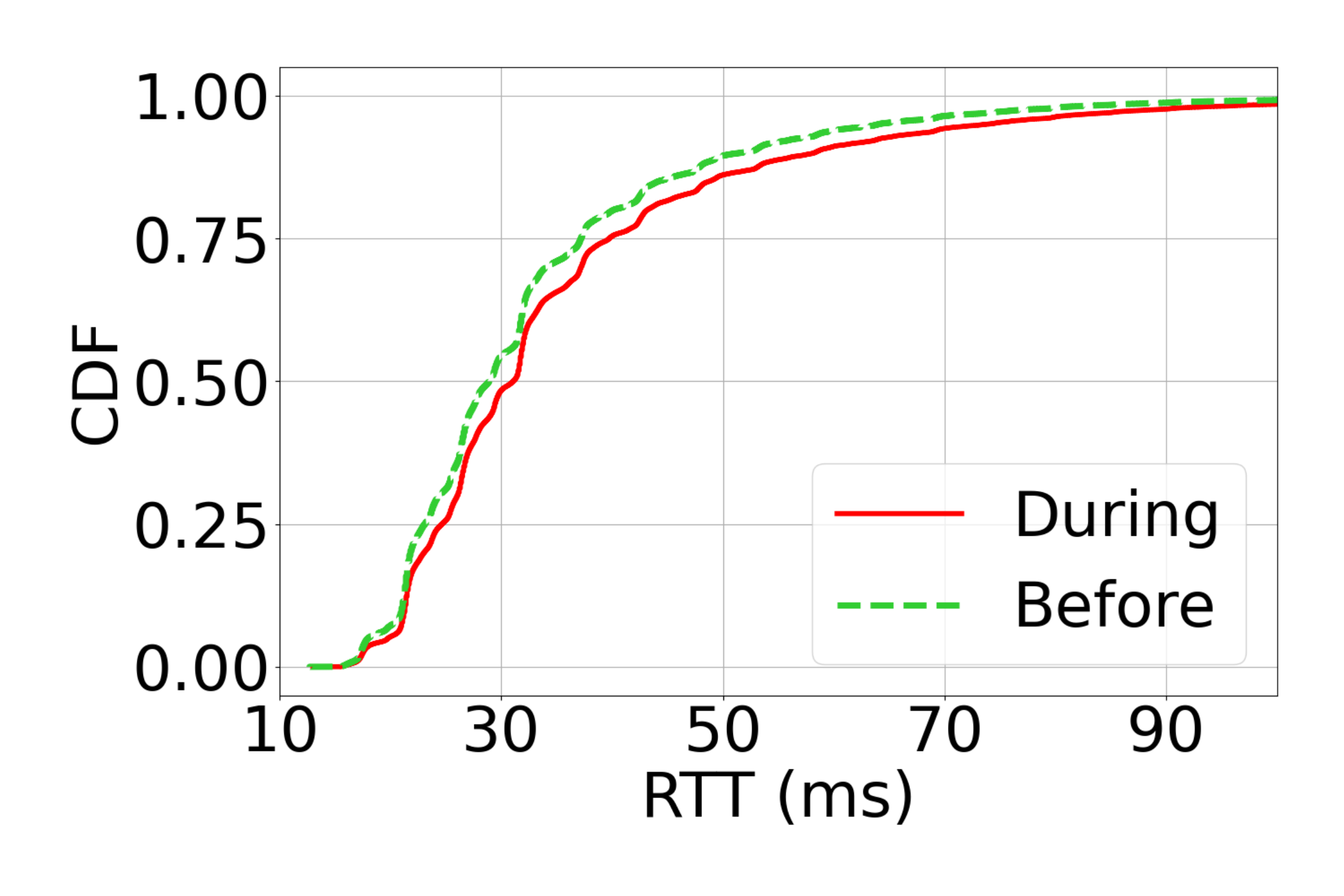}
        \caption{Seattle, UT-1, US}
    \end{subfigure}%
    \hfill
    \begin{subfigure}[t]{0.30\columnwidth}
        \centering
        \includegraphics[height=2.9cm, keepaspectratio]{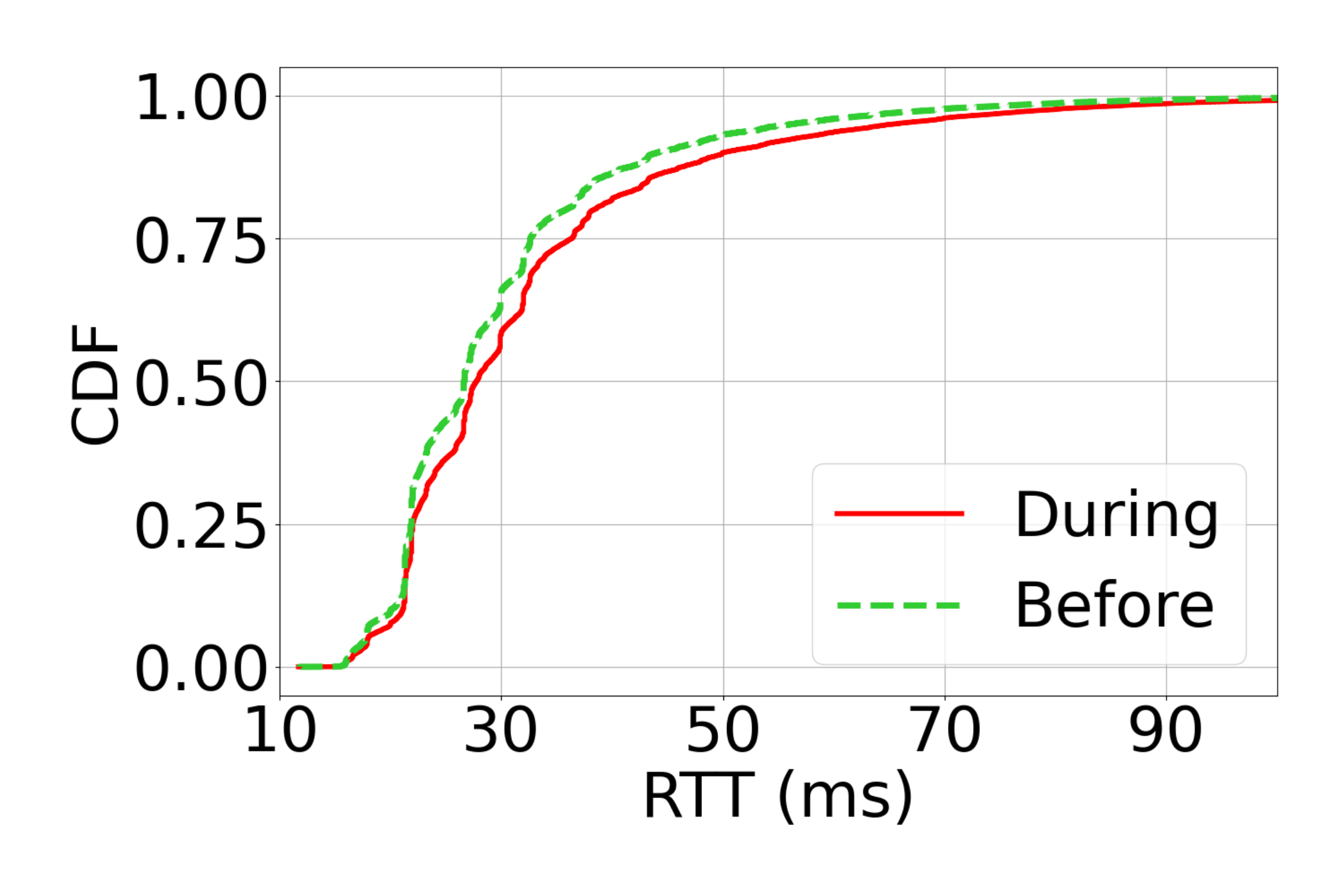}
        \caption{Seattle, UT-2, US}
    \end{subfigure}%

    \caption{Comparing the pre-storm (green dashed line) and during solar superstorm (red solid line) latency distribution, showing a notable right shift in latency distribution during the event across all the vantage points.}
    \label{fig:CDFpingRTTMay24}
\end{figure}

\begin{figure}
    \centering
    \begin{subfigure}[t]{0.30\columnwidth}
        \centering
        \includegraphics[height=2.9cm, keepaspectratio]{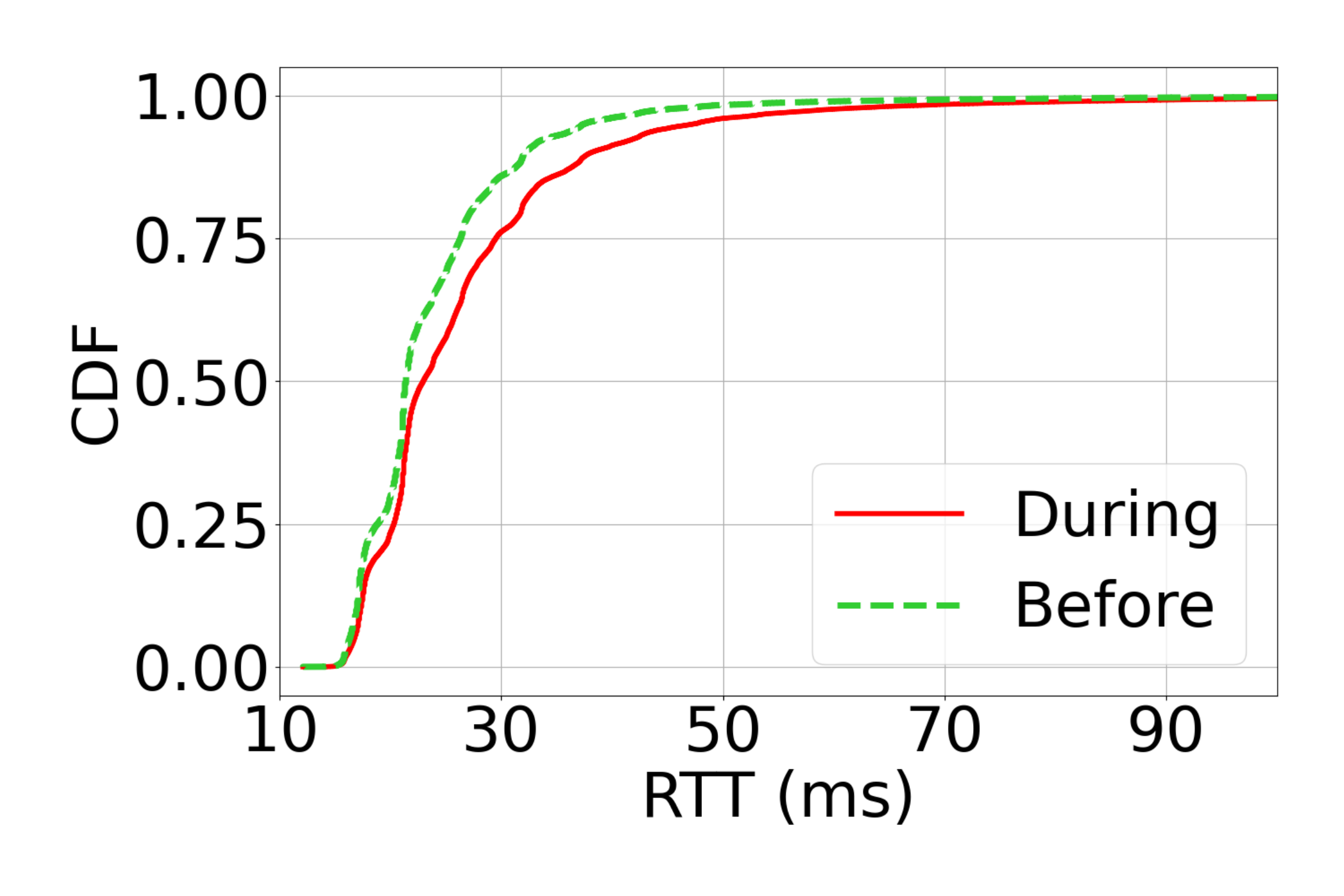}
        \caption{Bruhl, Germany}
    \end{subfigure}%
    \hfill
    \begin{subfigure}[t]{0.30\columnwidth}
        \centering
        \includegraphics[height=2.9cm, keepaspectratio]{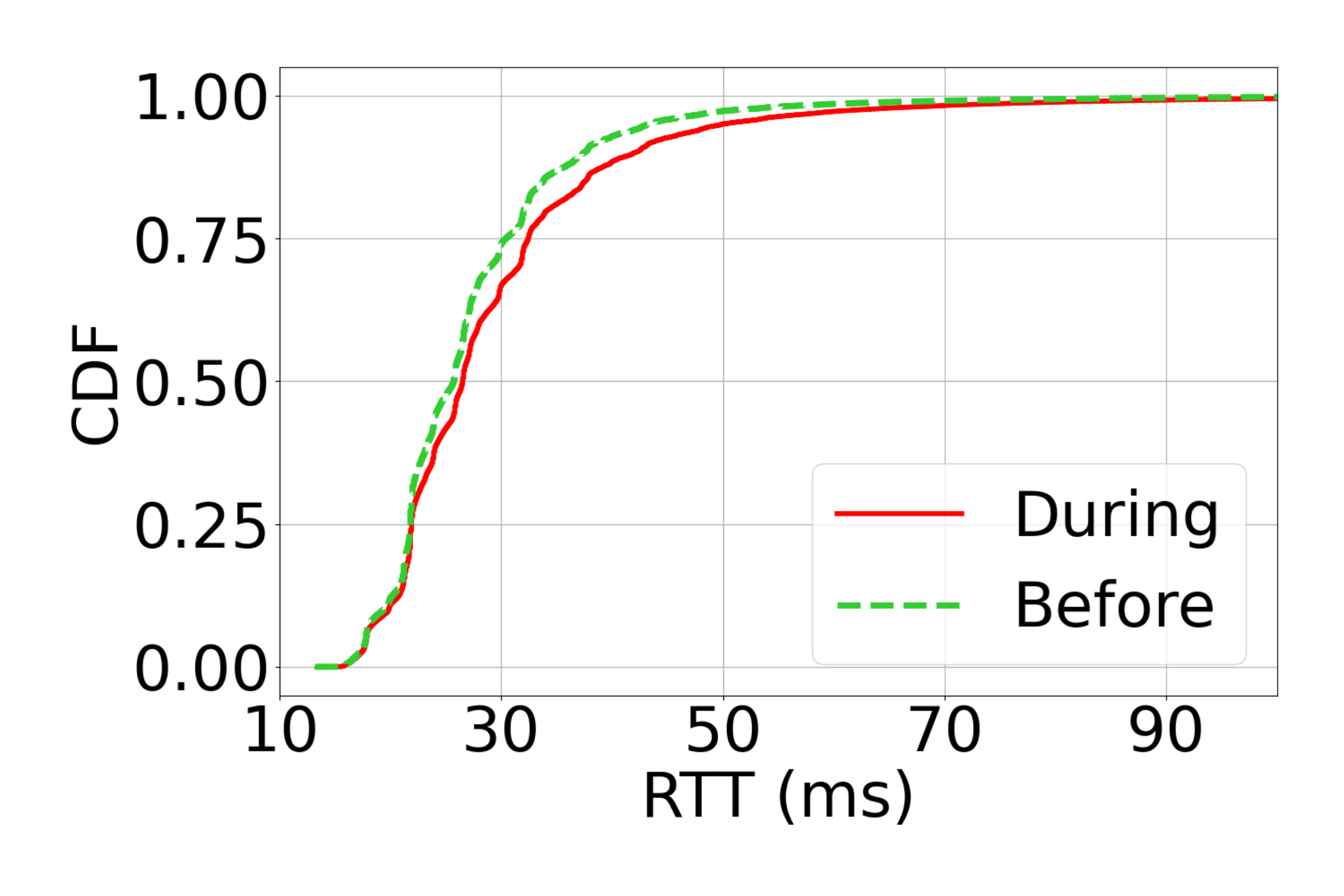}
        \caption{Frankfurt, Germany}
    \end{subfigure}%
    \hfill
    \begin{subfigure}[t]{0.30\columnwidth}
        \centering
        \includegraphics[height=2.9cm, keepaspectratio]{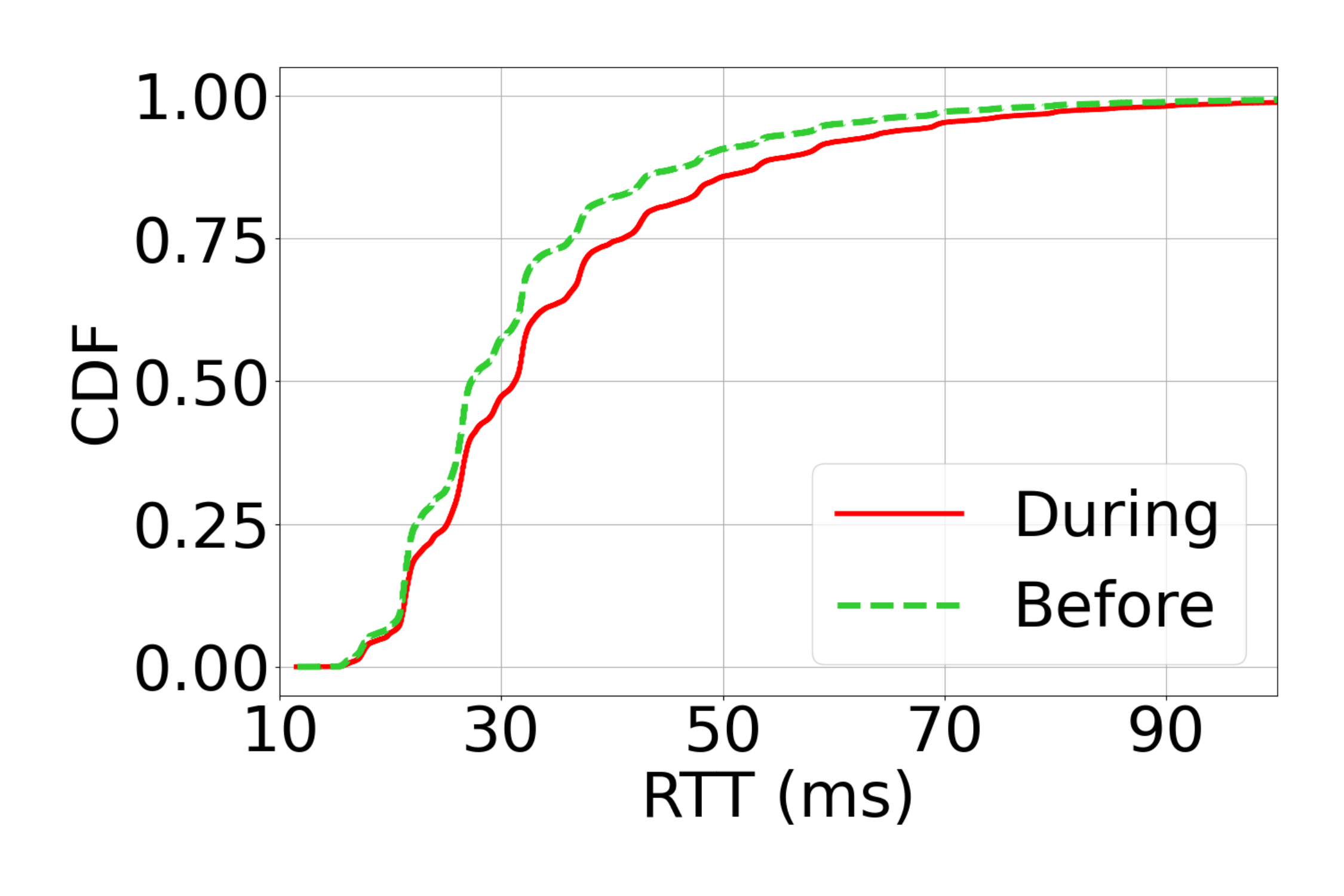}
        \caption{Victoria, Canada}
    \end{subfigure}%
    \hfill
    \begin{subfigure}[t]{0.30\columnwidth}
        \centering
        \includegraphics[height=2.9cm, keepaspectratio]{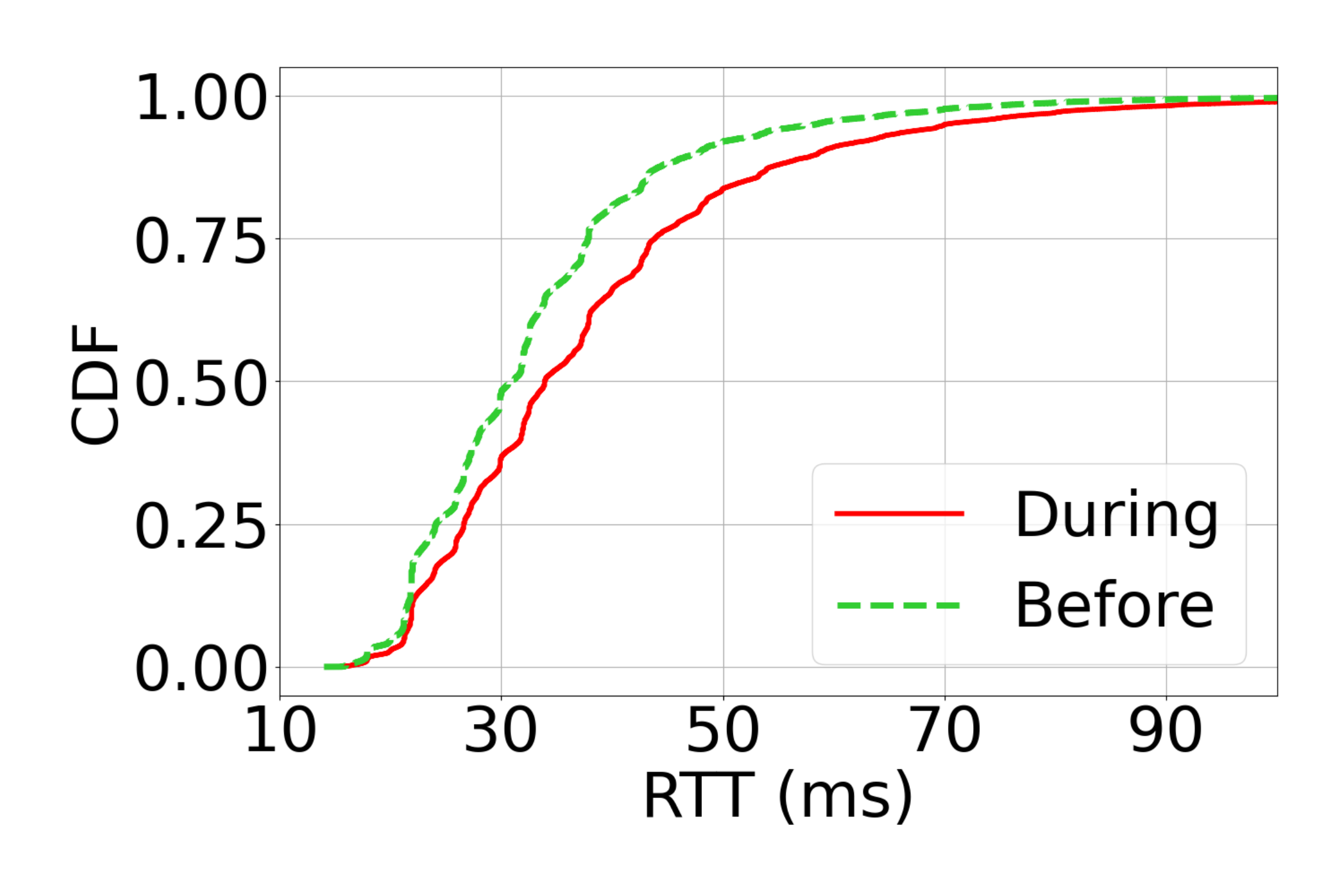}
        \caption{Calgary, Canada}
    \end{subfigure}%
    \hfill
    \begin{subfigure}[t]{0.30\columnwidth}
        \centering
        \includegraphics[height=2.9cm, keepaspectratio]{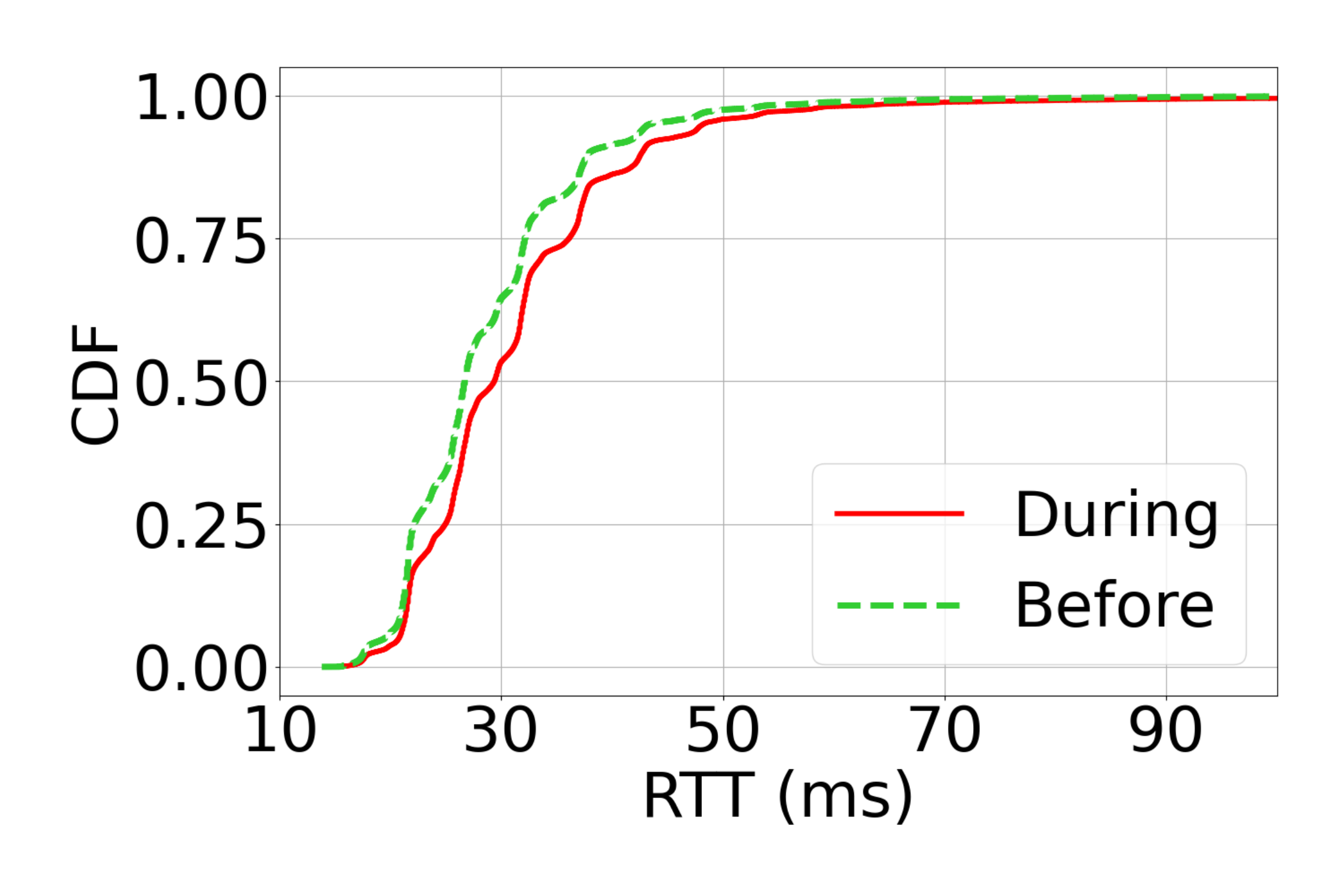}
        \caption{Ottawa IPv6, Canada}
    \end{subfigure}%
    \hfill
    \begin{subfigure}[t]{0.30\columnwidth}
        \centering
        \includegraphics[height=2.9cm, keepaspectratio]{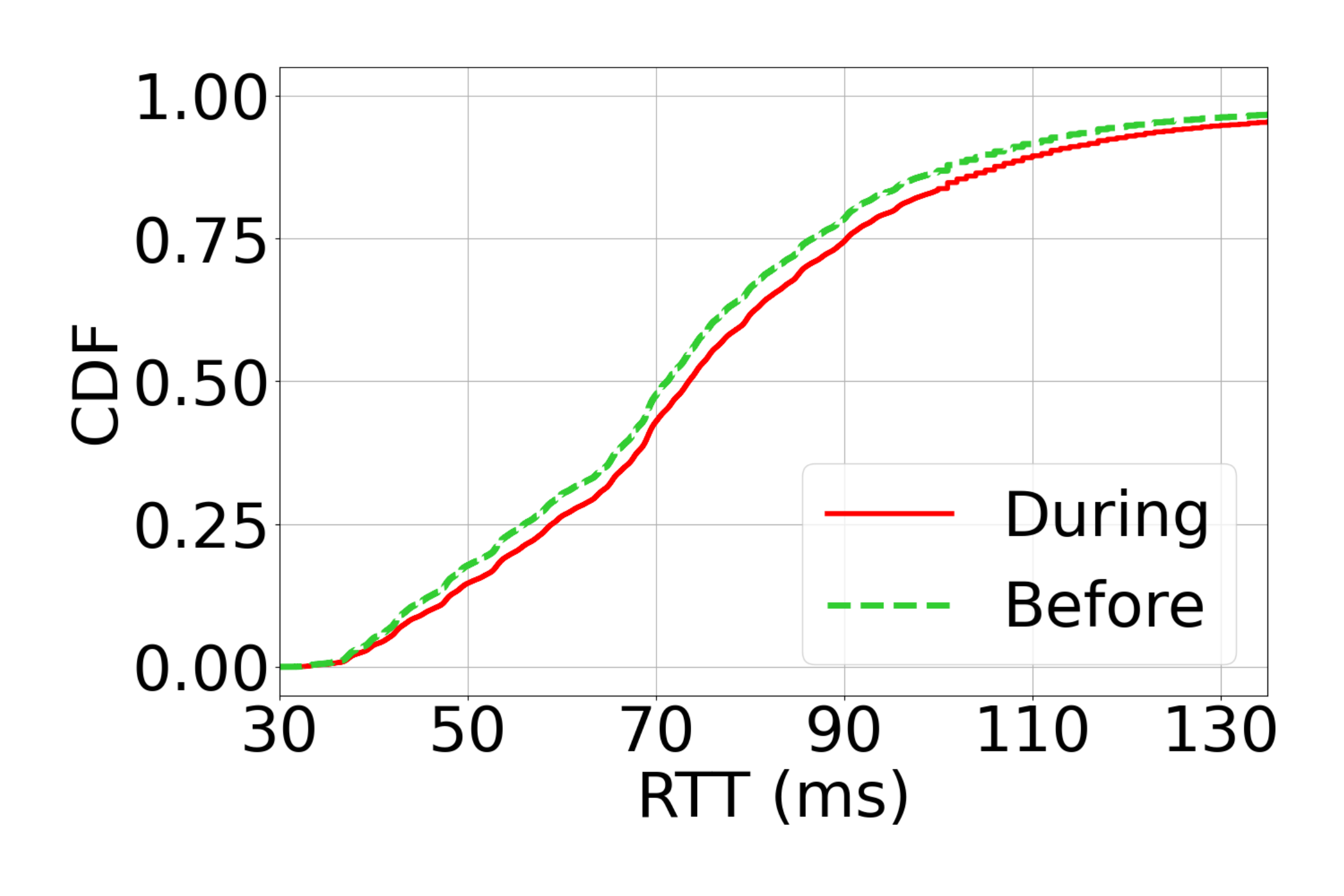}
        \caption{Ulukhaktok, Canada}
    \end{subfigure}%
    \hfill
    \begin{subfigure}[t]{0.30\columnwidth}
        \centering
        \includegraphics[height=2.9cm, keepaspectratio]{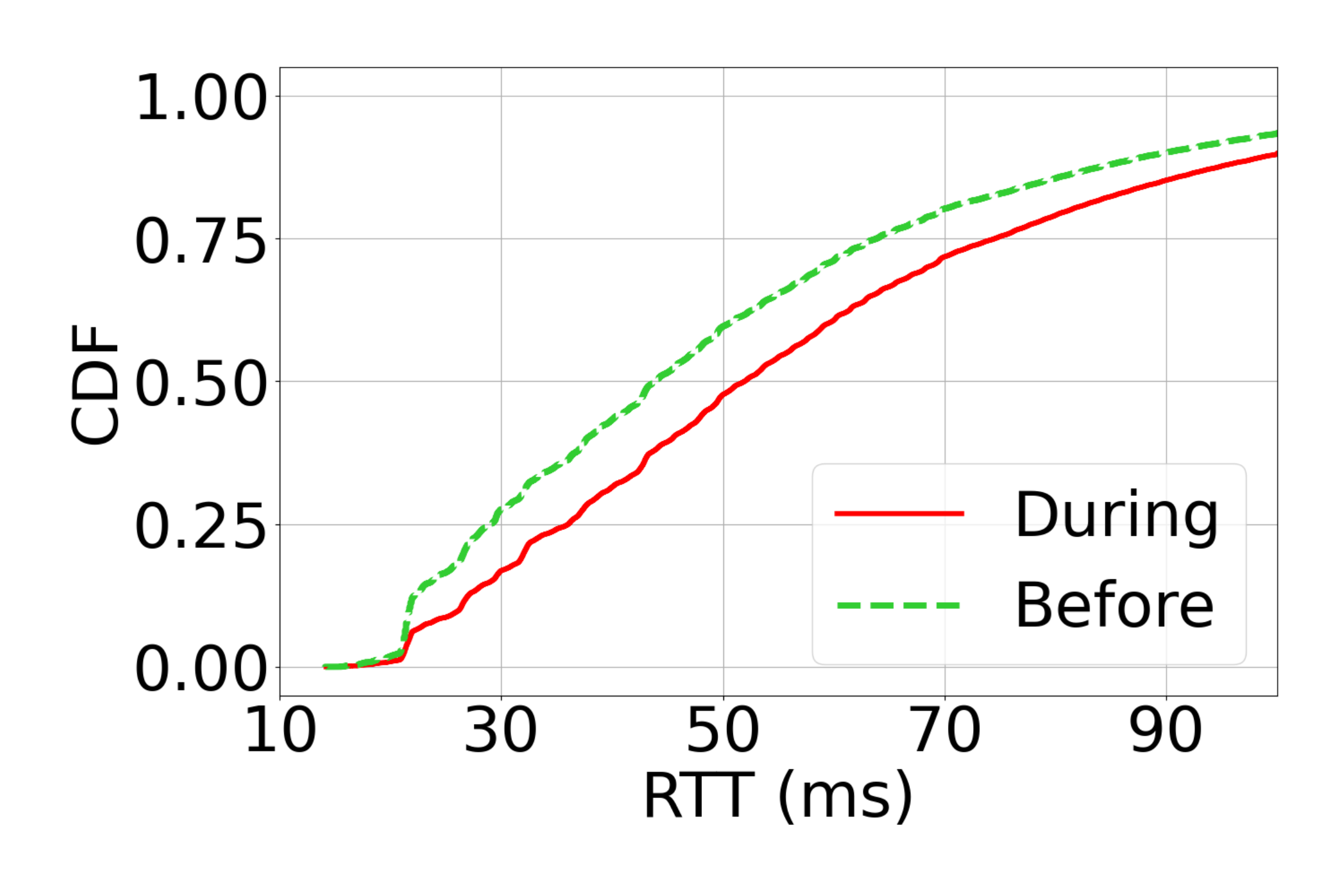}
        \caption{Dallas, US}
    \end{subfigure}%
    \hfill
    \begin{subfigure}[t]{0.30\columnwidth}
        \centering
        \includegraphics[height=2.9cm, keepaspectratio]{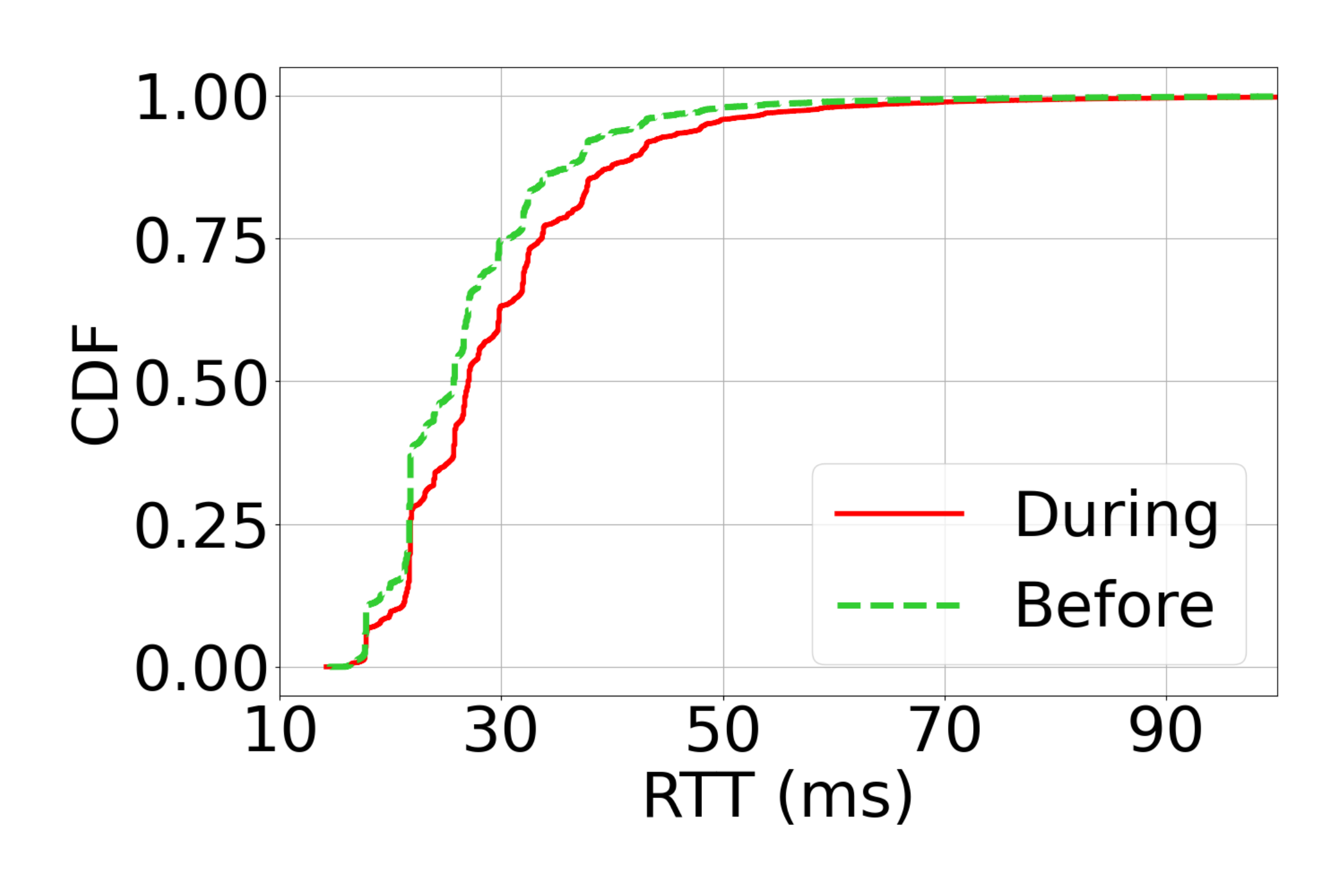}
        \caption{Salt Lake City, US}
    \end{subfigure}%
    \hfill
    \begin{subfigure}[t]{0.30\columnwidth}
        \centering
        \includegraphics[height=2.9cm, keepaspectratio]{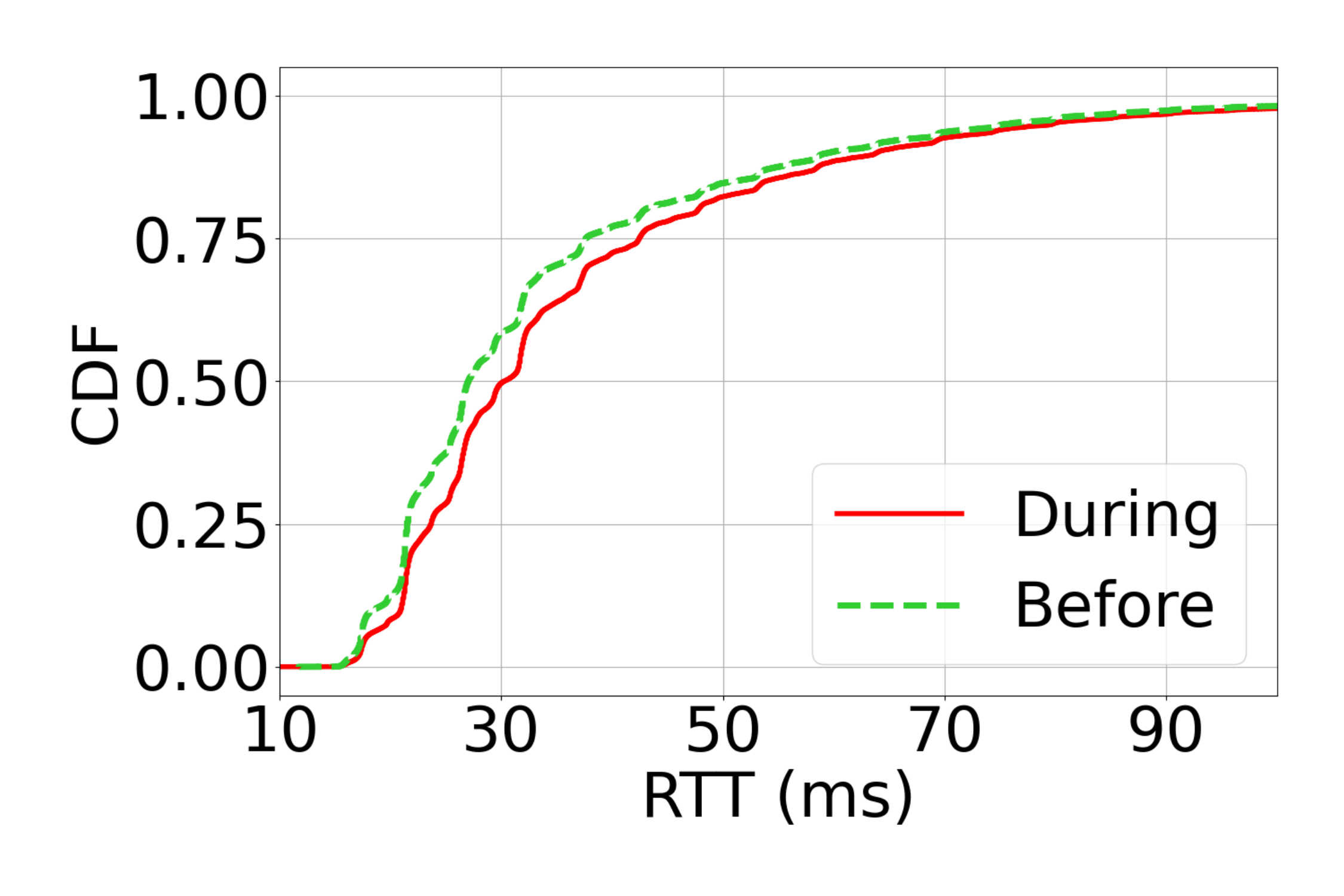}
        \caption{Seattle, US}
    \end{subfigure}%
    
    \caption{Comparing the pre-storm (green dashed line) and during October 2024 solar storm (red solid line) latency distribution, showing a similar right shift in latency distribution during the event across all the vantage points.}
    \label{fig:CDFpingRTTOct24}
\end{figure}

We examine how the latency distribution changes during a solar storm compared to the pre-storm scenario.
For that, we use a 10 ms cadence of RTT observations of each vantage point and plot two CDFs of RTT measurements from pre-storm (green dashed line) and during (red solid line) the storm window in Fig.~\ref{fig:CDFpingRTTMay24} and in Fig.~\ref{fig:CDFpingRTTOct24}. 
We use the same window as above: pre-storm from 3rd to 10th May and 1st to 7th October, and storm window from 11th to 13th May and 8th to 12th October, for the May 2024 and October 2024 solar storms, respectively. 
Notice the minimum latency recorded during the storm is similar to before the solar storm, and also notice a rightward shift in the latency distribution across all vantage points during solar storms.
This observation is exactly the same across all vantage points for both the May 2024 solar superstorm and the October 2024 solar storms in Fig.~\ref{fig:CDFpingRTTMay24} and in Fig.~\ref{fig:CDFpingRTTOct24}. 
This indicates clear degradation in latency during solar storms.
The Starlink connectivity experienced higher latency variability while minimum latency remain the same.
During the May 2024 solar superstorm, Frankfurt experienced maximum latency degradation, with median latency increasing by 10\% and reaching up to 40\% above the 80th percentile.
In Canada, we see up to 25\% latency inflation above the 80th percentile, with a 4\% change in median latency.
In the US, among three vantage points, Denver experiences the highest latency inflation, up to 15\% (10\%) above the 80th percentile (median).
During the October 2024 solar storms, the observations remained similar to those during the May 2024 solar superstorm.
In Germany, Bruhl experiences up to 36\% (7\%) latency inflation above the 80th percentile (median), whereas in Frankfurt, it is no more than 22\% (3\%).
In Canada, the highest inflation is up to 23\% (14\%) above the 80th percentile (median) observed in Calgary. 
In the US, Dallas has the highest inflation rate, reaching 23\%.

\subsection{Momentary outages}

\begin{figure}
    \centering
    \begin{subfigure}[t]{0.5\columnwidth}
        \centering
        \includegraphics[width=\columnwidth, keepaspectratio]{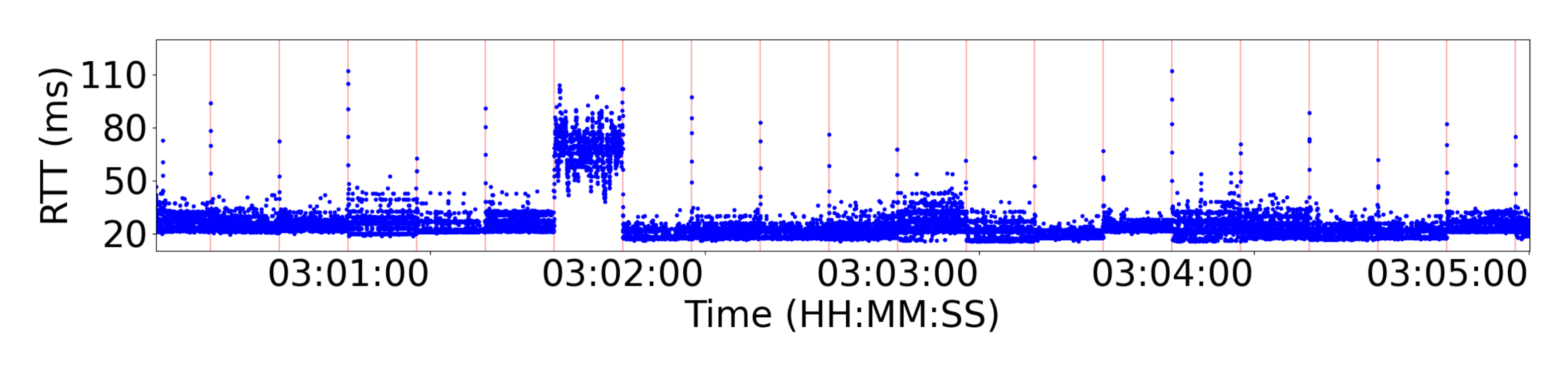}
        \caption{03:00 UTC to 03:05 UTC May 10, 2024}
    \end{subfigure}%
    \hfill
    \begin{subfigure}[t]{0.5\columnwidth}
        \centering
        \includegraphics[width=\columnwidth, keepaspectratio]{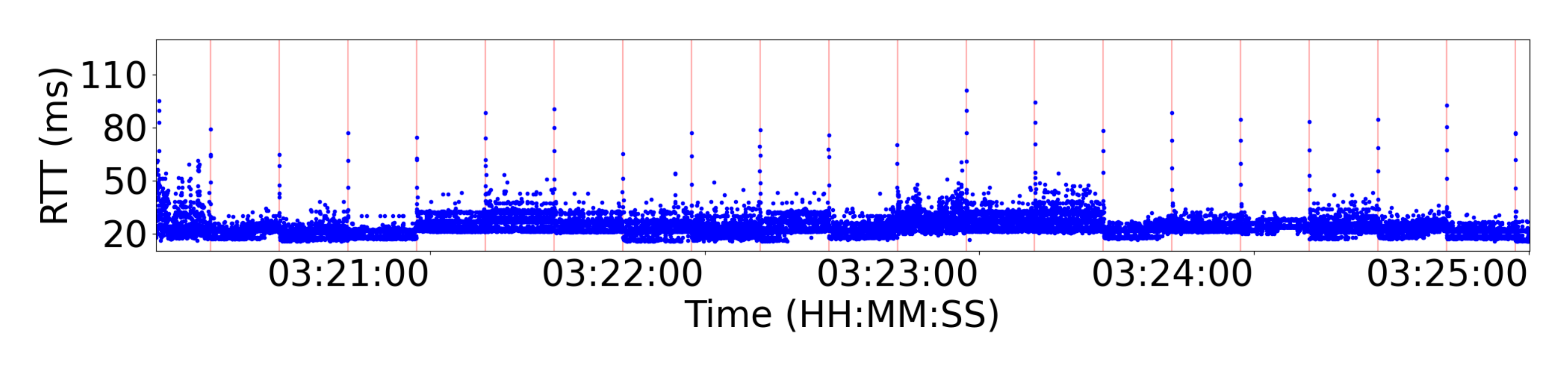}
        \caption{03:20 UTC to 03:25 UTC May 10, 2024}
    \end{subfigure}%
    \hfill
    \begin{subfigure}[t]{0.5\columnwidth}
        \centering
        \includegraphics[width=\columnwidth, keepaspectratio]{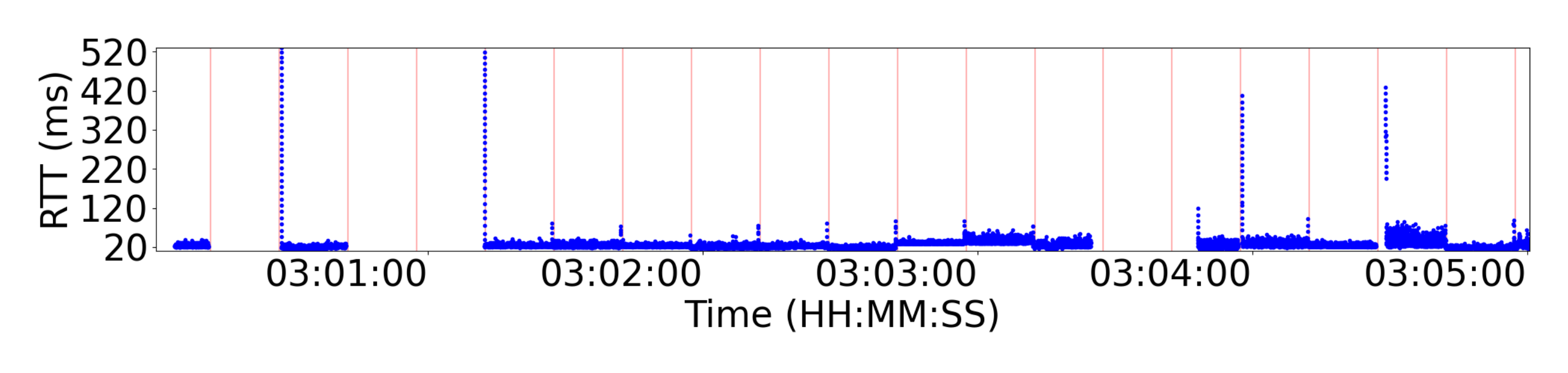}
        \caption{03:00 UTC to 03:05 UTC May 11, 2024}
    \end{subfigure}%
    \hfill
    \begin{subfigure}[t]{0.5\columnwidth}
        \centering
        \includegraphics[width=\columnwidth, keepaspectratio]{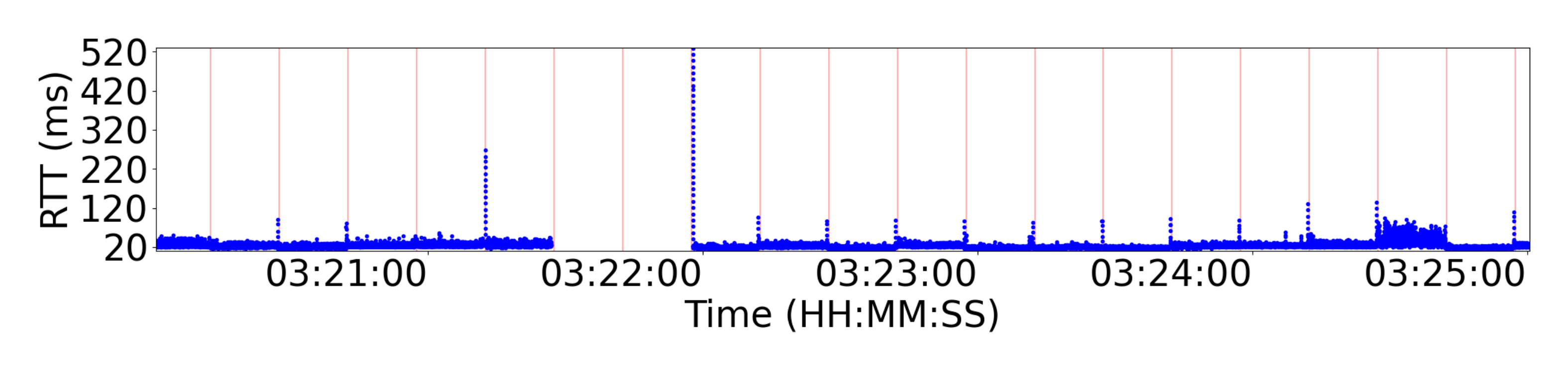}
        \caption{03:20 UTC to 03:25 UTC May 11, 2024}
    \end{subfigure}%

    \caption{Comparing the timeseries of individual latency observation (blue dots) from Frankfurt (c)-(d) during the peak of solar superstorm with (a)-(b) the previous day reveals repetitive short-lived outage of 10s of seconds, and during reconfiguration (red vertical lines), many latency spikes are 4X higher compared to the regular days. }
    \label{fig:ZoomScaleMsFrankfurtMay24}
\end{figure}

\begin{figure}
    \centering
    \begin{subfigure}[t]{0.5\columnwidth}
        \centering
        \includegraphics[width=\columnwidth, keepaspectratio]{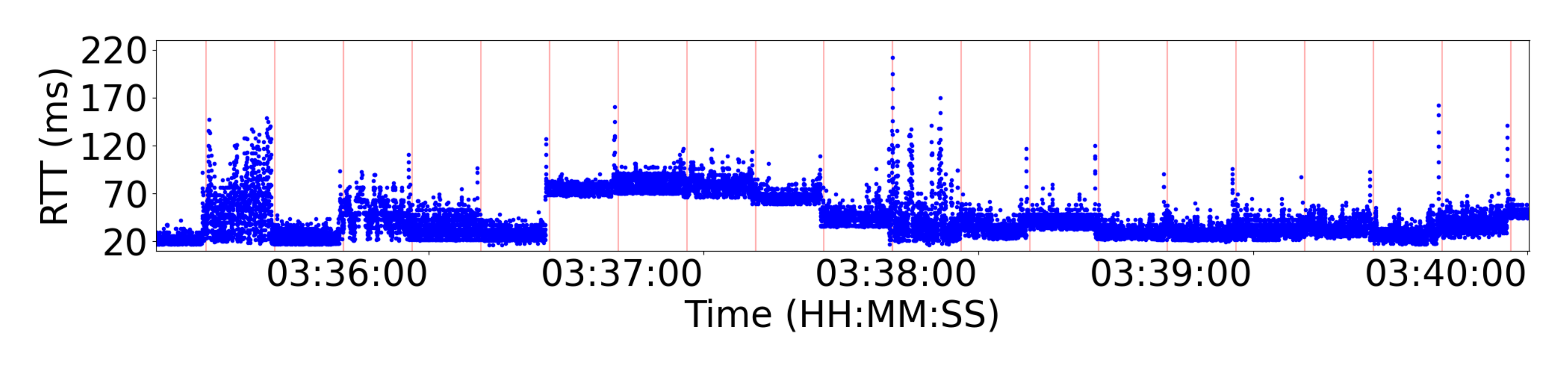}
        \caption{03:35 UTC to 03:40 UTC May 10, 2024}
    \end{subfigure}%
    \hfill
    \begin{subfigure}[t]{0.5\columnwidth}
        \centering
        \includegraphics[width=\columnwidth, keepaspectratio]{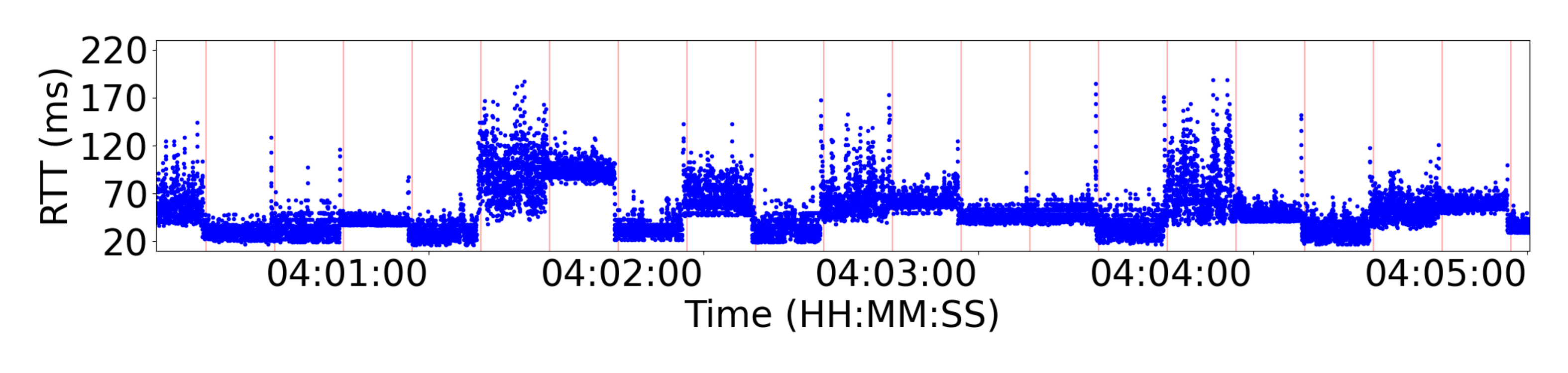}
        \caption{04:00 UTC to 04:05 UTC May 10, 2024}
    \end{subfigure}%
    \hfill
    \begin{subfigure}[t]{0.5\columnwidth}
        \centering
        \includegraphics[width=\columnwidth, keepaspectratio]{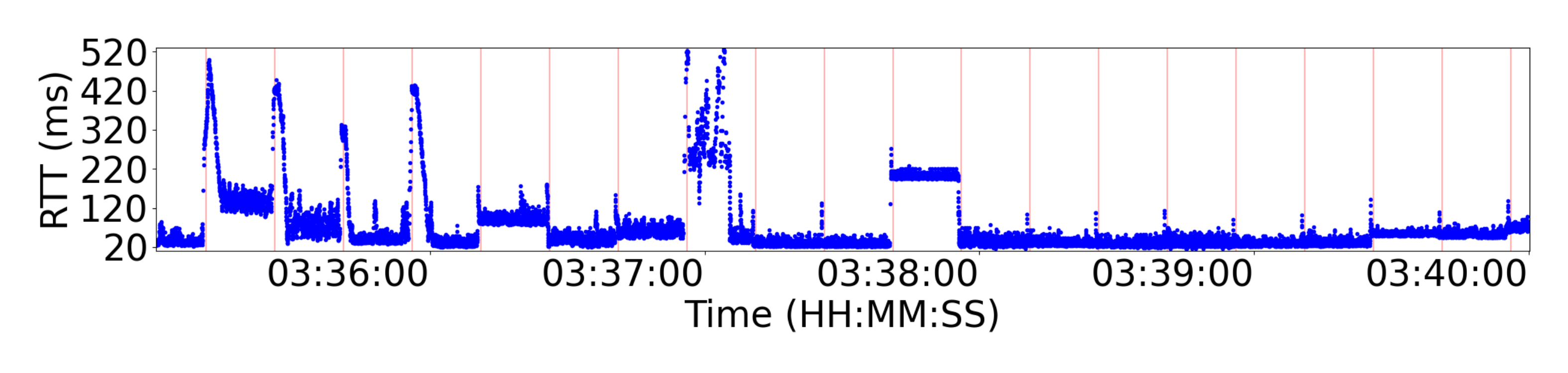}
        \caption{03:35 UTC to 03:40 UTC May 11, 2024}
    \end{subfigure}%
    \hfill
    \begin{subfigure}[t]{0.5\columnwidth}
        \centering
        \includegraphics[width=\columnwidth, keepaspectratio]{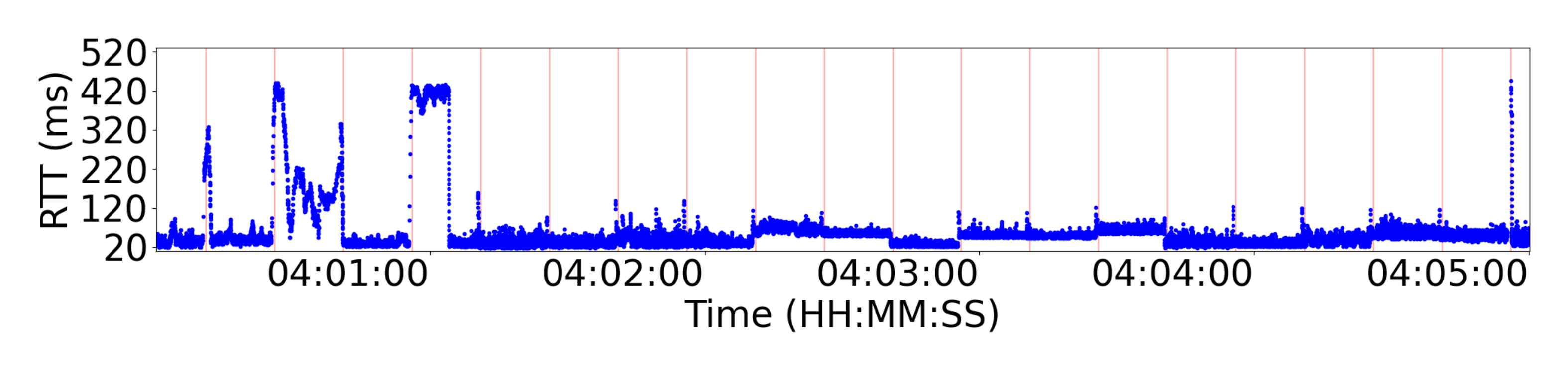}
        \caption{04:00 UTC to 04:05 UTC May 11, 2024}
    \end{subfigure}%
    
    \caption{Latency observations (blue dots) from Vancouver in panels (c)–(d) during the peak of the solar superstorm, compared with the previous day in panels (a)–(b), show similar latency spikes during reconfiguration events (red vertical lines). However, unlike Frankfurt, latency in Vancouver remains elevated for several seconds. }
    \label{fig:ZoomScaleMsVancouverMay24}
\end{figure}

When we magnify latency measurements to the millisecond scale, we uncover patterns obscured by statistical aggregation in earlier analyses. 
In Fig.~\ref{fig:ZoomScaleMsFrankfurtMay24}, we show the RTT observations in blue dots, and with vertical red lines highlighting the reconfiguration events occurring every 15 seconds~\cite{mohan2024multifaceted, izhikevich2024democratizing, MakingSenseLEO, 10623111}.
On 10 May 2024, a day before the onset of the solar superstorm, RTT measurements from Frankfurt are shown for two 5-minute windows from 03:00 to 03:05 UTC in Fig.~\ref{fig:ZoomScaleMsFrankfurtMay24}(a) and from 03:20 to 03:25 UTC in Fig.~\ref{fig:ZoomScaleMsFrankfurtMay24}(b). 
Notice that in a regular day during reconfiguration, the majority of RTT spikes remain below 120 ms.
In contrast, on 11 May 2024, during the peak of the solar superstorm, the same time windows in Fig.~\ref{fig:ZoomScaleMsFrankfurtMay24}(c)–(d) exhibit significantly higher RTT spikes during the reconfiguration exceeding 500 ms. 
Approximately a fourfold increase over the baseline on the previous day. 
We also observe short-lived outages lasting close to 15–30 seconds, and occasionally longer.
Note that we present only two examples within a 5-minute window for brevity.
Such spikes during reconfiguration and short outages, occurring in multiples of 15 seconds, have been observed repeatedly for hours. 
Similar behavior is observed in Vancouver. 
In Fig~\ref{fig:ZoomScaleMsVancouverMay24}(a)–(b) show two 5-minute windows on 10 May 2024, while the corresponding windows on 11 May 2024 during the peak of the solar superstorm are shown in Fig.~\ref{fig:ZoomScaleMsVancouverMay24}(c)–(d). 
During the storm, RTT spikes associated with reconfiguration are more than twice as high as on a typical day. 
Moreover, in Vancouver, elevated RTT persists for a few seconds, whereas in Frankfurt, spike durations are on the order of milliseconds.
We also analyzed latency measurements from other vantage points. 
However, we did not observe similar effects across most locations in the US. 

During the October 2024 solar storms, similar patterns were observed in Germany, though less frequently than the May 2024 superstorm. 
In Canada, such effects were negligible.
An exception is Dallas, where multiple outage windows, often in multiples of 15 seconds, have been observed since 1st October 2024. 
This suggests persistent connectivity issues, which likely contribute to the higher RTT variance observed in Fig.~\ref{fig:CDFpingRTTOct24}(g) compared to other US vantage points.

\subsection{Loss characteristics}

Using LENS's UDP-based \texttt{IRTT} measurements, we analyze how packet loss behavior changes during a solar storm compared to the usual days.

\subsubsection{Exploring the packet loss during solar storms:}

\begin{figure}
    \centering
    \begin{subfigure}[t]{0.5\columnwidth}
        \centering
        \includegraphics[width=\columnwidth, keepaspectratio]{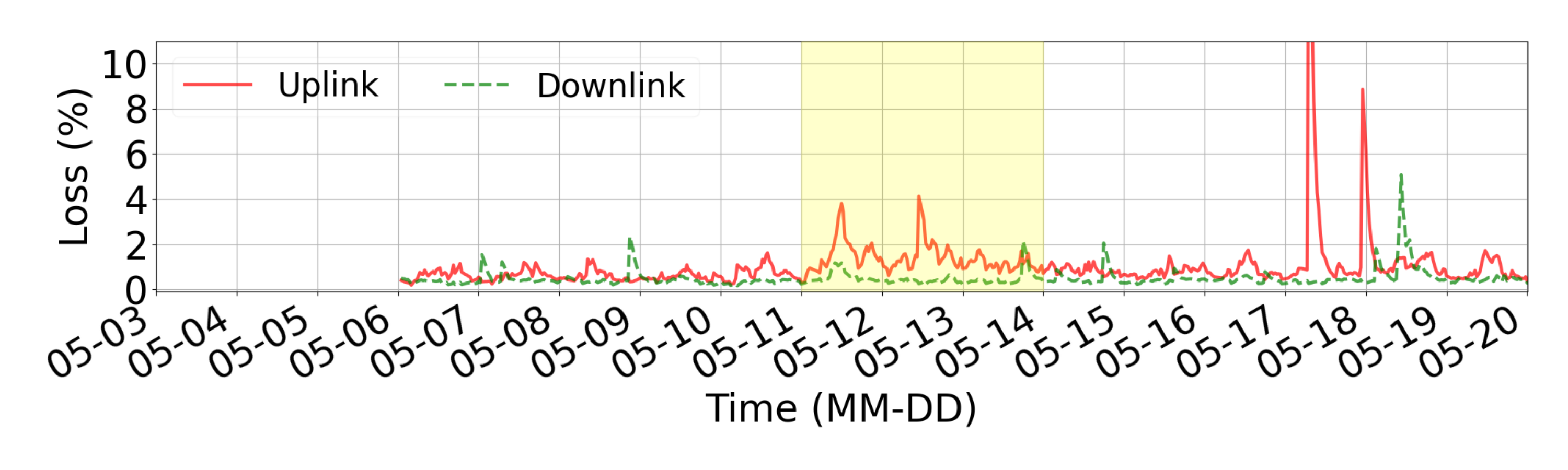}
        \caption{Frankfurt, Germany}
    \end{subfigure}%
    \hfill
    \begin{subfigure}[t]{0.5\columnwidth}
        \centering
        \includegraphics[width=\columnwidth, keepaspectratio]{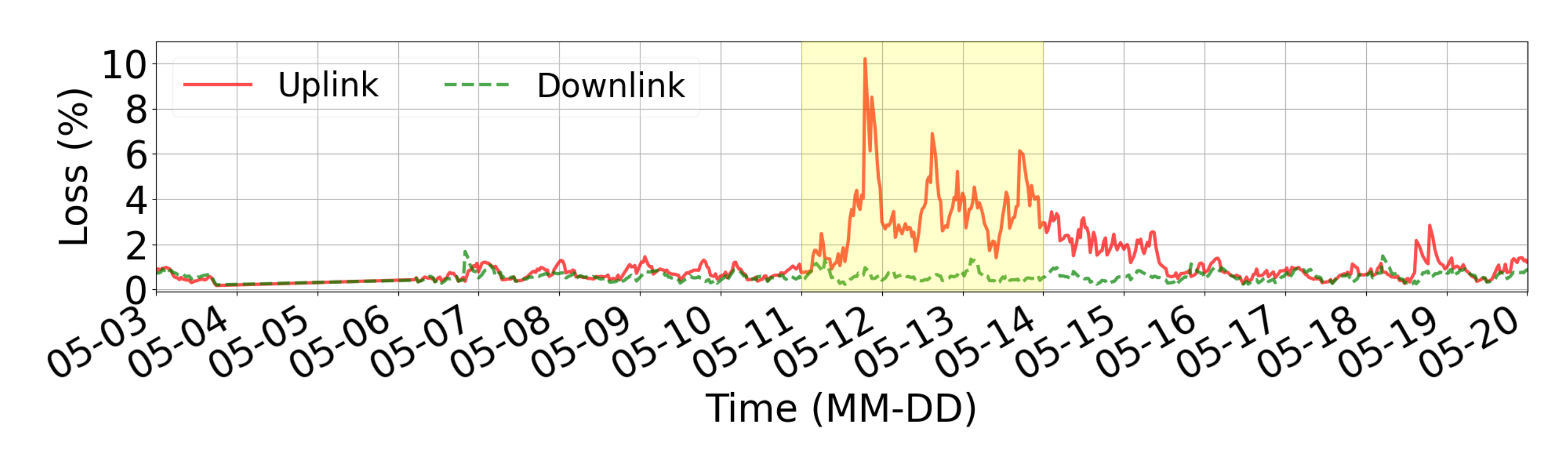}
        \caption{Vancouver, Canada}
    \end{subfigure}%
    \hfill
    \begin{subfigure}[t]{0.5\columnwidth}
        \centering
        \includegraphics[width=\columnwidth, keepaspectratio]{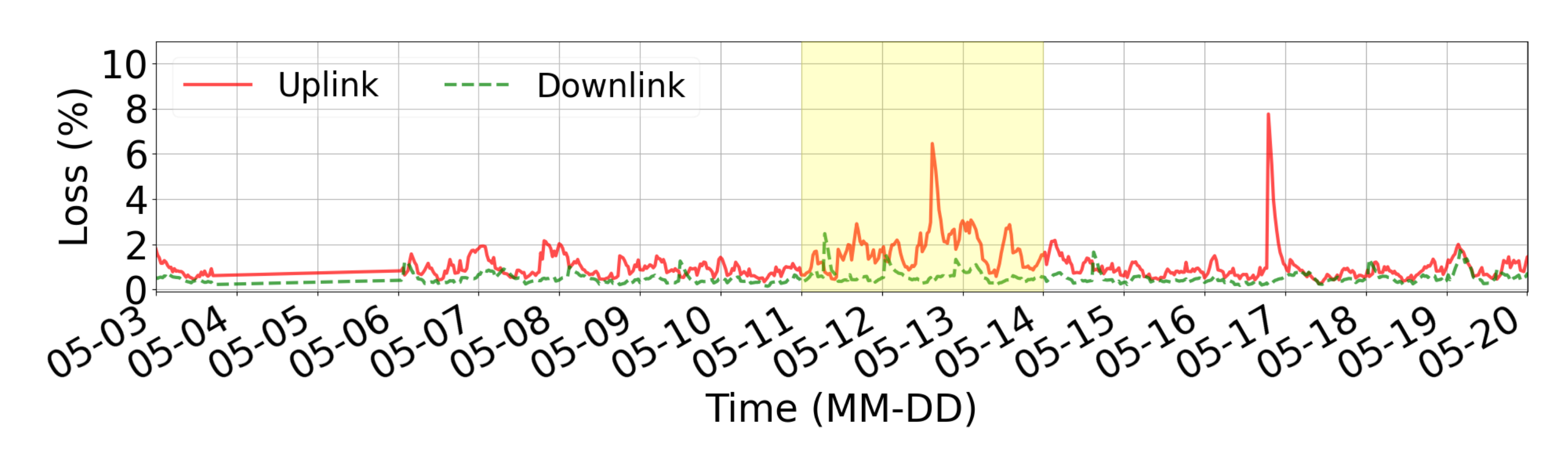}
        \caption{Victoria, Canada}
    \end{subfigure}%
    \hfill
    \begin{subfigure}[t]{0.5\columnwidth}
        \centering
        \includegraphics[width=\columnwidth, keepaspectratio]{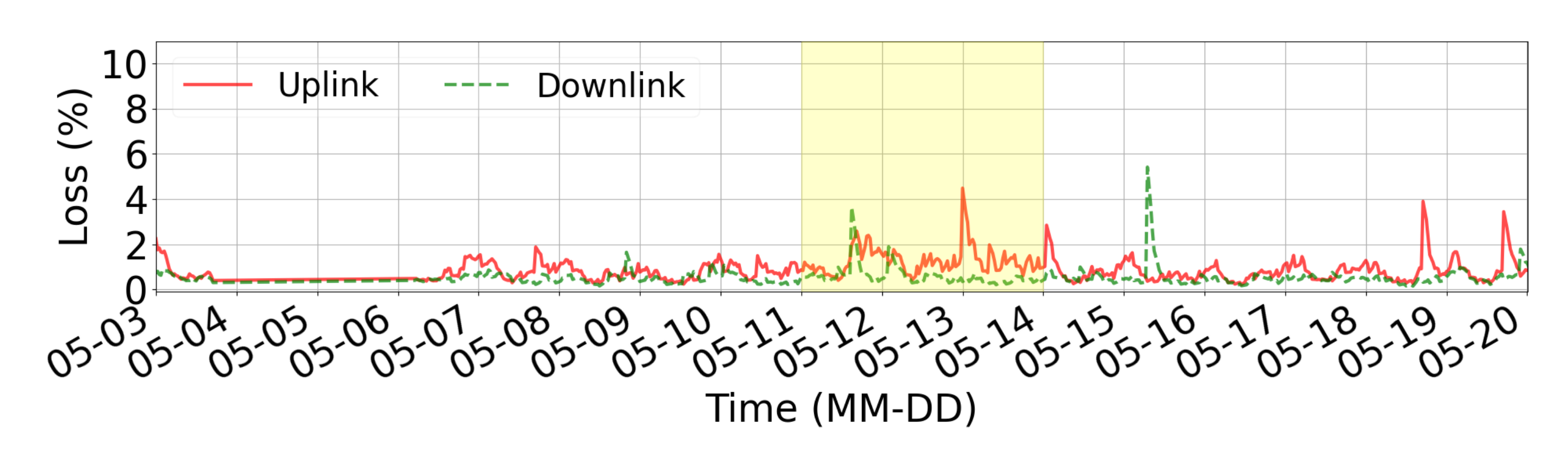}
        \caption{Seattle, US}
    \end{subfigure}%
    
    \caption{Packet loss in downlink (dashed green line) and uplink (solid red line) from four vantage points from (a) Germany, (b)-(c) Canada, and (d) the US during May 2024 solar superstorm (highlighted region) shows that the downlink is much more stable during such an event, while the uplink experiences higher losses across all the vantage points, reaching up to 10\%.  }
    \label{fig:LossIRTTMay24}
\end{figure}

\begin{figure}
    \centering
    \begin{subfigure}[t]{0.5\columnwidth}
        \centering
        \includegraphics[width=\columnwidth, keepaspectratio]{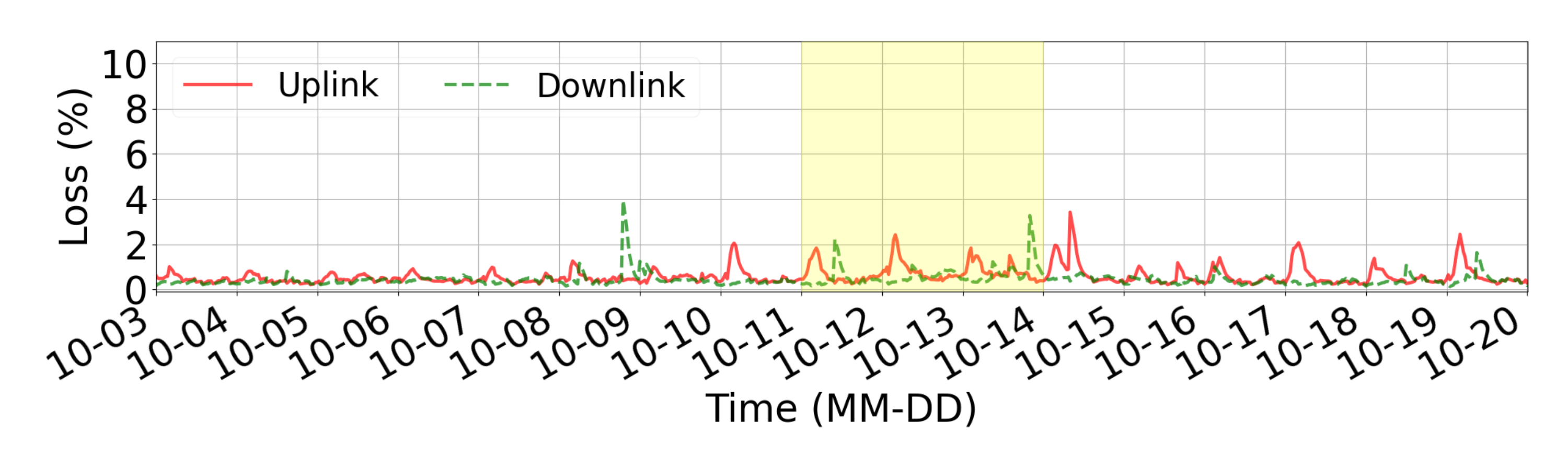}
        \caption{Frankfurt, Germany}
    \end{subfigure}%
    \hfill
    \begin{subfigure}[t]{0.5\columnwidth}
        \centering
        \includegraphics[width=\columnwidth, keepaspectratio]{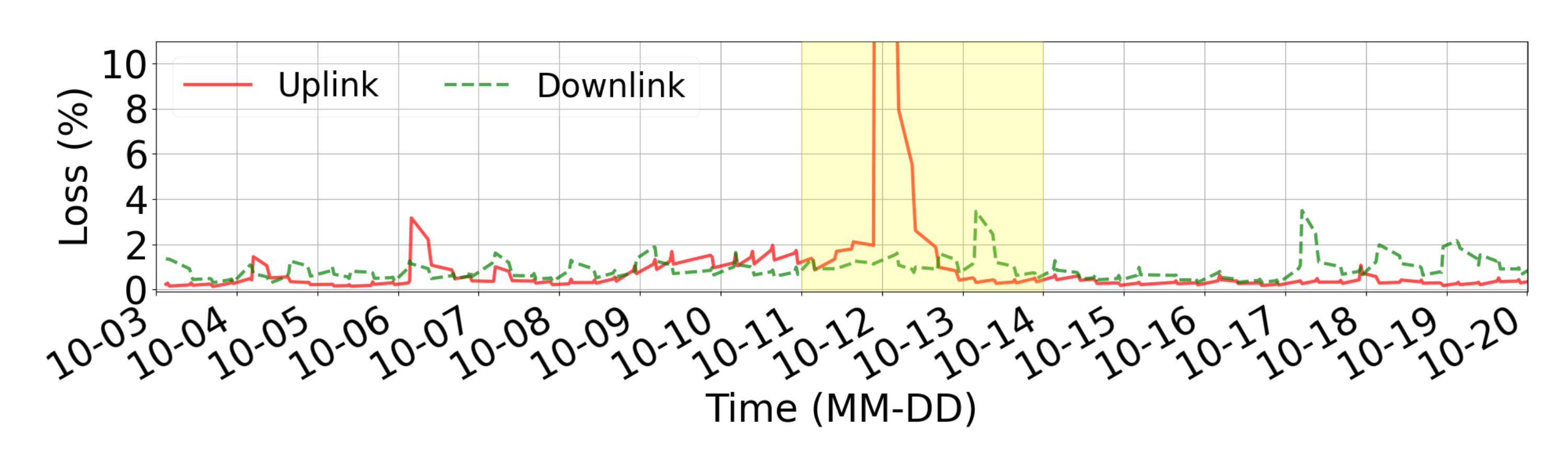}
        \caption{Seattle, US}
    \end{subfigure}%
    
    \caption{Packet loss in downlink (dashed green line) and uplink (solid red line) from two vantage points from Germany and the US during the October 2024 solar storm (highlighted region) shows similar implications in uplink, impact at (a) Frankfurt is limited, but (b) Seattle experiences 60\% loss in uplink for a while. }
    \label{fig:LossIRTTOct24}
\end{figure}

We calculate the loss percentage as the number of packets lost divided by the total number of packets sent over a 30-minute interval. 
We then compute the EWMA of the loss percentage over two periods: 3rd to 20th May and 3rd to 20th October, to examine how packet loss behavior changes during solar storm events.
In Fig.~\ref{fig:LossIRTTMay24}, we present the EWMA of the uplink and downlink loss percentages during the May 2024 solar superstorm, using four available vantage points from the LENs \texttt{IRTT} dataset. 
Notice that a baseline loss of up to 2\% is consistently observed across all vantage points in Germany, Canada, and the US. 
On 11th May, a clear increase in packet loss is visible across all the locations.
However, the magnitude of this increase varies significantly across these four locations.
In Frankfurt and Vancouver, downlink loss remains relatively stable, staying within 2\% of the baseline. 
In contrast, Victoria and Seattle experience higher downlink losses, reaching up to 3\% and 4\%, respectively.
On the other hand, uplink loss increases by up to 4\%, 10\%, 7\%, and 5\% in Frankfurt, Vancouver, Victoria, and Seattle, respectively. 
Among these, Vancouver experiences the most severe and prolonged impact.
The uplink loss percentage remains high for several days and returns to baseline on 16th May, five days after the event. 
At the other locations, uplink loss remains high for no more than 3 days.

For the October solar storm, data is available from only two vantage points in the LENs dataset: Frankfurt and Seattle. 
In these two vantage points, inflation in the loss percentage starts after 8th October. 
Frankfurt shows a relatively minor impact, with both uplink and downlink losses remaining within 4\%.
In contrast, Seattle experiences a significant degradation in uplink performance, with packet loss exceeding 20\% after 11th October.
Notably, the uplink appears more sensitive than the downlink and is highly susceptible to Solar events, thus showing more pronounced degradation during the storm.

\subsubsection{Quantifying loss inflation:}

\begin{figure}
    \centering
    \begin{subfigure}[t]{0.25\columnwidth}
        \centering
        \includegraphics[width=\columnwidth, keepaspectratio]{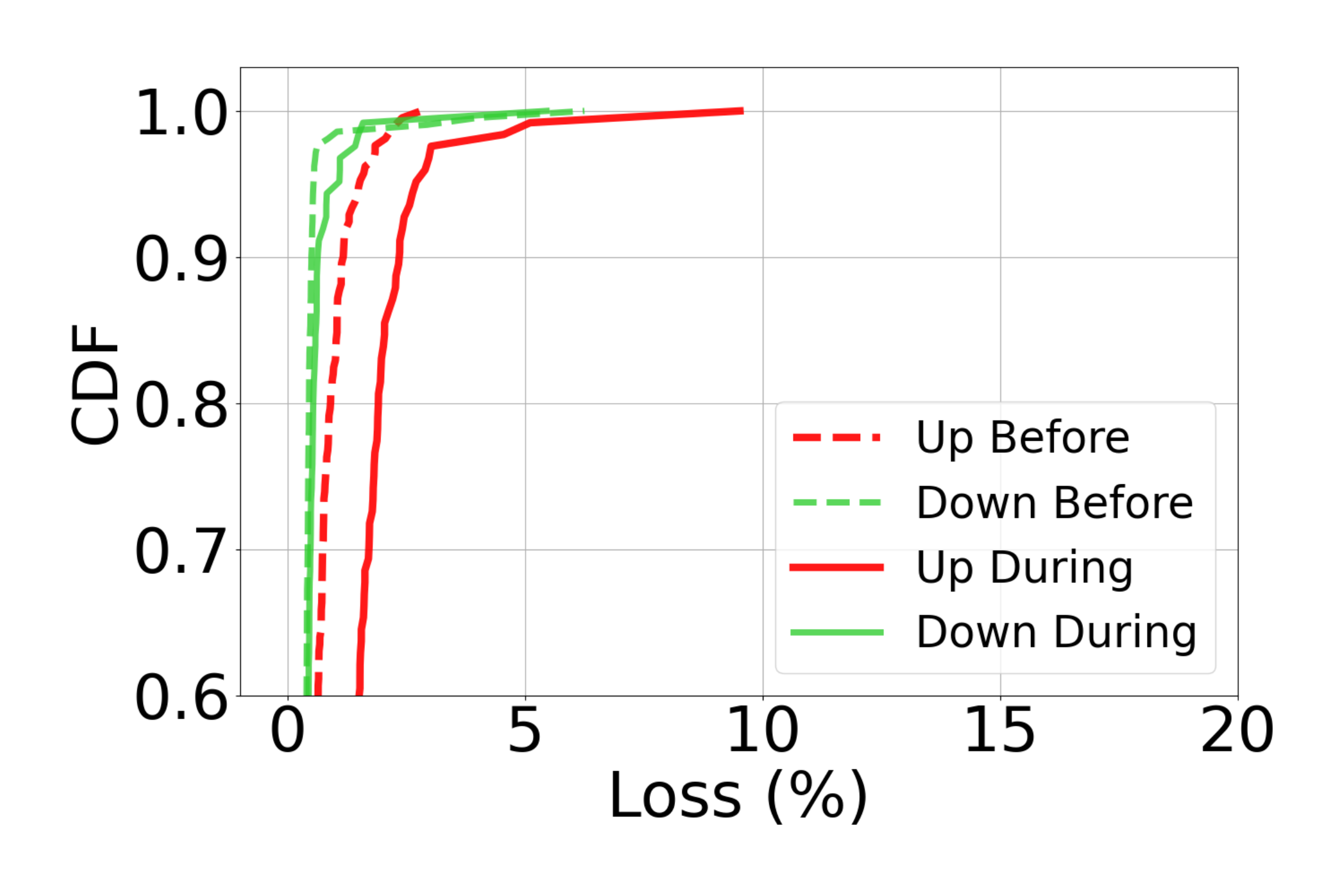}
        \caption{Frankfurt, Germany}
    \end{subfigure}%
    \hfill
    \begin{subfigure}[t]{0.25\columnwidth}
        \centering
        \includegraphics[width=\columnwidth, keepaspectratio]{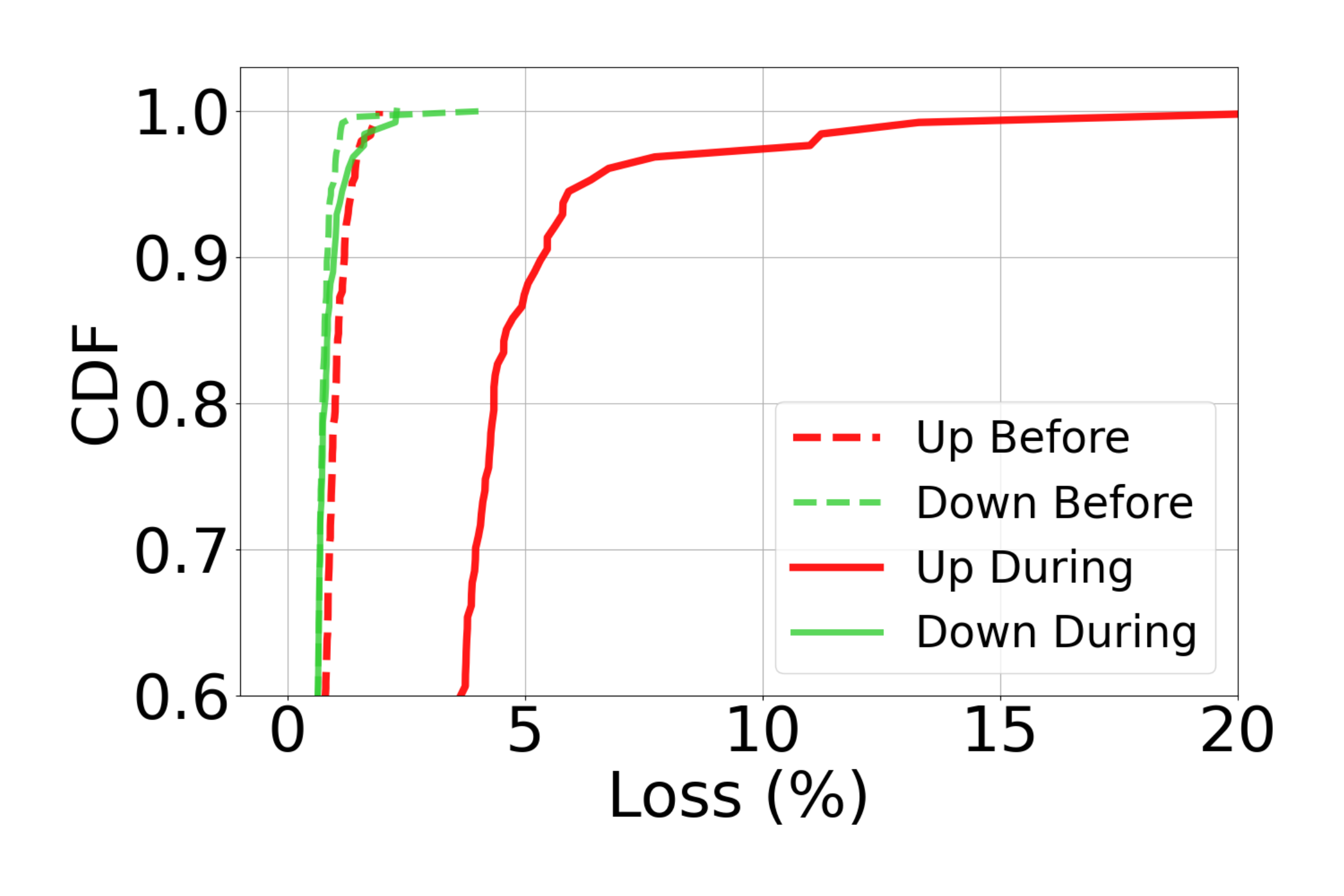}
        \caption{Vancouver, Canada}
    \end{subfigure}%
    \hfill
    \begin{subfigure}[t]{0.25\columnwidth}
        \centering
        \includegraphics[width=\columnwidth, keepaspectratio]{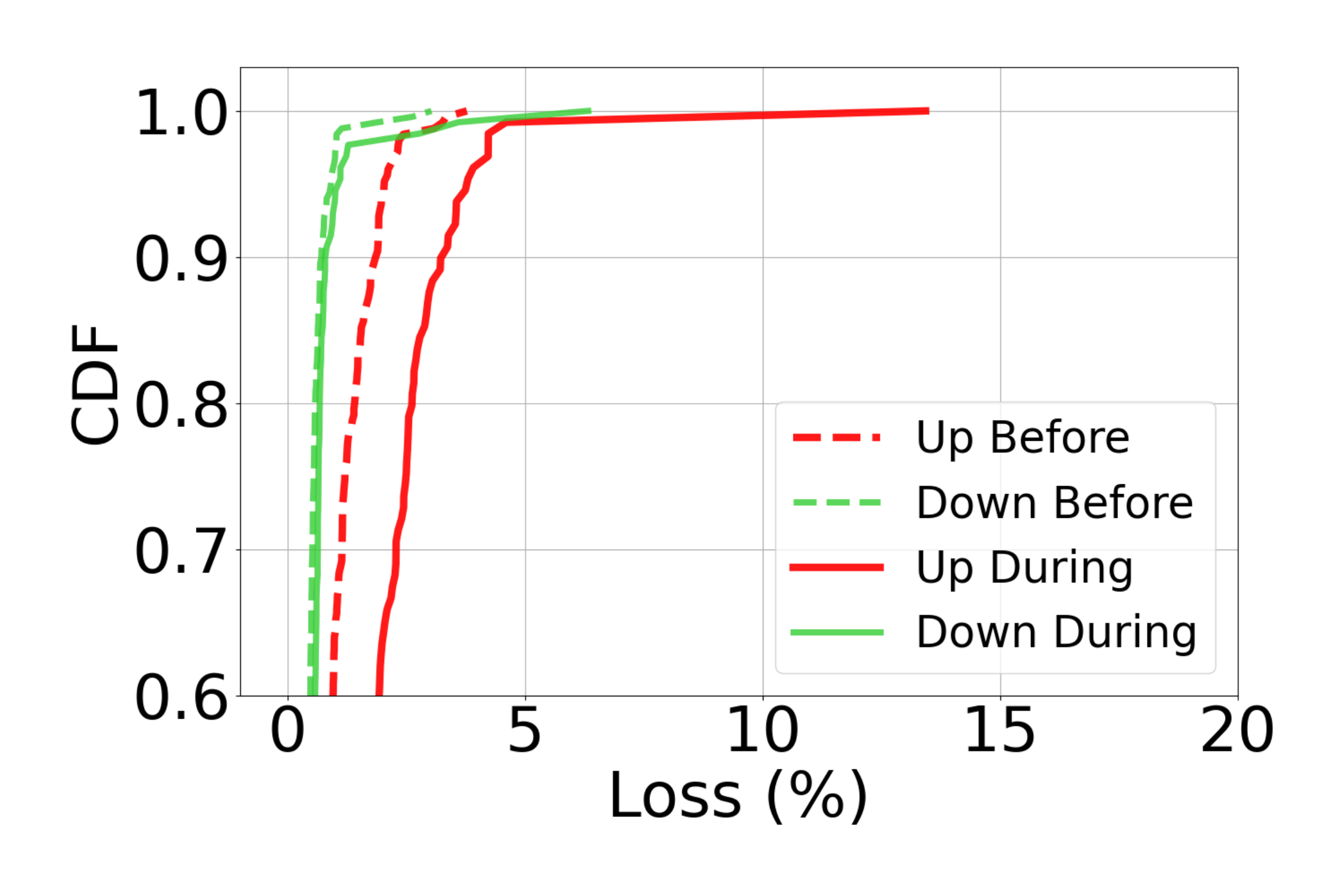}
        \caption{Victoria, Canada}
    \end{subfigure}%
    \hfill
    \begin{subfigure}[t]{0.25\columnwidth}
        \centering
        \includegraphics[width=\columnwidth, keepaspectratio]{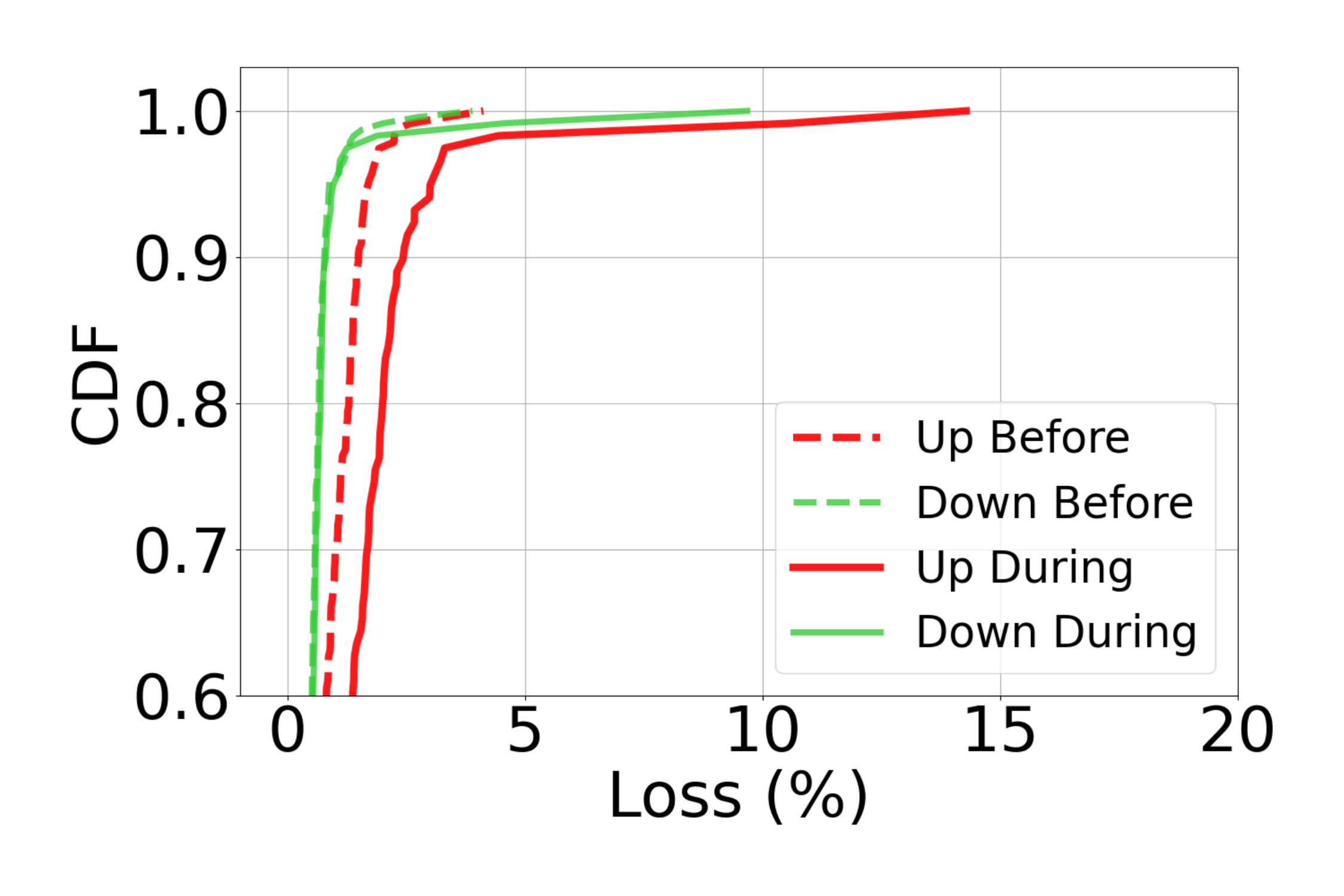}
        \caption{Seattle, US}
    \end{subfigure}%

    \caption{Comparing the pre-storm (dashed line) and during May 2024 solar superstorm (solid line) uplink (red) and downlink (green) packet loss showing the difference lies at the higher quantile. Downlink loss shows only minor changes above the 90th percentile, while uplink shows significant changes with long tails at all vantage points. }
    \label{fig:IRTTLossCDFMay24}
\end{figure}

\begin{figure}
    \centering
    \begin{subfigure}[t]{0.5\columnwidth}
        \centering
        \includegraphics[width=0.7\columnwidth, keepaspectratio]{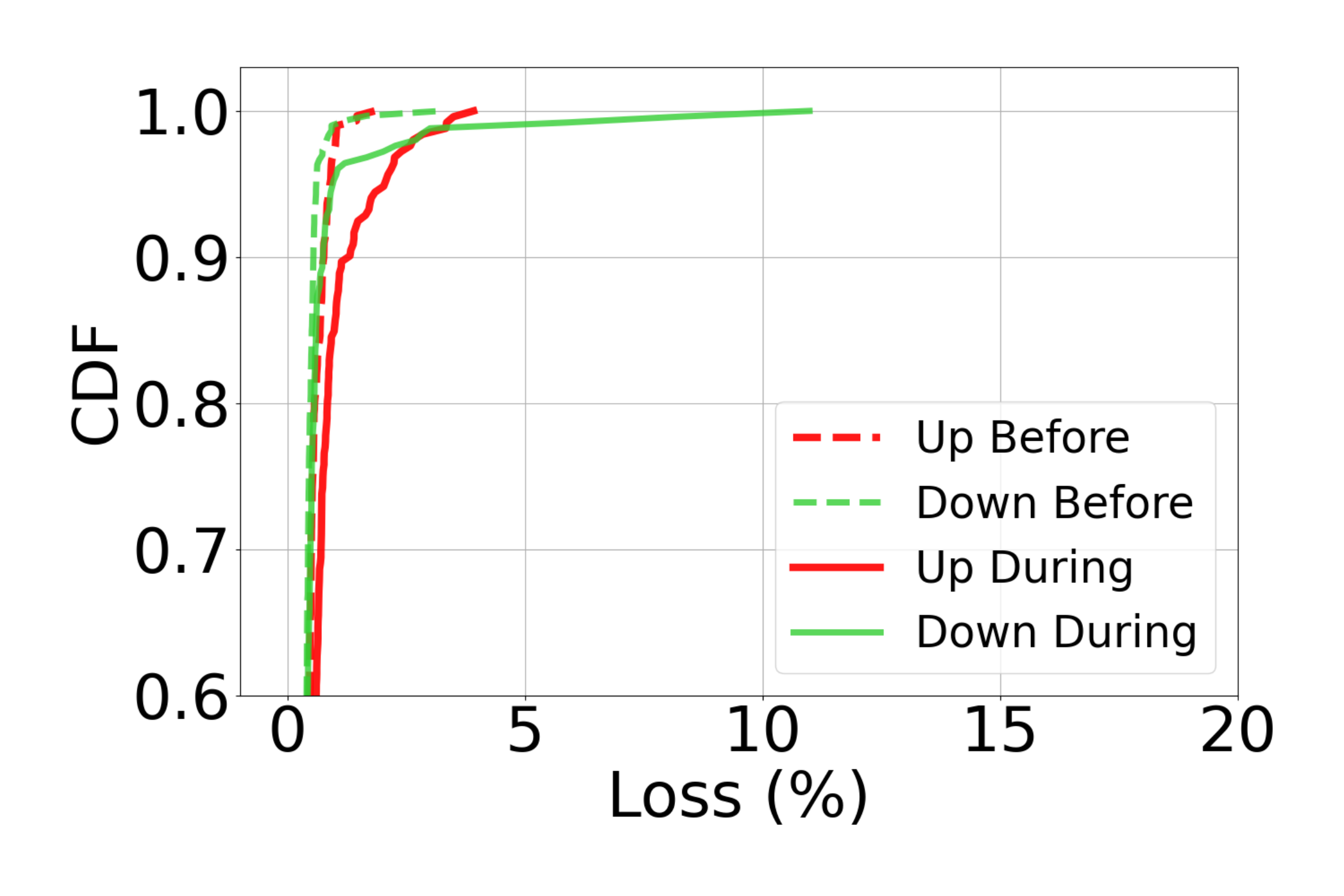}
        \caption{Frankfurt, Germany}
    \end{subfigure}%
    \hfill
    \begin{subfigure}[t]{0.5\columnwidth}
        \centering
        \includegraphics[width=0.7\columnwidth, keepaspectratio]{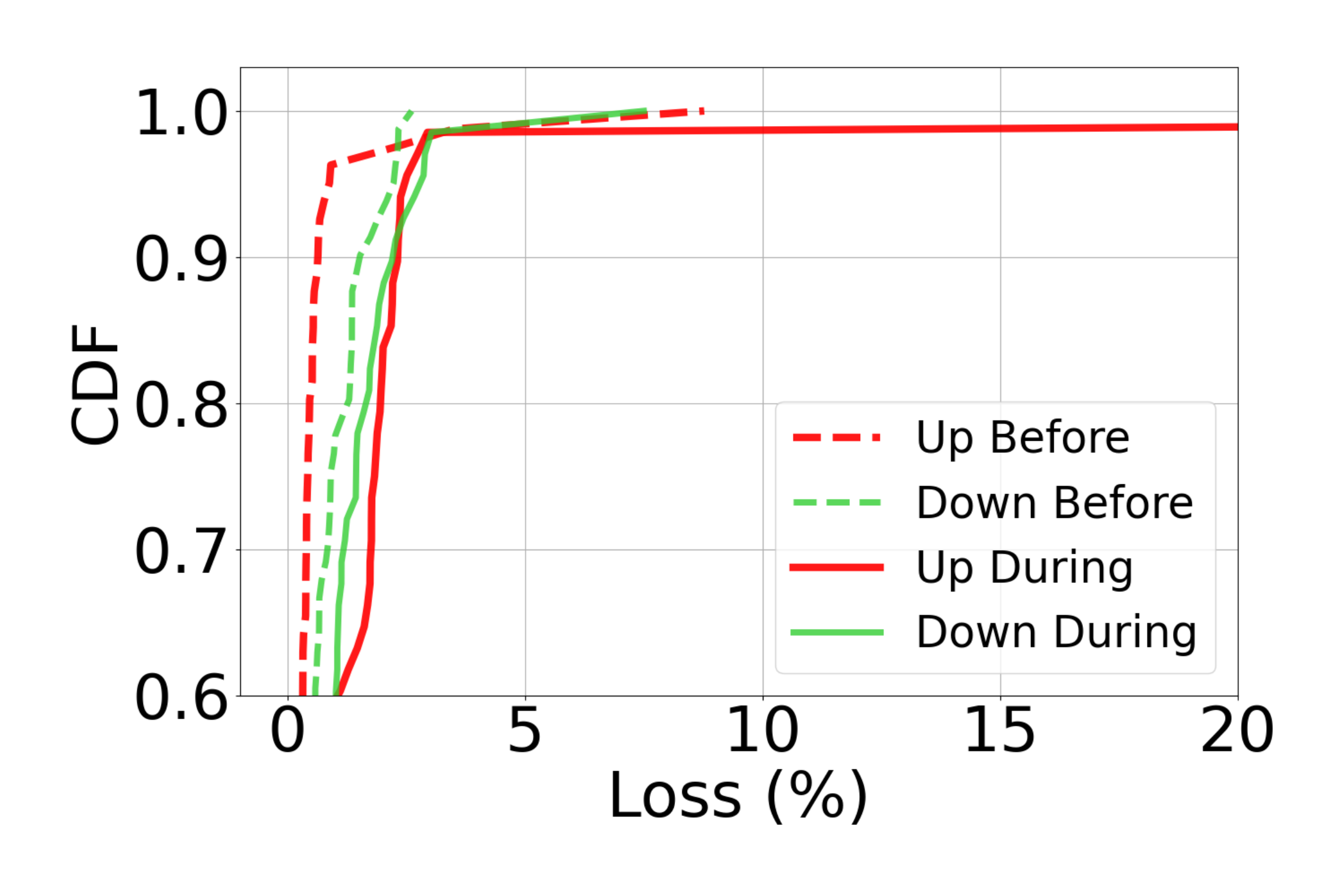}
        \caption{Seattle, US}
    \end{subfigure}%
    
    \caption{Comparing the pre-storm (dashed line) and during October 2024 solar storm (solid line) uplink (red) and downlink (green) packet loss, showing downlink has minor loss inflation, while uplink shows a significant increase in loss, specifically in (b) Seattle. }
    \label{fig:IRTTLossCDFOct24}
\end{figure}

Similar to the latency, we segregate 30-minute packet loss percentage observations into two windows of pre-storm: 3rd to 10th May, 1st to 7th October, and during storm: 11th to 13th May, 8th to 13th October, and plot the uplink and downlink loss CDF in Fig.~\ref{fig:IRTTLossCDFMay24}, and Fig.~\ref{fig:IRTTLossCDFOct24}, respectively.
Notice that in these figures, we use color red for the uplink and green for the downlink; we denote pre-storm with dashed lines and during storm with solid lines.
During the May 2024 solar superstorm, we see negligible changes in median packet loss. 
The main difference is observed at higher percentiles. 
Above the 90th percentile, in the downlink, we can observe a maximum of 3 to 5\% increase in loss in the tail of Victoria and Seattle.
Hence, the effect on the downlink is limited. 
A significant shift in uplink loss is observed across all vantage points, with increases of at least 6\%.
The tail of the uplink loss CDF at Victoria, Seattle, increased by 10\%, and in Vancouver it exceeded 20\%.

During the October solar storms in Fig.~\ref{fig:IRTTLossCDFOct24}, we see the downlink at Frankfurt experience a major hit, rising by 7\% above the 95th percentile, while the uplink shows a relatively lower impact, rising by 2\% above the 90th percentile.
In contrast, Seattle shows maximum degradation of increased loss of 5\% in downlink above the 95th percentile and beyond 60\% (up to 2\%) in uplink above the 95th percentile (within 90th percentile).

\begin{keybox}
\keynote 
During a solar storm, the Starlink segment of the network shows latency inflation up to 40\% at higher quantiles, distortion in the diurnal latency pattern, and increased latency variability.
Short-lived outages lasting 10s are also being observed.
The Starlink uplink seems more sensitive to such events than the downlink.
A noticeable rise in uplink packet loss up to 60\% while uplink loss remains within 3-5\%.
Implications are not globally consistent but are concentrated in specific regions.
\end{keybox}

\subsection{End user perceived Internet experience}

We use user-initiated speed test results from M-Lab and Cloudflare, and built-in ping measurements from RIPE Atlas probes, to analyze how the above-discussed implications are perceived in end-user Internet access.

\subsubsection{Preparing the measurement datasets:}

\begin{figure}
    \centering
    \begin{subfigure}[t]{0.50\columnwidth}
        \centering
        \includegraphics[width=\columnwidth, keepaspectratio]{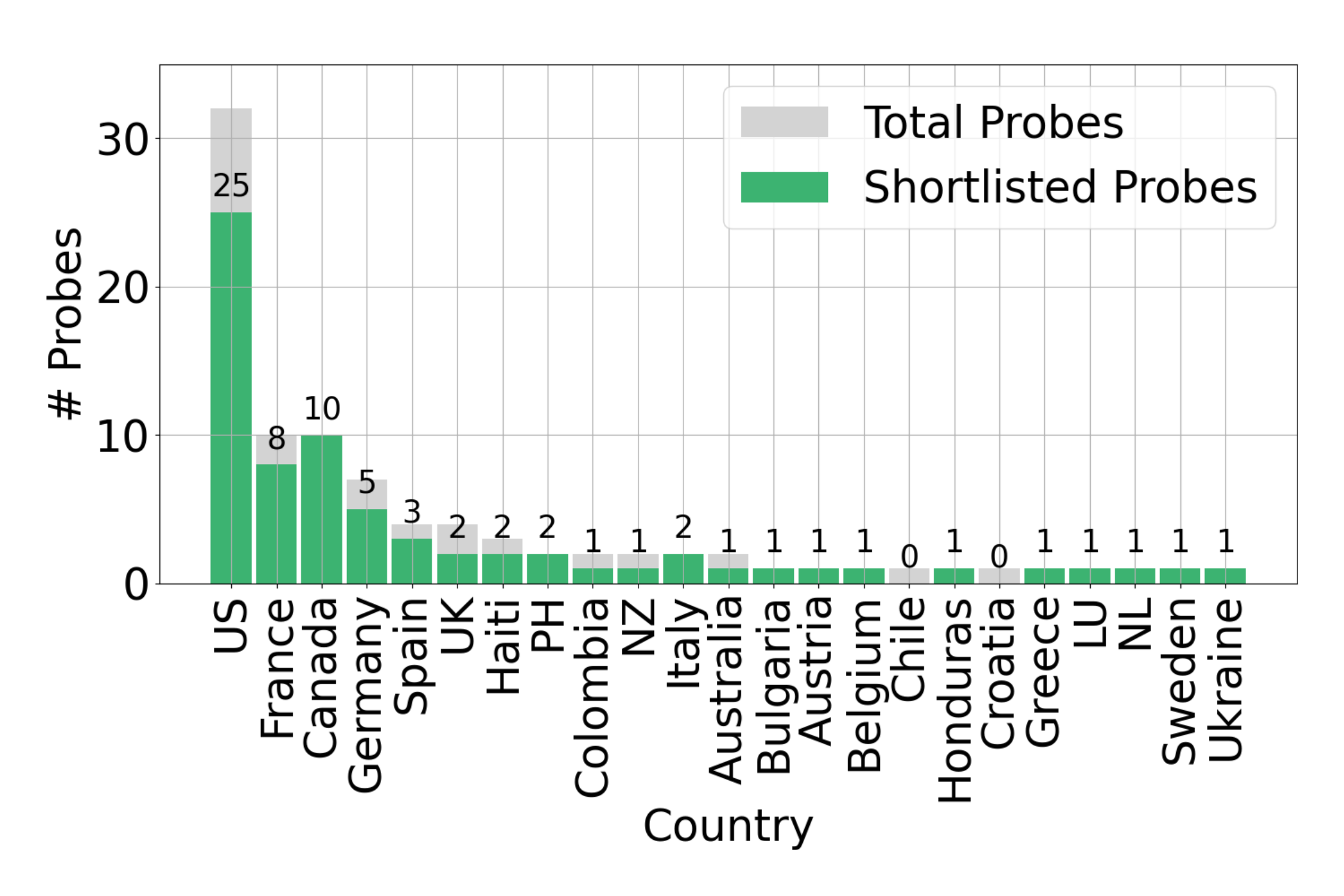}
        \caption{Shortlisted 71, out of 91 Probes, in May'24}
    \end{subfigure}%
    \hfill
    \begin{subfigure}[t]{0.50\columnwidth}
        \centering
        \includegraphics[width=\columnwidth, keepaspectratio]{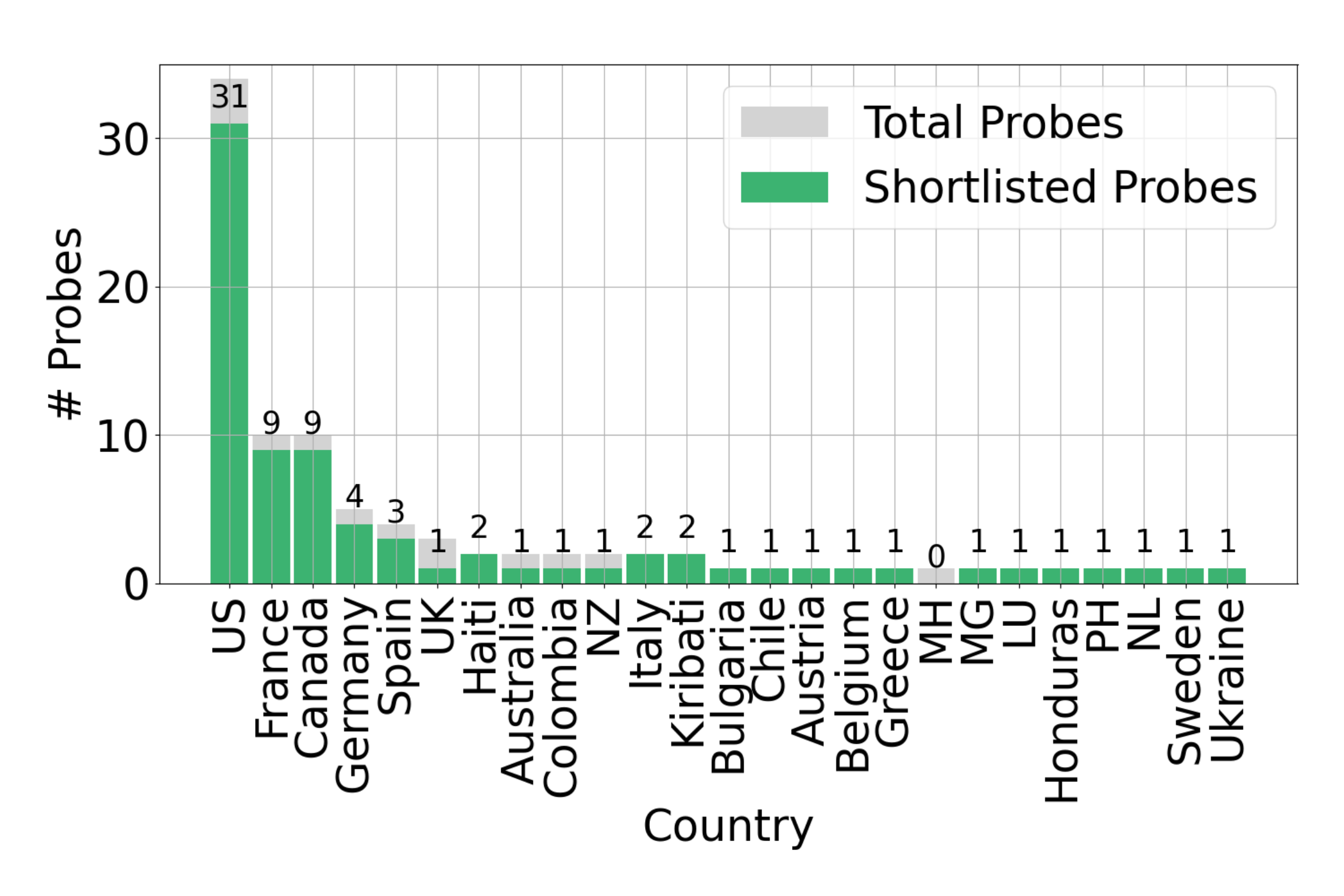}
        \caption{Shortlisted 78, out of 91 Probes, in Oct'24}
    \end{subfigure}%
    \caption{Country-wise available RIPE Atlas probes and shortlisted probes (i.e., not shifting RTTs in different levels over the observation window) for the analysis. [United States: US, United Kingdom: UK, New Zealand: NZ, Luxembourg: LU, Netherlands: NL, Philippines: PH, Marshall Islands: MH, Madagascar: MG,]}
    \label{fig:shortlistedRIPEAtlasProbes}
\end{figure}

The speed test results in the M-LAB and Cloudflare datasets, and the number of RIPE Atlas probes, are extremely skewed towards the US compared to the rest of the world.
To address this disparity, we treat each US state as a separate region, whereas for other countries, we treat each country as a region.
We ensure each region has at least one speed test per day; otherwise, we exclude that region from our analysis.
This turns out to be a negligible few hundred speed tests removed from a few thousand to a million speed test results.
Additionally, latency measurements from a few RIPE Atlas probes exhibit arbitrary shifts for days, as reported in prior work~\cite{basak2025investigation}.
We speculate that this is either due to a change in Starlink's egress ground station or to a change in DNS localization~\cite{bose2025investigating}.
However, we manually intervene to remove such probes and shortlist 71 and 78 probes out of the total 91 probes that do not fluctuate during our analysis window in May and October 2024.
Fig.~\ref{fig:shortlistedRIPEAtlasProbes} shows the country-wise number of probes we have shortlisted for our analysis.
In the following, we measure how overall Internet performance changes during solar storms and quantify region-wise degradation worldwide.

\subsubsection{Throughput:}

\begin{figure}
    \centering
    \begin{subfigure}[t]{0.25\columnwidth}
        \centering
        \includegraphics[width=\columnwidth, keepaspectratio]{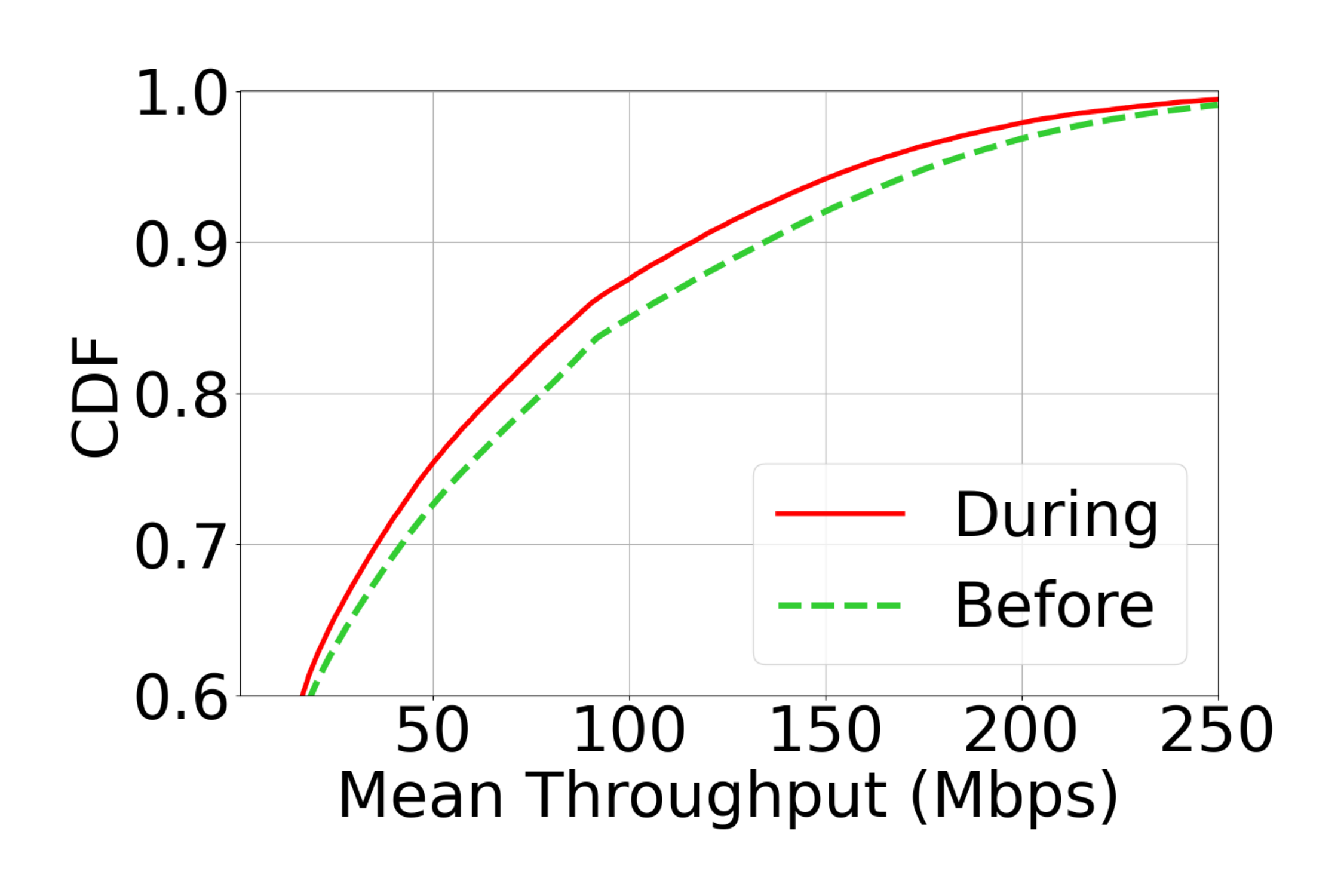}
        \caption{ndt7}
    \end{subfigure}%
    \hfill
    \begin{subfigure}[t]{0.25\columnwidth}
        \centering
        \includegraphics[width=\columnwidth, keepaspectratio]{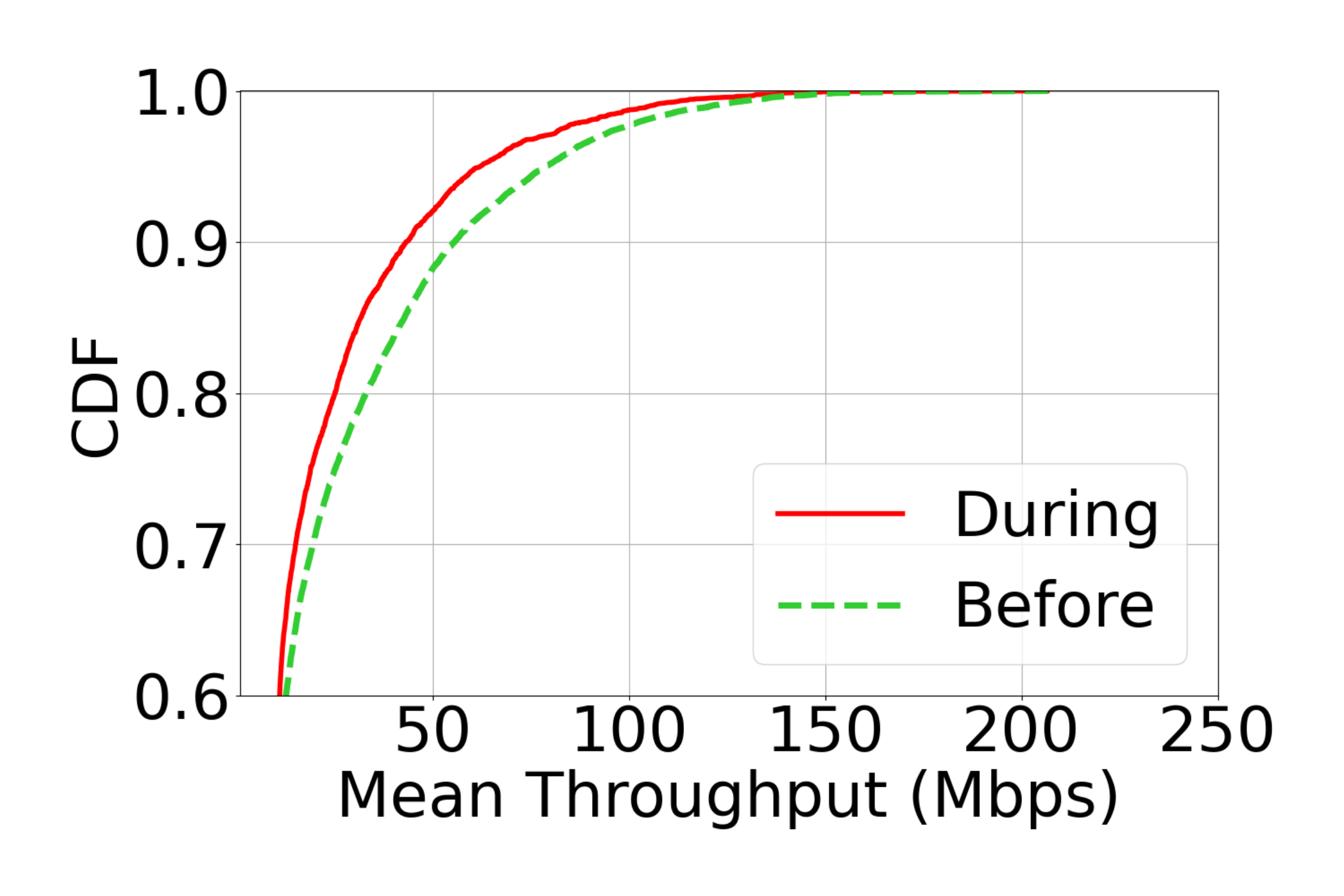}
        \caption{ndt5}
    \end{subfigure}%
    \hfill
    \begin{subfigure}[t]{0.25\columnwidth}
        \centering
        \includegraphics[width=\columnwidth, keepaspectratio]{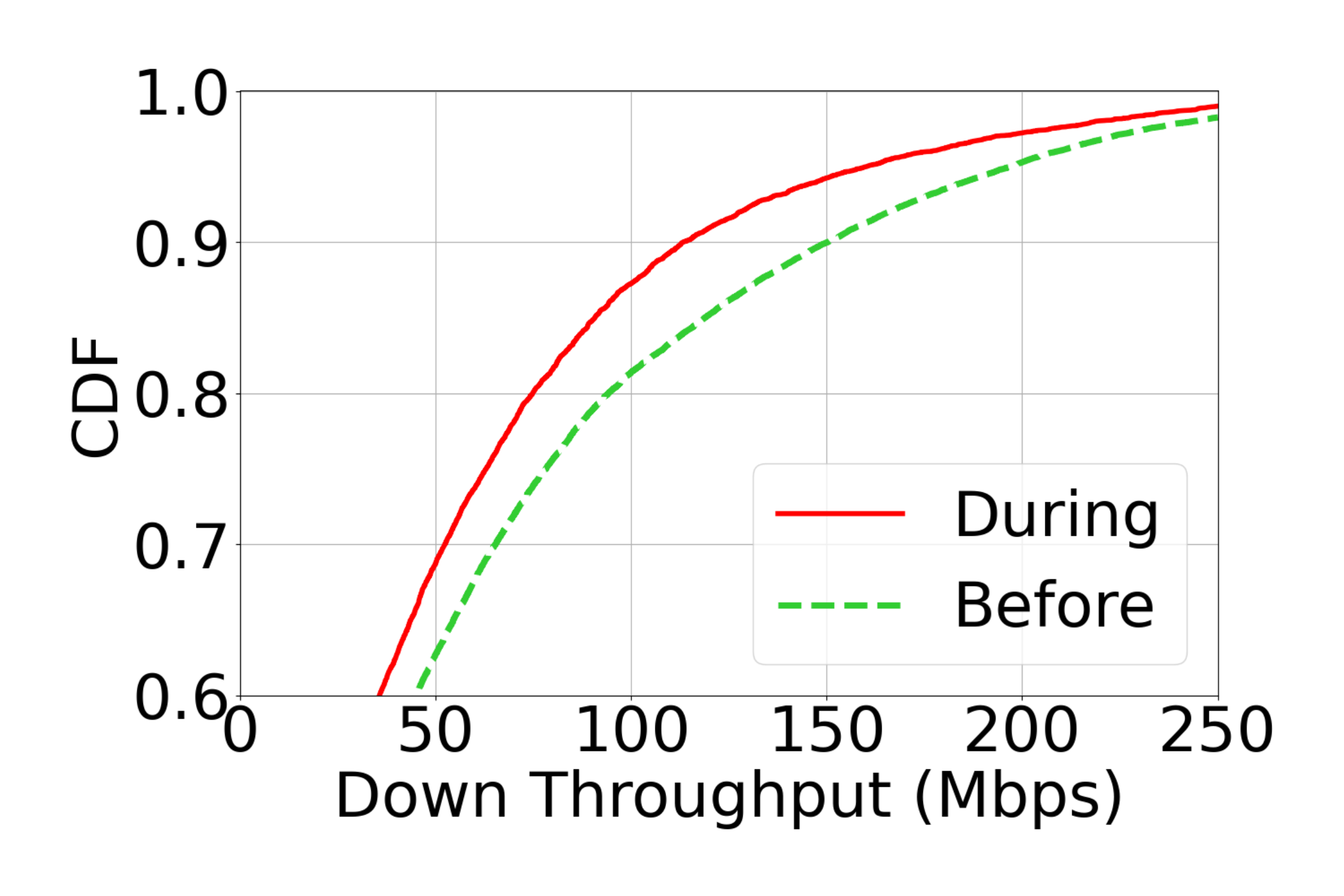}
        \caption{AIM downlink}
    \end{subfigure}%
    \hfill
    \begin{subfigure}[t]{0.25\columnwidth}
        \centering
        \includegraphics[width=\columnwidth, keepaspectratio]{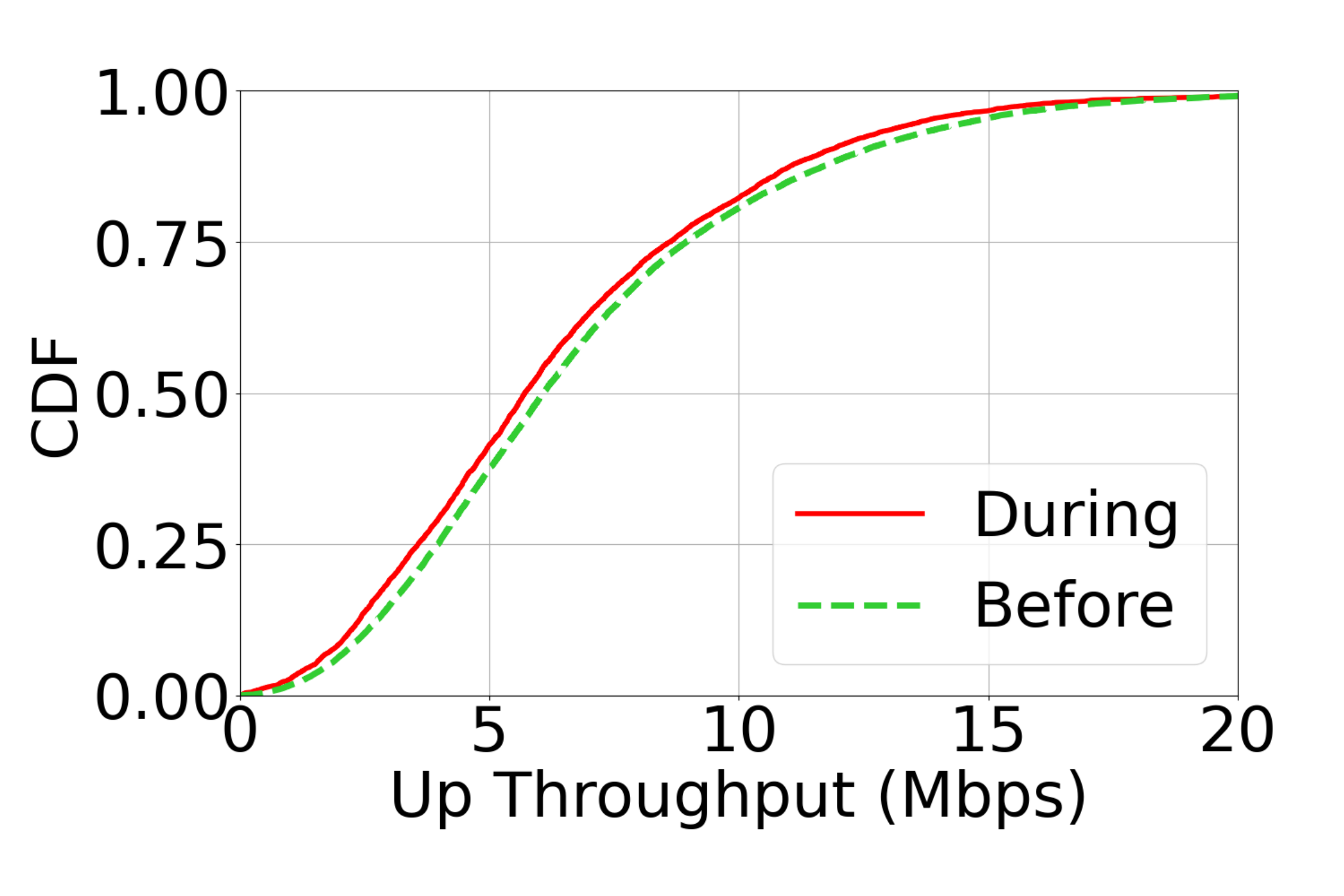}
        \caption{AIM uplink}
    \end{subfigure}%

    \hfill

    \begin{subfigure}[t]{0.25\columnwidth}
        \centering
        \includegraphics[width=\columnwidth, keepaspectratio]{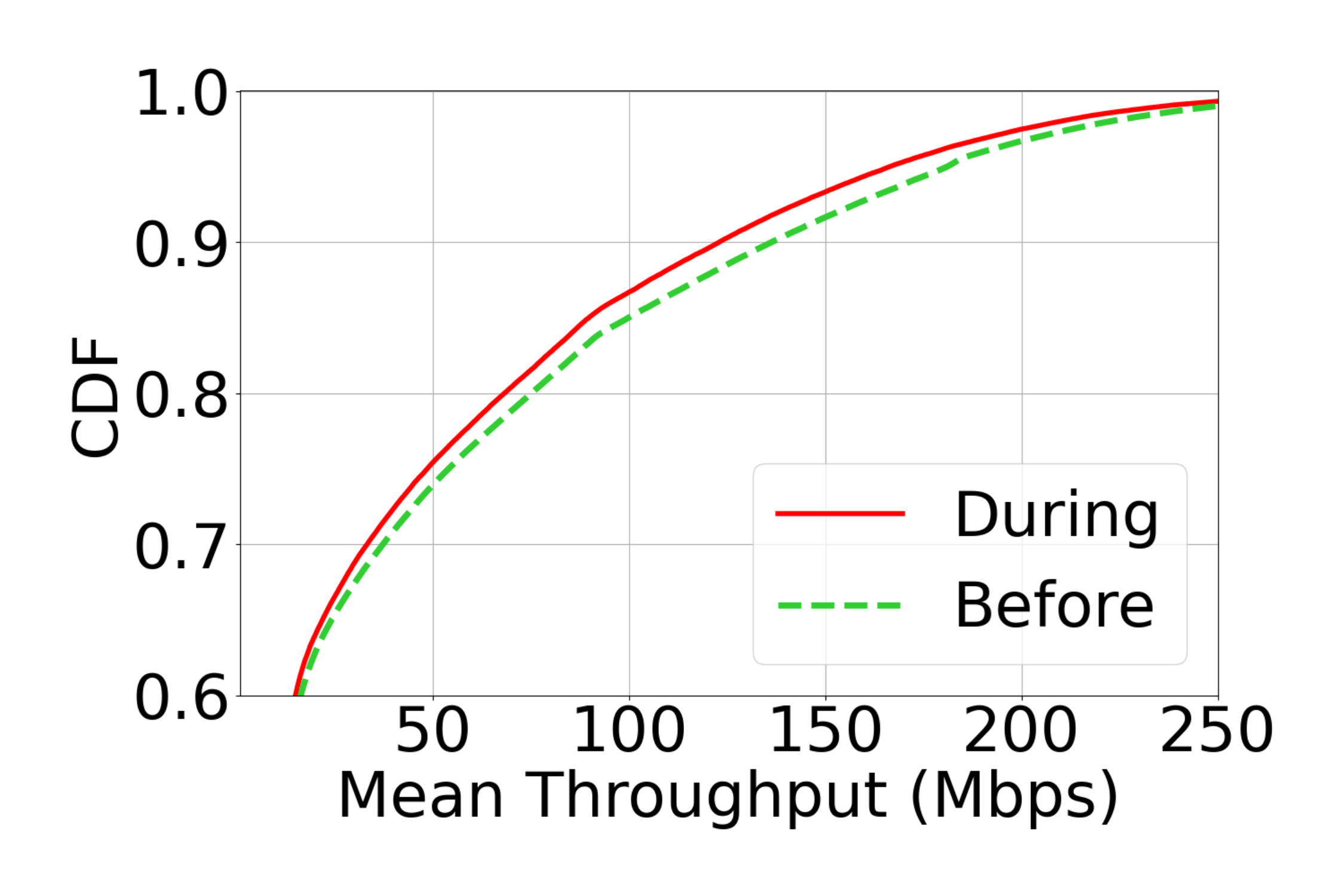}
        \caption{ndt7}
    \end{subfigure}%
    \hfill
    \begin{subfigure}[t]{0.25\columnwidth}
        \centering
        \includegraphics[width=\columnwidth, keepaspectratio]{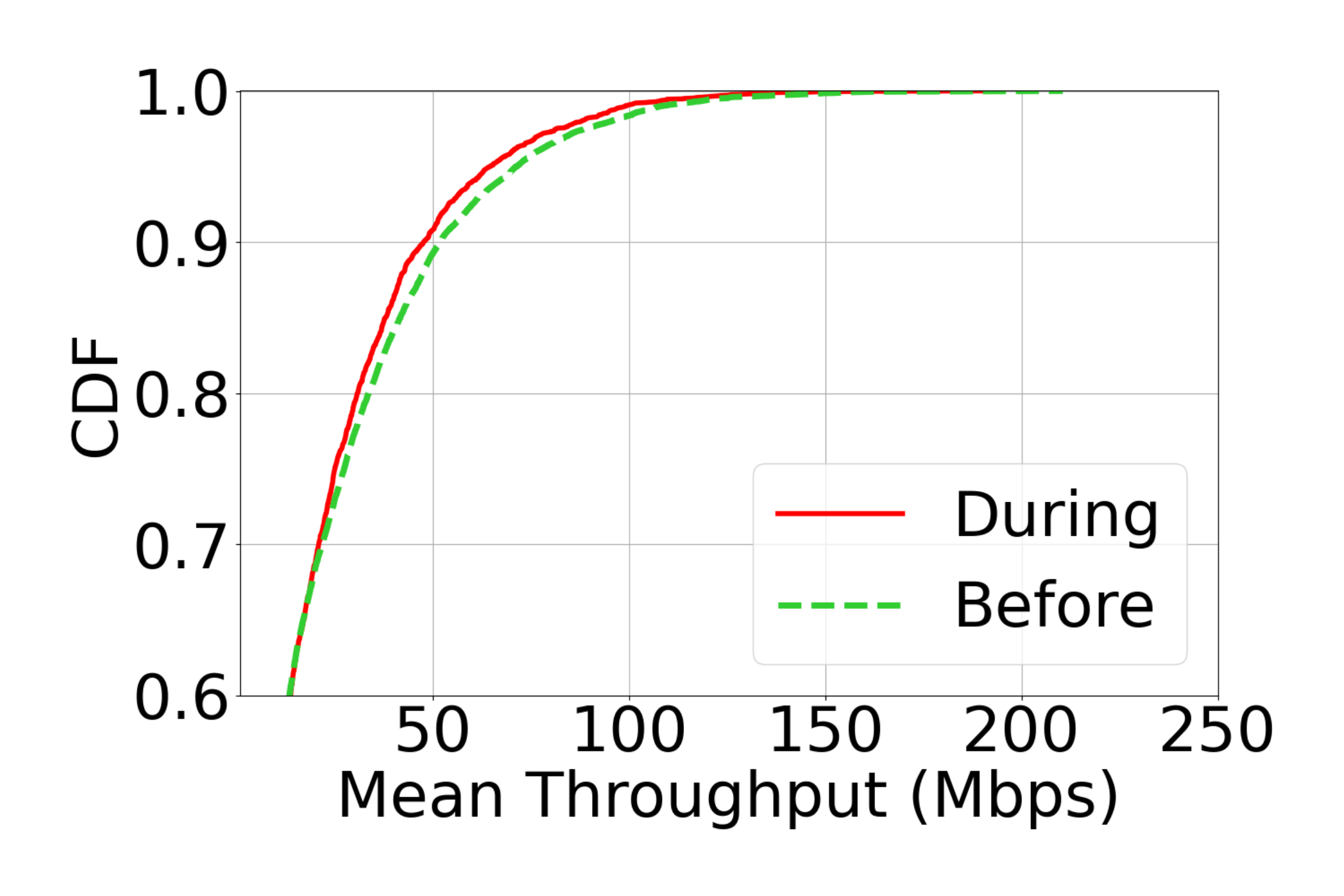}
        \caption{ndt5}
    \end{subfigure}%
    \hfill
    \begin{subfigure}[t]{0.25\columnwidth}
        \centering
        \includegraphics[width=\columnwidth, keepaspectratio]{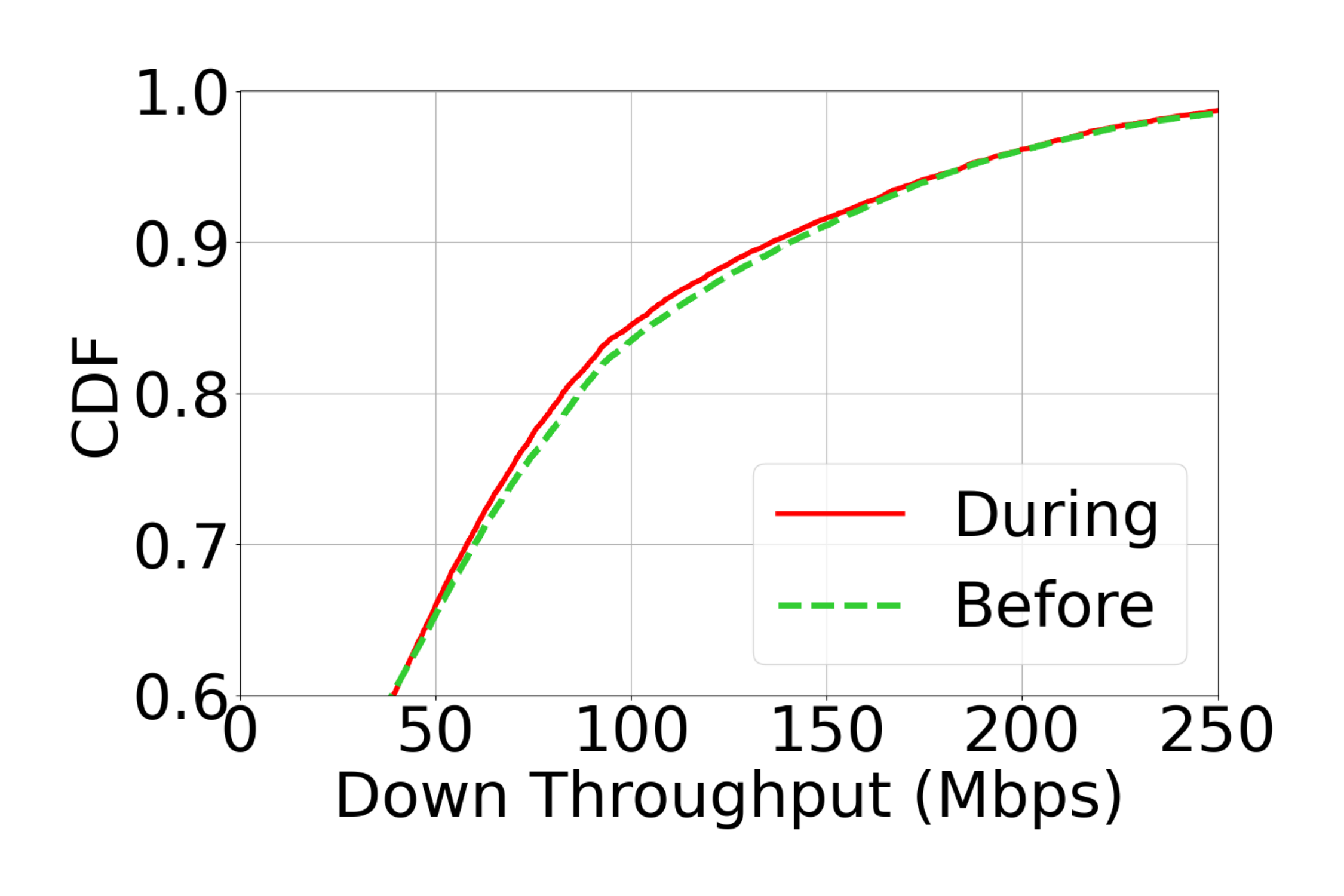}
        \caption{AIM downlink}
    \end{subfigure}%
    \hfill
    \begin{subfigure}[t]{0.25\columnwidth}
        \centering
        \includegraphics[width=\columnwidth, keepaspectratio]{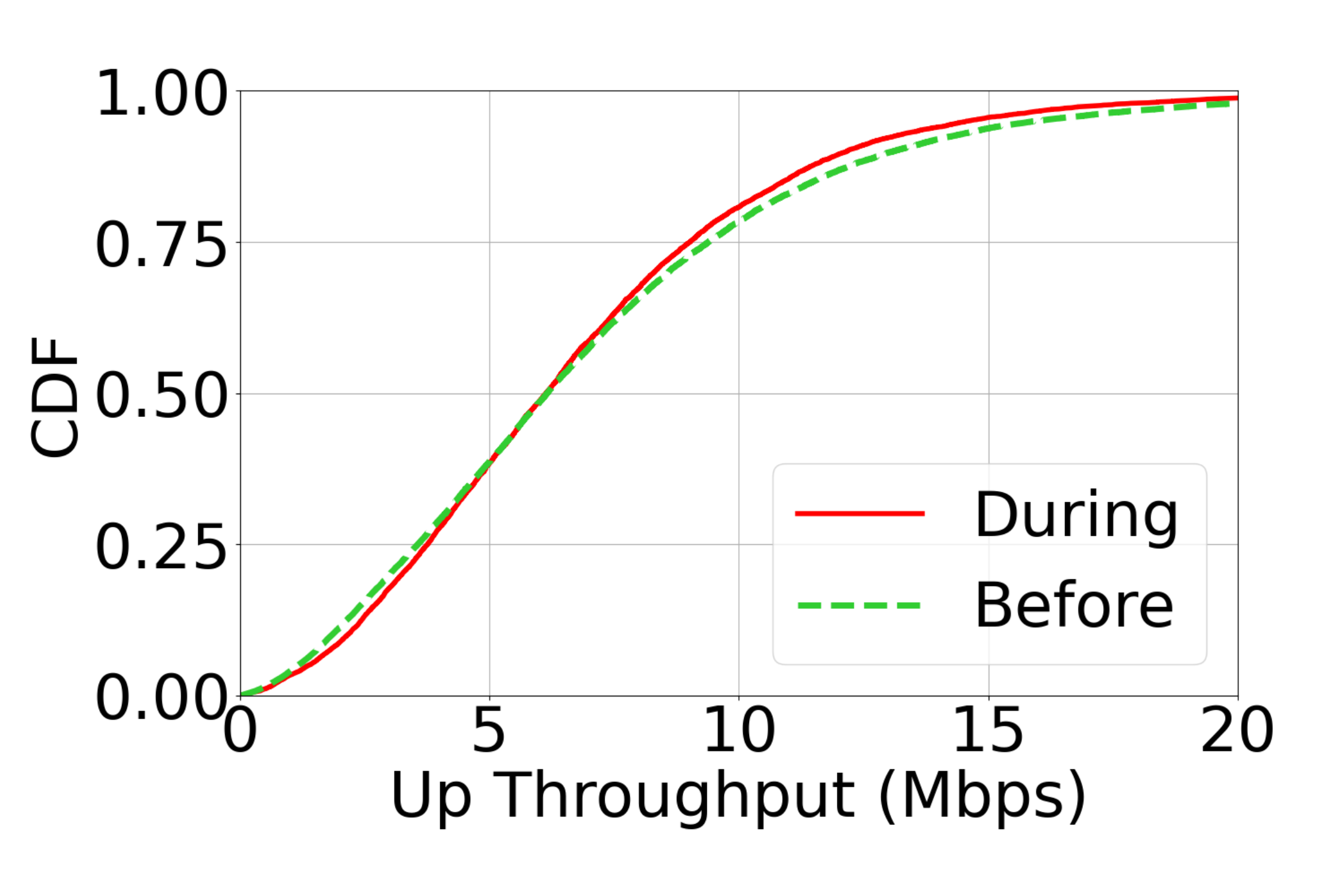}
        \caption{AIM uplink}
    \end{subfigure}%

    \caption{Overall throughput distribution of worldwide speed test results during (a)-(d) May 2024 solar superstorm and (e)-(h) October 2024 solar storm, indicating a notable drop in data rates during the solar storm.}
    \label{fig:userThroughputOverall}
\end{figure}

\begin{figure}
    \centering
    \begin{subfigure}[t]{0.25\columnwidth}
        \centering
        \includegraphics[width=\columnwidth, keepaspectratio]{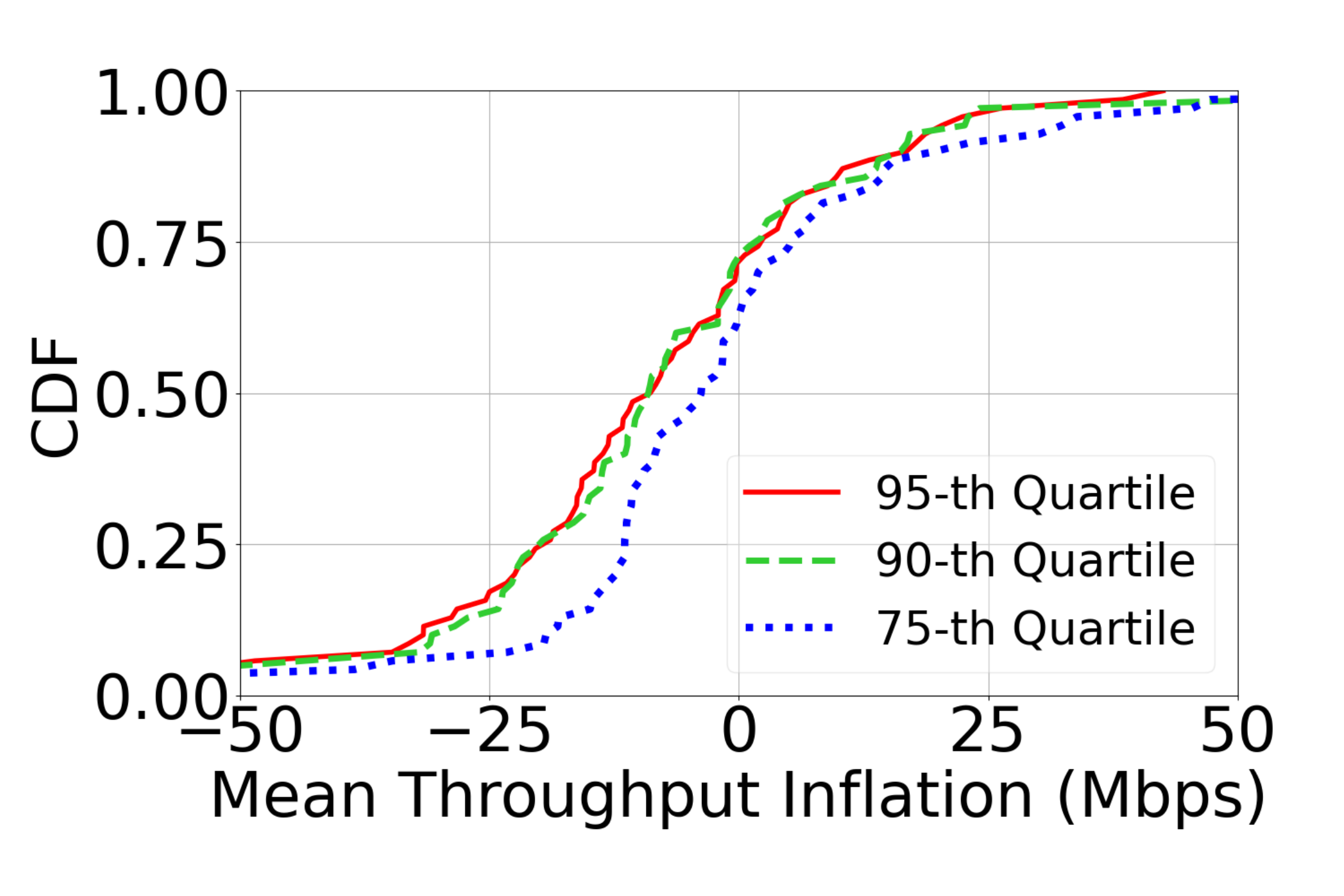}
        \caption{ndt7}
    \end{subfigure}%
    \hfill
    \begin{subfigure}[t]{0.25\columnwidth}
        \centering
        \includegraphics[width=\columnwidth, keepaspectratio]{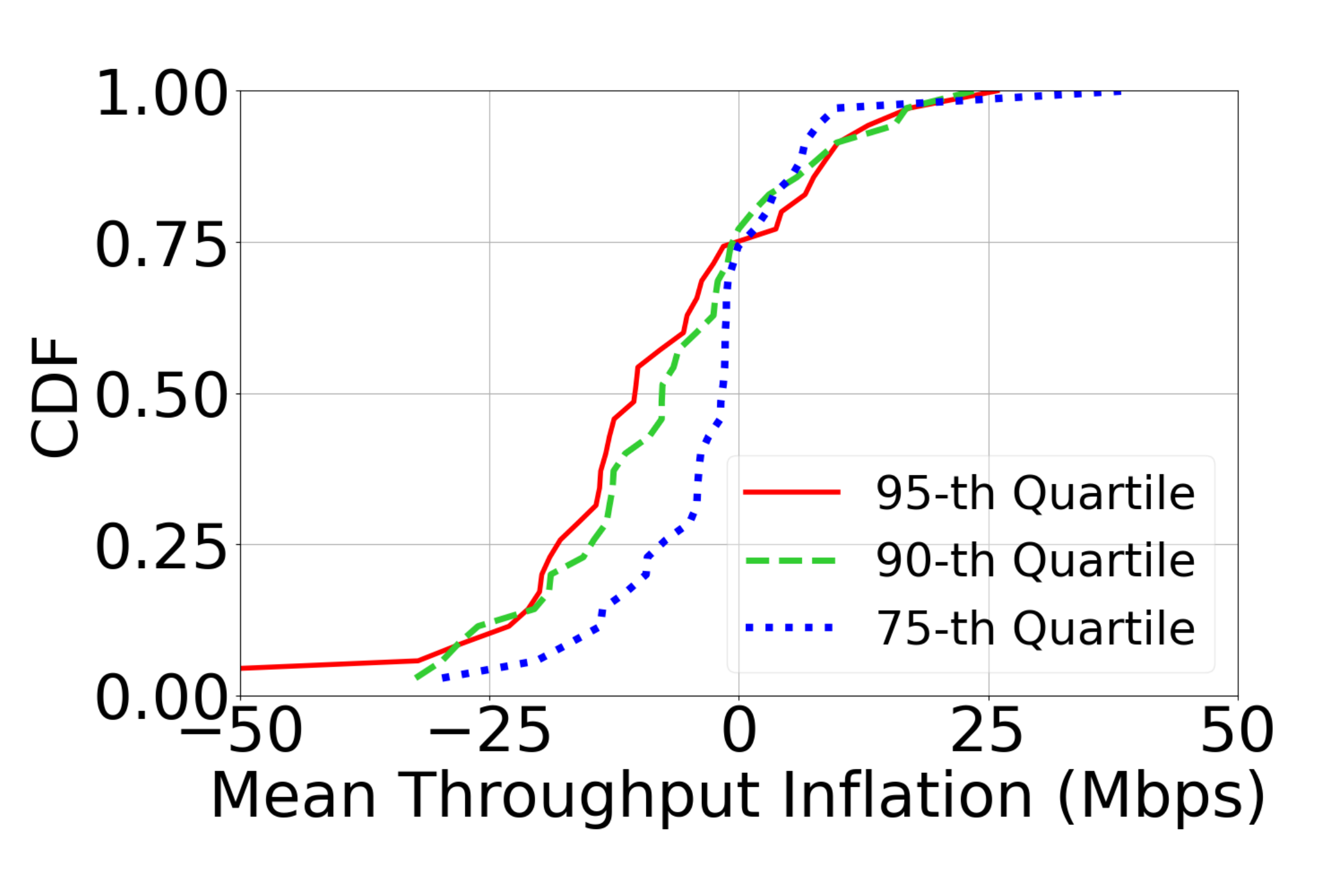}
        \caption{ndt5}
    \end{subfigure}%
    \hfill
    \begin{subfigure}[t]{0.25\columnwidth}
        \centering
        \includegraphics[width=\columnwidth, keepaspectratio]{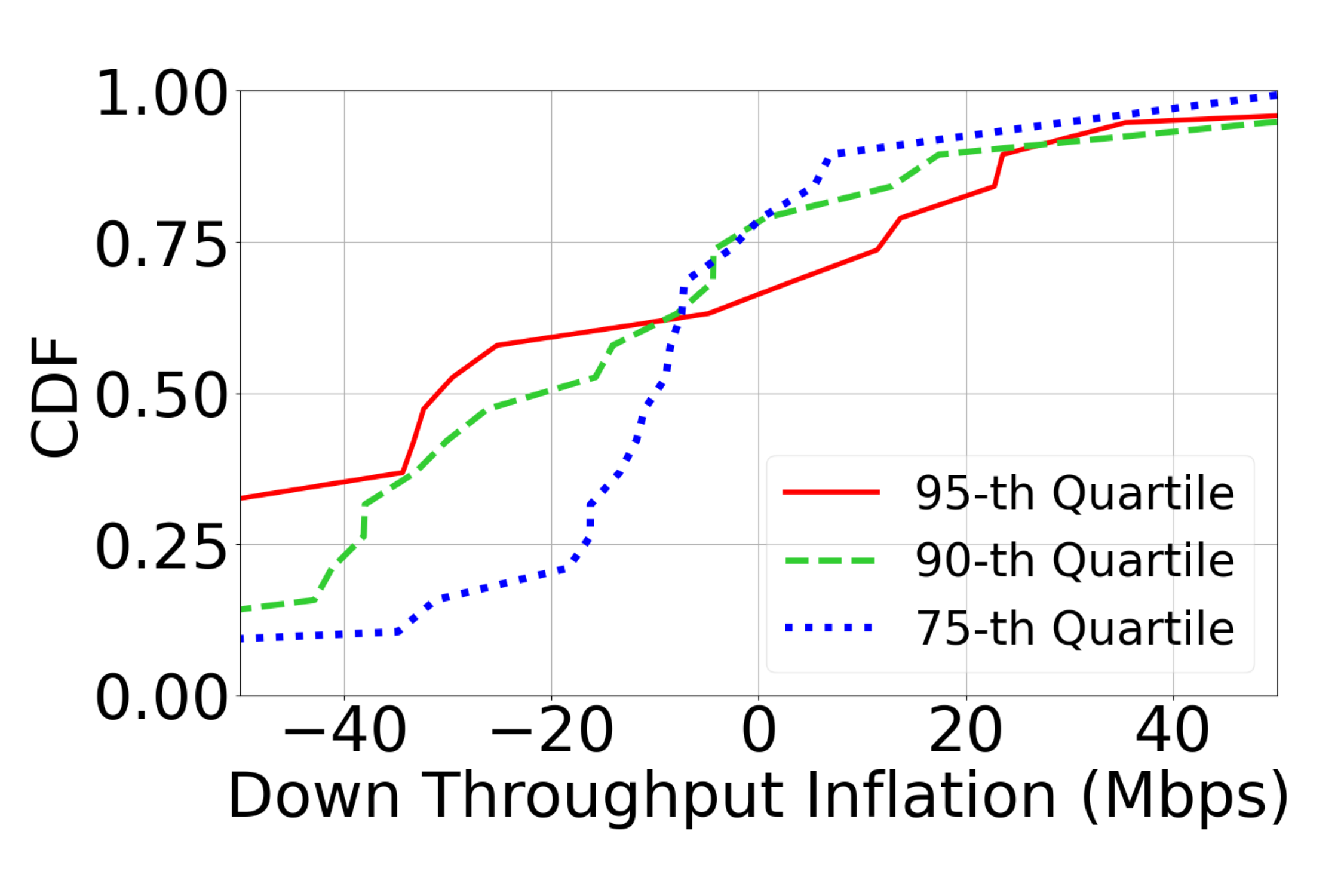}
        \caption{AIM downlink}
    \end{subfigure}%
    \hfill
    \begin{subfigure}[t]{0.25\columnwidth}
        \centering
        \includegraphics[width=\columnwidth, keepaspectratio]{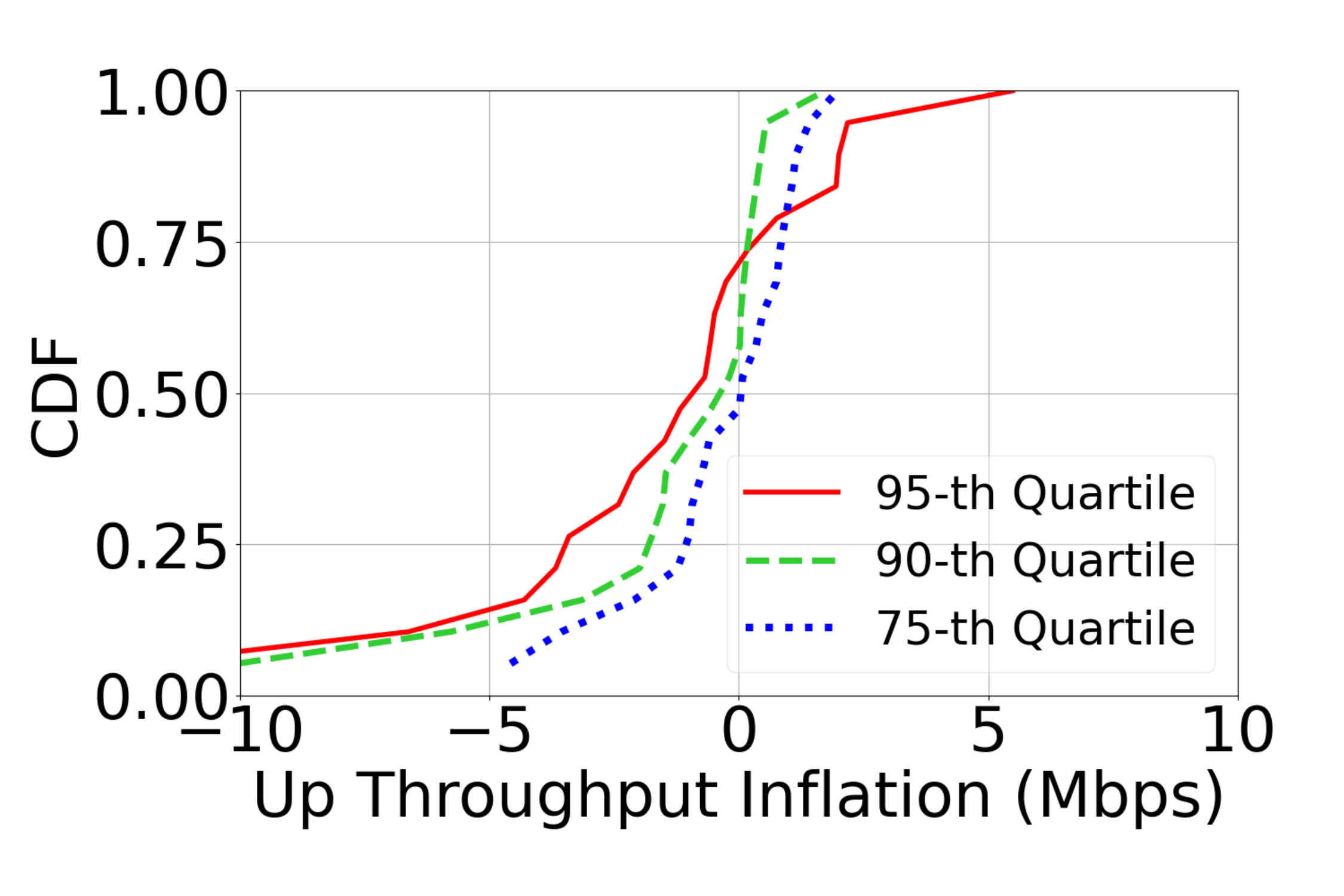}
        \caption{AIM uplink}
    \end{subfigure}%

    \hfill

    \begin{subfigure}[t]{0.25\columnwidth}
        \centering
        \includegraphics[width=\columnwidth, keepaspectratio]{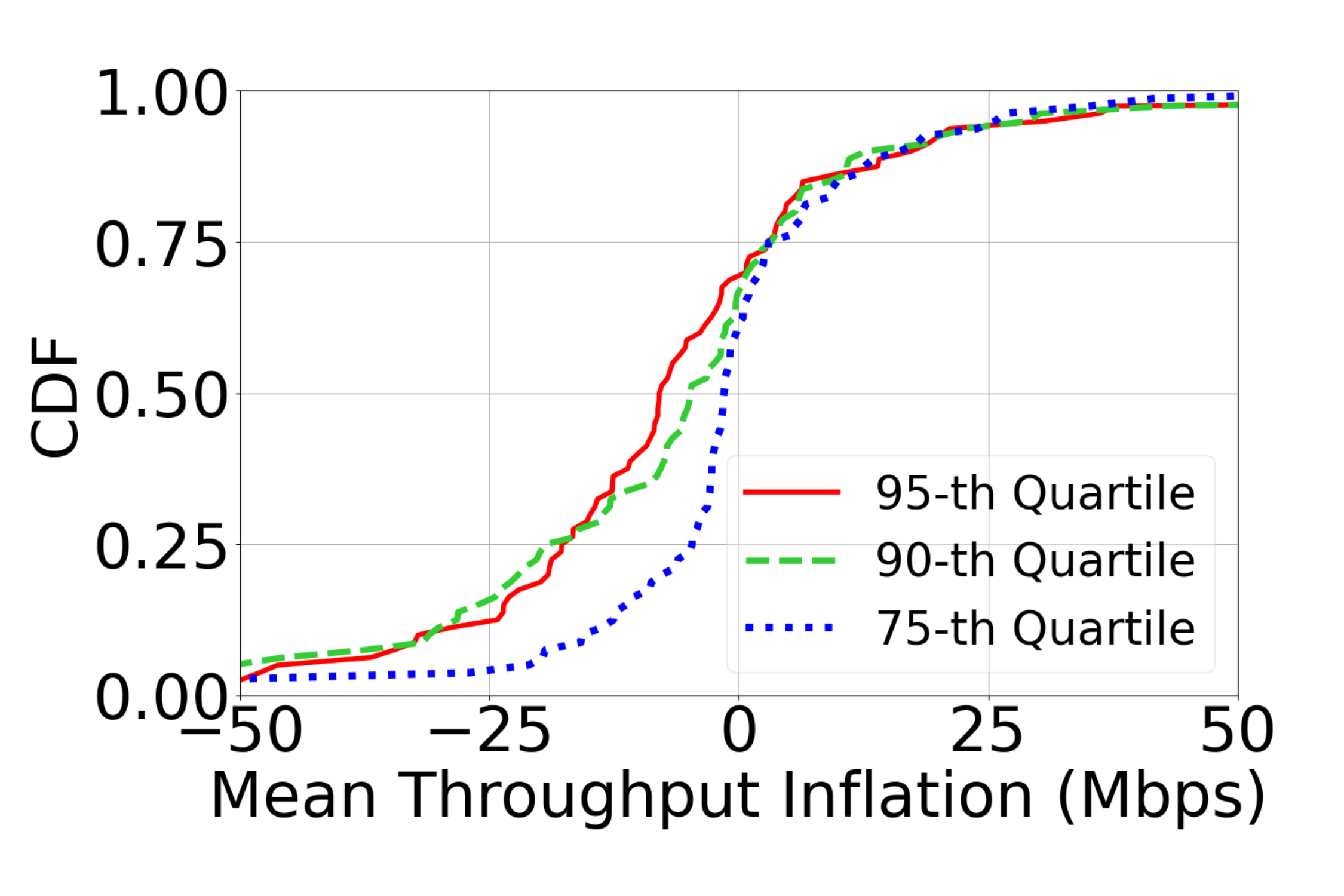}
        \caption{ndt7}
    \end{subfigure}%
    \hfill
    \begin{subfigure}[t]{0.25\columnwidth}
        \centering
        \includegraphics[width=\columnwidth, keepaspectratio]{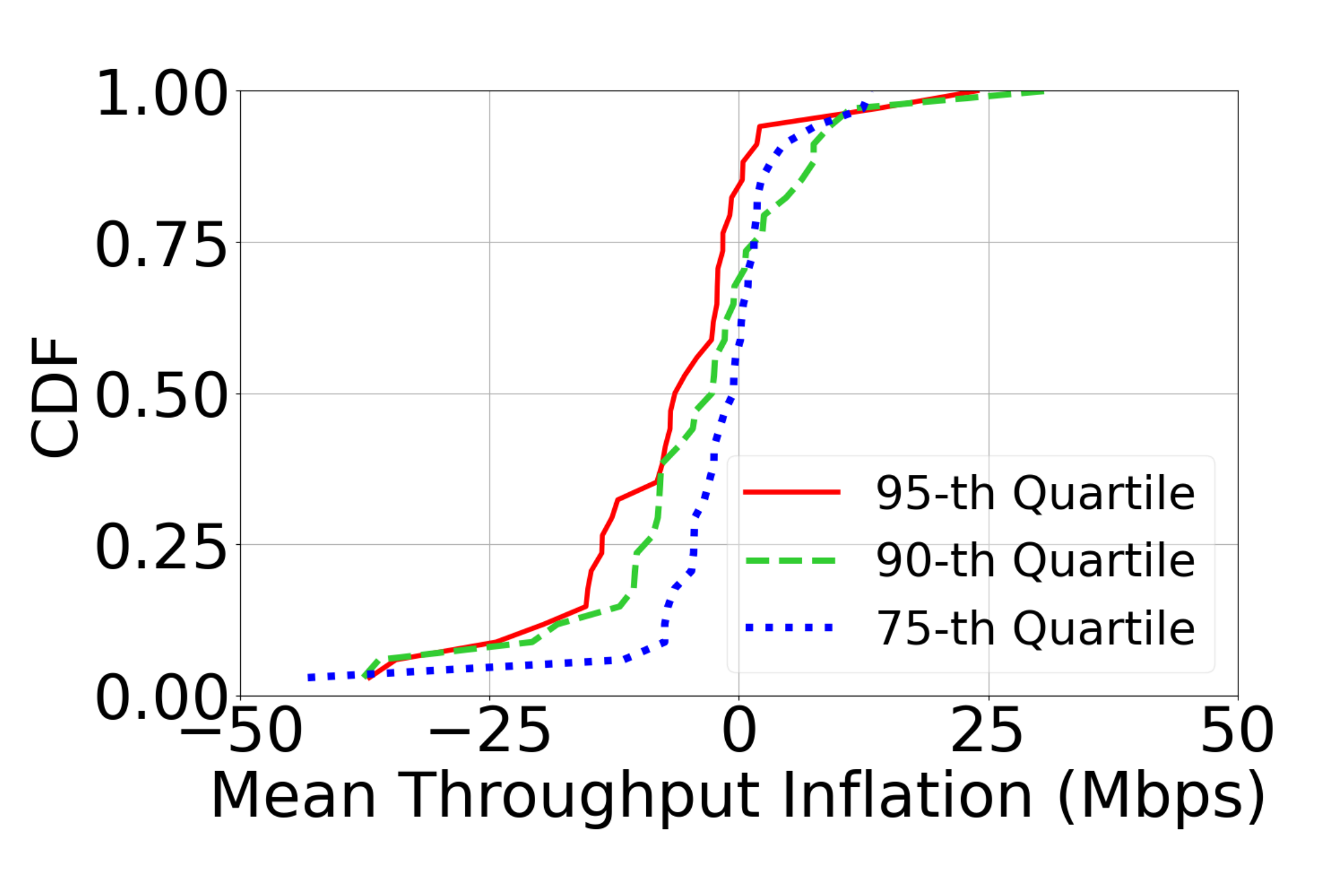}
        \caption{ndt5}
    \end{subfigure}%
    \hfill
    \begin{subfigure}[t]{0.25\columnwidth}
        \centering
        \includegraphics[width=\columnwidth, keepaspectratio]{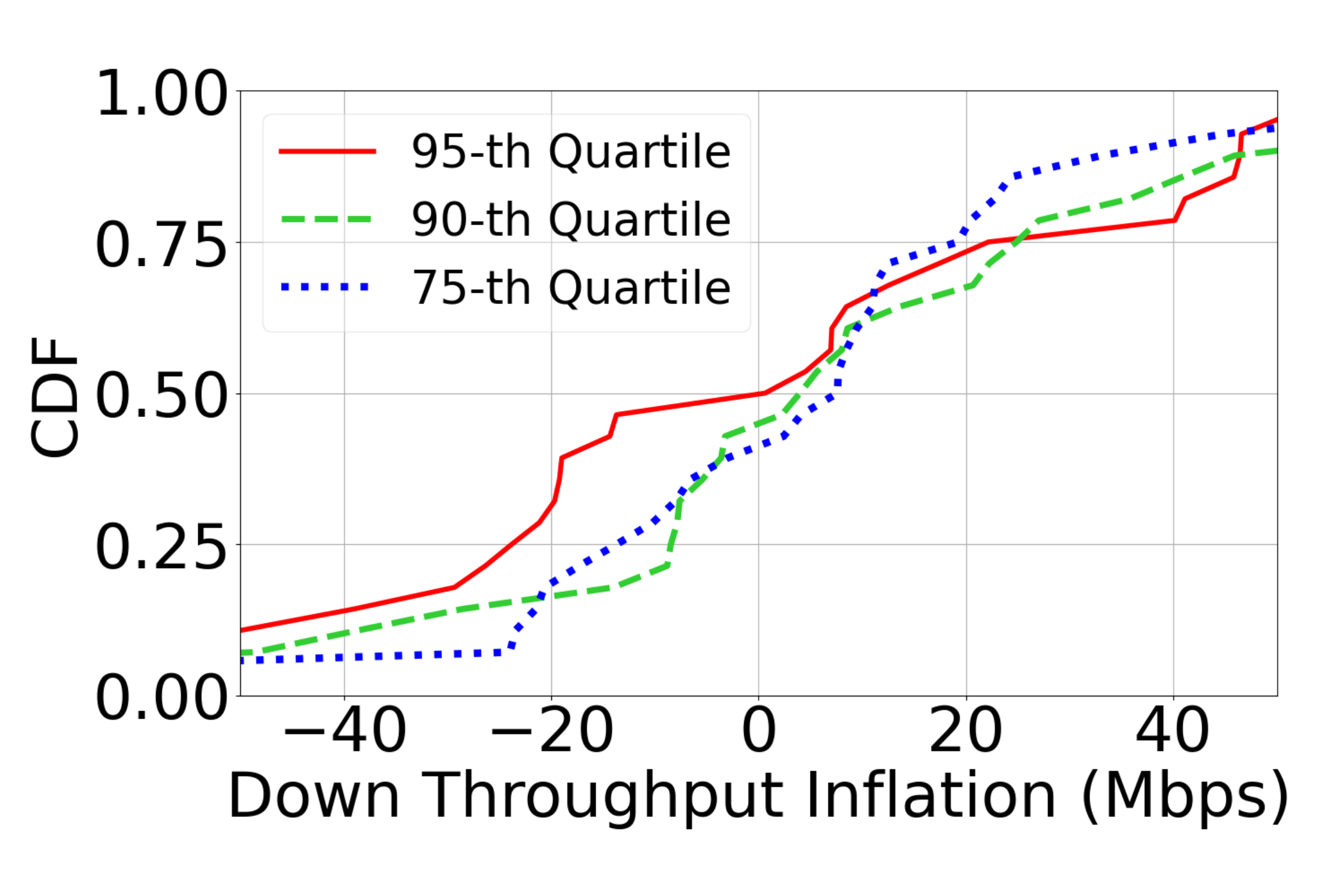}
        \caption{AIM downlink}
    \end{subfigure}%
    \hfill
    \begin{subfigure}[t]{0.25\columnwidth}
        \centering
        \includegraphics[width=\columnwidth, keepaspectratio]{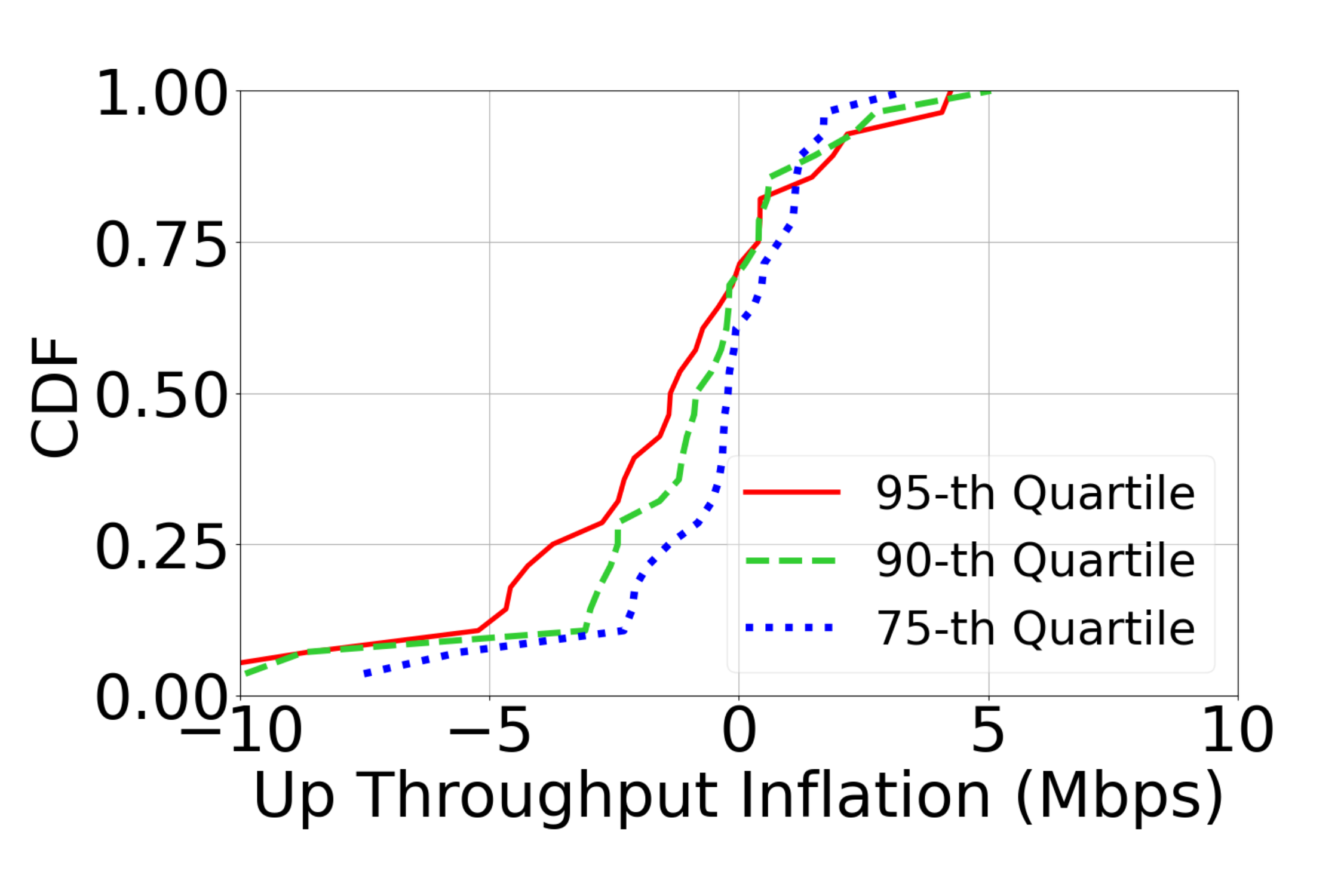}
        \caption{AIM uplink}
    \end{subfigure}%

    \caption{Distribution of region-wise throughput degradation from speed test results during (a)-(d) May 2024 solar superstorm and (e)-(h) October 2024 solar storm, indicating 75\% of the regions experienced a drop in data rates up to a few 10s of Mbps. }
    \label{fig:userThroughputInflation}
\end{figure}

In Fig.~\ref{fig:userThroughputOverall}, we plot the pre-storm and during-storm distribution of mean throughput from M-LAB and downlink and uplink throughput from Cloudflare AIM speed test results. 
Note that we use the same pre-storm and during-storm windows as the previous analysis.
During the May 2024 solar superstorm, observe in Fig.~\ref{fig:userThroughputOverall}(a)-(d) all the figures show a reduction in throughput above the 60th percentile, while the uplink is the exception, showing a negligible throughput decrease during the event.
Quantitatively, we observe drops of up to 18 Mbps (16 Mbps), 37 Mbps, and 1 Mbps in the M-LAB ndt7 (ndt5) mean throughput, the AIM downlink, and the AIM uplink, respectively. 
In contrast, during the October 2024 solar storm, the change in the throughput distribution is much smaller than the change during the May 2024 solar superstorm. 
Quantitatively, we observe drops to approximately 14 Mbps (6 Mbps) at the 95th percentile of the M-LAB ndt7 (ndt5) mean throughput. 
Approximately 5 Mbps at the 85th percentile AIM downlink, while negligible change in AIM uplink.

To look at region-specific details, we segregate the speed test results by region. 
Then, in each region, we calculate the quantile-wise throughput difference between the pre-storm and during-storm windows.
As we have already observed, the performance implication predominantly occurs at the higher quantile, while it is negligible below the median.
In Fig.~\ref{fig:userThroughputInflation}, we plot the distribution of the difference at the 75th, 90th, and 95th percentiles, where a negative value on the x-axis denotes the throughput degradation of the regions.
Notice that approximately 60-75\% of the regions worldwide experience a drop in throughput at the higher quantile. 
The statistics for both the May 2024 solar superstorm and the October 2024 solar storm are quite similar in terms of mean throughput and uplink throughput.
The AIM downlink during October 2024 in Fig.~\ref{fig:userThroughputInflation}(g) is an exception, showing a 50-50 split between regions experiencing throughput drops and hikes. 
We have relatively fewer speed test results in Cloudflare AIM across fewer geographical regions, which could lead to this exception.

\subsubsection{Latency and Jitter:}

\begin{figure}
    \centering
    \begin{subfigure}[t]{0.25\columnwidth}
        \centering
        \includegraphics[width=\columnwidth, keepaspectratio]{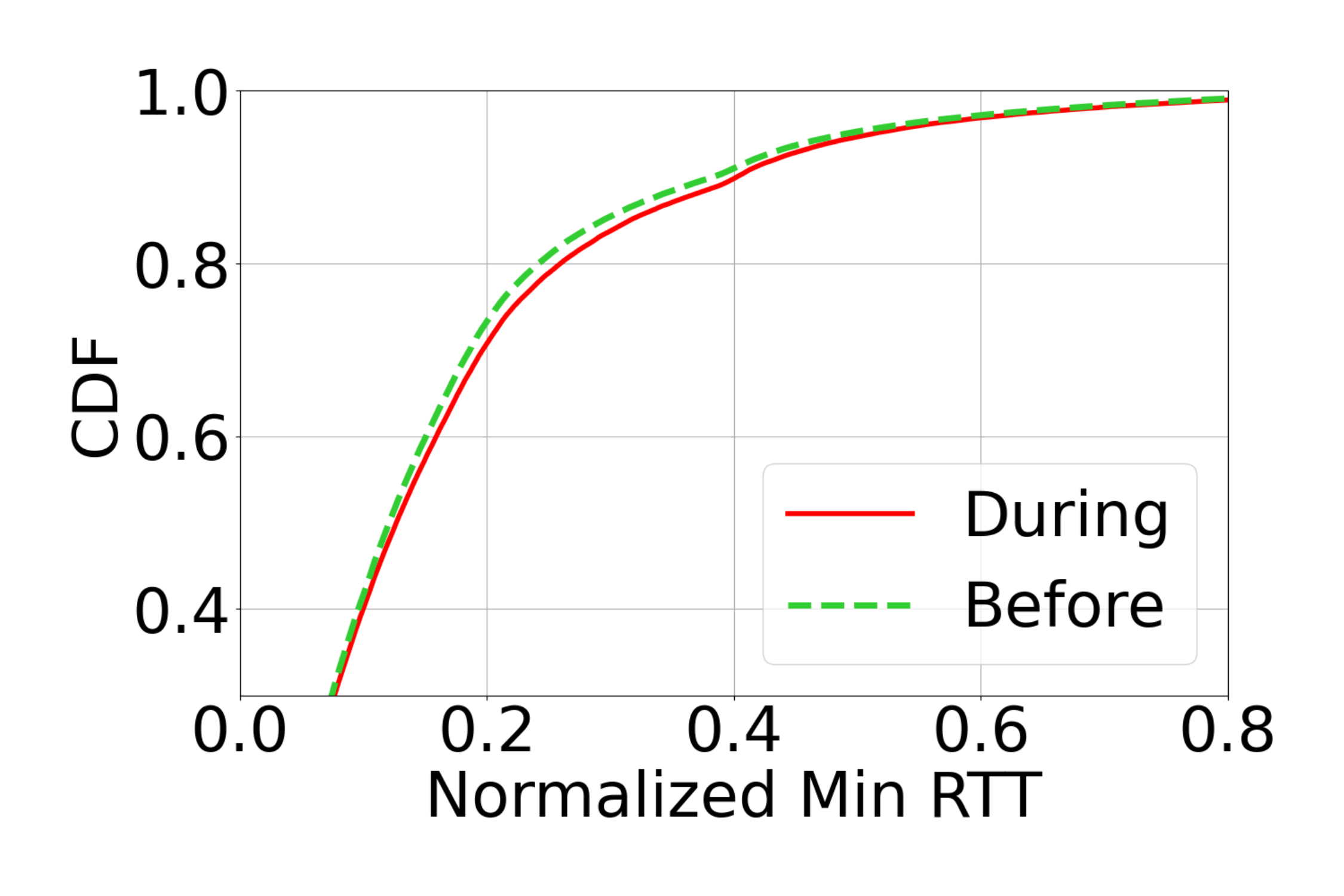}
        \caption{ndt7}
    \end{subfigure}%
    \hfill
    \begin{subfigure}[t]{0.25\columnwidth}
        \centering
        \includegraphics[width=\columnwidth, keepaspectratio]{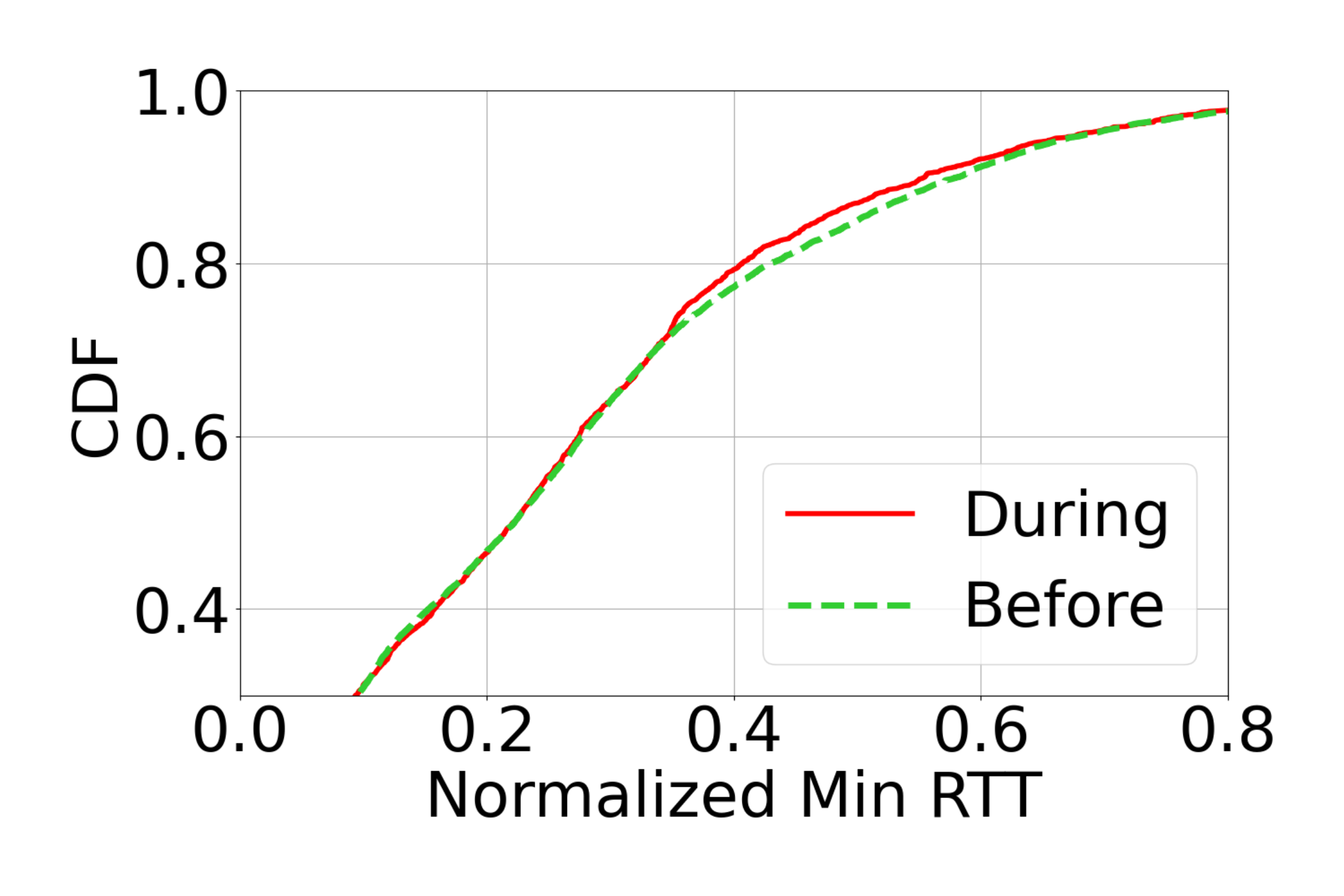}
        \caption{ndt5}
    \end{subfigure}%
    \hfill
    \begin{subfigure}[t]{0.25\columnwidth}
        \centering
        \includegraphics[width=\columnwidth, keepaspectratio]{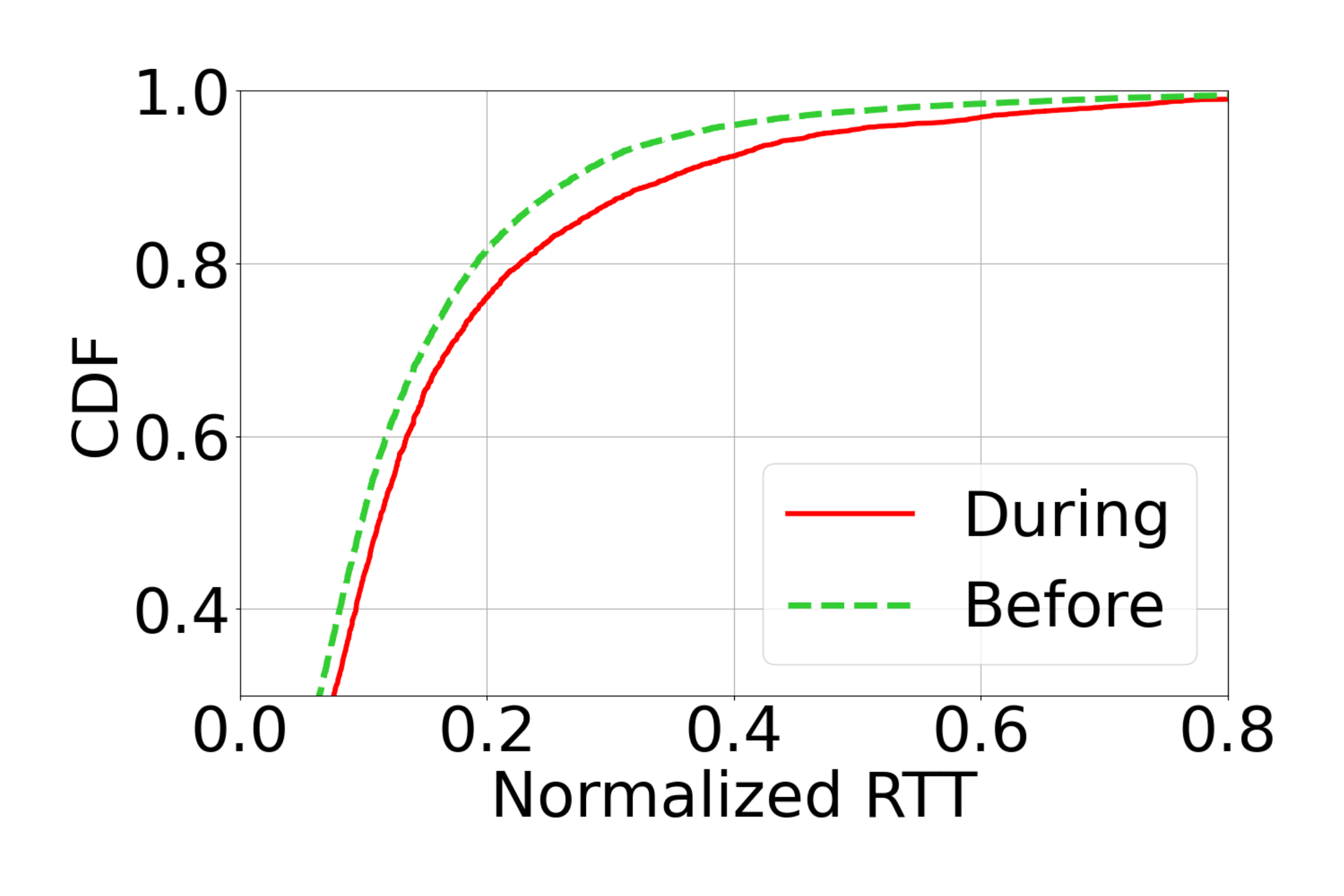}
        \caption{AIM}
    \end{subfigure}%
    \hfill
    \begin{subfigure}[t]{0.25\columnwidth}
        \centering
        \includegraphics[width=\columnwidth, keepaspectratio]{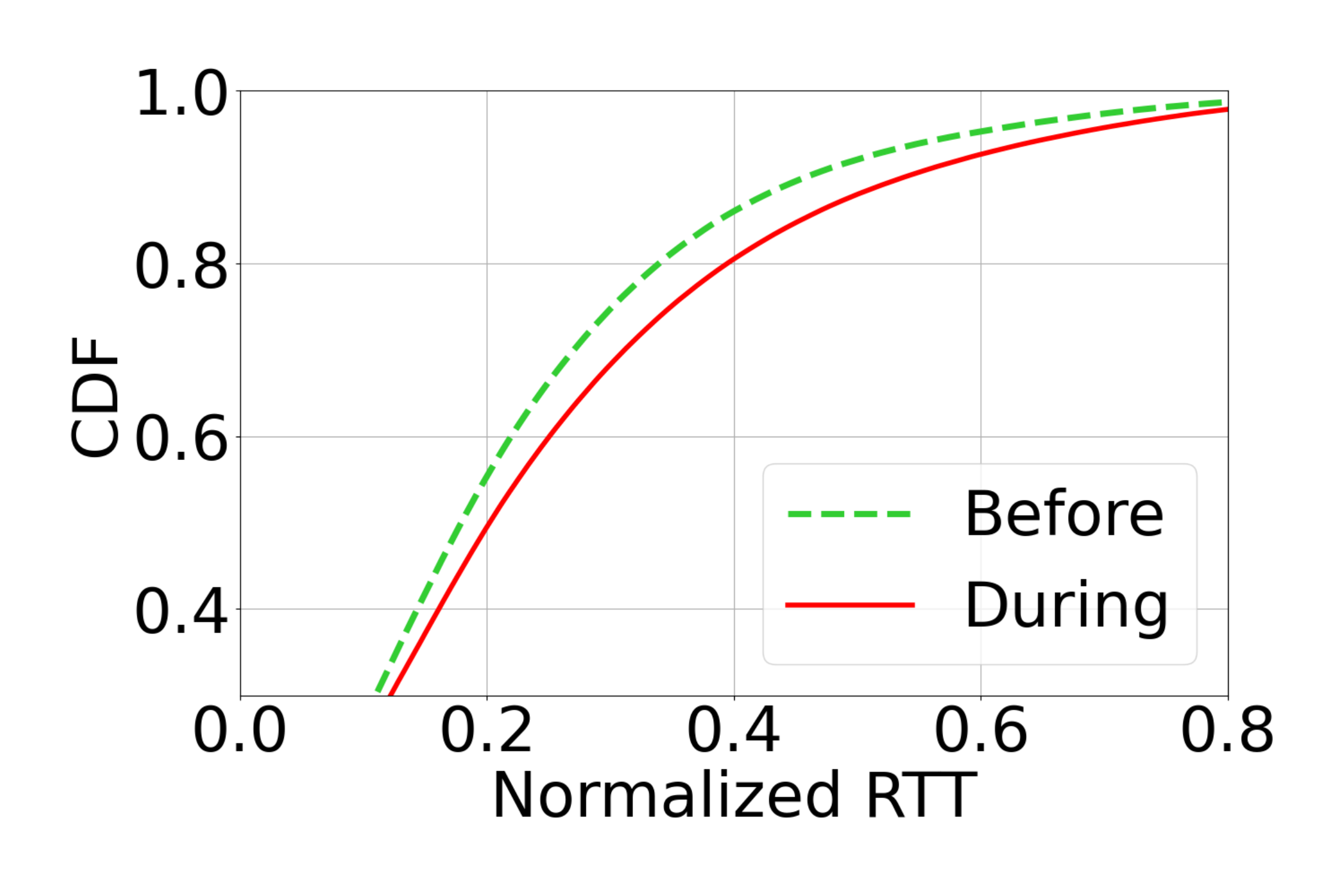}
        \caption{RIPE Atlas}
    \end{subfigure}%

    \hfill

    \begin{subfigure}[t]{0.25\columnwidth}
        \centering
        \includegraphics[width=\columnwidth, keepaspectratio]{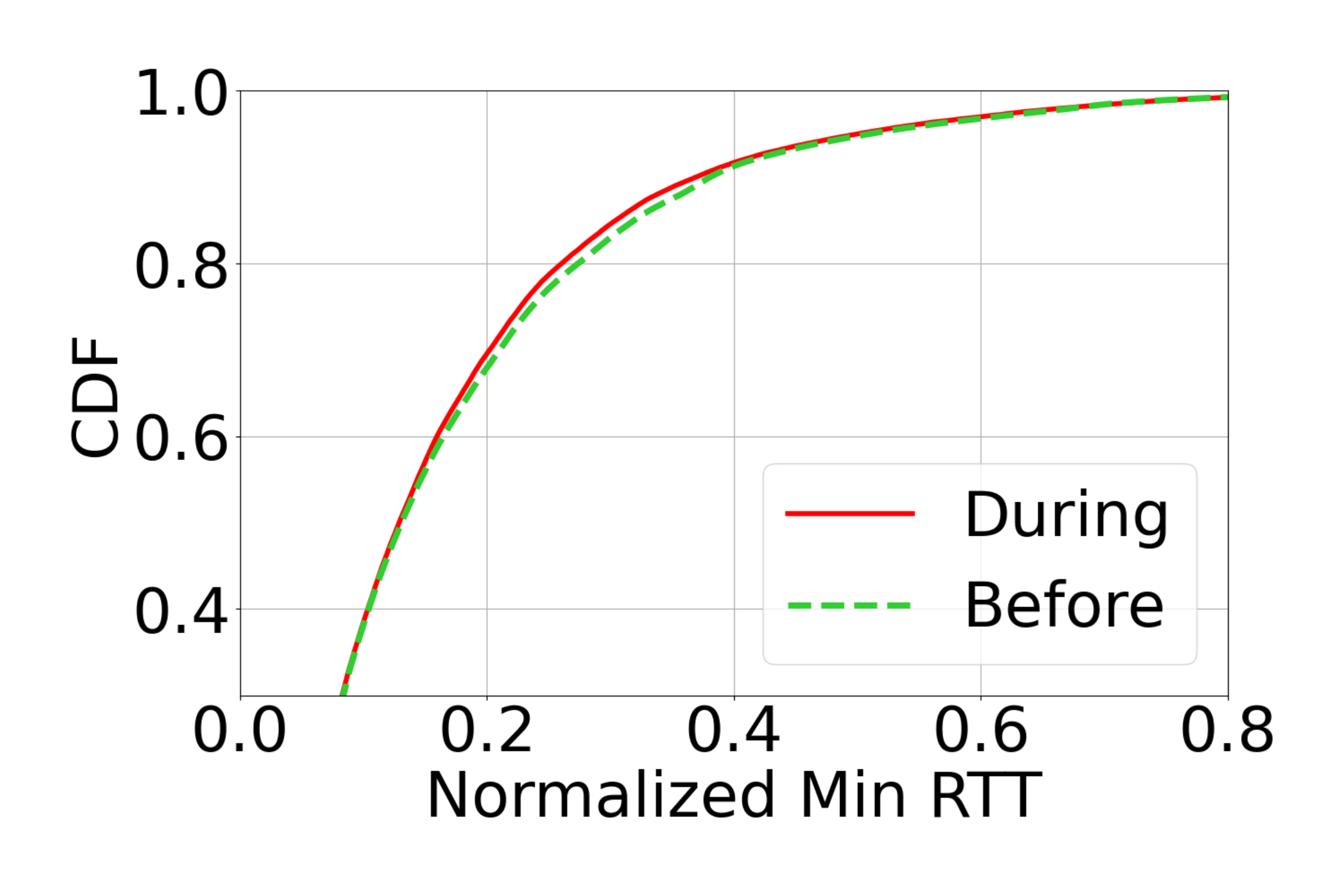}
        \caption{ndt7}
    \end{subfigure}%
    \hfill
    \begin{subfigure}[t]{0.25\columnwidth}
        \centering
        \includegraphics[width=\columnwidth, keepaspectratio]{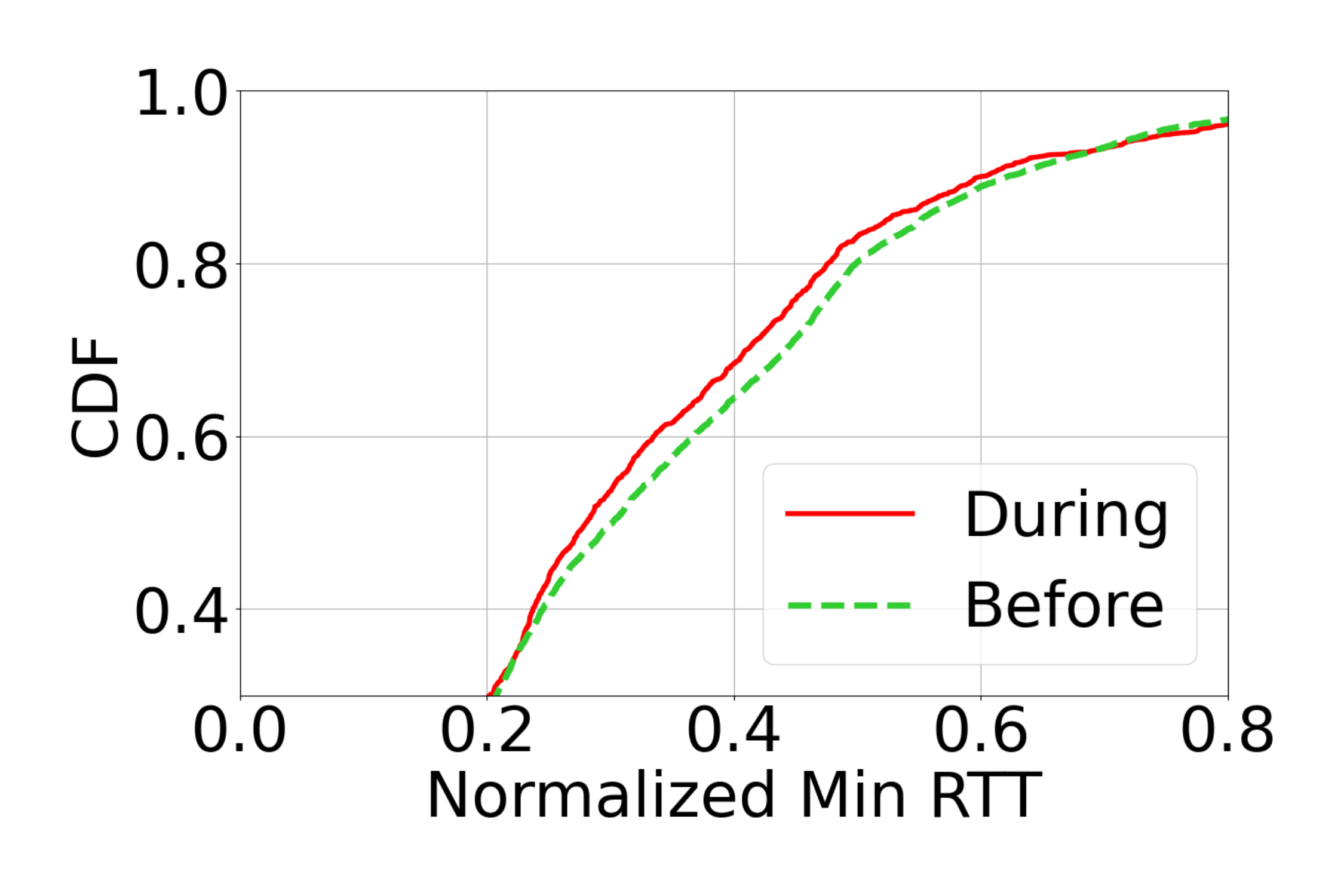}
        \caption{ndt5}
    \end{subfigure}%
    \hfill
    \begin{subfigure}[t]{0.25\columnwidth}
        \centering
        \includegraphics[width=\columnwidth, keepaspectratio]{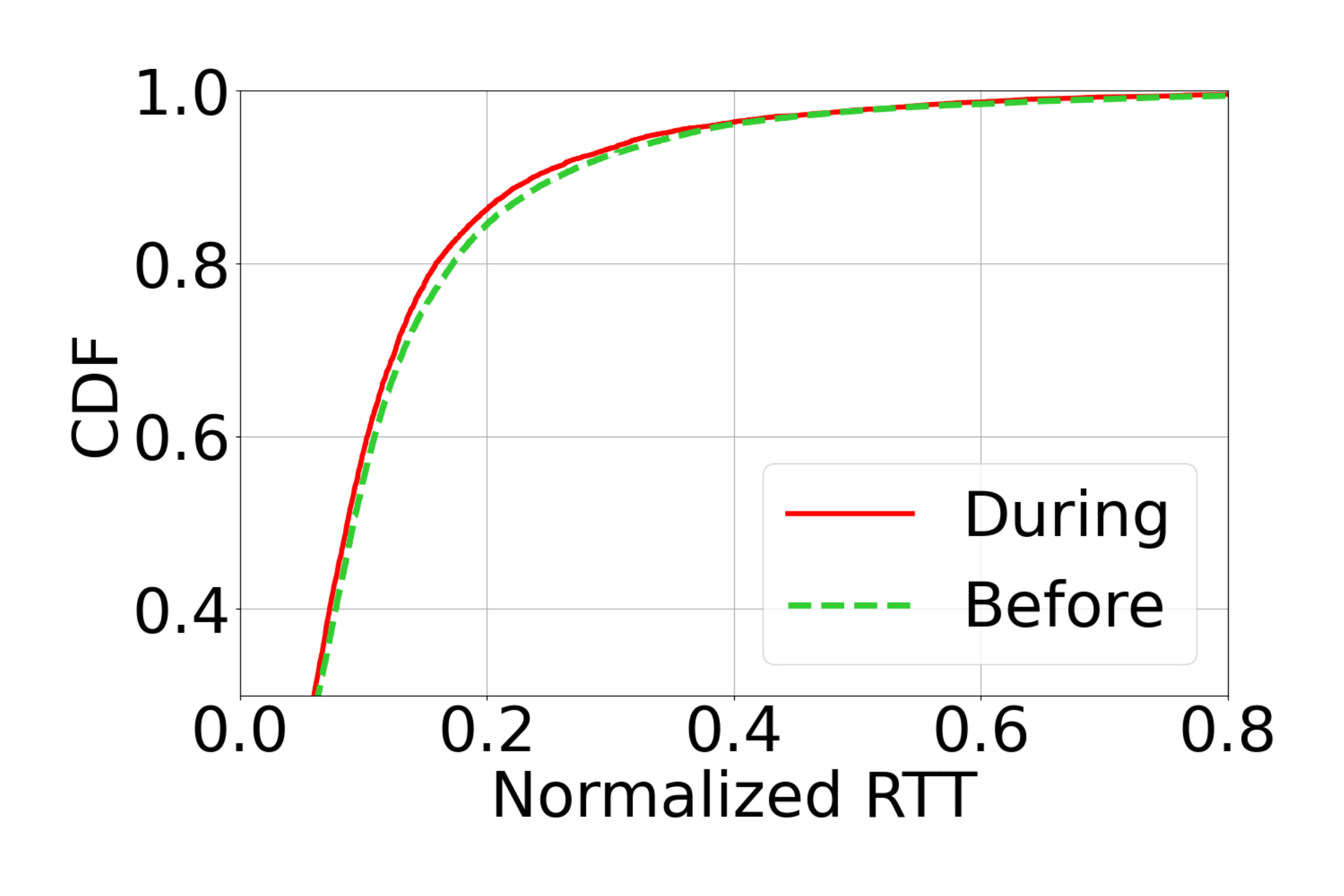}
        \caption{AIM}
    \end{subfigure}%
    \hfill
    \begin{subfigure}[t]{0.25\columnwidth}
        \centering
        \includegraphics[width=\columnwidth, keepaspectratio]{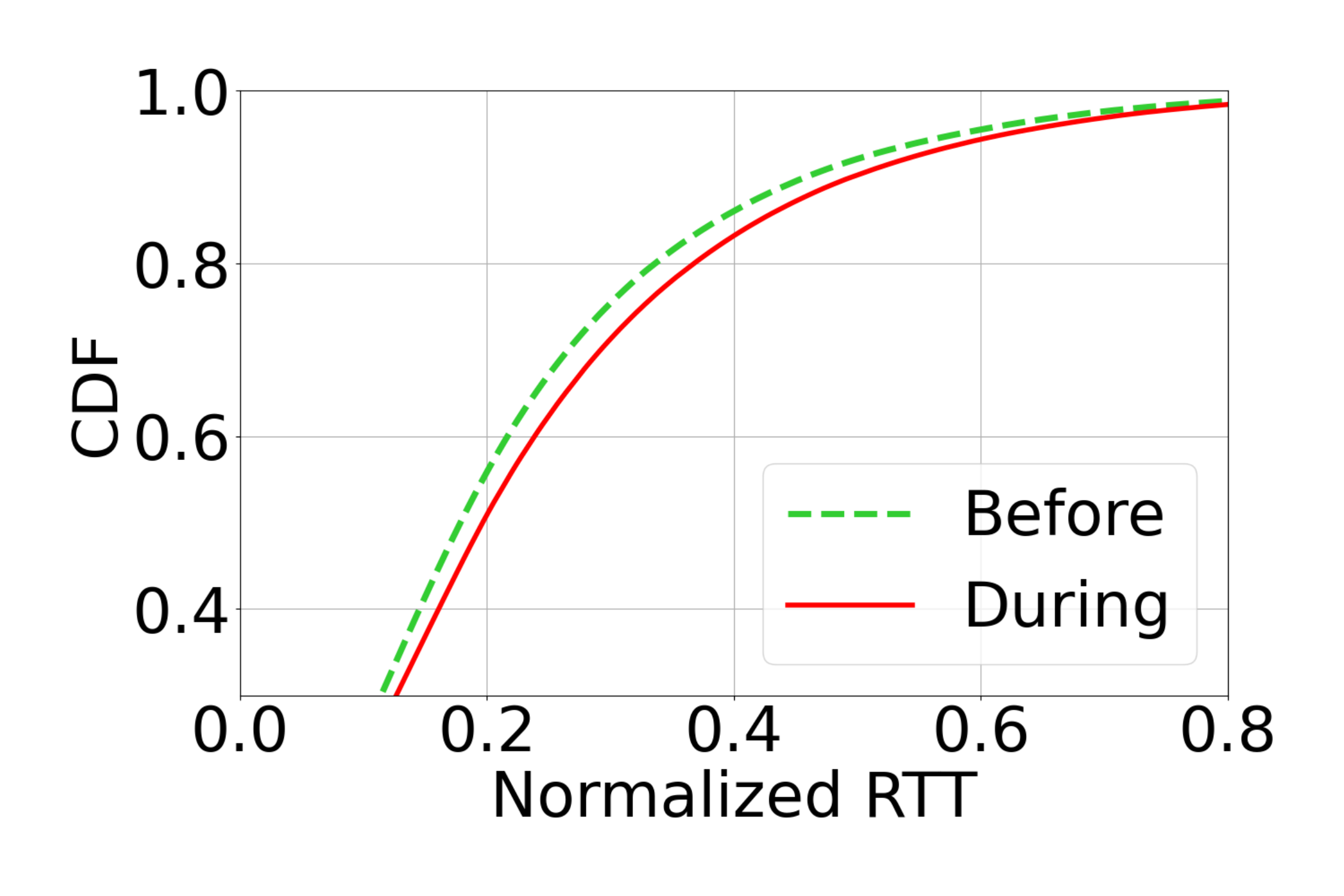}
        \caption{RIPE Atlas}
    \end{subfigure}%

    \caption{Overall normalized latency distribution of worldwide speed test results during (a)-(d) May 2024 solar superstorm and (e)-(h) October 2024 solar storm, showing end user might experience minor inflation in latency during the solar superstorm, while during the October solar storm would not feel much difference.}
    \label{fig:userLatencyOverall}
\end{figure}

\begin{figure}
    \centering
    \begin{subfigure}[t]{0.25\columnwidth}
        \centering
        \includegraphics[width=\columnwidth, keepaspectratio]{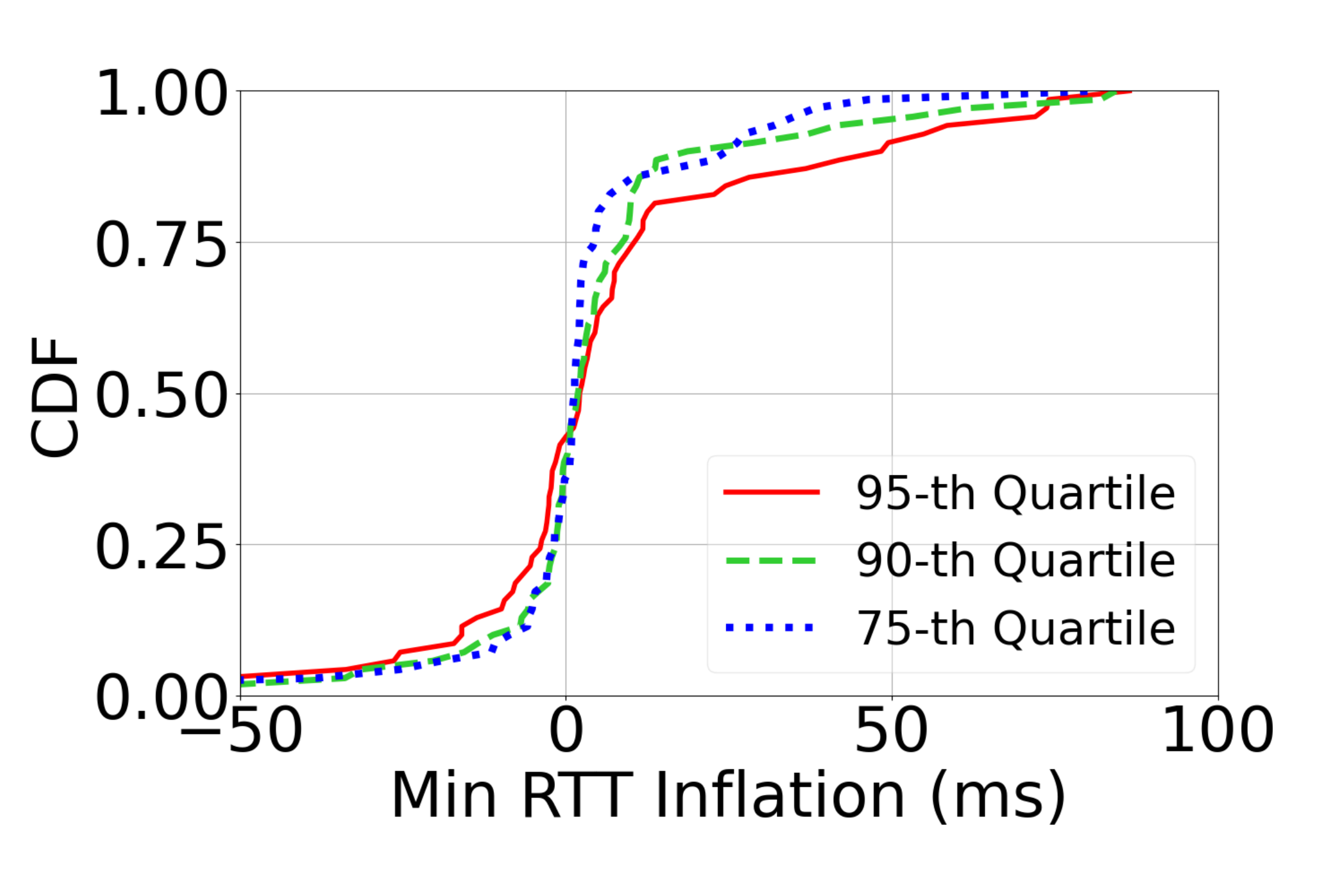}
        \caption{ndt7}
    \end{subfigure}%
    \hfill
    \begin{subfigure}[t]{0.25\columnwidth}
        \centering
        \includegraphics[width=\columnwidth, keepaspectratio]{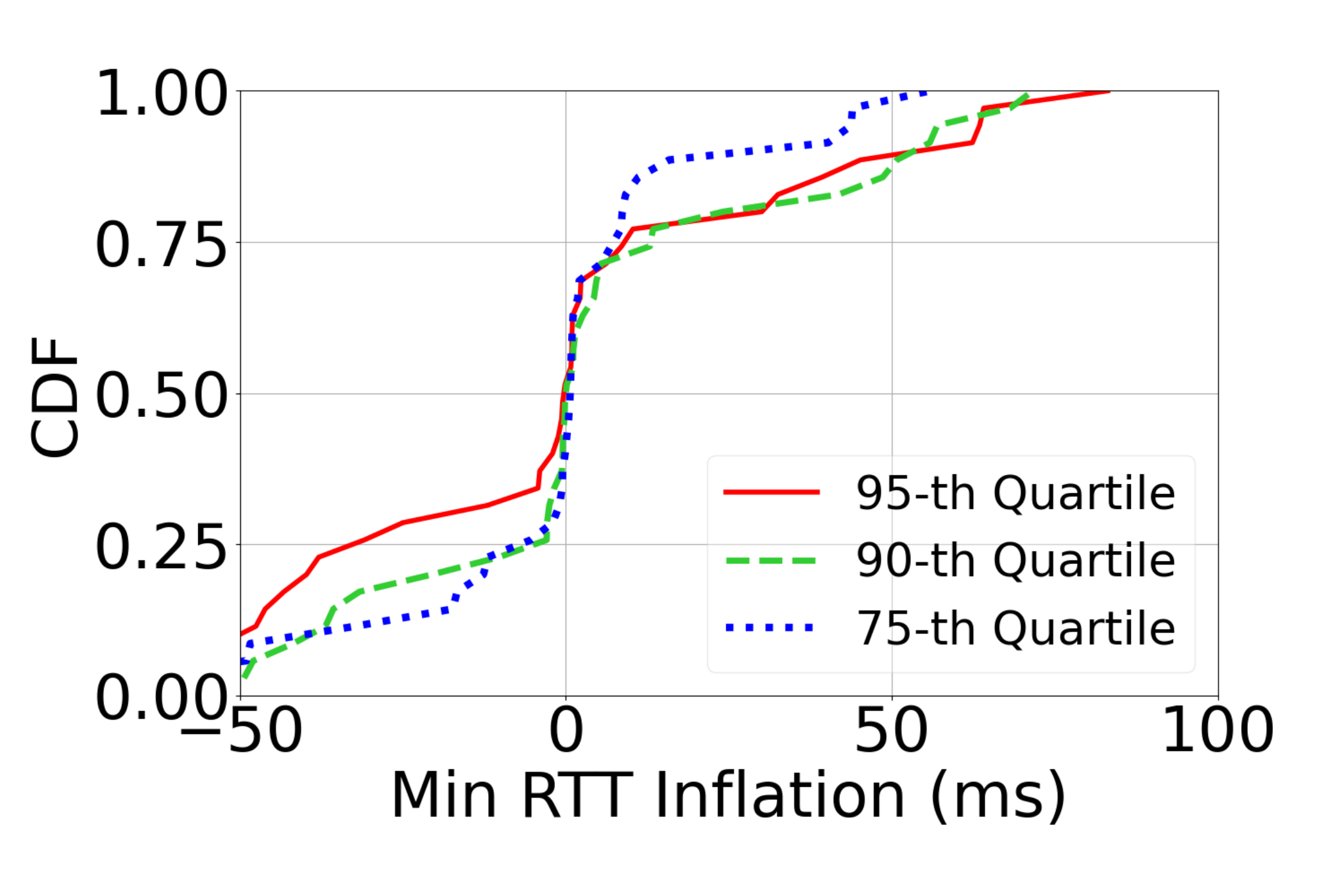}
        \caption{ndt5}
    \end{subfigure}%
    \hfill
    \begin{subfigure}[t]{0.25\columnwidth}
        \centering
        \includegraphics[width=\columnwidth, keepaspectratio]{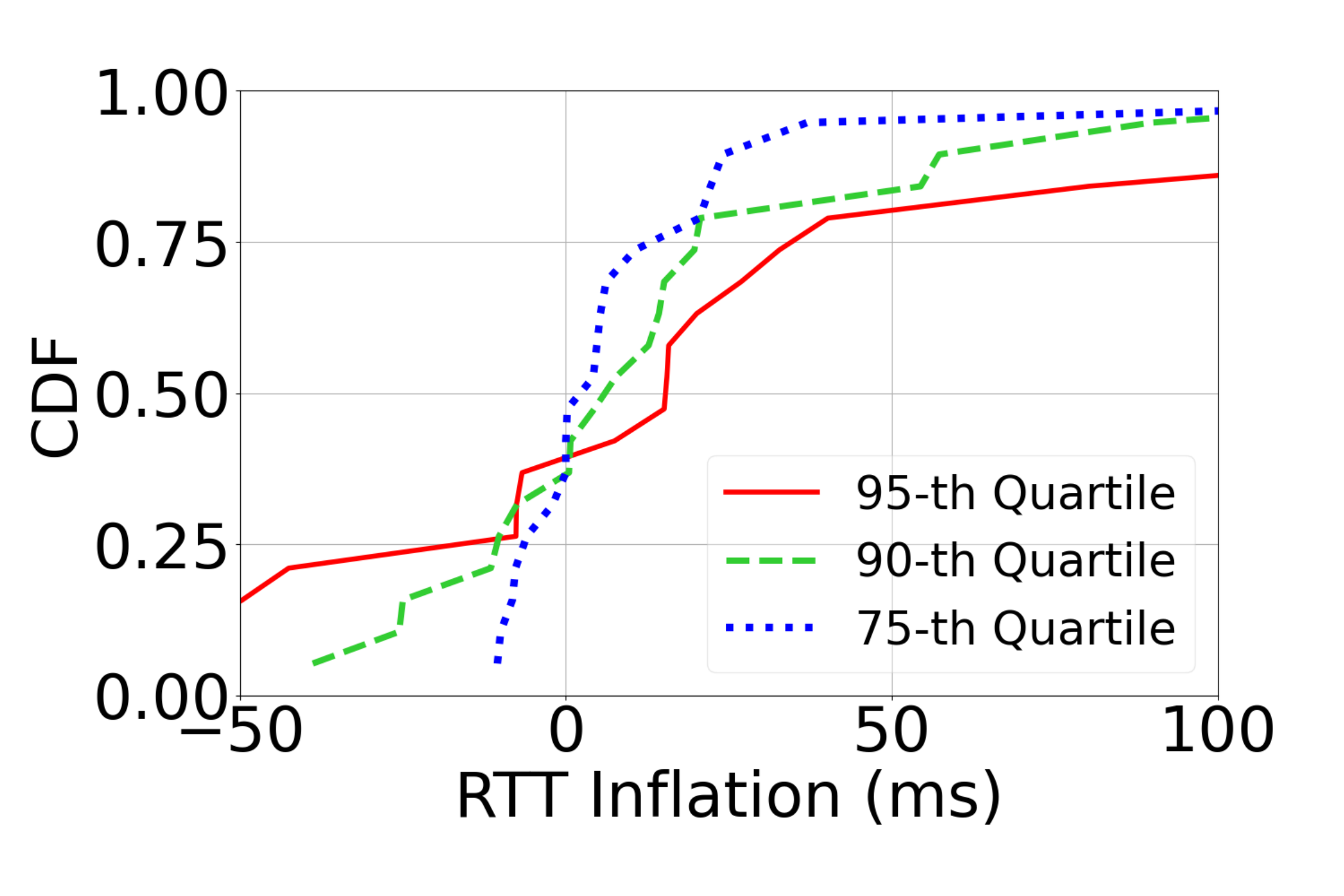}
        \caption{AIM}
    \end{subfigure}%
    \hfill
    \begin{subfigure}[t]{0.25\columnwidth}
        \centering
        \includegraphics[width=\columnwidth, keepaspectratio]{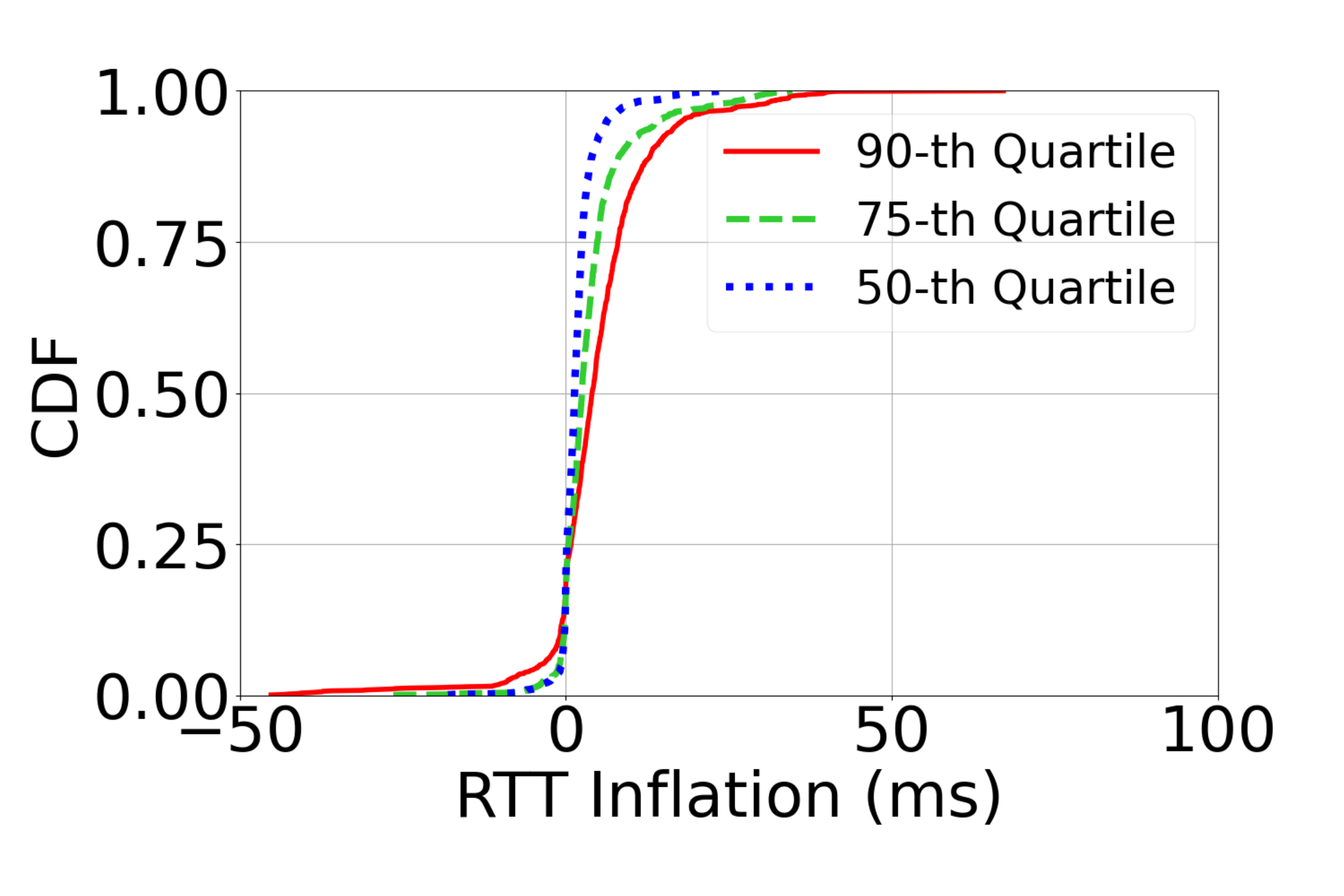}
        \caption{RIPE Atlas}
    \end{subfigure}%

    \hfill

    \begin{subfigure}[t]{0.25\columnwidth}
        \centering
        \includegraphics[width=\columnwidth, keepaspectratio]{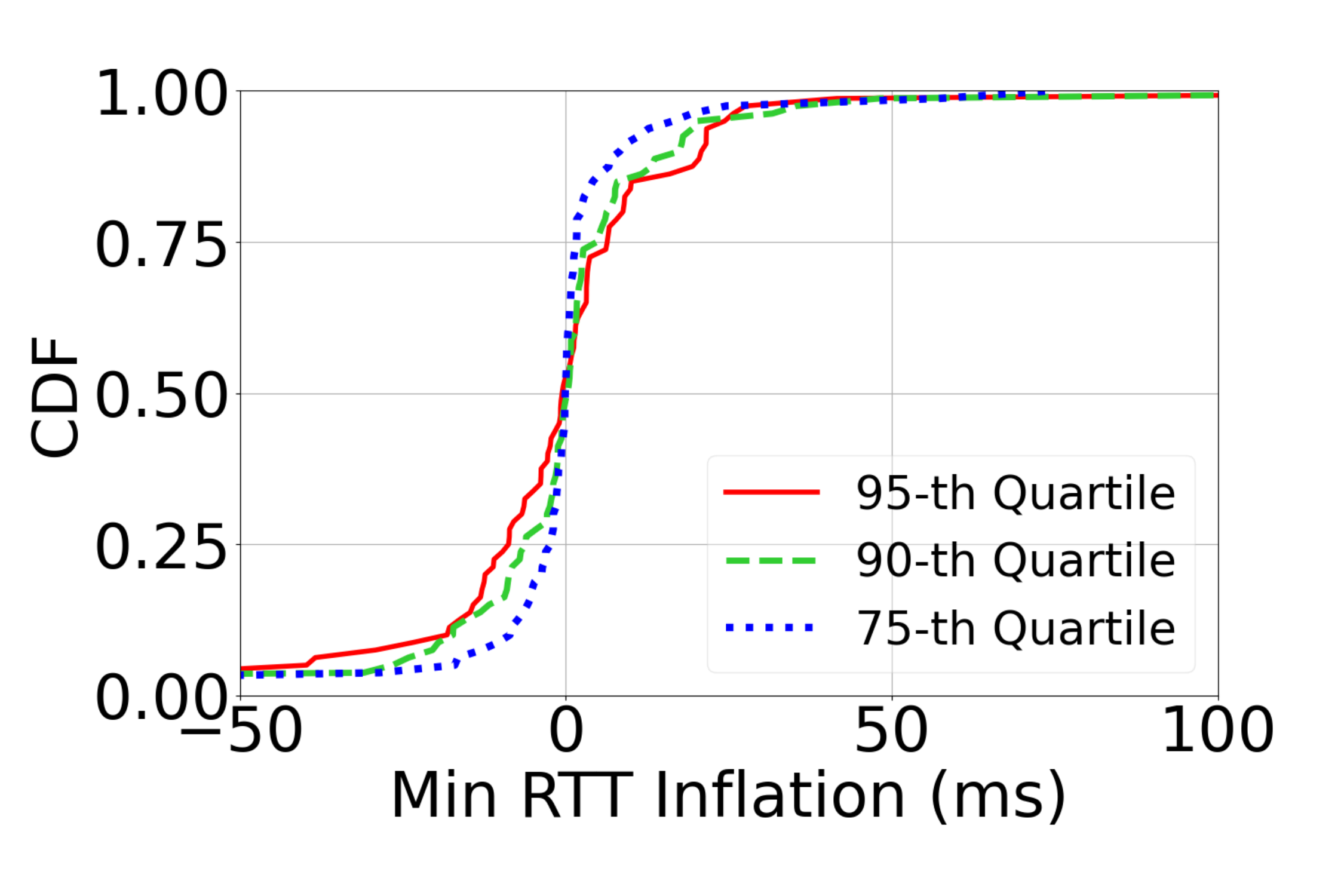}
        \caption{ndt7}
    \end{subfigure}%
    \hfill
    \begin{subfigure}[t]{0.25\columnwidth}
        \centering
        \includegraphics[width=\columnwidth, keepaspectratio]{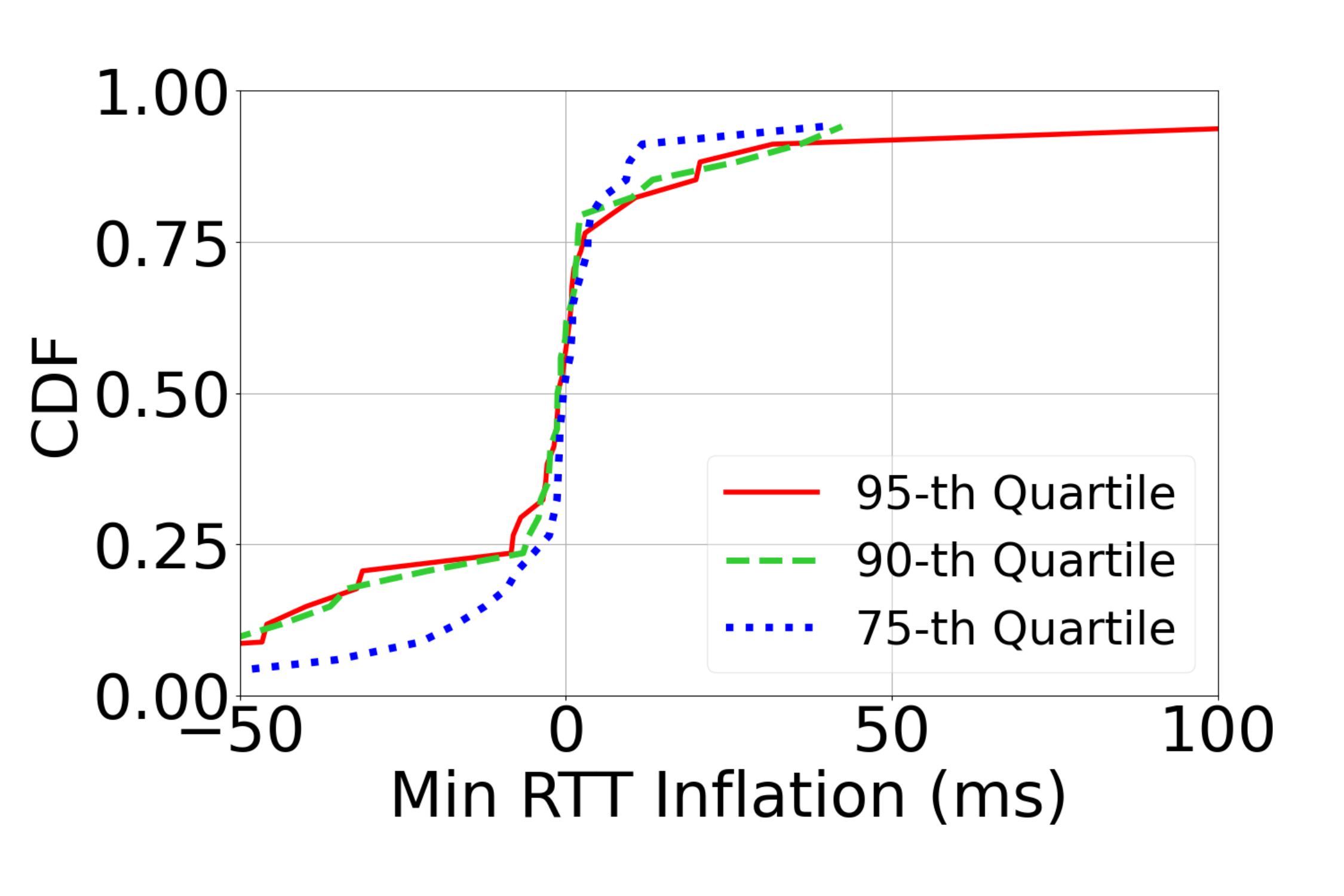}
        \caption{ndt5}
    \end{subfigure}%
    \hfill
    \begin{subfigure}[t]{0.25\columnwidth}
        \centering
        \includegraphics[width=\columnwidth, keepaspectratio]{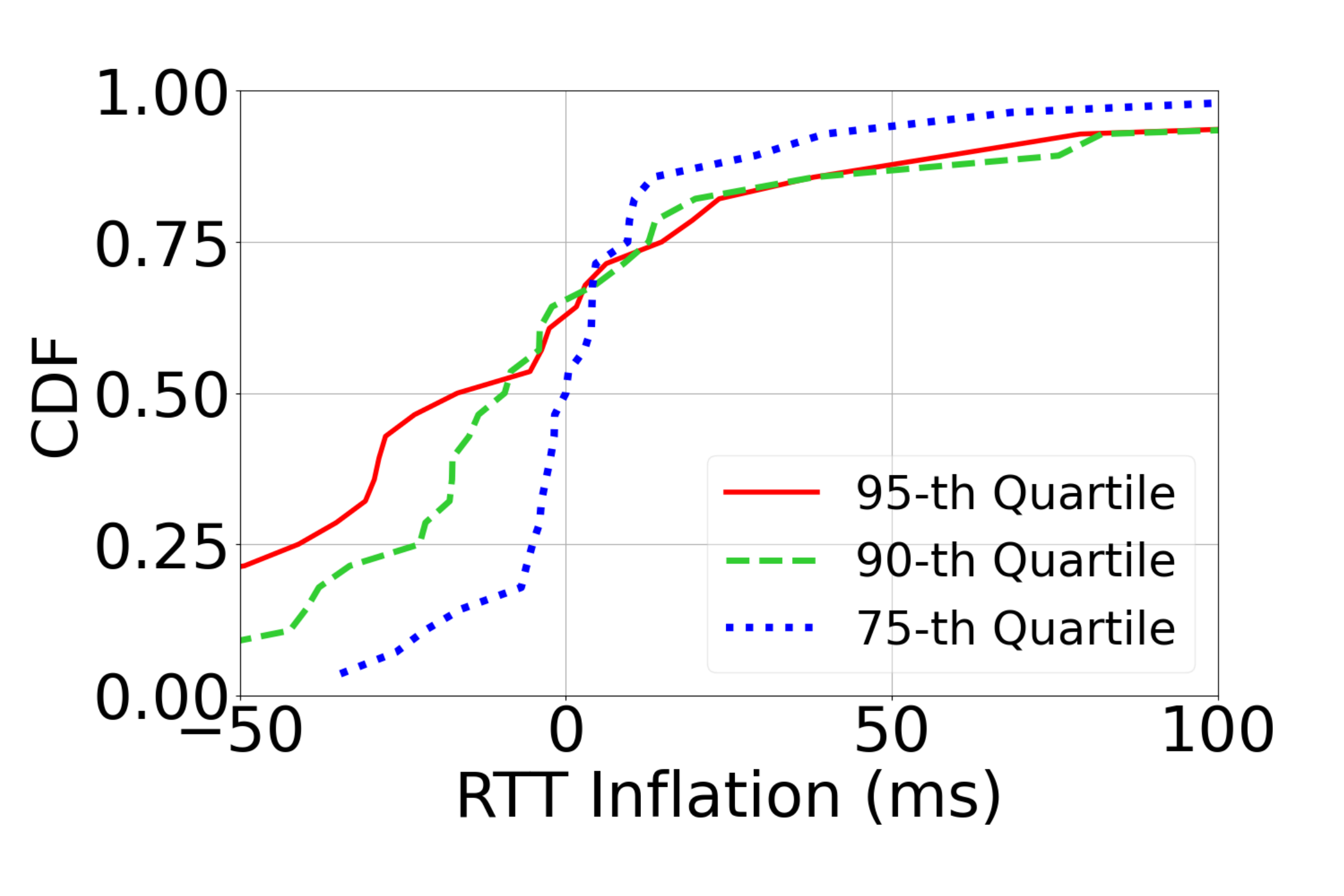}
        \caption{AIM}
    \end{subfigure}%
    \hfill
    \begin{subfigure}[t]{0.25\columnwidth}
        \centering
        \includegraphics[width=\columnwidth, keepaspectratio]{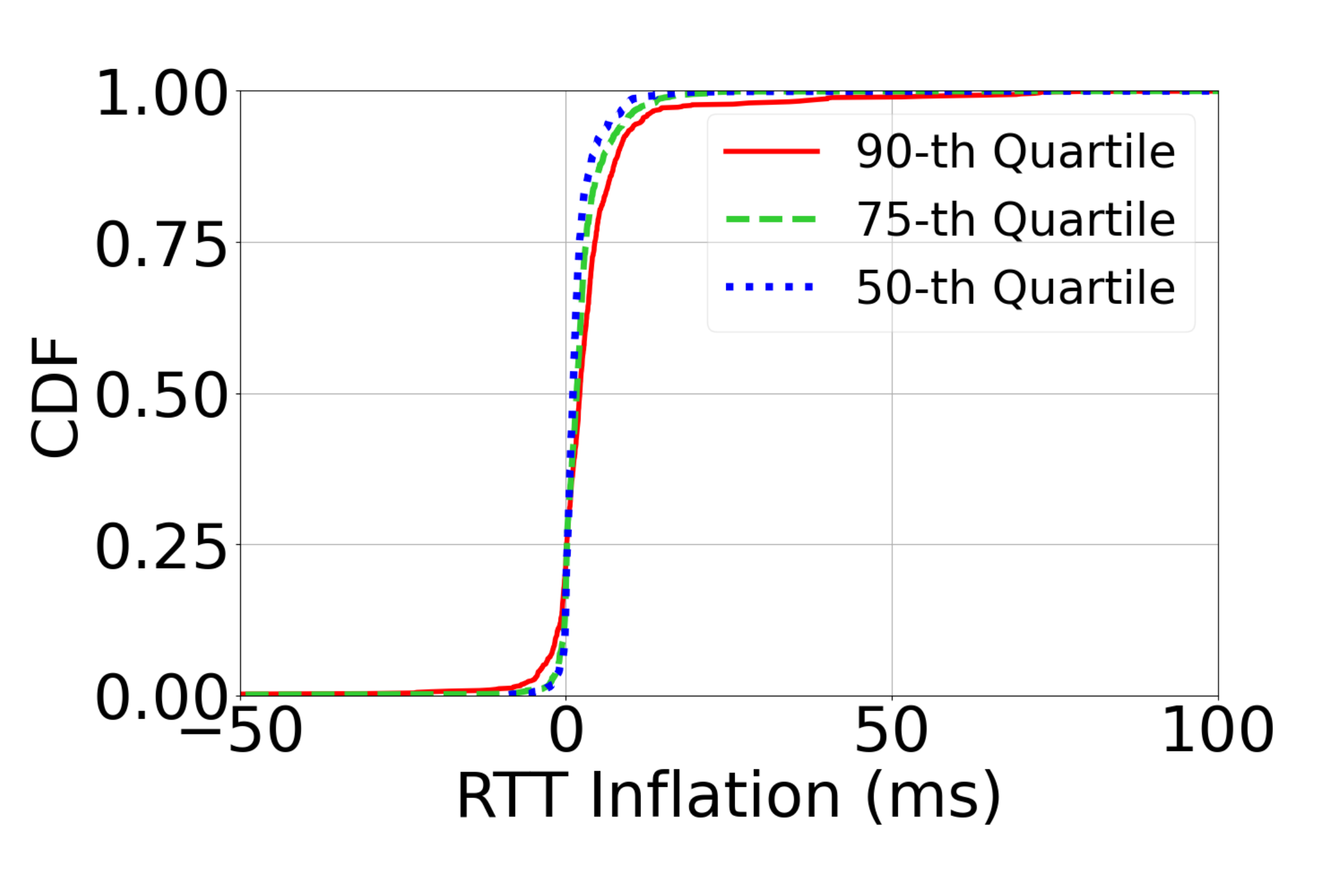}
        \caption{RIPE Atlas}
    \end{subfigure}%

    \caption{Distribution of region-wise latency inflation from speed test results during (a)-(d) May 2024 solar superstorm and (e)-(h) October 2024 solar storm, indicating user in 25\% of the region could experience higher latency of 10s of ms during the solar superstorm, while during the October solar storm, the distribution is symmetric. }
    \label{fig:userLatencyInflation}
\end{figure}

\begin{figure}
    \centering
    \begin{subfigure}[t]{0.25\columnwidth}
        \centering
        \includegraphics[width=\columnwidth, keepaspectratio]{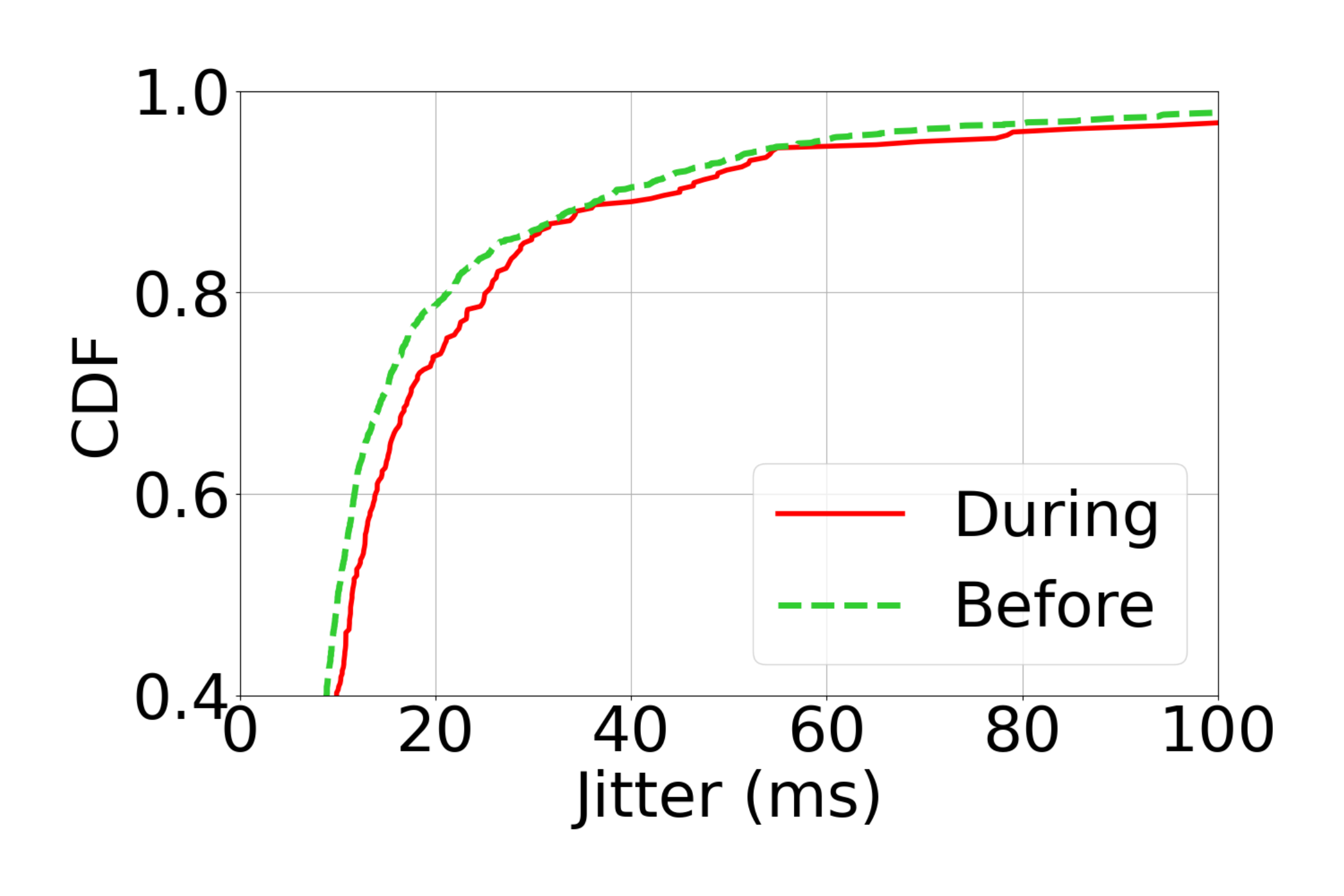}
        \caption{AIM, May'24}
    \end{subfigure}%
    \hfill
    \begin{subfigure}[t]{0.25\columnwidth}
        \centering
        \includegraphics[width=\columnwidth, keepaspectratio]{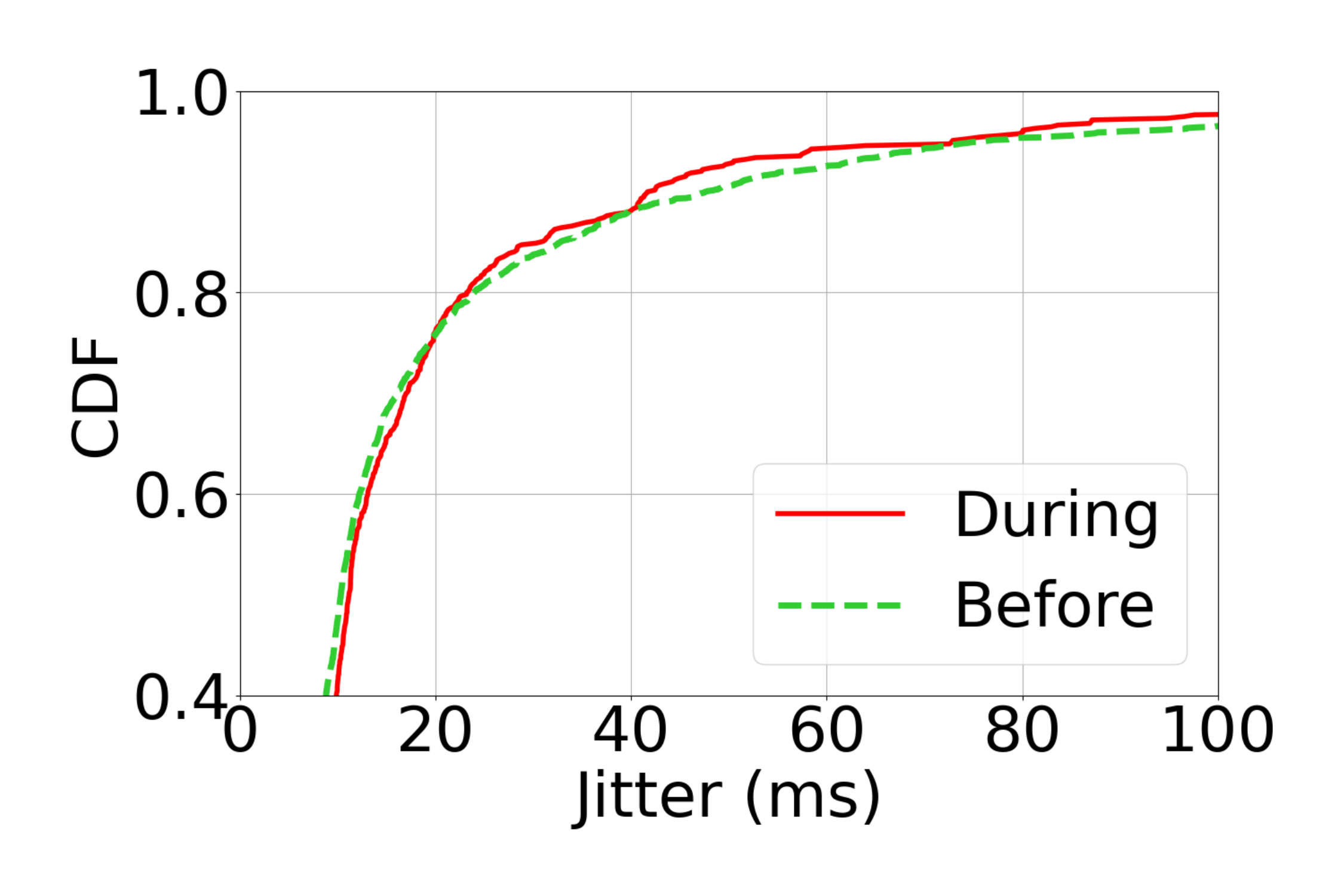}
        \caption{AIM, Oct'24}
    \end{subfigure}%
    \hfill
    \begin{subfigure}[t]{0.25\columnwidth}
        \centering
        \includegraphics[width=\columnwidth, keepaspectratio]{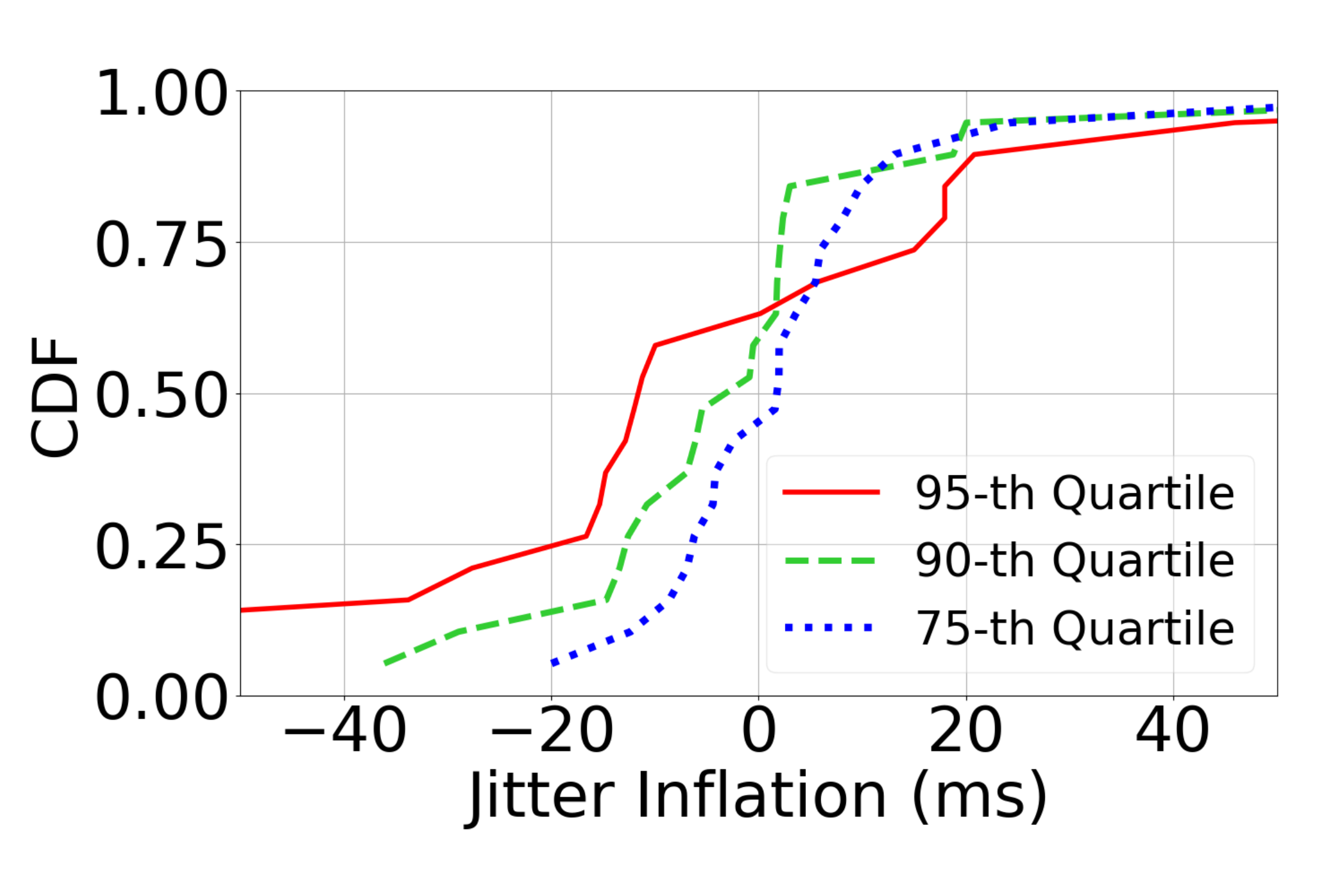}
        \caption{AIM, May'24}
    \end{subfigure}%
    \hfill
    \begin{subfigure}[t]{0.25\columnwidth}
        \centering
        \includegraphics[width=\columnwidth, keepaspectratio]{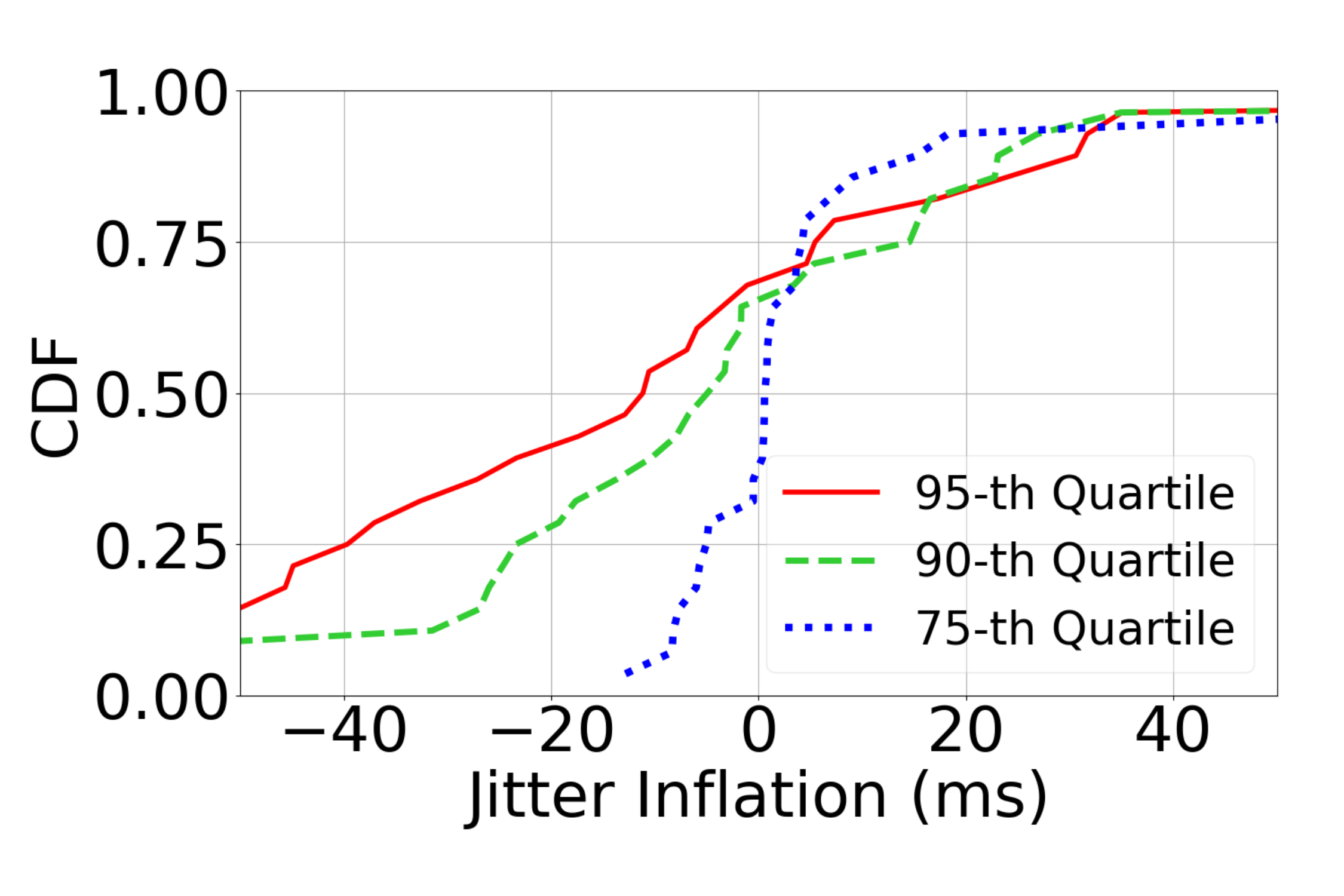}
        \caption{AIM, Oct'24}
    \end{subfigure}%

    \caption{Overall jitter distribution of worldwide speed test results during (a) May 2024 solar superstorm shows a negligible increase, (b) October 2024 solar storm closely overlaps, and (c)-(d) do not show any regional pattern. }
    \label{fig:userJitter}
\end{figure}

The latency profile over Starlink connectivity drastically varies worldwide. 
To address the exceptional diversity in latency measurements, we group speed tests by region, remove extreme outliers (>99th percentile), and normalize latency observations.
For RIPE Atlas probes, we do the same for each probe-destination pair.
In Fig.~\ref{fig:userLatencyOverall}, we plot the distribution of normalized latency observations for the same pre-storm and during-storm windows of May 2024 and October 2024 solar storms.
Notice that during the May 2024 solar superstorm, we observe up to 7\%, 8\%, and 15\% inflation in ndt7 min RTT, Cloudflare AIM RTT, and RIPE Atlas probe RTT, respectively.
The distribution in ndt5 min RTT closely overlaps, showing no significant changes during the solar storm.
In contrast, during October 2024, we do not see any latency implications in M-LAB or Cloudflare AIM.
However, in RIPE Atlas probes, we observe up to 10\% RTT inflation during a solar storm.

In Fig.~\ref{fig:userLatencyInflation}, we show the region-wise changes using the same approach as throughput.
Here, a positive value on the X axis indicates latency inflation in the regions.
Notice that the nature of MLAB ndt and Cloudflare AIM distributions is inherently different from the RIPE Atlas probe distribution.
The latency measurements in MLAB ndt and Cloudflare AIM rely on TCP/UDP and HTTP/WebSocket statistics collected when an end user voluntarily initiates a web-based speed test~\cite{cloudflareHowTestWorks, ndt}, which reflect the application-level experience.
In contrast, RIPE Atlas probes rely on periodic ICMP \texttt{pings} to measure the end-to-end latency~\cite{ripeTechnicalDetails}, indicating the network-level implications.
During the May 2024 solar superstorm, in M-LAB ndt7 Fig.~\ref{fig:userLatencyInflation}(a), we see users in 20\% of the regions experience a major RTT inflation.
Cloudflare AIM in Fig.~\ref{fig:userLatencyInflation}(c) also shows similar statistics, while Fig.~\ref{fig:userLatencyInflation}(b) does not give such an indication.
In Fig.~\ref{fig:userLatencyInflation}(d), we see 75\% probe destination pairs experience RTT inflation during the storm.
During the October 2024 solar storm, we observe almost negligible changes in M-LAB and Cloudflare AIM, while in Fig.~\ref{fig:userLatencyInflation}(h), we see that around 50\% of probe destination pairs experience slight RTT inflation.

We also look at jitters in Cloudflare AIM speed test results in Fig.~\ref{fig:userJitter}.
We observe a small increase in jitter of 5 ms at the 75th percentile in the overall distribution during the May 2024 event, whereas we do not see a similar change in October 2024.
Consequently, we do not see any regional trends in Fig.~\ref{fig:userJitter}(c)-(d).

\subsubsection{Packet loss:}

\begin{figure}
    \centering
    \begin{subfigure}[t]{0.25\columnwidth}
        \centering
        \includegraphics[width=\columnwidth, keepaspectratio]{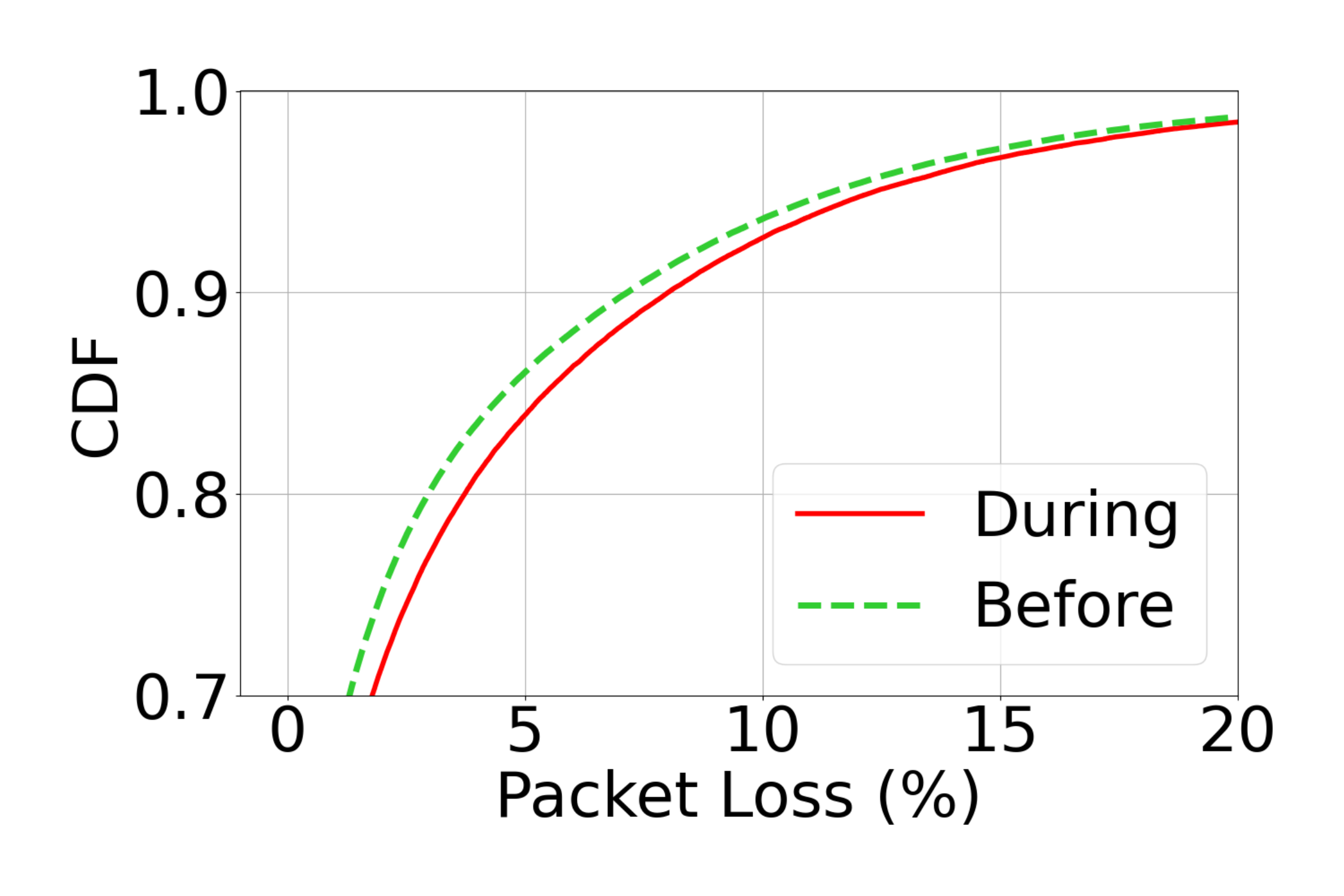}
        \caption{ndt7}
    \end{subfigure}%
    \hfill
    \begin{subfigure}[t]{0.25\columnwidth}
        \centering
        \includegraphics[width=\columnwidth, keepaspectratio]{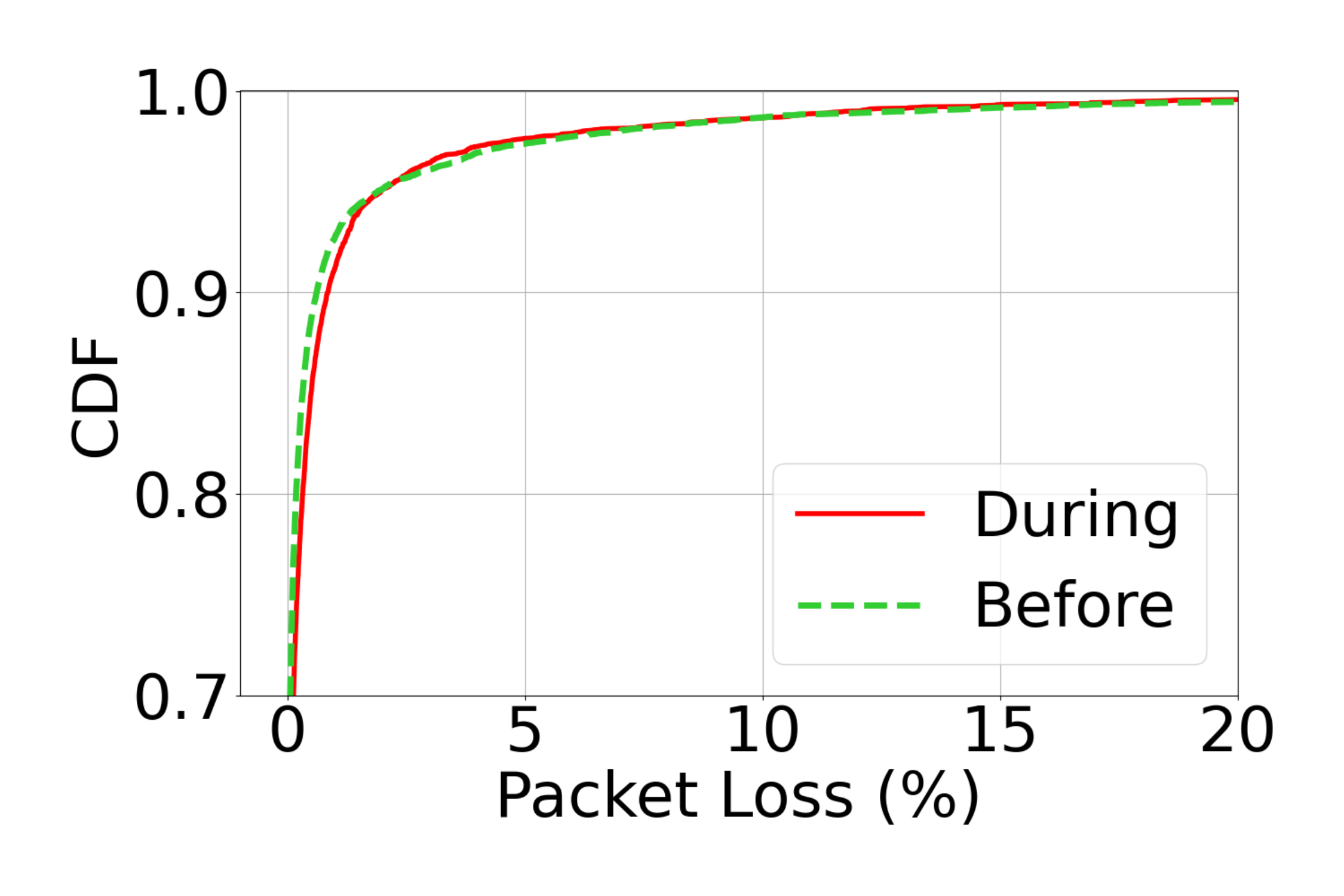}
        \caption{ndt5}
    \end{subfigure}%
    \hfill
    \begin{subfigure}[t]{0.25\columnwidth}
        \centering
        \includegraphics[width=\columnwidth, keepaspectratio]{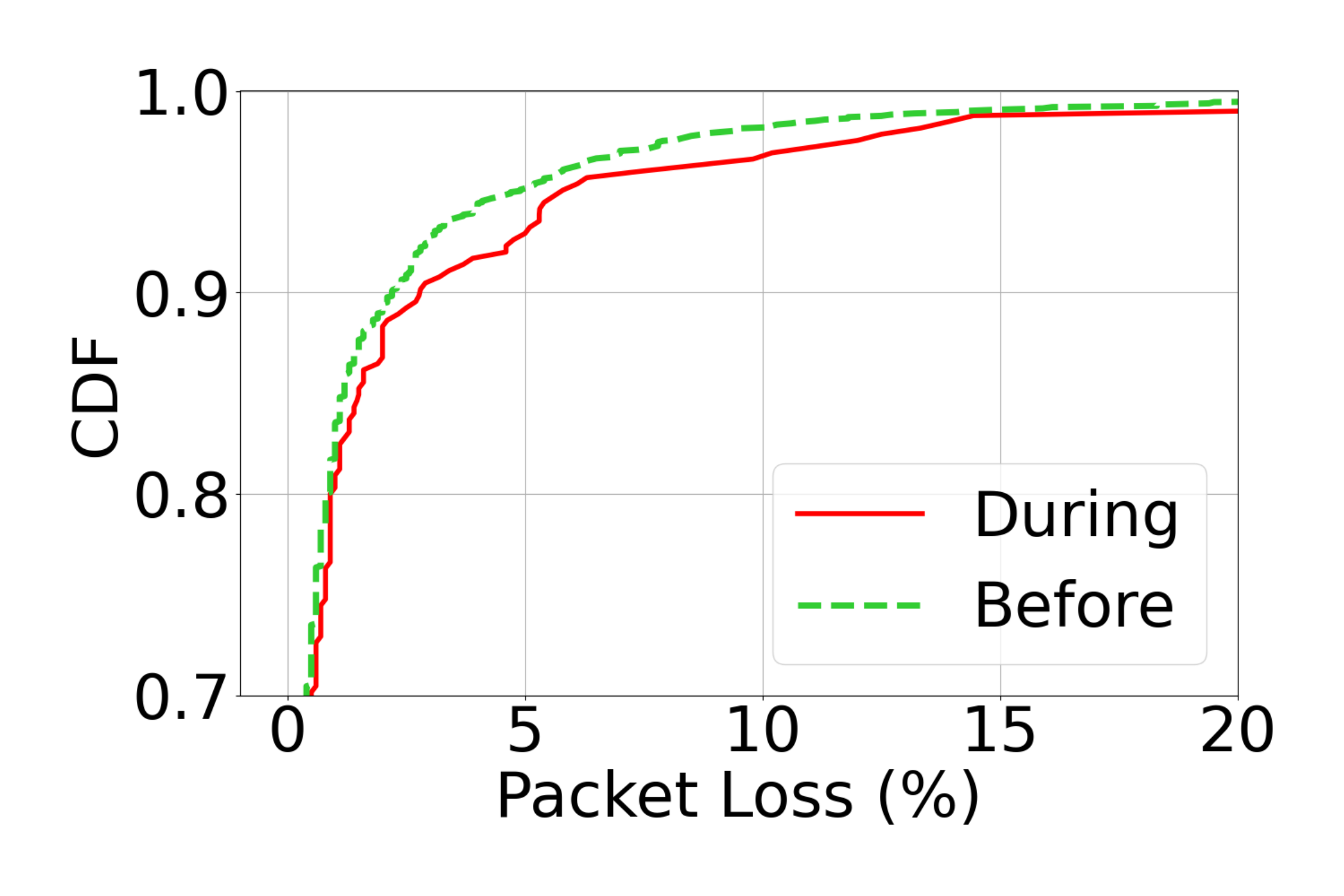}
        \caption{AIM}
    \end{subfigure}%
    \hfill
    \begin{subfigure}[t]{0.25\columnwidth}
        \centering
        \includegraphics[width=\columnwidth, keepaspectratio]{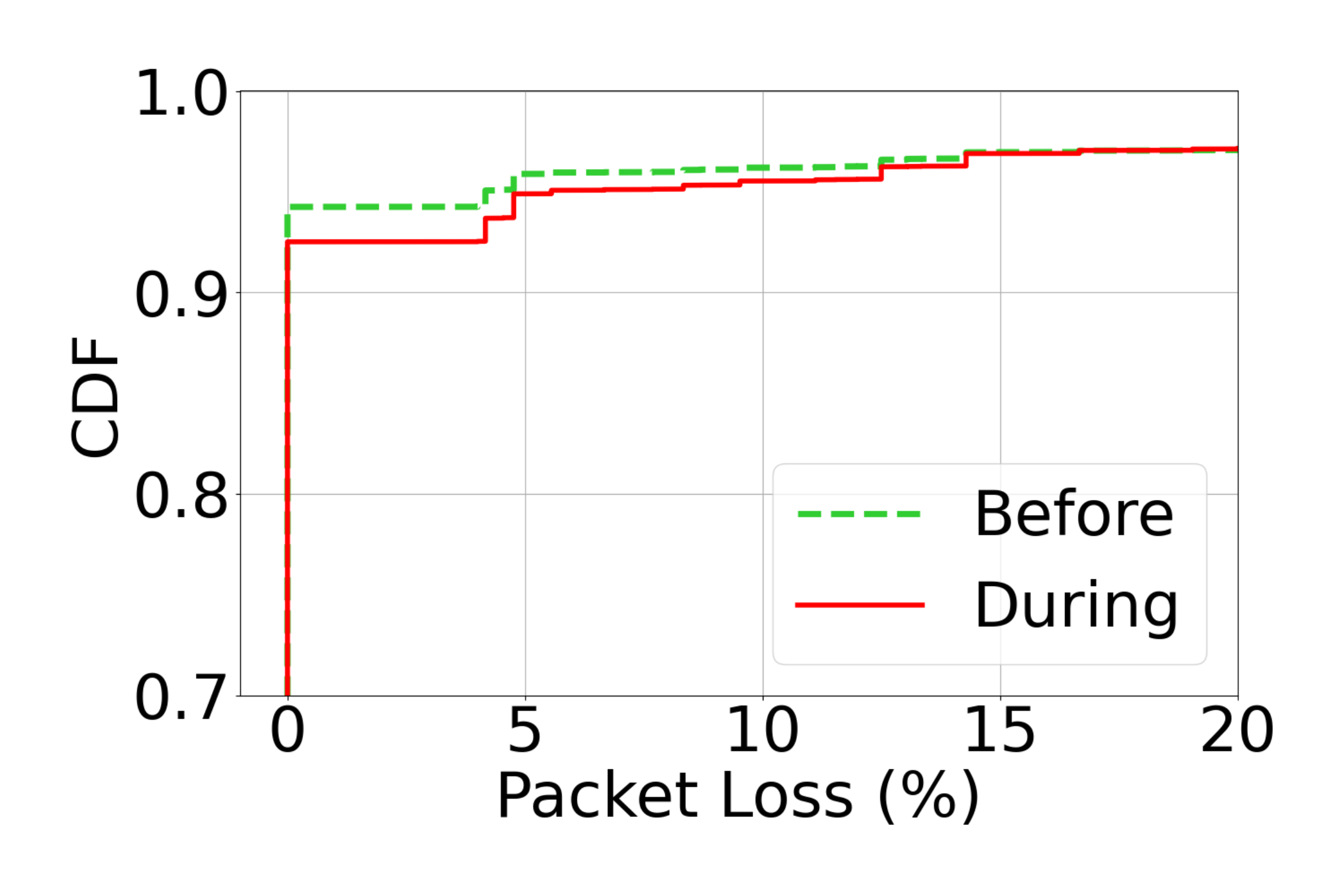}
        \caption{RIPE Atlas}
    \end{subfigure}%

    \hfill

    \begin{subfigure}[t]{0.25\columnwidth}
        \centering
        \includegraphics[width=\columnwidth, keepaspectratio]{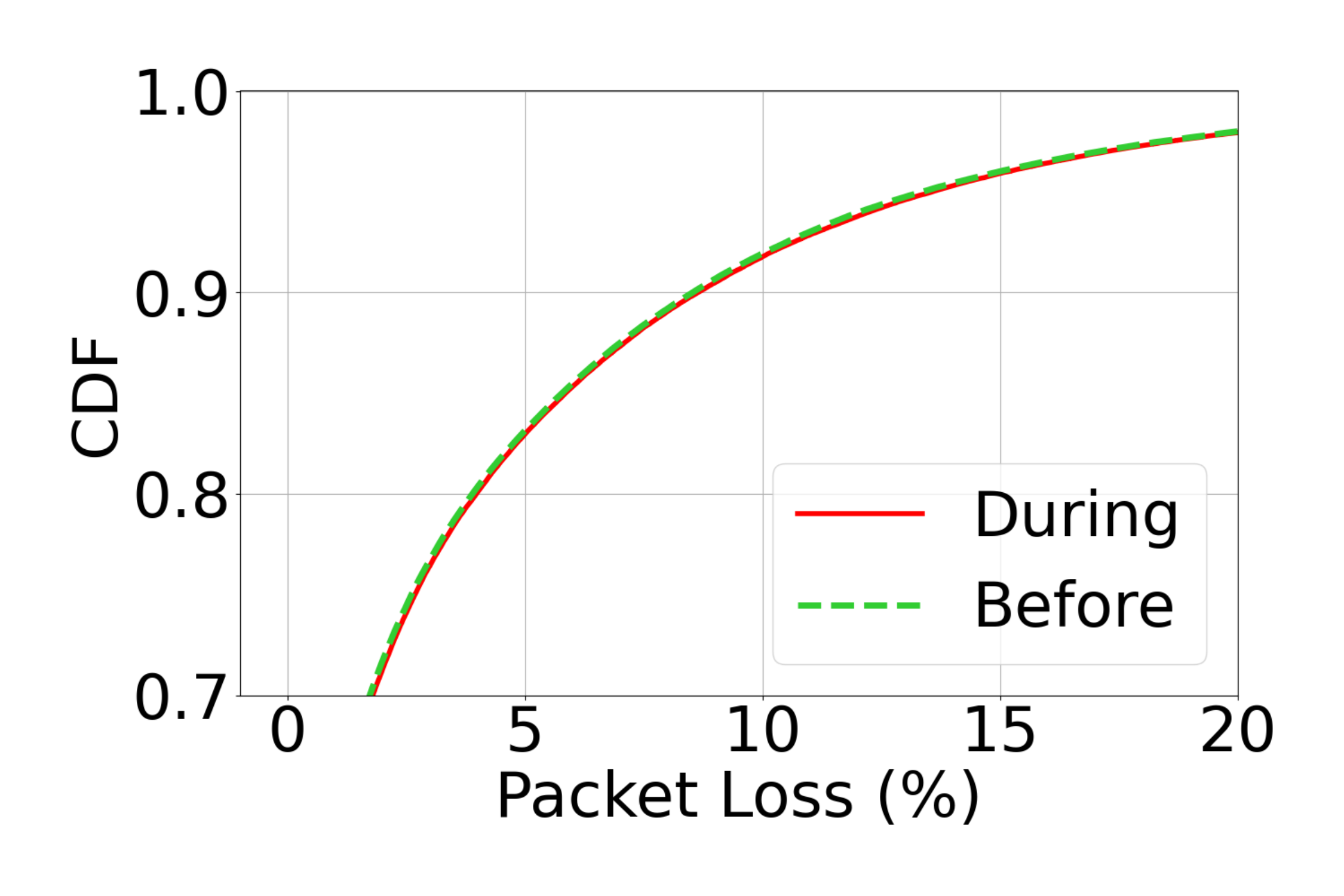}
        \caption{ndt7}
    \end{subfigure}%
    \hfill
    \begin{subfigure}[t]{0.25\columnwidth}
        \centering
        \includegraphics[width=\columnwidth, keepaspectratio]{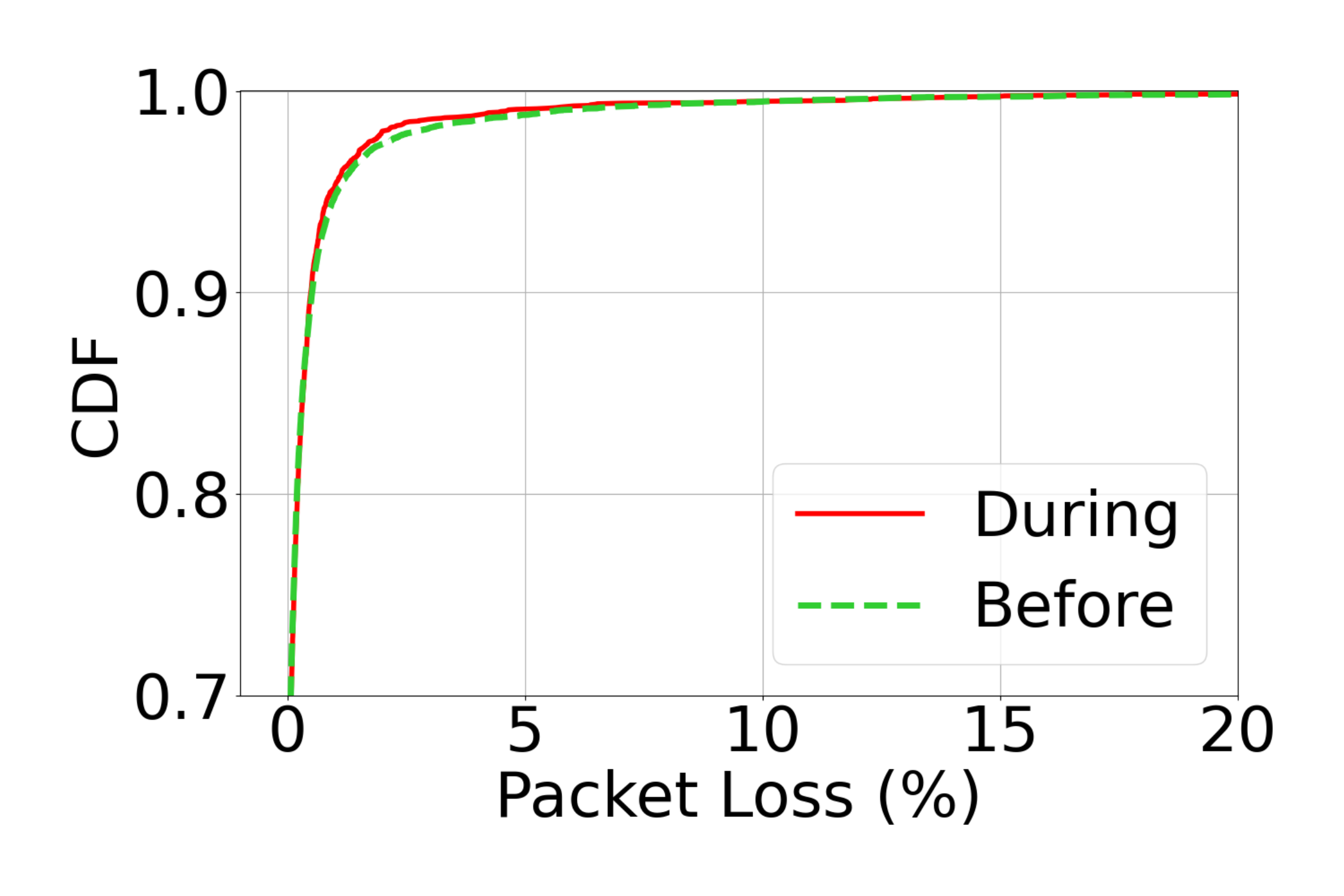}
        \caption{ndt5}
    \end{subfigure}%
    \hfill
    \begin{subfigure}[t]{0.25\columnwidth}
        \centering
        \includegraphics[width=\columnwidth, keepaspectratio]{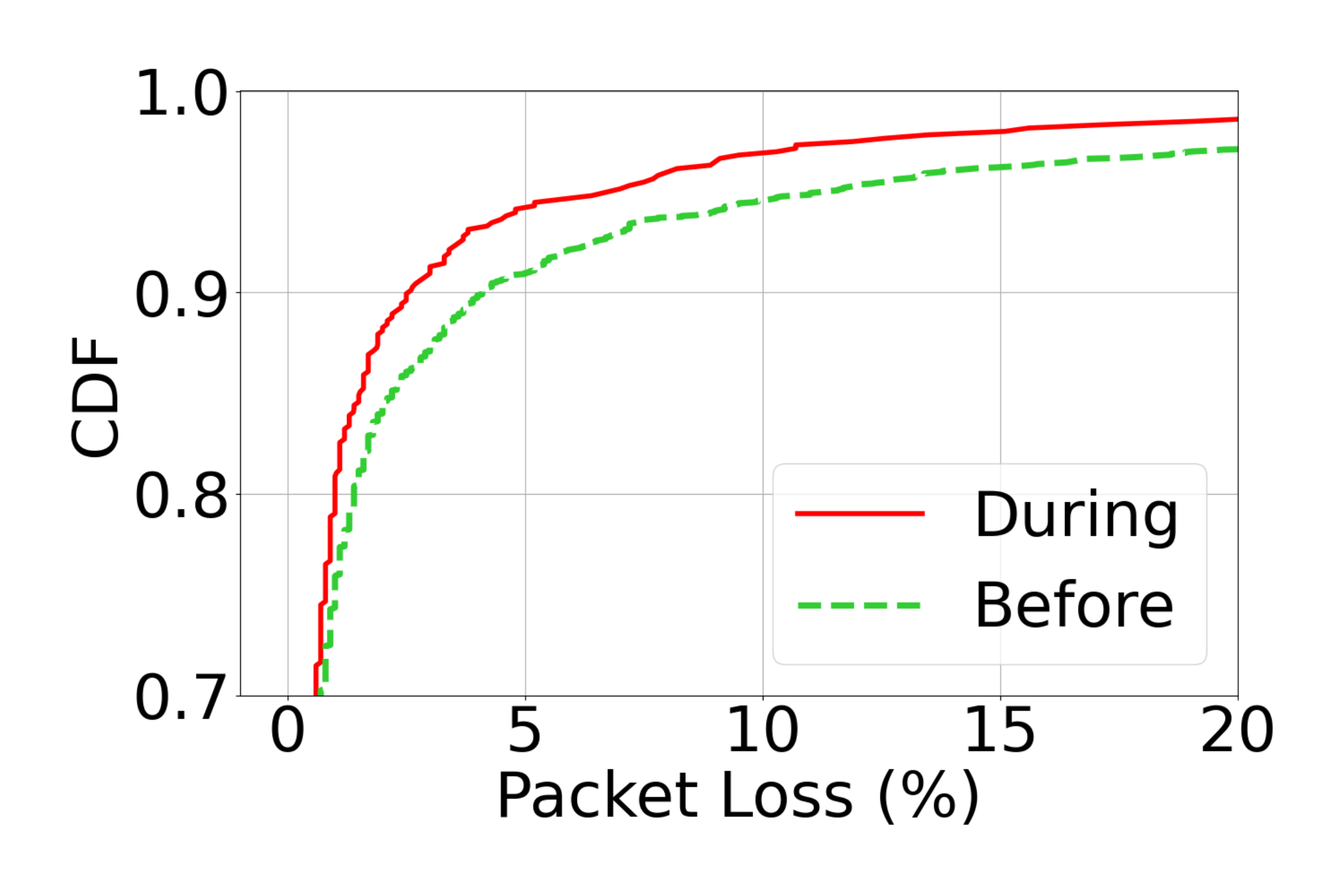}
        \caption{AIM}
    \end{subfigure}%
    \hfill
    \begin{subfigure}[t]{0.25\columnwidth}
        \centering
        \includegraphics[width=\columnwidth, keepaspectratio]{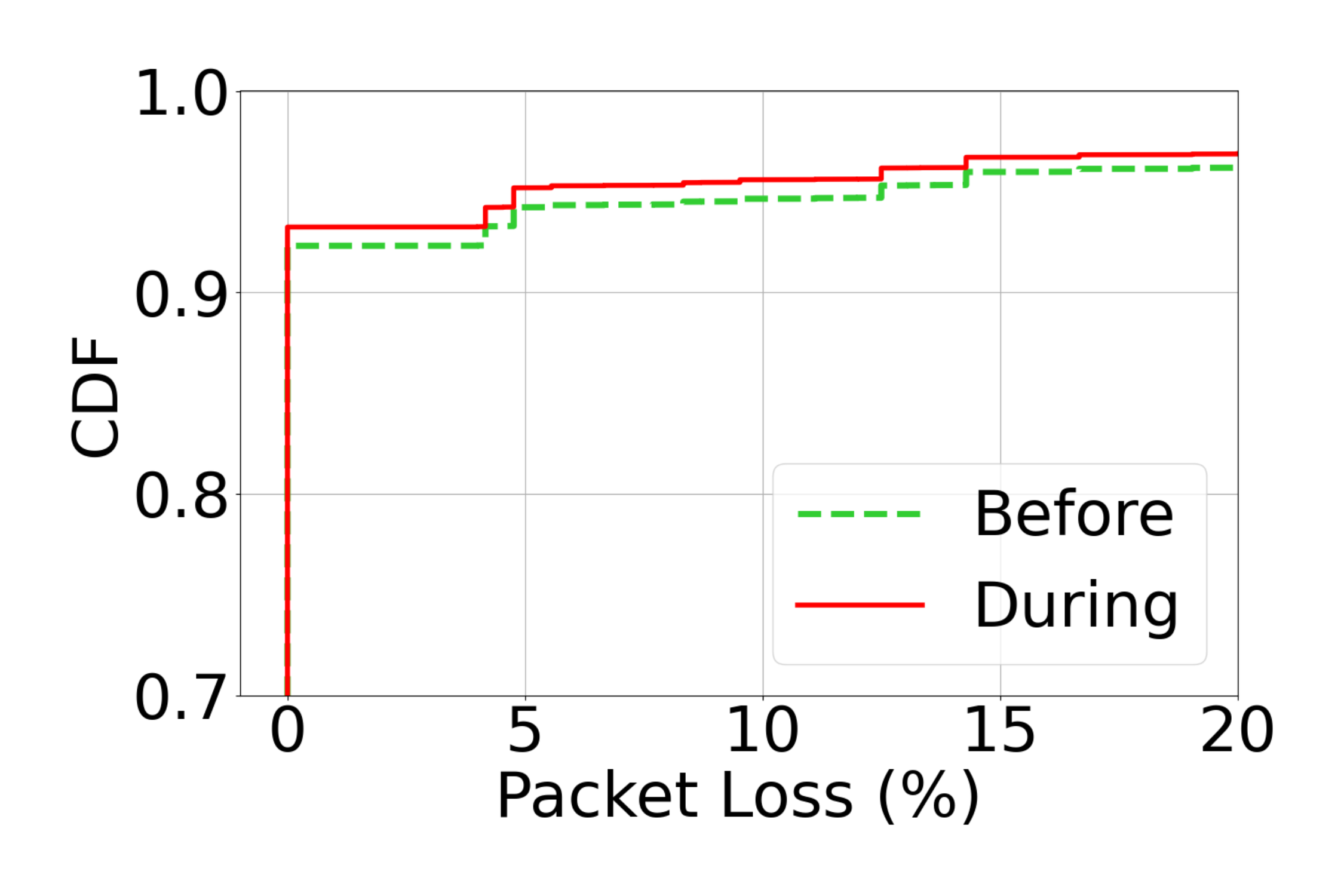}
        \caption{RIPE Atlas}
    \end{subfigure}%

    \caption{Overall packet loss distribution of worldwide speed test results during (a)-(d) May 2024 solar superstorm and (e)-(h) October 2024 solar storm, showing noticeable inflation in loss during the solar superstorm, while during the October solar storm, it mostly overlaps. (g) Cloudflare AIM speed test record shows an exception. }
    \label{fig:userLossOverall}
\end{figure}

\begin{figure}
    \centering
    \begin{subfigure}[t]{0.25\columnwidth}
        \centering
        \includegraphics[width=\columnwidth, keepaspectratio]{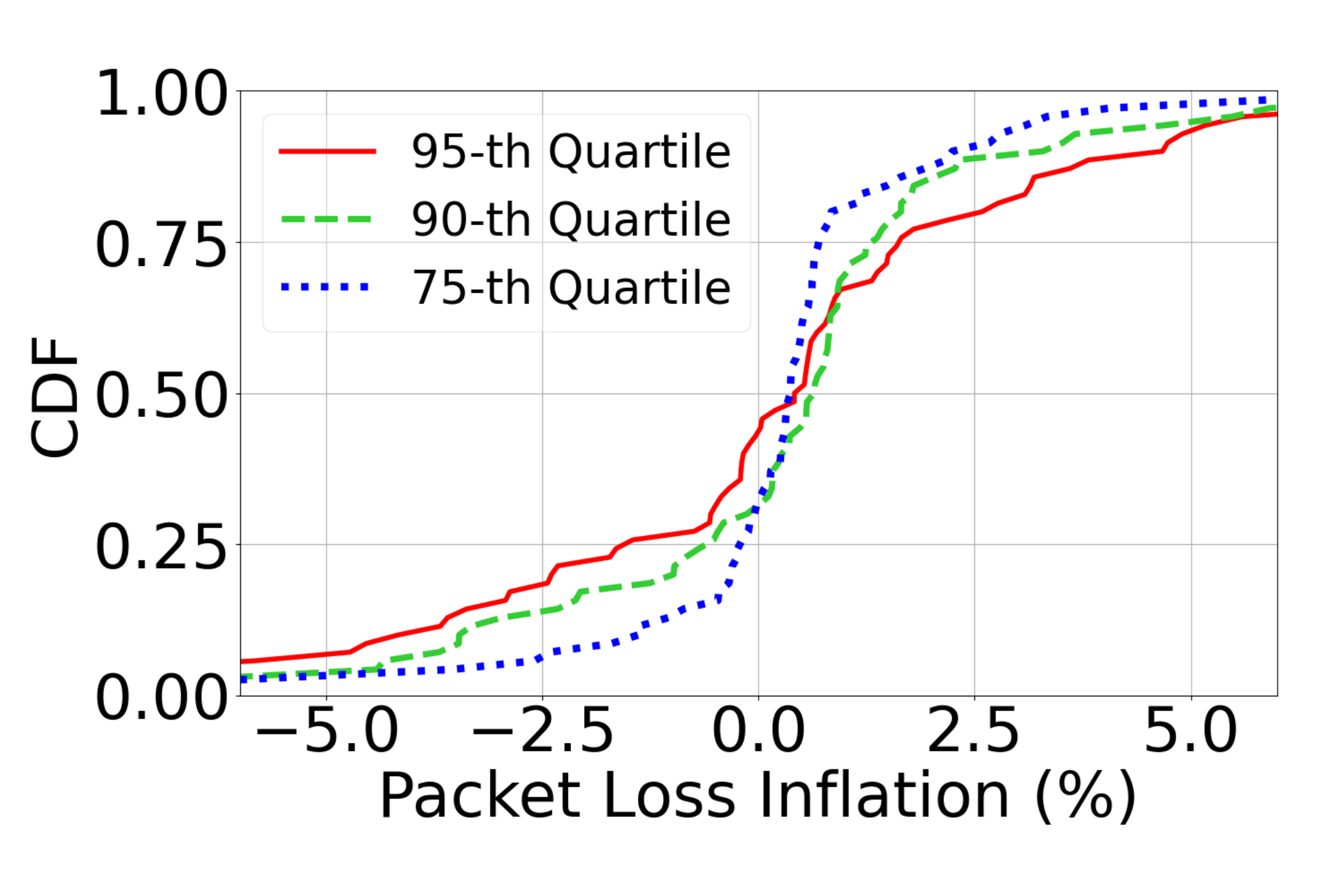}
        \caption{ndt7}
    \end{subfigure}%
    \hfill
    \begin{subfigure}[t]{0.25\columnwidth}
        \centering
        \includegraphics[width=\columnwidth, keepaspectratio]{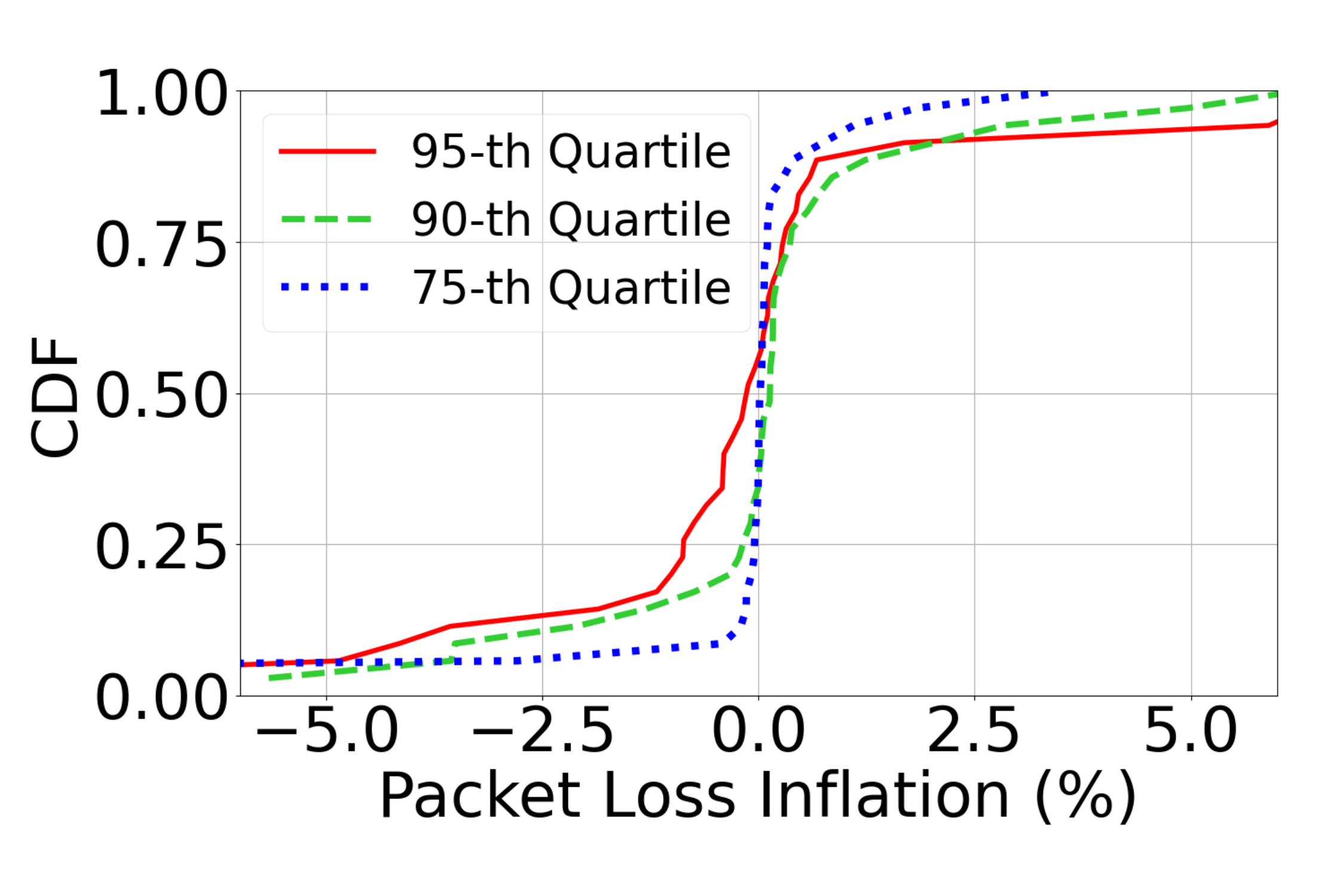}
        \caption{ndt5}
    \end{subfigure}%
    \hfill
    \begin{subfigure}[t]{0.25\columnwidth}
        \centering
        \includegraphics[width=\columnwidth, keepaspectratio]{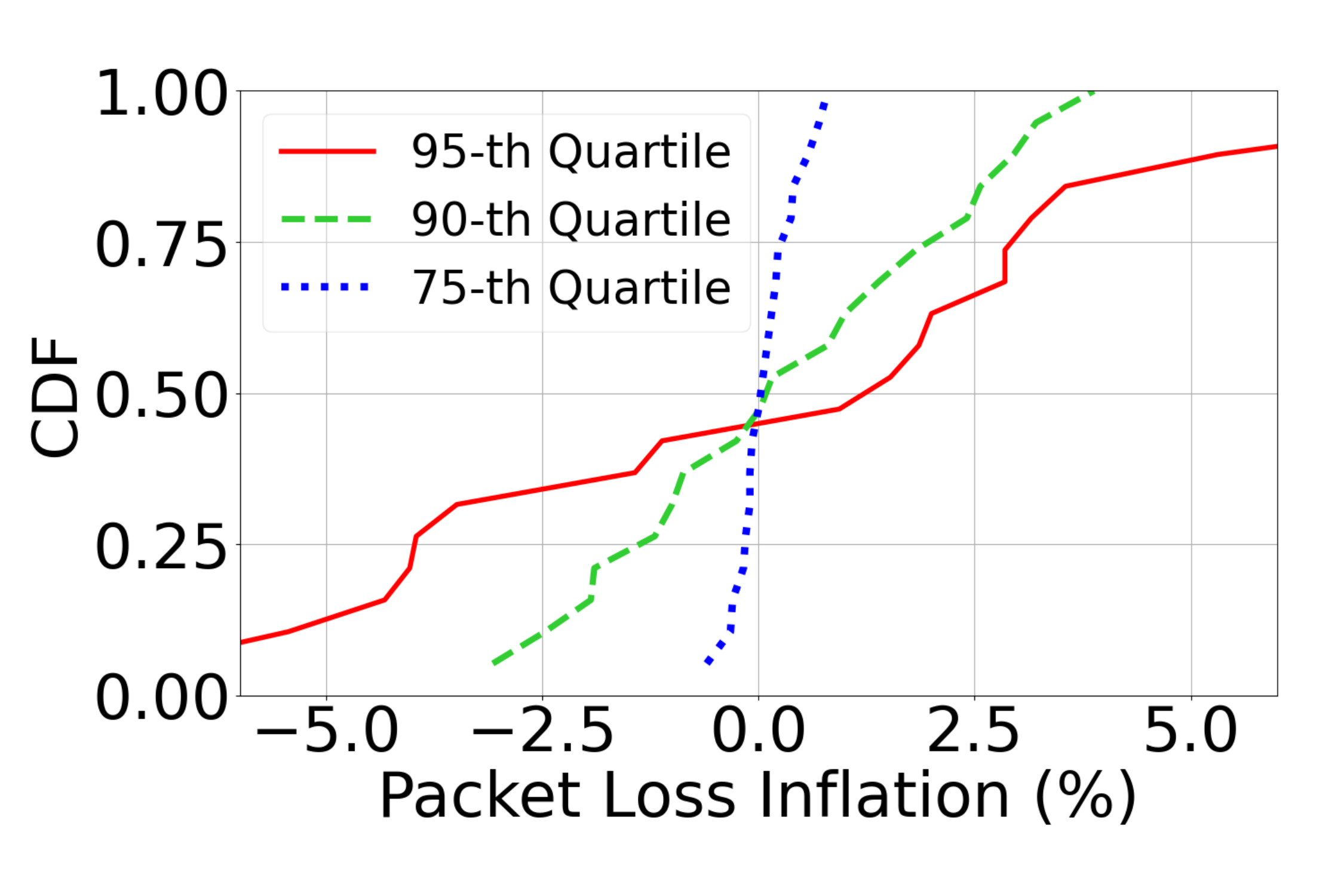}
        \caption{AIM}
    \end{subfigure}%
    \hfill
    \begin{subfigure}[t]{0.25\columnwidth}
        \centering
        \includegraphics[width=\columnwidth, keepaspectratio]{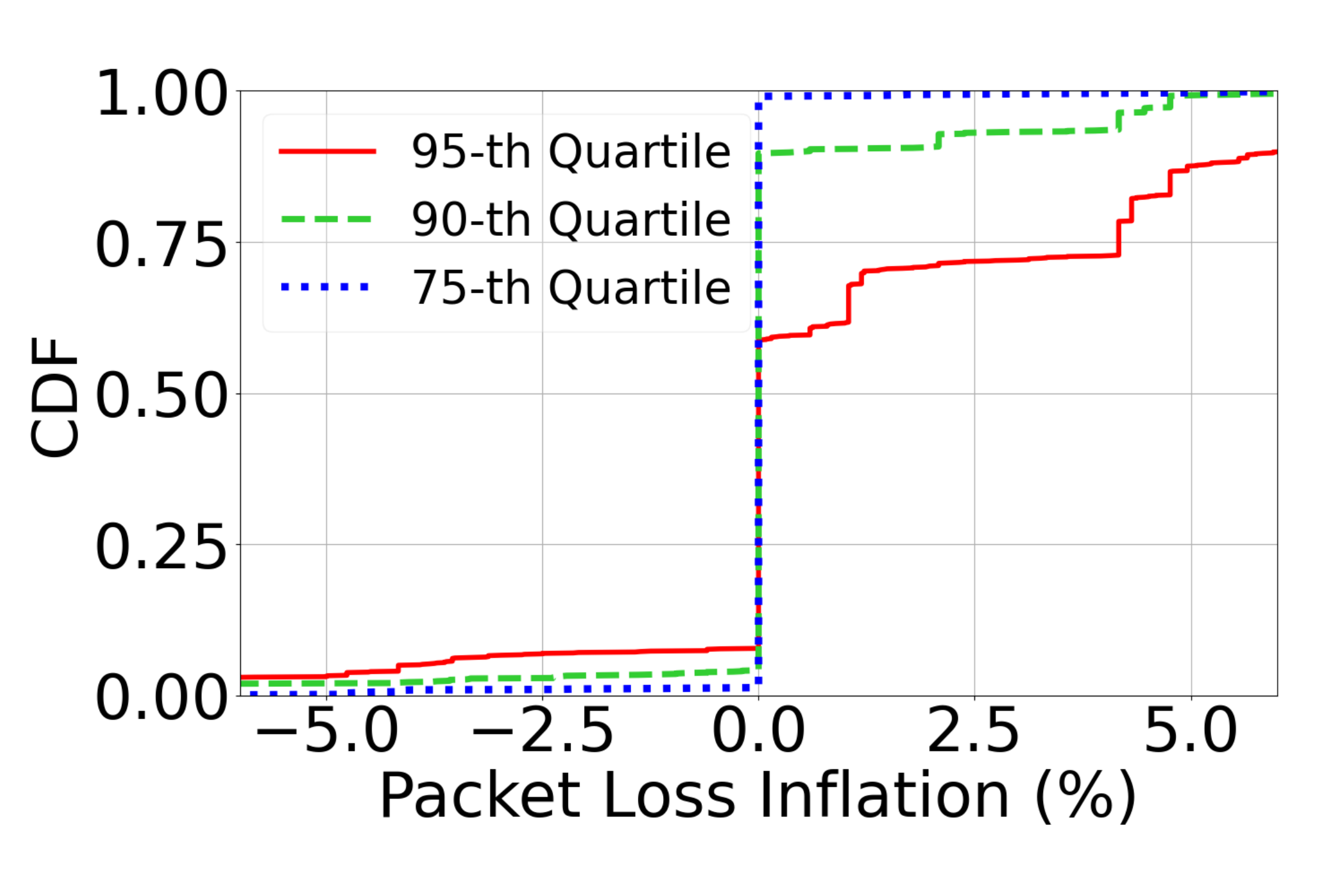}
        \caption{RIPE Atlas}
    \end{subfigure}%

    \hfill

    \begin{subfigure}[t]{0.25\columnwidth}
        \centering
        \includegraphics[width=\columnwidth, keepaspectratio]{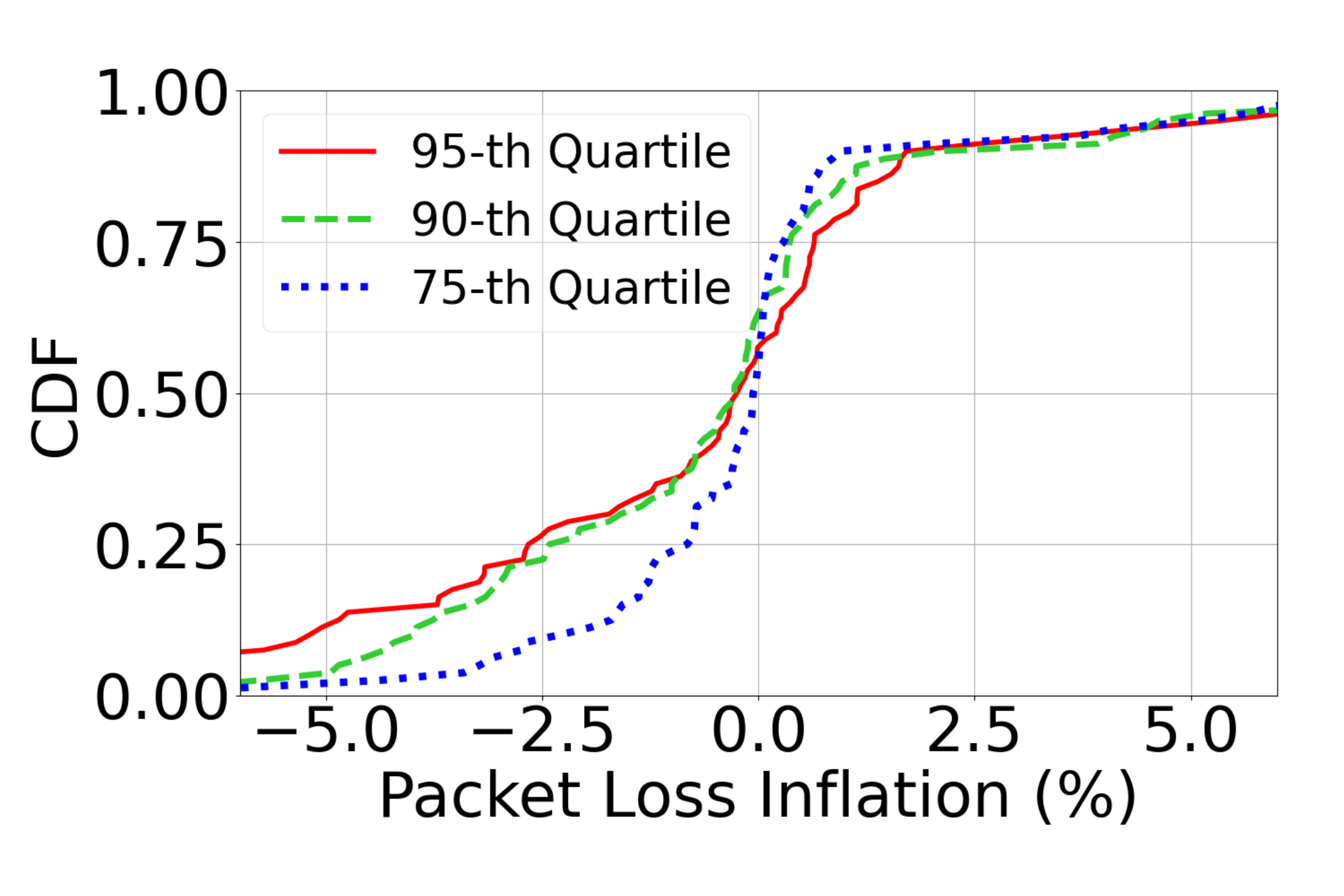}
        \caption{ndt7}
    \end{subfigure}%
    \hfill
    \begin{subfigure}[t]{0.25\columnwidth}
        \centering
        \includegraphics[width=\columnwidth, keepaspectratio]{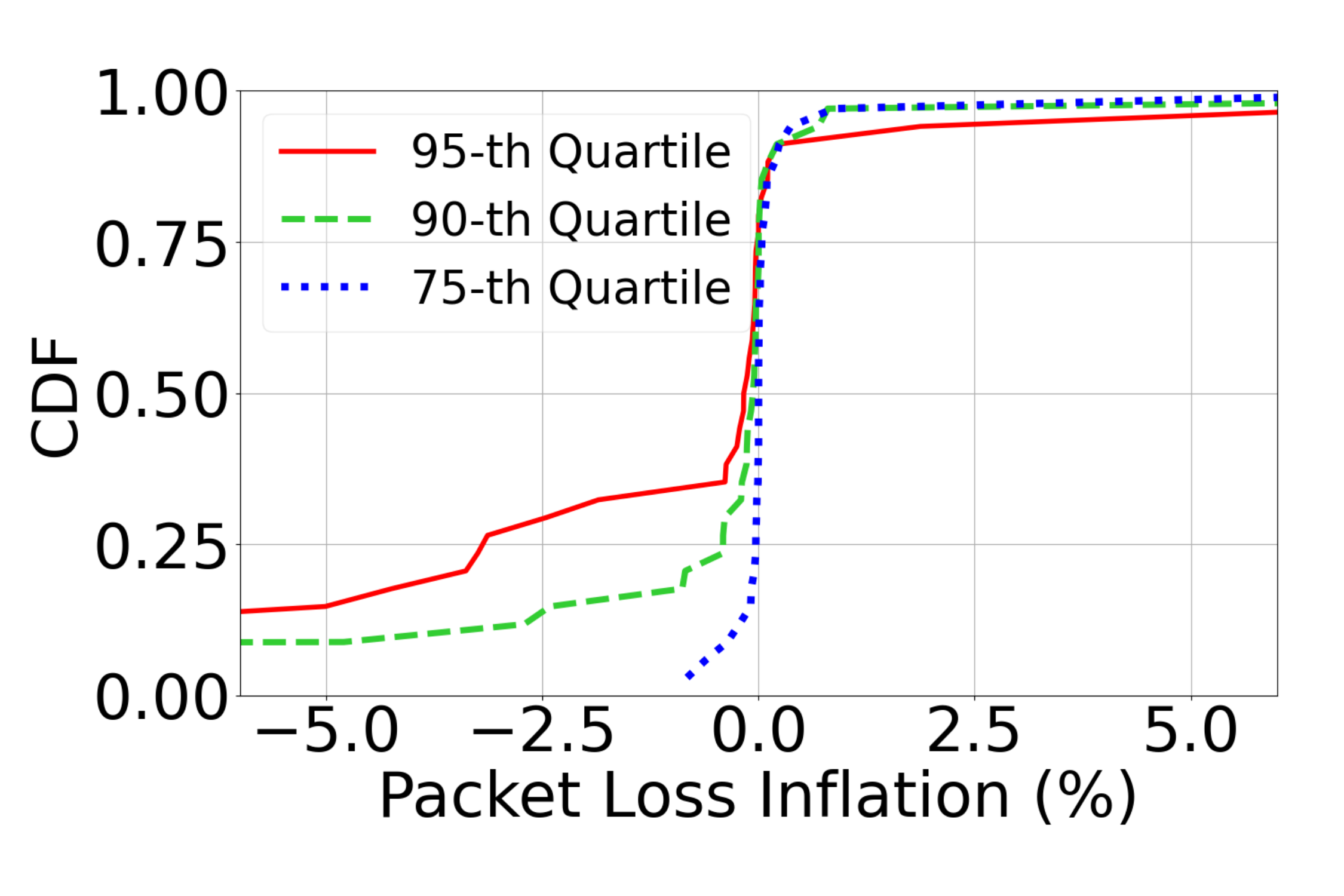}
        \caption{ndt5}
    \end{subfigure}%
    \hfill
    \begin{subfigure}[t]{0.25\columnwidth}
        \centering
        \includegraphics[width=\columnwidth, keepaspectratio]{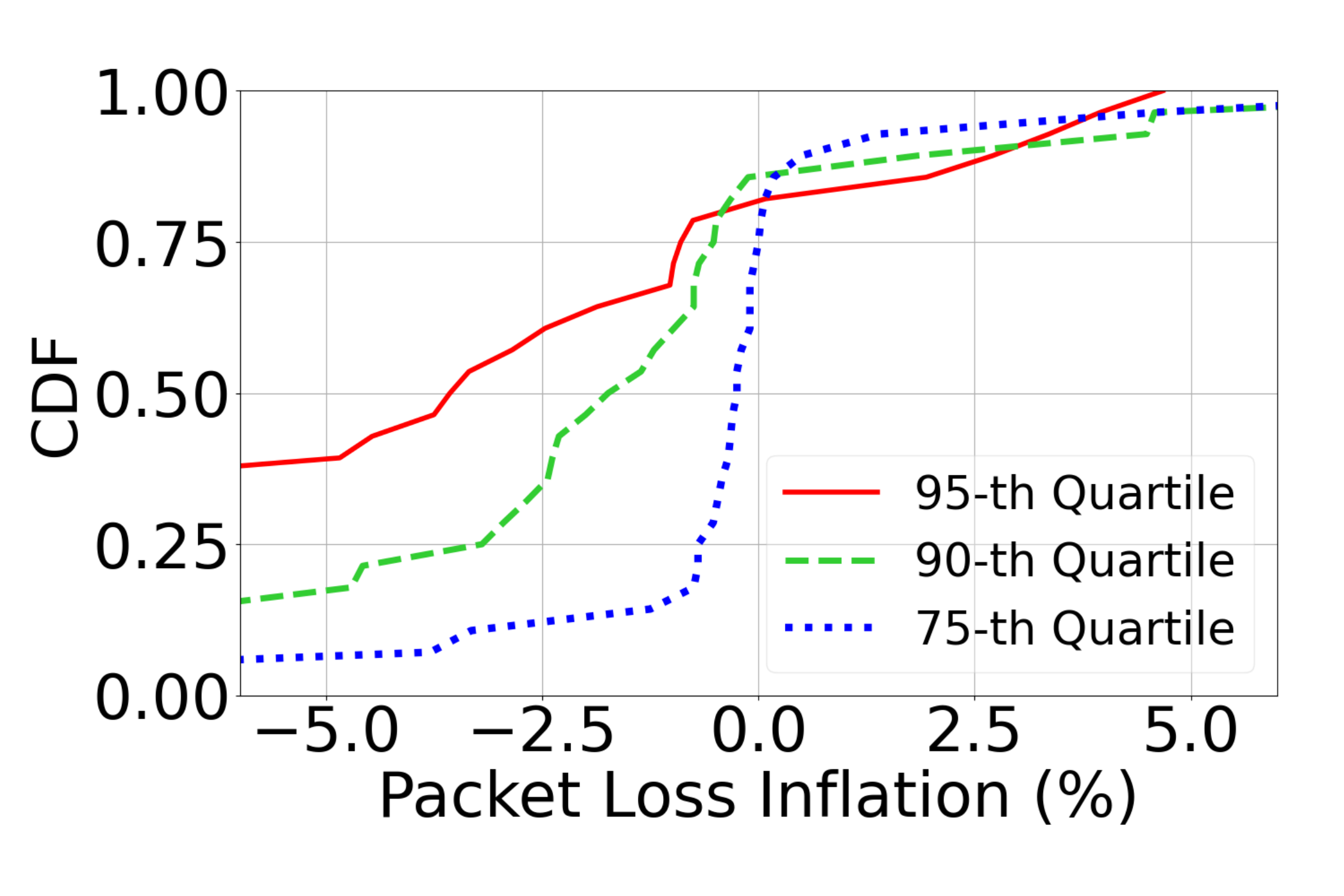}
        \caption{AIM}
    \end{subfigure}%
    \hfill
    \begin{subfigure}[t]{0.25\columnwidth}
        \centering
        \includegraphics[width=\columnwidth, keepaspectratio]{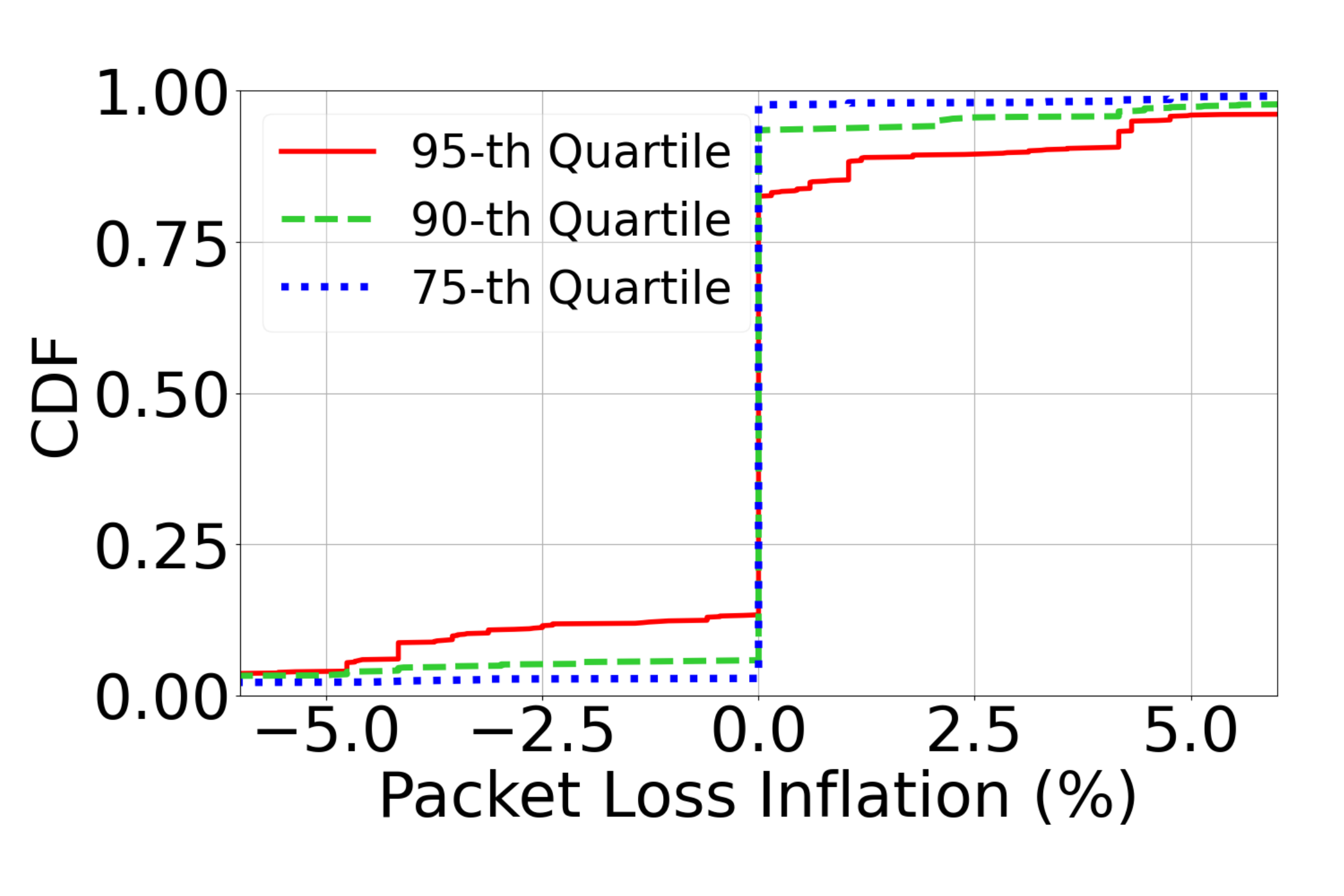}
        \caption{RIPE Atlas}
    \end{subfigure}%

    \caption{Distribution of region-wise loss inflation from speed test results during (a)-(d) May 2024 solar superstorm and (e)-(h) October 2024 solar storm showing (a) 60\% of the regions experience a subtle inflation in the loss, (d) a few probe and destination pairs experience more than 2\% inflation in loss. (b)-(c) remain symmetric. (e)-(h) No such notable indication during the October 2024 solar storm. }
    \label{fig:userLossInflation}
\end{figure}

In Fig.~\ref{fig:userLossOverall}, plot the overall packet loss distribution from M-LAB ndt, Cloudflare AIM, and RIPE Atlas probes.
Note that RIPE Atlas probes do not report the packet loss percentage explicitly.
We calculate the loss percentage as the number of packets lost over the total number of packets sent over a 30-minute window. 
During the May 2024 solar superstorm, we observe packet loss inflation of 0.8\%, 1\%, and 4\% (at 93 percentile) on M-LAB ndt7, Cloudflare AIM, and RIPE Atlas probes.
We see overlap in M-LAB ndt5.
During the October 2024 solar storm, all distributions overlap closely, indicating a negligible impact, except for Cloudflare AIM.

We show the region-wise changes in Fig.~\ref{fig:userLossInflation}, where we see around 60\% of regions in M-LAB ndt7 experience loss inflation during the May 2024 solar superstorm, and around 10\% and 30\% of probe destination pairs experience more than 2\% loss inflation above the 90th and 95th percentile. 
M-LAB ndt5 and Cloudflare AIM do not indicate any particular regional trends.
During the October 2024 solar storm, there were no regional trends.

\begin{keybox}
\keynote 
An end user, depending on their geographic region, might experience a temporary drop in the data rate of 10s of Mbps during a solar storm.
There is clear evidence of subtle latency and packet loss inflation.
However, the magnitude of inflation does not look severe in the publicly sourced datasets. 
Users might not notice a change in web browsing or video streaming, but could experience connectivity issues in interactive applications like video conferencing and cloud gaming due to transient outages, latency, or loss spikes we observed in the Starlink segment.
\end{keybox}

\section{Limitations}
\label{sec:limitations}

In this section, we point out a few limitations we encountered while studying the orbital decay and connectivity degradation of LEO satellites during solar events.

\parab{Detecting in orbit maneuver} - 
In this analysis, we reveal the Starlink fleet management strategies and correlate altitude change statistics with geospatial upper-atmospheric density conditions to explain the systematic pattern. 
However, the scope of this methodology is limited to the overall trend analysis.
It cannot tell which satellite is maneuvering, exactly when. 
This limitation prevents the analysis of regional network implications corresponding to the satellite's physical dynamics above those regions.

\parab{Invisible origin of network implications} - 
In our analysis, we observe a transient impact on the Starlink segment of the network.
It is highly likely that the observed implications are due to Starlink satellites.
However, because the network layer is invisible, it cannot be validated.

\parab{Limited insight into user experience} - 
A user-initiated speed test conducted over a large geographic area at an arbitrary time provides evidence of a probable degradation in the user experience.
However, quantifying end-user experience degradation requires continuous, systematic measurement of interactive user application performance during such solar events.


\section{Conclusion}
\label{sec:conclusion}

In this work, we present \projectname{}, an open-source tool that leverages multimodal datasets from diverse sources to trace how solar eruptions propagate to Earth and trigger solar storms. 
It enables in-depth analysis of LEO satellite trajectory variations and the resulting implications for LEO network connectivity during such events.
Using \projectname{}, we conduct a detailed study of two major recent events: 
(i) the May 2024 solar superstorm and 
(ii) the October 2024 solar storm. 
Our analysis shows that, during these events, the density of Earth’s upper atmosphere increased by up to 20 and 13 times over baseline levels, respectively.
While Starlink typically performs batch-wise orbit corrections to maintain its satellite fleet, during the May 2024 superstorm, it experienced rapid altitude decay and responded in real time. 
The system temporarily raised satellite altitudes slightly above their long-term operational altitude to counteract increased atmospheric drag until conditions stabilized.
We further investigate the root cause of the `W' pattern in satellite-altitude variations observed in prior studies of the May 2024 superstorm~\cite{DeepDiveSolarStorms}. 
Our findings indicate that geospatial differences in atmospheric density between day and night are the primary drivers of this pattern. 
This effect persists throughout the year and shifts continuously in synchronization with the Earth's position around the Sun. 
During solar storms, the pattern becomes more pronounced due to amplified atmospheric density.
In addition, we identify several network-level impacts, including inflation latency and loss, distortions in diurnal latency patterns, and short-lived outages. 
Our analysis reveals that Starlink uplinks are more susceptible to these disruptions than downlinks. Although end-user performance shows noticeable reductions in data rates and modest increases in latency and packet loss, these effects can still degrade the quality of experience for interactive applications.

\begin{acks}

This research is funded by the Prime Minister's Research Fellowship (PMRF) from the Ministry of Education, Government of India.
Simulation results have been provided by the Community Coordinated Modeling Center (CCMC) at Goddard Space Flight Center through their publicly available simulation services (https://ccmc.gsfc.nasa.gov). 
The NCAR Thermosphere-Ionosphere- Electrodynamics General Circulation Model (TIE-GCM) was developed by Wenbin Wang at the University Corporation for Atmospheric Research (UCAR).

\end{acks}

\bibliographystyle{ACMReferenceFormat}
\bibliography{references}

\end{document}